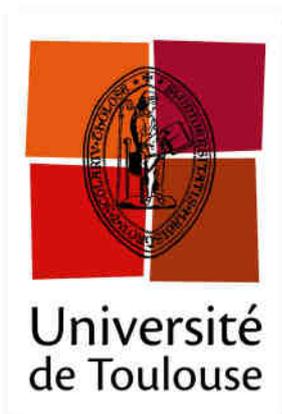

# THÈSE

En vue de l'obtention du

## DOCTORAT DE L'UNIVERSITÉ DE TOULOUSE

**Délivré par :** *l'Université Toulouse 3 Paul Sabatier (UT3 Paul Sabatier)*

**Présentée et soutenue le** *25/09/2017* **par :**
**Jason CHAMPION**

**Photoevaporation des disques protoplanétaires par les photons UV d'étoiles massives proches : observation de proplyds et modélisation**

### JURY

| | | |
|---|---|---|
| Maryvonne GERIN | Directeur de Recherche | Président du Jury |
| Yann ALIBERT | Maître de Conférence | Rapporteur |
| Emilie HABART | Chargé de Recherche | Examinateur |
| Karine DEMYK | Directeur de Recherche | Examinateur |
| Emmanuel CAUX | Directeur de Recherche | Examinateur |
| Olivier BERNE | Chargé de Recherche | Directeur de Thèse |

**École doctorale et spécialité :**
    *SDU2E : Astrophysique, Sciences de l'Espace, Planétologie*
**Unité de Recherche :**
    *Institut de Recherche en Astrophysique et Planétologie (UMR 5277)*
**Directeur de Thèse :**
    *Olivier BERNE*
**Rapporteurs :**
    *Maryvonne GERIN* et *Yann ALIBERT*

# Résumé

Les disques protoplanétaires entourant les jeunes étoiles sont les embryons des systèmes planétaires. A différentes phases de leur évolution, ils peuvent subir d'importantes pertes de masse par photoévaporation : des photons énergétiques, issus de l'étoile centrale ou d'une étoile voisine, chauffe le disque qui perd en masse sous l'échappement des particules. Cependant, ce mécanisme et la physique sous-jacente n'ont que peu été contraints par les observations. Les objectifs de cette thèse sont d'étudier la photoévaporation dans le cas particulier où elle est due à des photons FUV, d'identifier les principaux paramètres physiques (densité, température) et processus (chauffage et refroidissement) impliqués, et d'estimer son impact sur l'évolution dynamique des disques. L'étude repose sur le couplage observations – modélisations des disques photoévaporés par les photons UV en provenance d'étoiles massives proches. Ces objets, appelés "proplyds", ont leur disque entouré d'une large enveloppe nourrie des flots de photoévaporation.

A l'aide d'un modèle 1D d'une région de photodissociation, j'ai développé un modèle pour l'émission dans l'infrarouge lointain des proplyds. Ce modèle a été utilisé pour interpréter les observations, issues principalement de *Herschel*, pour quatre proplyds. Il apparait que les conditions physiques en surface de leur disque sont similaires: une densité de l'ordre de $10^6$ cm$^{-3}$ et une température d'environ 1000 K. Cette température est maintenue par un équilibre dynamique : si la surface se refroidit, la perte de masse diminue et l'enveloppe se réduit. L'atténuation UV produite par l'enveloppe diminue alors et le disque, recevant plus de photons UV, chauffe. La majorité du disque peut s'échapper sous forme de flots de photoévaporation avec des taux de perte de masse de quelques $10^{-7}$ M$_\odot$ yr$^{-1}$ ou plus, en accord avec les observations précédentes des traceurs du gaz ionisé.

A la suite de ce travail, j'ai développé un modèle hydrodynamique 1D pour étudier l'évolution dynamique d'un disque en photoévaporation par un champ de rayonnement externe. Le code inclut l'évolution visqueuse due à la turbulence et des prescriptions pour différents types de photoévaporation, incluant la prescription dans le cas défini par les observations et modèles décrits précédemment. Dans ce cas, deux régions du disque évoluent différemment. L'évolution dynamique du disque externe, où les flots de photoévaporation se développent, est dominée par la photoévaporation. Le disque externe se dissipe dans un temps de $10^5$ ans, laissant un disque tronqué. Le disque interne, où la gravitation retient les flots, est aussi impacté en raison d'un transfert de masse vers le disque externe en photoévaporation, mais l'accroissement de la perte de masse est faible. Le disque interne, dont l'évolution est dominée par la viscosité, peut survivre quelques $10^6$ ans. Finalement, j'ai effectué une étude statistique en utilisant ce modèle et montré qu'il était capable de reproduire des observations d'un grand échantillon de disques dans Orion où la photoévaporation externe est supposée modifier la fonction de masse des disques.

D'après mes résultats, la photoévaporation externe est très efficace dans le disque externe. A moins que des planètes puissent se former en moins de $10^5$ ans, leur formation semble difficile ici. Le disque interne est un environnement beaucoup plus favorable. Les modèles de synthèse de populations planétaires, étudiant les impacts à l'échelle d'une grande population, sont une perspective prometteuse de ce travail car ils peuvent offrir la possibilité d'étudier l'évolution de systèmes planétaires en formation et irradiés par les étoiles massives voisines. Dans le futur proche, l'arrivée du *James Webb Space Telescope* et sa haute résolution spatiale dans l'infrarouge, en synergy avec ALMA dans le submillimétrique, vont permettre de sonder plus directement les propriétés des flots de photoévaporation et d'avoir une vision locale de ses effets.

# Abstract

Protoplanetary disks are found around young stars, and represent the embryonic stage of planetary systems. At different phases of their evolution, disks may undergo substantial mass-loss by photoevaporation: energetic photons from the central or a nearby star heat the disk, hence particles can escape the gravitational potential and the disk loses mass. However, this mechanism, and the underlying physics regulating photoevaporation, have not been well constrained by observations so far. The aims of this thesis are to study photoevaporation, in the specific case when it is driven by far-UV photons, to identify the main physical parameters (density, temperature) and processes (gas heating and cooling mechanisms) that are involved, and to estimate its impact on the disk dynamical evolution. The study relies on coupling observations and models of disks being photoevaporated by UV photons coming from neighbouring massive star(s). Those objects, also known as "proplyds", appear as disks surrounded by a large cometary shaped envelope fed by the photoevaporation flows.

Using a 1D code of the photodissociation region, I developed a model for the far-IR emission of proplyds. This model was used to interpret observations, mainly obtained with the *Herschel Space Observatory*, of four proplyds. We found similar physical conditions at their disk surface: a density of the order of $10^6$ cm$^{-3}$ and a temperature about 1000 K. We found that this temperature is maintained by a dynamical equilibrium: if the disk surface cools, its mass-loss rate declines and the surrounding envelope is reduced. Consequently, the attenuation of the UV radiation field by the envelope decreases and the disk surface, receiving more UV photons, heats up. Most of the disk is thus able to escape through photoevaporation flows leading to mass-loss rates of the order of $10^{-7}$ M$_\odot$ yr$^{-1}$ or more, in good agreement with earlier spectroscopic observations of ionised gas tracers.

Following this work, I developed a 1D hydrodynamical code to study the dynamical evolution of an externally illuminated protoplanetary disk. This code includes the viscous evolution driven by turbulence and prescriptions of different photoevaporation mechanisms, including a prescription for external photoevaporation based on the observational results and models discussed above. I found that, in this case, two regions of the disk evolve differently. The dynamical evolution of the outer disk, where escape flows can develop, is dominated by the photoevaporation. The outer disk is found to be dissipated within $10^5$ years, leaving the disk truncated. The inner disk, where the gravitational field avoid photoevaporation flows, is also impacted due to a mass transfer in the outer disk that is photoevaporated, but the enhancement of mass loss is not significant. The dynamical evolution of the inner disk is rather dominated by viscosity and it may survive a few $10^6$ years. Finally, I conducted a statistical analysis using this model, and showed that it is able to reproduce observations of a large sample of disks in Orion where external photoevaporation is believe to modify the disk mass function.

According to my results, external photoevaporation is very efficient in the outer disk, hence, unless planets can form in less than $10^5$ years, it is unlikely that they can form there. The inner disk may be a more favorable environment since it may survive several $10^6$ years. Plan-etary population synthesis models, that study impacts at the scale of a large population, are a promising perspective of this work since they offer the possibility to study the evolution of forming planetary systems irradiated by nearby massive stars. In the near future, the arrival of the *James Webb Space Telescope* and its high spatial resolution in the infrared, in synergy with ALMA in the submillimeter range, will permit to probe the properties of the photoevaporation flows more directly and to get a local view of its effects.

**Key words:** protoplanetary disk – photoevaporation – proplyd – photodissociation region – infrared observation – hydrodynamical simulation

# Contents

# List of Tables

# List of Figures

# List of Acronyms

| Acronym | Definition |
|---------|------------|
| ALMA    | Atacama Large Millimeter/sub-millimeter Array |
| BG      | Big Grain |
| EUV     | Extreme Ultraviolet (10 nm to 91.2 nm) |
| FUV     | Far UltraViolet (91.2 nm to 200 nm) |
| FWHM    | Full Width at Half Maximum |
| HIFI    | Heterodyne Instrument for the Far Infrared |
| HST     | Hubble Space Telescope |
| IMF     | Initial Mass Function |
| IR      | InfraRed (750 nm to 1 mm) |
| IRAC    | InfraRed Array Camera |
| ISM     | InterStellar Medium |
| ISRF    | InterStellar Radiation Field |
| JWST    | James Webb Space Telescope |
| MC      | Molecular Cloud |
| MIPS    | Multiband Imaging Photometer and Spectrometer |
| MRI     | Magneto-Rotational Instability |
| ONC     | Orion Nebula Cluster |
| PACS    | Photodetector Array Camera and Spectrometer |
| PAH     | Polycyclic Aromatic Hydrocarbon |
| PDR     | PhotoDissociation Region |
| SED     | Spectral Energy Distribution |
| SPIRE   | Spectral and Photometric Imaging Receiver |
| UV      | UltraViolet |
| VSG     | Very Small Grain |
| YSO     | Young Stellar Object |

# Introduction (français)

Ces dernières années, les annonces de découvertes de systèmes planétaires se sont succédé. Il n'est désormais plus surprenant de trouver de nouvelles exoplanètes, et leur présence, probablement universelle, entre progressivement dans l'esprit collectif. L'engouement qui suit chaque annonce porte alors plutôt sur l'aspect atypique du nouveau système, la potentielle habitabilité de quelques unes de ses planètes, ou bien la ressemblance de l'une d'elle avec la Terre. Aucun des systèmes exoplanétaires ne semble pourtant être le jumeau de celui dans lequel nous vivons. Le Système Solaire est en effet très ordonné : une série de petites planètes rocheuses, une ceinture d'astéroïdes, puis une série de géantes de gaz ou de glace. Le fait de ne pas retrouver cette organisation ailleurs a remis en question les scénarios de formation des systèmes planétaires, basées uniquement sur l'observation du Système Solaire. L'intérêt s'est très vite porté sur les embryons de systèmes planétaires, c'est-à-dire sur les disques de matière qui entourent les jeunes étoiles et que l'on nomme disques protoplanétaires. En plus de la découverte concomitante des premières exoplanètes, l'observation de ces disques s'offre réellement à nous depuis l'avènement du télescope spatial Hubble dans les années 1990 ou, plus récemment, grâce au réseau de radiotélescopes ALMA permettant de sonder plus encore les détails de ces objets.

Tout n'est pas rose dans le développement de ces embryons. En se tournant vers la nébuleuse d'Orion, l'une des plus proches régions de formation d'étoiles, les premiers disques observés par Hubble semblaient être attaqué depuis l'extérieur. En effet, des étoiles massives naissent aussi au sein de cette nébuleuse. En véritable monstres célestes, ces étoiles consomment tellement vite leur réserve d'hydrogène qu'elles émettent en retour une grande quantité d'énergie, sous forme de rayonnement ultraviolet, très énergétique et parfois destructeur. S'il n'est pas question pour un disque de survivre autour d'une telle étoile, les disques des étoiles voisines, plus communes et pouvant ressembler au Soleil, sont également affectées. L'arrivée des photons UV à leur surface provoque un échauffement important. Si la température augmente suffisamment, une partie de la matière peut s'échapper. C'est ce que l'on appelle la photoévaporation. Un disque subissant de la photoévaporation externe, c'est-à-dire à cause d'une étoile extérieure au système, comme dans ce cas, perd alors sa masse à une vitesse bien plus importante que dans le cas d'un développement normale. La vie de l'embryon du système planétaire s'en trouve alors écourtée si bien que l'avortement de la formation des planètes semble inévitable.

Pourtant, répondre à la question du futur de ces disques en particulier n'est pas si simple. Dans un tel cas, la naissance ou non du système planétaire se joue à peu de choses, et les connaissances que nous avons aujourd'hui ne permettent pas de véritablement trancher. L'originalité du sujet de cette thèse est de coupler l'apport de nouvelles observations, permettant de sonder les disques subissant une photoévaporation externe, avec de la simulation numérique afin d'en prédire l'évolution, et de juger la faisabilité de la formation planétaire en leur sein. Cette thèse se découpe en trois parties. La première introduit la vision actuelle que nous avons de la formation d'un système planétaire (chapitre 1), et ce que nous connaissons aujourd'hui des disques protoplanétaires, leurs propriétés et leur évolution générale (chapitre 2). La seconde partie traite de l'étude de la région du disque irradiée par les rayons UV, que l'on nomme "région de photodissociation". La physique qui gouverne ces régions est d'abord présentée (chapitre 3), afin de comprendre ensuite les observations qui ont été menées et analysées (chapitre 4). Le couplage de ces observations avec la modélisation permet de sonder ces objets et d'y étudier le phénomène de photoévaporation (chapitre 5). La dernière partie décrit ce qui fait évoluer un disque protoplanétaire (chapitre 6), comment on peut le mettre en équation et le simuler numériquement (chapitre 7), puis enfin ce que cela nous apprend sur le devenir d'un disque soumis à de la photoévaporation externe (chapitre 8).

# Introduction (english)

In recent years, there have been many announcements of planetary system discoveries. It is no longer surprising to find new exoplanets, and their presence, probably universal, gradually enters the collective spirit. The enthusiasm that follows each announcement is then rather related to the atypical aspect of the new system, the potential habitability of some of its planets, or the resemblance of one of them with the Earth. However, none of the exoplanetary systems seems to be the twin of the one in which we live. The Solar System is very orderly: a series of small rocky planets, an asteroid belt, and then a series of gas or ice giants. The failure to trace this organisation elsewhere has challenged the scenarios of formation of planetary systems, based originally on the observation of the Solar System solely. The interest has rapidly focused on the embryos of planetary systems, that is to say on the disks of matter which surround the young stars and which are called protoplanetary disks. In addition to the concomitant discovery of the first exoplanets, the possibility to observe these disks has really been offered to us since the advent of the Hubble Space Telescope in the 1990s or, more recently, with the network of radio-telescopes ALMA which permits to probe even more their details.

Things are not looking so good in the development of these embryos. Turning to the Orion nebula, one of the closest star-forming regions, the first disks observed by Hubble seemed to be attacked from the outside. Indeed, massive stars are also present within this nebula. As true celestial monsters, these stars consume their hydrogen stock so fast that they emit in return a large amount of energy, in the form of ultraviolet radiation, very energetic and sometimes destructive. If this is hard for a disk to survive around such a star, the disks of the neighbouring stars, more common and similar to the Sun, are also affected. The impinging of the UV photons at their surface causes an important heating. If the temperature rises sufficiently, some of the material can escape. This is called photoevaporation. A disk undergoing external photoevaporation, that is to say, because of a star outside the system as in this case, will then loses its mass at a much higher rate than in the case of a normal development. The life of the planetary system embryo is thus shortened so that the abortion of the planet formation seems inevitable.

Yet, answering the question of the future of these disks in particular is not so simple. In such a case, the birth of a planetary system could go either way, and the knowledge we have today does not allow us to really state on its feasibility. The originality of this thesis is to combine the contribution of new observations, allowing to probe disks under external photoevaporation, with numerical simulations in order to predict their evolution, and to judge the feasibility of the planet formation within them. This thesis is divided into three parts. The first introduces the current vision of the formation of a planetary system (chapter 1), and what we today know about protoplanetary disks, their properties and their general evolution (chapter 2). The second part deals with the study of the region of the disk irradiated by UV photons, known as the "photodissociation region". The physics that govern these regions are firstly presented (chapter 3), in order to understand then the observations that have been carried out and analysed (chapter 4). The coupling of these observations with the models permits to probe these objects and to study the mechanism of photoevaporation (chapter 5). The last part describes which processes drive the protoplanetary disk evolution (chapter 6), how they can be put into an equation and numerically simulated (chapter 7), and finally what this study teaches us on the future of a disk subjected to external photoevaporation (chapter 8).

# Part I

# Formation and evolution of planetary systems

# Table of Contents

# Chapter 1

# General scenario: from a nebula to a planetary system

## 1.1 Historical context

### 1.1.1 Ancient times: first steps towards a theory of the planets

Offering the opportunity to nourish the scientific curiosity, or to simply enjoy the view, the Moon and the planets have always been present in the night sky. Archeology has taught us that astronomy, the natural science that studies celestial objects, was already present during prehistory. Originally of a practical nature, the lunar phases were followed in order to maintain the cycles of religious and popular events, as well as for agricultural and hunting needs. The oldest written trace of planet identification was achieved by ancient Babylonian astronomers in the 2nd millennium BC. Sadly, only fragments of their astronomy, engraved in tablets, have survived through the ages. Even so, the surviving fragments show that the Babylonians set the fundamentals of scientific astronomy, and were the first civilisation known to possess a simple, but functional, theory of the planets.

The major steps in the history of astronomy were then made by the ancient Greek civilisation (starting in the millennium BC). They noted that a few objects, or lights in the sky, moved differently in comparison with the motion of most of the stars. Not only different, their movements were quite irregular compared to them. Consequently, they were called πλανητες αστερες (read "planetes asteres"), meaning "wandering stars", that finally gives the word "planet" used today. They were seven: the Moon, the planets Mars, Mercury, Jupiter, Venus, Saturn and the Sun, our star. At this moment, they had no idea of what kind of objects they were. Each one was associated to a divinity supposed to reign successively in a manner that engender the order of the days in a week from which their names are derived[1]. More than an example of an ancient Greek astronomy remnant in our every day life, this was a big step toward the understanding of our planetary system structure, with the fact that all the "planets" are moving in roughly the same line in the sky, today called the "ecliptic". Another advance was made by Aristarchus of Samos ($\sim 310 - 230$ BC), an ancient Greek mathematician and very clever astronomer, who was able to estimate roughly the relative sizes and distances of the Moon and the Sun. He noticed that the Moon moves, with respect to the background stars, with a velocity of one Moon diameter per hour. Considering that the Moon eclipses in the shadow of the Earth last three hours at most, he deduced that the Moon should be about three times smaller than the Earth which is a quite correct result. From other geometrical considerations on the lunar phases, he estimated that the Sun was further away from the Earth than the Moon is. According to their same apparent size, he deduced that the Sun was much larger than the Moon and the Earth. While not correct in the number, this relative order was right. Based on that discovery, it appeared logical to him that the small bodies turn around the largest. He thus presented the first known model of the Solar System, our planetary system, that placed the Sun at the centre of the known Universe with the Earth orbiting around it. The Sun was here supposed to be a star like the others in the night sky, as suggested before him by Anaxagoras ($\sim$ 510 – 428 BC). In this model, the "fixed stars" should be rotating around the Sun too, and not the Earth. As a consequence, from Earth, we should see relative motion between stars because of the motion of the Earth on its orbit. This apparent motion called "parallax" can not be observe with the naked eyes. According to Aristarchus, this absence of visible motion could be explained if stars are very far away from the Earth. However, at that time, the simplest explanation for the absence of parallax was that the Earth does not move and is effectively at the centre of the Universe. Because of this argument and religious considerations (mostly), the model of Aristarchus was rejected in favour of the geocentric theories (Fig. 1.1) of Aristotle ($\sim$ 384 – 322 BC) and Ptolemy ($\sim$ 100 – 170), which led for many centuries.

---

[1] In French/English, the order is: Moon → Lundi/Monday, Mars → Mardi/Tuesday, Mercury → Mercredi/Wednesday, Jupiter → Jeudi/Thursday, Venus → Vendredi/Friday, Saturn → Samedi/Saturday, Sun → Dimanche/Sunday.

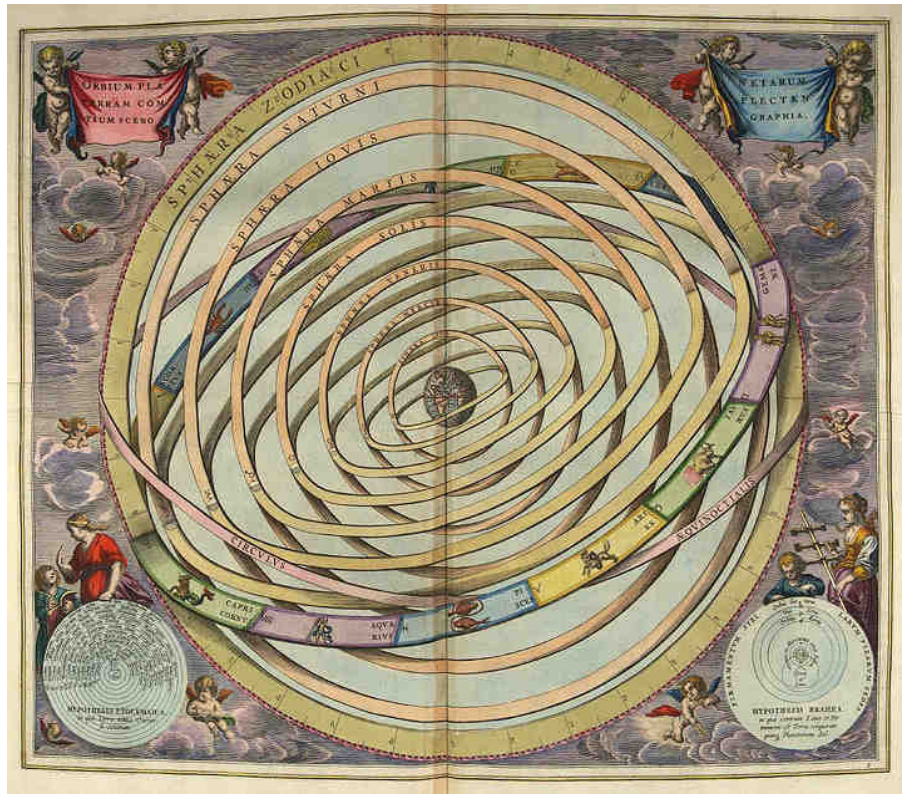

Figure 1.1: Illustration of the Ptolemaic orbits, from *Harmonia Macrocosmica* (1661) by Andreas Cellarius.

### 1.1.2 Renaissance and modern ages: structure of the Solar System

The renaissance period (14th to 17th century) marks the end of many centuries of a relatively slow development of astronomy, both observationally and theoretically. Galileo Galilei (1564 – 1642) is considered to be the father of observational astronomy since, in 1609, he built the first telescope designed to be pointed to the sky. In 1610, he discovered four moons orbiting around Jupiter, nowadays known as the "Galilean satellites". This discovery led Galileo to convince himself that there is no reason for the Earth to be at the center of the Universe, and made him one of the greatest defender of a soon growing theory: heliocentrism.

Heliocentrism is the theory which places the Sun at the center of the Universe with every planet orbiting around it, including the Earth. This was the forgotten idea of Aristarchus of Samos, revived in Europe thanks to Nicolaus Copernicus (1473 – 1543). At first around 1510, Copernicus secretly spread this theory by making available to friends of him a manuscript named "Commentariolus", describing his ideas about the heliocentric hypothesis based on seven basic observations and assumptions. Because he did not want to risk the scorn to which he would expose himself on account of the novelty and incomprehensibility of his theses, or even worse, he resisted to openly publish his views. The book "De revolutionibus orbium coelestium" ("on the revolutions of the celestial spheres") presenting his major work was finally published the year of his death, in 1543. This event represents the point of no return in the replacement of the geocentric theory by the heliocentrism. In the following decades, it became more and more accepted among scientists. Tycho Brahe (1546 – 1601) conducted precise observations of the motion of the planet Mars and showed that this planet was not orbiting the Earth in a circle but, instead, orbits the Sun following an ellipse with our star at one focus. From those observations, Johannes Kepler (1571 – 1630) was convinced by heliocentrism and then derived his famous three laws of planetary motions:

**First law** The orbit of a planet is an ellipse with the Sun at one of the two foci;

**Second law** A line segment joining a planet and the Sun sweeps out equal areas during equal intervals of time;

**Third law** The square of the orbital period of a planet is proportional to the cube of the semi-major axis of its orbit.

Sir Isaac Newton (1642 – 1726) later showed, in his book "Philosophiae Naturalis Principia Mathematica" ("Mathematical Principles of Natural Philosophy"), published in 1687, that the Kepler's laws are a consequence of his own laws of motion and law of universal gravitation. Modern science was born there, with the aim to decrypt the laws that rule the world. Yet, while the Solar System structure was globally understood, its origin remained a mystery.

### 1.1.3 Enlightenment on the origin of the Solar System

The initial step towards a theory of the Solar System origin was the general acceptance of heliocentrism by the scientific community by the end of the 17th century. The term "Solar System" comes from that period. The French philosopher and mathematician René Descartes (1596 – 1650) was the first to propose a scenario for the origin of the Solar System in his book "Le Monde" ("The World"), published in 1664 after his death. He proposed that the Universe is filled with vortices of swirling particles from which the Sun and the planets have emerged (Fig. 1.2, left). If we today know that the motion of matter is mostly driven by the law of gravitation, Descartes, by trying to explain the observed circular motion of planets with vortices, and the famous painter Van Gogh, by imagining a turbulent Universe (Fig. 1.2, right), were on the right track since, as we will see later in Sect. 1.4.1 and Sect. 6.2, vortices/turbulence may be involved at some points.

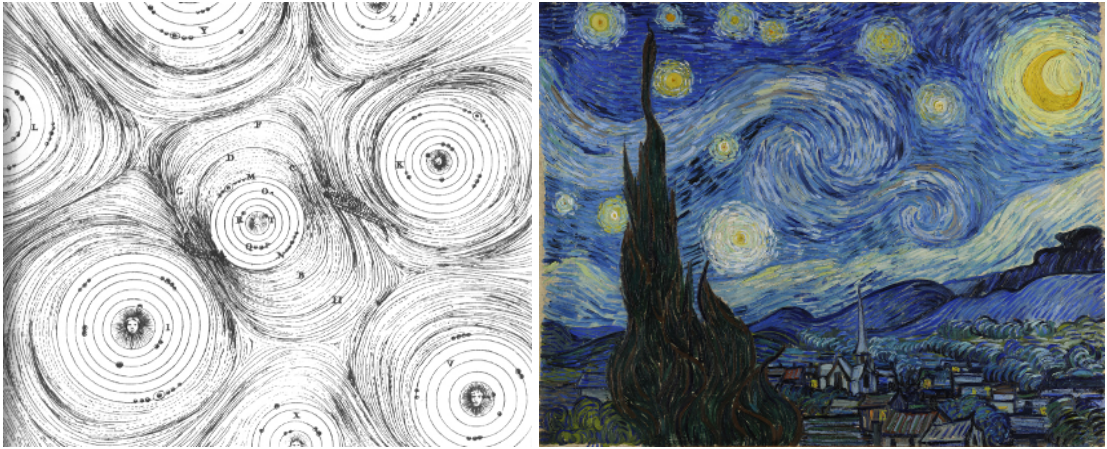

Figure 1.2: Illustration of the vortices theory for the planet formation. Left: drawing of the Descartes' vortices (credit: library of the Paris Observatory, France). Right: the famous painting "Starry Night" (1889) by Vincent Van Gogh (currently belonging to the collection of the Museum of Modern Art in New York, USA) curiously close to that idea.

In 1749, Georges-Louis Leclerc (1707 – 1788), also known as the "Comte de Buffon", proposed that the planets were formed, either from a collision of a comet (the true nature of a comet was not known at this period) onto the Sun, or by a removal of matter from a neighbouring passing star because of the Sun gravitational force. In both scenarios, matter is left around the Sun and may condense in planets when cooling. This theory was in competition with another contemporary one: the nebular hypothesis, initially proposed by Emanuel Swedenborg (1688 – 1772) and elaborated independently by the philosopher Immanuel Kant (1724 – 1804) in 1755 and the scientist Pierre-Simon Laplace (1749 – 1827) in 1796. They proposed that the Solar System could have formed out of a nebula (an interstellar cloud as already observed at this period but not defined precisely) that collapsed into a star, and that the remaining material gradually settled into a flat disk orbiting it which then condensed to form the planets. This theory explains why all the planets have quasi-coplanar and prograde orbits (the motion on the "ecliptic" line observed by ancient Greek astronomers). Since this nebular hypothesis is not perfect, there were many alternatives scenarios proposed but, after more than two centuries of new observations of the Solar System, including the discovery of two ice giant planets (Uranus by William Herschel in 1781 and Neptune by Urbain Le Verrier in 1846), this model is still the most popular scenario among astronomers who continue to endeavour to improve it. In the last decades, this theory developed for the Solar System, had to face the discovery of numerous extrasolar planets which brought a lot of surprising informations.

### 1.1.4 Contemporary times: discovery of extrasolar planets

During the renaissance period, the heliocentrism theory was in the mind of scientist and philosopher as well as the fact that the Sun was probably similar to other stars in the sky. The possibility to imagine other worlds in the Universe, around those other stars, was thus inevitable. Giordano Bruno (1548 – 1600), an Italian philosopher, believed in that possibility and wrote, in the book

"De l'infinito universo et mondi" ('On the infinite universe and worlds") published in 1584, that "This space we declare to be infinite... In it are an infinity of worlds of the same kind as our own". As many other scientists after him, he was right.

An extrasolar planet, or exoplanet, is a planet that orbits a star other than the Sun. The first detection of such an object was made in 1988 (Campbell et al., 1988). However, the very first confirmation of an exoplanet orbiting a main-sequence star (that is to say, a "living" star with characteristics close to the Sun) was made in 1995, at the *Observatoire de Haute-Provence* in France, when a giant planet was found in a four-day orbit around a nearby star called 51 Pegasi (Mayor and Queloz, 1995). At the time of writing, July 14 2017, there are 3631 confirmed exoplanets localised in 2712 planetary systems with 611 of them that are multiple planet systems.[2]

The scenario of planet formation built up before 1995 was only based on the 8 planets of the Solar System. Since then, exoplanets have showed us that we are actually in a particular system rather than in a generic one. Therefore, the scenario is in continuous evolution, and many recent changes/advances have been made to explain the large diversity in the planetary systems. The main surprise was that the structure of inner terrestrial planets (by order from the Sun: Mercury, Venus, Earth and Mars) followed by a series of giant planets (Jupiter, Saturn, Uranus and Neptune) is typical of our system. Indeed, a lot of "hot Jupiters/Neptunes" or "super-Earths" have been found, meaning respectively a gas/ice giant, or a big terrestrial planet, close to their host star. At the same time, distant gas giant planets, that we do not observe in our system, have also been discovered. The recently discovered TRAPPIST-1 system illustrates the difference of structure from one system to another (Fig. 1.3).

In the following sections of this chapter, I develop the broad outlines of the current theory to form a planetary system around a low-mass star ($M_* < 8 \ \mathrm{M}_\odot$) from an initial "nebula".

## 1.2 Pre-stellar phase: collapse of a molecular cloud in the interstellar medium

Observations on galactic scales (our own or other galaxies) showed that the star formation rate is highly correlated with the presence of molecular gas (Leroy et al., 2008; Bigiel et al., 2008). If the medium filling a galaxy, called the InterStellar Medium (ISM), is mainly ionised or atomic because of radiation, there are some places where the density is high enough ($> 10 \ \mathrm{M}_\odot/\mathrm{pc}^2$, see Lee et al., 2012) to protect the gas, and permit the formation and persistence of molecules. Those places form what is called a Molecular Cloud (MC). There are many types of molecular clouds spanning a mass range over many orders of magnitude, with the most of the mass located in the large clouds (Roman-Duval et al., 2010). The large range in size and mass of molecular clouds led to a specific terminology: from the largest "giant molecular clouds", then follows "molecular clouds", "clumps", and "dense cores". Table 1.1 gives some of their properties. All of these structures are gravitationally bound and the smaller structures are embedded in the larger ones as Russian dolls. Generally, and especially for massive molecular clouds, the magnetic field is not strong enough to support cloud against gravitational collapse (Crutcher, 2012). The star formation process occurs on time scales which are much larger than pure free-fall collapse (ratio about 1%, see Krumholz et al., 2012), and is so slow that only a small fraction of the initial mass will be converted into stars before the gas dispersion. This is because the fragmentation and collapse, that goes from clouds to clumps, dense cores and finally protostellar cores, are also subject to opposing processes such as thermal pressure, turbulence and magnetic field.

---

[2]Those numbers may rapidly evolve, the reader is thus referred to the websites http://exoplanet.eu and/or http://exoplanets.org to obtain an update, and eventually more details on those numerous "worlds".

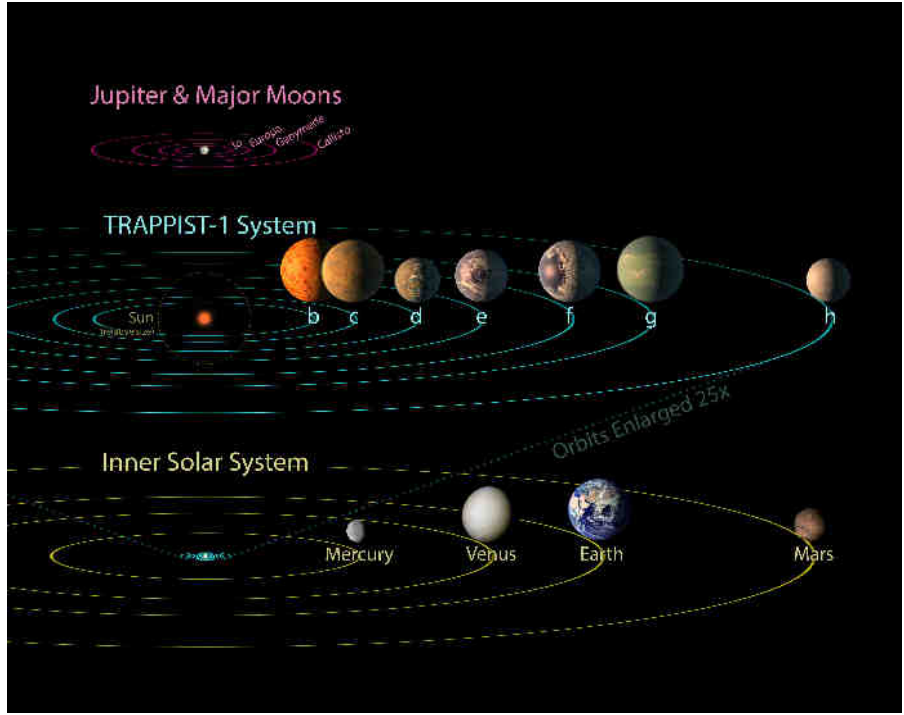

Figure 1.3: TRAPPIST-1 Comparison to Solar System and Jovian Moons. Image credit: NASA/JPL-Caltech.

Table 1.1: Physical properties of molecular clouds

| Category | Mass ($M_\odot$) | Size (pc) | Density $n_{H_2}$ ($cm^{-3}$) |
|---|---|---|---|
| Giant Molecular Cloud | $10^5 - 10^7$ | $50 - 200$ | $10 - 10^3$ |
| Molecular Cloud | $10^3 - 10^5$ | $5 - 50$ | $10^2 - 10^4$ |
| Clump | $10 - 10^4$ | $0.5 - 5$ | $10^3 - 10^5$ |
| Dense core | $0.1 - 10^3$ | $0.05 - 0.5$ | $10^4 - 10^6$ |
| Protostellar core | $0.01 - 10^2$ | $0.01 - 0.1$ | $10^5 - 10^6$ |

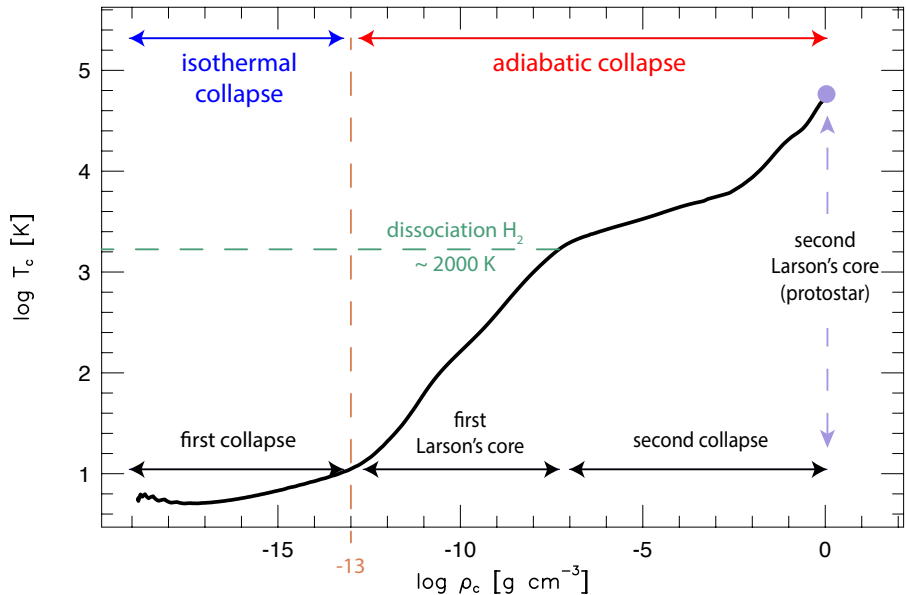

Figure 1.4: Temperature as a function of the density at the center of a cloud core during the collapse phases according to the model of Masunaga and Inutsuka (2000). Figure adapted from the original one in their paper.

Let us consider a molecular cloud that has collapsed to form an unstable dense core that will lead to a single star. The collapse is described in Fig. 1.4. There is a first collapse phase for which the energy from gas compression can be released by dust emission. This first phase is thus isothermal. When the density reaches about $10^{13}$ g cm$^{-3}$, the emission is trapped in the core so its temperature raises and the collapse becomes adiabatic. The collapse eventually forms an hydrostatic core, also known as the first Larson's core (Larson, 1969), which is a few astronomical units in size and its mass is about 1% of the initial dense core mass. When the first core temperature reaches about 2000 K, the dihydrogen molecules dissociate. Since this reaction is endothermic (it needs energy), the temperature is rising much slower in this second collapse. When the central density reaches a stellar density (about 1 g cm$^{-3}$), a protostar with stellar dimensions, also known as the second Larson's core, is formed in hydrostatic equilibrium. The protostar will then strongly accrete matter from the collapsing envelope though a disk (see following sections). At $10^6$ K in its center, the protostar will cross the "birth-line" and, by starting the fusion of deuterium, will become a pre-main sequence star. When the center reaches $10^7$ K, and that fusion of hydrogen can start, the object becomes a "true" star by joining the main sequence of the Hertzsprung-Russel diagram (see App. A). The star is formed but there is another concomitant story to tell. Let us take a closer look at what happens for the "disk" linking the core's envelope and the newly born star.

## 1.3 The disk phases

Dense core precursors of protostars can live as long as a million of years but may evolve quickly as it starts to collapse. Numerical models of both magnetic and nonmagnetic collapsing molecular cores indicate that a disk surrounding the protostar form rapidly within $10^4$ years after the collapse has been initiated (Yorke et al., 1993; Hueso and Guillot, 2005). Such a disk is naturally created because of angular momentum conservation during the collapse. To classify the Young Stellar Objects (YSO) according to their evolution, one method is to follow where the mass is currently located: in the envelope surrounding the protostar, the forming disk or the protostar itself. It is hard to directly measure the mass present in each component (but possible today), so indirect methods are used instead. A classification based on the spectral energy distribution of dust emission in the infrared is generally used since Lada and Wilking (1984). Those authors have highlighted different classes, namely class I, II and III, formalised then by Lada (1987), and corresponding to different stages in the evolution of the disk. An earlier phase with no disk, named class 0, corresponding to the collapsing core mentioned earlier in this chapter, was added by Andre et al. (1993) A visible classification is also used in parallel, where a Classical T Tauri Star (CTTS) is defined as a young stellar object with high Hα and UV emission because of a high accretion onto the star (corresponding roughly to class II objects in IR), and a Weak-lined T Tauri Star (WTTS) when no or very weak emission is seen (corresponding roughly to class III objects in IR). The observational infrared classification as well as the visible one are consistent with the theoretical view of a rotating collapsing core (Adams et al., 1987). A summary of their definitions are given in Table 1.2 and the scheme of this scenario is depicted in Fig. 1.5.

Table 1.2: Classification of young stellar objects

| Class | SED slope | Mass localisation | Observational characteristics |
|---|---|---|---|
| 0 | no emission | $M_{\text{env}} > M_{\text{star}} > M_{\text{disk}}$ | Sub-mm, no optical or NIR emission |
| I | $\alpha_{\text{IR}} > 0$ | $M_{\text{star}} > M_{\text{env}} \approx M_{\text{disk}}$ | Generally optically obscured |
| II | $-1.6 < \alpha_{\text{IR}} < 0$ | $M_{\text{disk}}/M_{\text{star}} \approx 1\%, M_{\text{env}} \approx 0$ | Accreting disk, strong Hα and UV |
| III | $\alpha_{\text{IR}} < -1.6$ | $M_{\text{disk}}/M_{\text{star}} \ll 1\%, M_{\text{env}} \approx 0$ | Passive disk, no or very weak accretion |

### 1.3.1 Protostellar disk: formation of the central star

The mass of the envelope core decreases by almost one order of magnitude between the class 0 and I YSO (Young et al., 2003). During this transition, the central protostar gains half of its final mass in only a few percents of the class 0 and class I lifetime (corresponding thus to a few $10^4$ years, see Evans et al., 2009). Laughlin and Bodenheimer (1994) suggested that this rapid evolution is possible if the disk is gravitationally unstable because of a high disk-to-protostar mass ratio. Consequently, this instability engenders sporadic burst of high accretion (Zhu et al., 2009) that occur about 10 times during the formation of a typical low-mass star (Hartmann and Kenyon, 1996). A protostellar disk is thus a transient massive and unstable disk, through which the mass is rapidly transported from the envelope to the protostar.

### 1.3.2 Protoplanetary disk: precursor of planetary bodies

After the transfer of the mass from the envelop to the (proto-)star, which has now probably started is life of star by getting onto the main-sequence, a remaining disk is left with a small mass compare to the one of the host star, about 1% only. This mass ratio makes the disk more stable allowing it to evolve more slowly (this kind of object and its evolution will be describe in more details in Sect. 2) and eventually form planets.

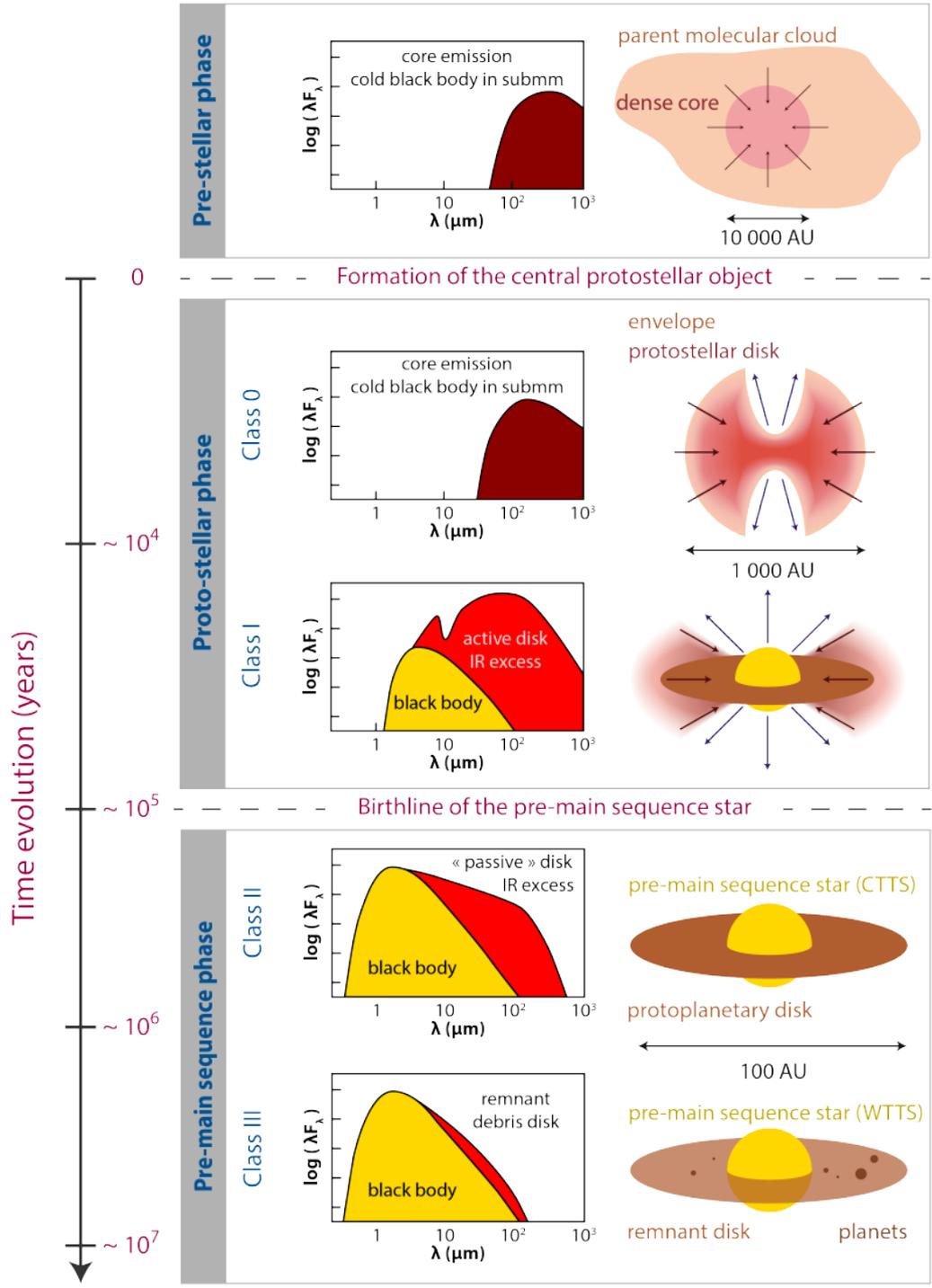

Figure 1.5: Evolutionary sequence for low-mass stars seen from observations. Personal adaptation from classical figures.

## 1.4 Planetary system formation and evolution

### 1.4.1 Planet formation by the classic core accretion scenario

Safronov (1972) introduced the bases of the core accretion scenario, then revisited by Pollack et al. (1996), to explain the formation of planets within protoplanetary disk. This classic scenario, supposed valid both for terrestrial and giant planets, implies to form initial terrestrial cores inside the disk to finally produces the planets following these steps:

- dust coagulation and formation of planetesimals;
- formation of planetary embryos and growth of the solid cores of giant planets;
- runaway gas accretion by the solid cores to form giant planets;
- the assembly of the planetary embryos to terrestrial planets.

I will develop each step in the following sections.

**Formation of planetesimals**

While dust grains represent only 1% of the newly formed disk mass, they represents the initial bricks for planet formation. The evolution of these solid particles in the protoplanetary disk is governed by transport and collisional processes, which are linked to each other. In a disk, solids tend to follow the keplerian motion while the gas, subject to pressure, is slowed and is so found in a sub-keplerian motion. As everyone may experience in his life by riding a bicycle through the air, whenever there is a difference in velocity between an object and the surrounding gas, the drag force acts towards eliminating this difference. As the rider may be slowed by the air, the gas tends to slow the particles in the disk. Since the particles are more or less coupled with the gas, depending on their size (the smallest being the most coupled), they do not all react in a same way. From this simple concept of size dependent drag forces, this implies that the trajectory and the velocity of a particle is depending on its size. Having different velocities means that particles will collide with each other, leading to two main consequences: a sticking/growth or a shattering/destruction of the involved particles.

The initial stage of planet formation, in this core accretion scenario, involve the growth from sub-micrometer sized dust grains to km-sized bodies, called planetesimals, that are gravitationally bound. To understand how particles grow along so many order of magnitude in mass, one needs to understand at which velocities particles collide, how often that happen and what is the outcome of the collision. All of these three points depend on the particle-particle velocities. There are various contributions to particle mean relative velocities in protoplanetary disks. For small-small particle, the dominant source is the Brownian motion (random motion of particles suspended in a fluid resulting from their collision with the fast-moving atoms or molecules of it). For the others, it is a mix between turbulence, vertical settling, azimuthal and radial drift (all of that caused by the drag force) and yield maximum impact velocities of a few 10 m s$^{-1}$. The main collisional outcomes are:

**Sticking** Hit-and-stick collisions.

**Bouncing** Particles bouncing off each other without changing their mass but possibly causing compaction.

**Erosion/Cratering** Smaller projectile removes mass from larger target, possibly shattering itself.

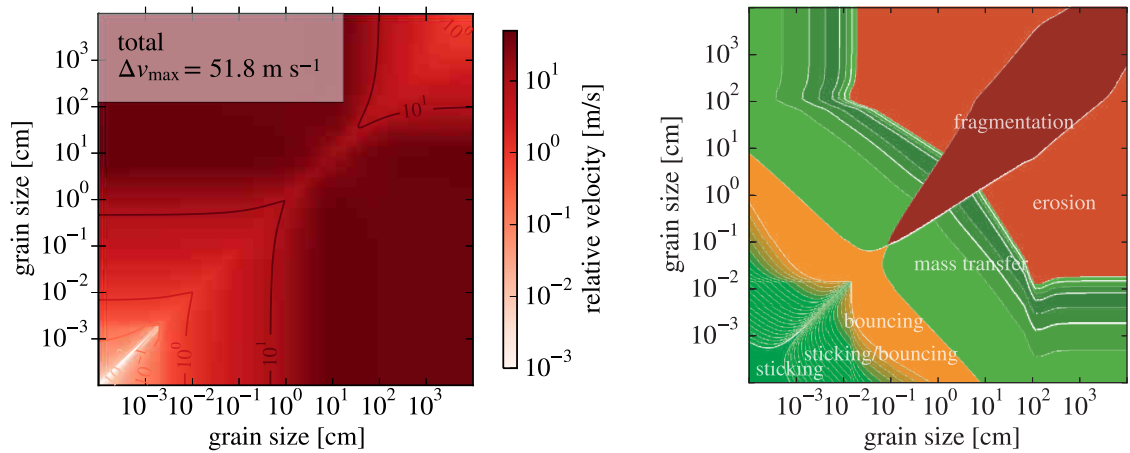

Figure 1.6: Total relative particle-particle velocity (left) and mean collisional outcomes (right) for silicate grains from (Windmark et al., 2012) as expected for a minimum-mass solar nebula disk (Weidenschilling, 1977b) at 1 AU for the corresponding velocities. For the right panel, green regions denote net growth of the larger collision partner, red mass loss, and orange denotes mass-neutral bouncing collisions. Figures adapted from Birnstiel et al. (2016)

**Mass transfer** Smaller projectile comes to pieces after colliding with target and deposits some fraction of its mass on it.

**Fragmentation** Complete destruction of the particles leading to a fragment population, generally distributed in a power-law size distribution.

The final outcome will depend on the impact velocity, the impact parameter (perpendicular distance between the path of a projectile and the center of the target), the grain sizes and their compositions. It is hard to take all of that into account but a simplified model can give a good estimate of how the bulk of particle distribution will evolve (Fig. 1.6).

Those processes permit the formation of macroscopic solid particles from submicrometer dust grains, but not easily because there are a few barriers to overcome. The first one applies mainly for the smallest particle for which the repulsion between two charged particles can avoid the collision: the charge barrier (Okuzumi, 2009). Following collisions of equal-sized particles for example (cf. the diagonal on Fig. 1.6), we can see that, when particles grow, their sticking probability decreases and bouncing becomes the dominant collisional outcome. The situation where particles stop growing and become stuck at the sticking/bouncing transition, at a few tenths of millimetres, was termed the bouncing barrier (Zsom et al., 2010). When particles get bigger, they finally have to face the fragmentation barrier (Brauer et al., 2008), with most of collisions being destructive. At every moment, but especially for biggest particles, the drift toward the central star is important and tends to be more effective in removing particles than particle growth can make them bigger (Weidenschilling, 1977a). However, these obstacles can be overcome. Because of a velocity dispersion around the mean relative velocity, the bouncing barrier can be crossed, but the fragmentation can also occurs more rapidly in consequence. The main convincing solution is to consider particles that are farther than the snowline in the disk (particular distance from the central protostar where it is cold enough for volatile compounds such as water to condense into solid ice grains), so that they are not only made of silicate but silicate and water ice. The increase of surface energy, about ten times, brought by the hydrogen bounds of the water ice might help, particularly to avoid the fragmentation barrier, but also the bouncing barrier by reducing the compression of aggregates.

Besides collisional growth previously discussed, the planetesimal formation can be done, or help, by gravitational collapses related to dynamical effects that are able to cause strong local concentrations of dust particles. One of them is the formation of axisymmetric overpressure regions (balanced between pressure gradient and Coriolis force) called pressure bumps. Simulations of turbulence show that large-scale variations in the strength of the turbulence lead to the spontaneous formation of pressure bumps at the largest scales of the turbulent flow. These pressure bumps are quasi-stable and live for hundreds of orbits before they disassemble and reform. Dust particle migrate to the centre of the bump so that particle concentrations may reach a threshold above which very large planetesimals can form by the subsequent gravitational contraction and collapse phases (Johansen et al., 2011). Another mechanism could be the vortex trapping. Here, azimuthally elongated vortices, created directly from a flow instability, can also trap dust particles (e.g. Lyra and Lin, 2013). A last mechanism could be the streaming instability (Youdin and Goodman, 2005) that happens when the gas-to-dust mass ratio approached 1 in the midplane of the disk. This time, the drag force is driven by the dust so that the gas is easily accelerated to the keplerian velocity, the drag force and the radial drift are then reduced, leading to a pill-up of particles.

When objects bound by their own gravitational attraction (as opposed to their surface and material binding forces for smaller objetcs) have successfully been formed, called planetesimals, which typically occurs at sizes above several kilometres (e.g. Benz and Asphaug, 1999), the next growth phases will be driven by the gravitation.

**From planetesimals to protoplanets**

The conventional mechanism to make a planetesimal to grow is by accreting other planetesimals through collisions. Let us introduce the important definition of the collisional cross section,

$$\sigma = \pi R^2 \left(1 + \frac{v_{\text{esc}}^2}{v_{\text{rel}}^2}\right), \tag{1.1}$$

where $v_{\text{rel}} \approx \sqrt{2}v$ with $v$ the velocity dispersion, $v_{\text{esc}} = \sqrt{2GM/R}$ the escape velocity of the planetesimal with mass $M$ and radius $R$. The evolution from protoplanetesimals to protoplanets goes through different phases defined here and illustrated in Fig. 1.7:

**Ordered growth** If planetesimals' masses are small, what is generally true initially, $v_{\text{esc}} \ll v_{\text{rel}}$, and the cross-section is simply geometric, i.e. $\sigma = \pi R^2$. In this first phase, the planetesimals of all sizes grow at the same rate.

**Runaway growth** As masses increase, the gravitational focusing (the term $v_{\text{esc}}^2/v_{\text{rel}}^2$) increases and enhances the cross-section. Since this enhancement is more efficient for more massive planetesimals, the growth of the larger bodies dramatically runaways over smaller ones. As a result, few massive bodies stand out over the rest of the planetesimals population (see e.g. numerical study of Wetherill and Stewart, 1989).

**Oligarchic growth** At late stages of the runaway growth, planetary embryos become sufficiently large to start interacting with each other so that the overall dynamics is now going to be dominated by these few bodies, named the "oligarchs". They increase the velocity dispersion of the neighbouring planetesimals, reducing the efficiency of gravitational focusing and slowing down their growth. In this regime (Kokubo and Ida, 1998), neighbouring oligarchs growth at similar rates maintaining similar masses. The overall outcome at this stage is a bi-modal distribution of an embryo-planetesimal system.

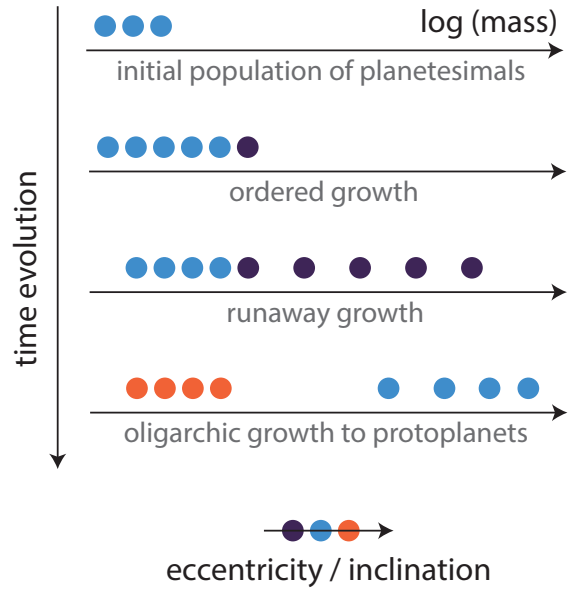

Figure 1.7: Simple illustration of the mass and orbital stability evolution of planetesimals through the different growing phases to form protoplanets.

The oligarchic growth (and other processes) led to a disk depleted of gas.

**Giant planet formation**

In this core accretion scenario, a giant planet starts with a massive solid core (or protoplanet) of a mass around 10 $M_{\text{Earth}}$. This allows the core to directly accrete a gaseous atmosphere from the disk of gas. It seems difficult to form a such core in the inner disk before the snowline (about 3 – 5 AU for the young Solar System) so that giant planet formation should takes place farther away.

Pollack et al. (1996) have revisited the bases of the core accretion scenario and developed the case of giant planet. The Fig. 1.8 illustrates the evolution they depicted, updated by the model of Mordasini et al. (2012). First, a rapid build-up of the core occurs until the isolation mass is reached (phase labeled I), then a plateau phase follows (phase labeled II), characterised by a slow increase of the envelope mass which allows further core growth, and then the transition to gas runaway accretion is observed. Shortly after the crossover point, when the core and envelope masses are the same (start of phase III), the gas accretion rate increases strongly because of beginning runaway accretion. When the accretion rate is maximal, the detached phase (labeled D) begins and the collapse of the envelope starts. This collapse is very rapid but is still an hydrostatic contraction. When the final mass of the planet is almost reached and the accretion drops, this marks the beginning of the evolutionary phase (labeled E). This evolutionary path is possible as long as the disk of gas is still there. If a planet, as Jupiter, has done its runaway phase and is finally composed mostly of gas, we call it a "gas giant planet". If the planet growth stops, because of the dispersal of the disk gas for example, while the icy and rocky core dominate the planet mass, we call it an "ice giant planet", as Uranus and Neptune for the Solar System. On the other side, if this accretion lasts long enough to form a "giant" with a mass greater than about 13 Jupiter masses, it will form a brown dwarf. More than a planet, a brown dwarf is a sub-stellar object since such a mass allow the fusion of deuterium. At 75 – 80 Jupiter masses, the object is able to fuse ordinary hydrogen, so that this value represents the limit between a brown dwarf and a star. It is however very unlikely that a second star may be formed in a disk by this mean.

**Terrestrial planet formation**

At the end of the "oligarchic growth" and the giant planets formation, when the disk of gas has been dissipated, many (probably hundreds) protoplanets are left. They do not accrete anymore since they have taken all that was around them. With no more gas to stabilise the orbits by damping the random velocities from dynamical friction (Kenyon and Bromley, 2006), and in the presence of few giant planets, the remaining protoplanets will increase their eccentricity and velocity dispersion, and will eventually collide through giant impacts. Protoplanets are now massive enough to survive those impacts. In this last phase, called the "chaotic growth", lasting about 100 millions of years, terrestrial planets may be formed. It sets the mass and composition of the planets and, if no other big event happens, the final orbital architecture of the planetary system. The Moon was likely formed by one of the latest impacts on the proto-Earth.

### 1.4.2 Need for complementary scenarios: pebbles accretion and gravitational instabilities

While explaining the formation of planets from the beginning, the core accretion scenario has to face some troubles, mainly to form giant planets. It is initially necessary to form a massive core of 10 – 15 Earth masses to trigger the runaway accretion of a gaseous envelope, and this take a lot of time. If this could be overcome by a high planetesimal density after the snowline, the pebbles accretion paradigm is now more and more used to enhanced significantly the accretion rates by implying centimetre-sized, named "pebbles", to metre-sized objects that are highly coupled with the gas. Still, this scenario fails in explaining very distant giant exoplanets that have been discovered recently via direct imaging, as around β Pictoris b (Lagrange et al., 2010) and the HR 8799 system (Marois et al., 2010) where orbital separation reaches up to 70 AU. One possible alternative to form giant planets at large orbital separation is by gravitational instabilities (Boss, 1997), which likely operate, in parallel with the core accretion, in the outer disk during the early stages of disk evolution.

In the pebbles accretion scenario, the accretion of mm- to cm- sized grains can significantly enhanced the core growth (e.g. Lambrechts and Johansen, 2012). Those pebbles are highly coupled with the gas of the disk via aerodynamic drag so the relative velocity can be strongly damped during gravitational encounters and the cross section can be significantly enhanced. The formulation of pebbles accretion has been incorporated in global models of planet formation in the last years (e.g. Bitsch et al., 2015). It was found that this scenario overcomes many of the challenges in the formation of ice and gas giants in evolving disk, including the formation of the cores of giant planets in the Solar System and for extrasolar planets at large separations.

Development of spirals inside a disk may form unstable over-densities if the the self-gravity becomes higher than the pressure and thermal forces. Two criterions must be fulfilled to allow such instabilities: the stability criterion of Toomre must favour the self-gravitational potential of the disk gas rather than the enthalpy of the disk gas (Toomre, 1964); the cooling time of disk gas must be smaller than the dynamical time needed to do an orbital period (Gammie, 2001). A massive disk and/or a cold region (outer region mainly) is necessary. If those conditions are fulfilled, the formation of the giant planet will start by a quasi-static compression of the overdensity. At this stage, the hydrogen is molecular but when the core reach about 2000 K, the dihydrogen will dissociate (similar evolution than a dense core collapse in a molecular cloud) and a rapid dynamic collapse is initiated. The last phase of compression is quasi-static again while the planet is cooling. If the core accretion scenario needs a few million of years to form a giant planet, gravitational instabilities could reduce this time to a few orbital periods only. This scenario do not allow the formation of giant close the star (not under 100 AU for the Solar System according to Boley 2009) as the two previously cited conditions can not be fulfilled.

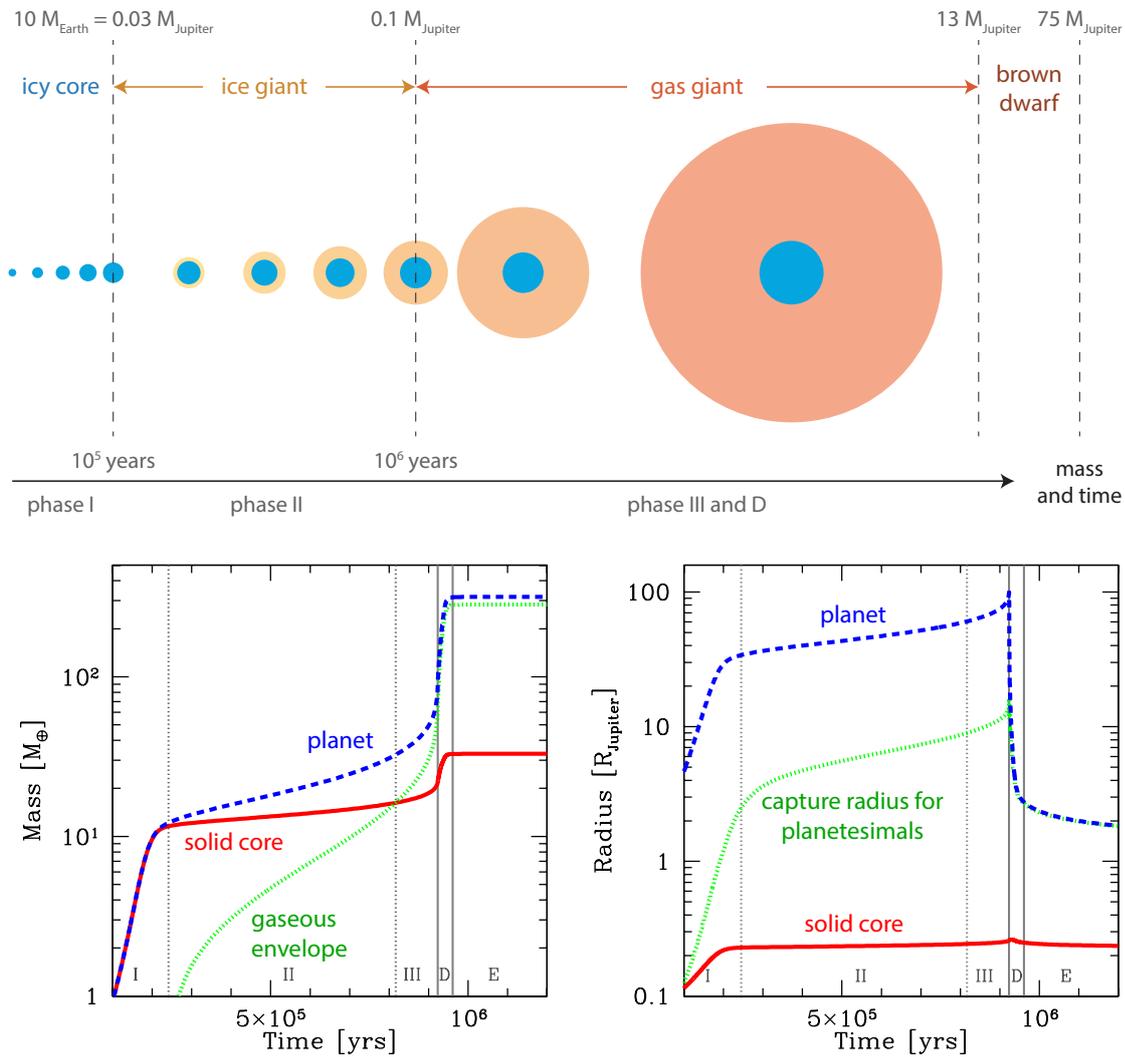

Figure 1.8: Illustration of the formation of a giant planet following the growth of the core, the gas envelope and the whole planet. Top: schematic illustration of the different phases leading to the formation of an ice giant or a gas giant plant (or possibly a brown dwarf). Bottom: simulation of the in situ formation of Jupiter at 5.2 AU. Gray vertical lines show the major phases, labeled in the top left panel: I, II, III during the attached phase, D for the detached phase, and E for the evolutionary phase. The left panel shows the evolution of the core mass (red solid line), the envelope mass (green dotted line), and the total mass (blue dashed line). The right panel shows the evolution of the core radius (red solid line), the total radius (blue dashed line), and the capture radius for planetesimals (green dotted line). Figure adapted from Mordasini et al. (2012).

The core accretion scenario, likely enhanced by the pebbles accretion, and the gravitational instabilities are thus two compatible scenarios that may be involved together in the formation of a planetary system but in different regions of the disk and within a different timescale.

### 1.4.3 Orbital evolution of a planetary system

The final structure of a planetary system depends on the planet formation process but not only. Indeed, the planets, during and after formation interact with other objects of the planetary system (central star or disk) that make their orbital properties evolve mainly by:

**Planet-disk interactions (during formation)** They damp eccentricities and inclinations efficiently, and affect the planet semi-major axis by migration;

**Planet-planet interactions (during and after formation)** They pump eccentricities and inclinations, and affect the planet semi-major axis through scattering events.

After formation, the interactions of the planets with the host star, or the debris disk, can also make the orbital properties evolved.

The presence of gas is thus critical to stabilise or destabilise the orbits and to make planets move closer or farther to the star. There are two kinds of torques that can move a planet inside the disk: the wake torque caused by spiral density waves, and the corotation torque caused by an exchange of angular momentum of the gas in the horseshoe region on the planet's orbit. The wake torque is generally negative (the inner wake is positive while the outer is negative and dominates), producing an inward migration. The corotation torque is, however, generally positive (also the combination of a positive and a negative component so the sign may change in some conditions) leading to an outward migration. The combination of the two, for different kind of planets, leads to different types of migration:

**Type I migration** It applies to low-mass planets (up to few tens of Earth masses) and since it is a competition between the two kind of torques, there are some locations known as "planet traps" when both torques focus planet in a stable location.

**Type II migration** Massive planets (at least a few Earth masses) can open a gap in the disk because of the inner and outer wakes. If the gap is sufficiently important (for planet more massive than Jupiter), only the wake torque is left and an inward migration runs on timescales longer than $10^4 - 10^5$ years. This migration can be considerably slowed down when the planet mass becomes larger than the mass of the gas outside the planet gap. Contrary as what is often thought, the gap does not force the planet to follow the viscous evolution of the disk and the gas can cross the gap (Dürmann and Kley, 2015).

**Type III migration** If a partial gap is open (consequence of a Saturn-mass planet - about 100 Earth masses) and that the dynamical corotation torque dominates, this lead to an inward or outward runaway migration before a saturation phase.

As migration, many mechanisms drive the final architecture of planetary systems. The great diversity of extrasolar planetary systems, showed by (mis-)aligned hot Jupiters and the many multiple super-Earths systems, illustrates that plurality of mechanisms. If it is difficult to derive a unique general theory due to the broad range of processes that echoes the diversity at the end, it seems that the broad lines are here well defined. The evolution of protoplanetary disk also follow different paths leading to different kind of planetary systems at the end. The state-of-the-art about protoplanetary disk will be developed in the next chapter.

# Chapter 2

# The protoplanetary disk phase

Protoplanetary disks (Fig 2.1) are found around young stars as a consequence of a collapsing rotating core. After a very rapid phase of accretion, moving the mass from the envelope dense core to the protostar, the remaining disk left, i.e. the protoplanetary disk, will possibly give birth to a planetary system. However, this is a race against time since the lifetime of a disk is of the order of million years, comparable to the theoretical time expected to form planets in the classic core accretion scenario. In this chapter, I describe the typical properties of a protoplanetary disk (Sect. 2.1), the processes driving its evolution (Sect. 2.2), and focus mainly on one of them: photoevaporation (Sect. 2.3).

## 2.1 Disk properties

### 2.1.1 Composition

Disks form from dense molecular cores in the ISM (Sect. 1.3) so that their composition, in terms of gas and dust, should be initially close to the one of the ISM. In this medium, dust is mainly composed of sub-micrometer-sized amorphous silicates, generally olivine and pyroxene, and an admixture of graphite grains and Polycyclic Aromatic Hydrocarbons (or PAHs, Draine, 2003). Disks start from that composition but molecules freeze-out from the gas phase onto the grain surfaces to produce ice mantles, the grains agglomerate by collisions, and are processed via thermal annealing. Dust in a disk have thus a much higher crystallinity fraction and a larger size than in the ISM (e.g. McClure et al., 2010).

The bulk of the disk is cold and dense so that gas is molecular and $H_2$ dominates. Dihydrogen is difficult to observe but there are many other species that one can detect. The molecule CO has the highest abundance after $H_2$, and a large dipole moment leading to strong emission, so that it is often used to trace molecular gas. Organic molecules have been detected in disks (e.g. Lahuis et al., 2006) and are supposed to be produced by the sublimation of icy grain mantles. OH and $H_2O$ are also present (e.g. Carr and Najita, 2008). However, the disk is not fully molecular since atomic or ionised species have also been detected, mainly in regions subject to energetic photons (e.g. Acke et al., 2005; Pascucci and Sterzik, 2009; Mathews et al., 2010). Disk chemistry is actually very rich, and the reader is thus referred to Bergin (2011) for a more deep-in review.

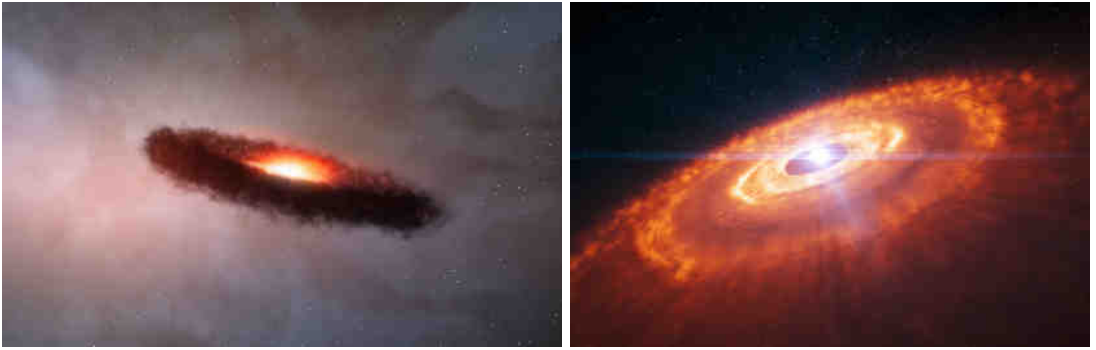

Figure 2.1: Artistss impression of protoplanetary disks. Left: disk of dust and gas similar to silhouette disks seen in visible with HST. Image credit: ALMA (ESO/NAOJ/NRAO) / M. Kornmesser (ESO). Right: disk similar to the ones observed recently via sub-millimeter dust emission using ALMA, revealing the complex structure of the disk with concentric rings of gas and gaps indicating planet formation. Image credit: ESO/L. Calçada.

### 2.1.2 Total masses

Disk masses are generally best determined by submillimetre or millimetre wavelength observations of the dust. Under the assumption than the continuum emission is optically thin, which is true except in the innermost regions of the disk where the density is very high, the optical depth $\tau_\nu$ is given as the integral of the dust opacity $\kappa_\nu$, times the density $\rho$ along the line of sight $s$,

$$\tau_\nu = \int \rho \kappa_\nu ds = \kappa_\nu \int \rho ds = \kappa_\nu \Sigma, \qquad (2.1)$$

where $\Sigma$ is the projected total surface density (gas and dust included). This is a direct observational way to estimate the total mass but also the mass distribution if the observation is sufficiently spatially resolved. A commonly used prescription for the dust opacity is the one from Beckwith et al. (1990) where

$$\kappa_\nu = 0.1 \left(\frac{\nu}{10^{12} \text{ Hz}}\right)^\beta \text{ cm}^2 \text{ g}^{-1}. \qquad (2.2)$$

The absolute value as well as the power-law index $\beta$ are related to the size distribution and composition of the dust grains (e.g. Ossenkopf and Henning, 1994). The reader should also note that since only the dust is observed while the objective is to measure the total gas plus dust mass, this normalization implicity includes a gas-to-dust mass ratio of 100. This is initially true since this represents the mean value for the ISM but this can evolve with the disk, especially at some particular locations.

A few large surveys have been conducted to measure the disk mass distribution, as in Taurus-Auriga (Beckwith et al., 1990; Andrews and Williams, 2005), in the ρ Ophiuchus cloud (Andre and Montmerle, 1994; Andrews and Williams, 2007b), or in the Orion nebula cluster (Mann and Williams, 2010). For Taurus-Auriga and Ophiuchus, it has been found that the median mass of Class II YSO disks is 5 Jupiter's mass (or $5 \times 10^{-3}$ M$_\odot$), with a median disk-to-star mass ratio of 0.9% (at the previous stage of Class I, so for protostellar disk, values are respectively $1.5 \times 10^{-2}$ M$_\odot$ and 3.8%, see Table 4 in Andrews and Williams 2007b). Using a logarithmic mass interval, the disk mass distribution (Fig. 2.2) is rather flat with a steep decline beyond $5 \times 10^{-2}$ M$_\odot$, for all regions. However, in Orion, where photoevaporation is probably highly efficient because of a very intense UV radiation field, the lowest mass disks are not present.

It is logical that a higher central star mass will have a larger protoplanetary disk mass since it requires more material to pass through the initial protostellar disk. Fig. 2.3 compiles (sub)millimeter measurements of Class II YSO disks from the literature. It confirms that disk mass tends to be lower around low-mass stars such that the ratio of the disk-to-star ratio is around 1%, close to the median of the detections, but exhibits a large scatter. This is correct from brown dwarfs (substellar object) to Herbig Ae/Be (most massive part of the low-mass stars) stars, where all disks have a value of this ratio between 0.1 and 10%, but this does not work for O stars (massive stars), probably because of rapid disk dissipation by photoevaporation or different star-formation mechanisms. Since massive stars are out of the range of this thesis, the reader is refer to the review of Zinnecker and Yorke (2007).

### 2.1.3 Surface density

The surface density profile (the mass distribution with the disk radius) can be estimated from spatially resolved observations of disks. It has been generally modelled by a power-law, $\Sigma \propto r^{-p}$, with the index $p$ between 0 and 1.5 (e.g. Lay et al., 1997). This kind of profile, with a simple truncation, are hard to reconcile with observations at the outer edge of disks, especially because of a difference between the sizes observed in gas and dust emission (see Dutrey et al., 1996; Isella et al., 2007). When looking at disks in the Orion nebula, McCaughrean and O'dell (1996) also exclude such pure power-laws and showed that an exponential decay at the outer edge was required. Hughes et al. (2008) showed that the outer edge problem was also solved by this solution. Since, this kind of exponentially tapered profile is also consistent with physical models of accretion disks (e.g. Hartmann et al., 1998), the profile generally taken into account rather follows the form

$$\Sigma \propto r^{-p} e^{\left(-r^{2-p}\right)}. \tag{2.3}$$

### 2.1.4 Radius

Disk sizes are not easy to measure because the outer parts do not emit strongly because of their generally low temperatures. However, they are still efficient absorbers. In consequence, it is possible to directly measure the extension of disks by observing disks in silhouettes, that is to say in absorption by contrast of a luminous background. Vicente and Alves (2005) have used this to conduct direct size determination of silhouette disks in the Orion nebula cluster that are seen against the optically bright HII region (i.e. ionised region of the cluster) in background. Excluding outliers, they found radii ranging from 50 to 194 AU for 22 disks and inferred a median radius about 75 AUs.

To date, contrary to the disk mass, we do not have enough measurements of disk sizes on a large range of star masses to establish any dependency.

### 2.1.5 Height

Protoplanetary disks are not flat but rather flared with a height increasing with the radius. This was initially suggested by Kenyon and Hartmann (1987) before being directly visible in *Hubble* images of disks seen in silhouette (e.g. Padgett et al., 1999). The pressure scale height $H$ gives an idea of the disk height (at a factor a few) and is depending on the competition between the disk thermal pressure and the vertical component of the stellar gravity so that

$$H = \frac{c_\mathrm{S}}{\Omega_\mathrm{K}}, \tag{2.4}$$

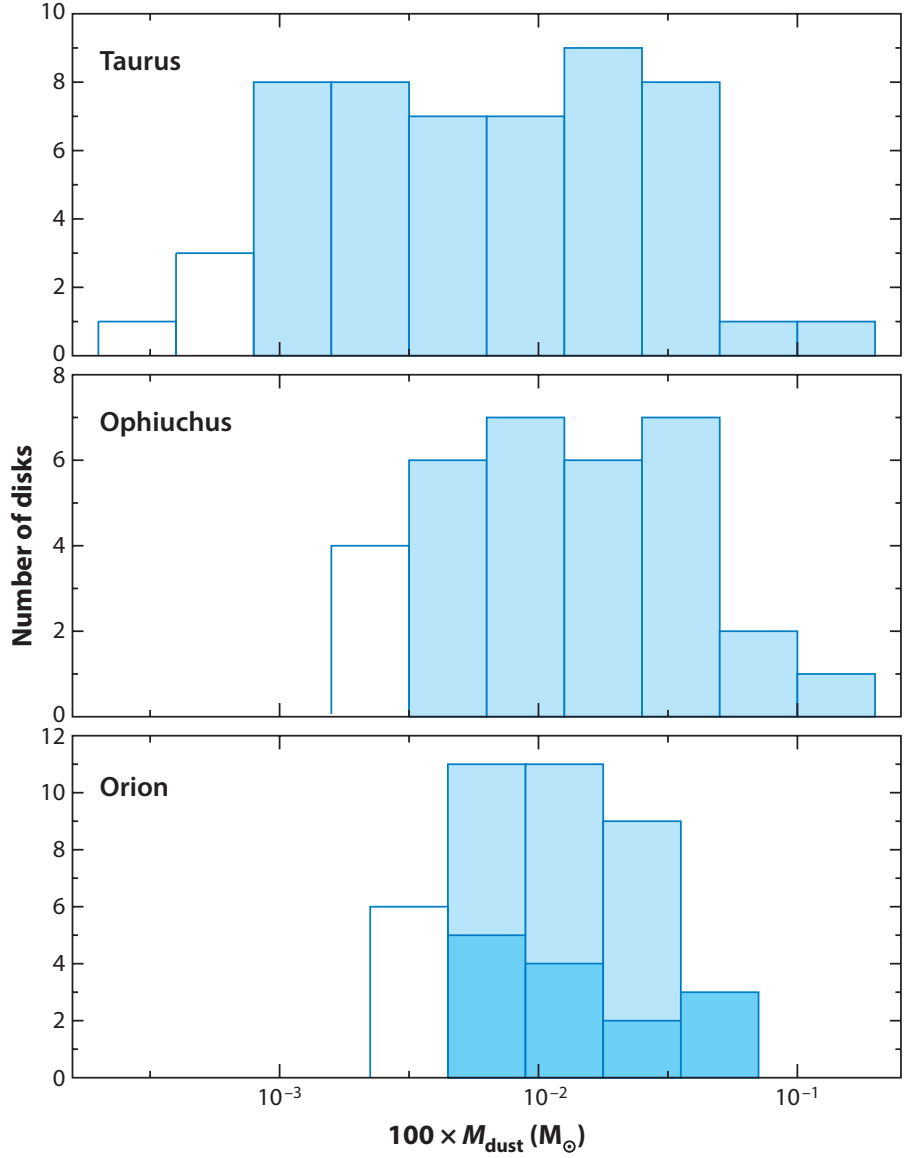

Figure 2.2: Distribution of protoplanetary (Class II YSO) disk masses in Taurus, Ophiuchus and Orion. The dust masses are derived from millimeter fluxes and extrapolated to a total mass assuming a maximum grain size of 1 mm, a characteristic temperature of 20 K, and an interstellar gas-to-dust ratio of 100 from data of Andrews and Williams (2007b); Mann and Williams (2010). The filled bars show the range where the millimeter measurements are complete for each region. The darker blue bars in the Orion histogram show the disks with projected distances 0.3 pc beyond the O6 star, $\Theta^1$ Ori C, in the Trapezium Cluster. Figure from Williams and Cieza (2011).

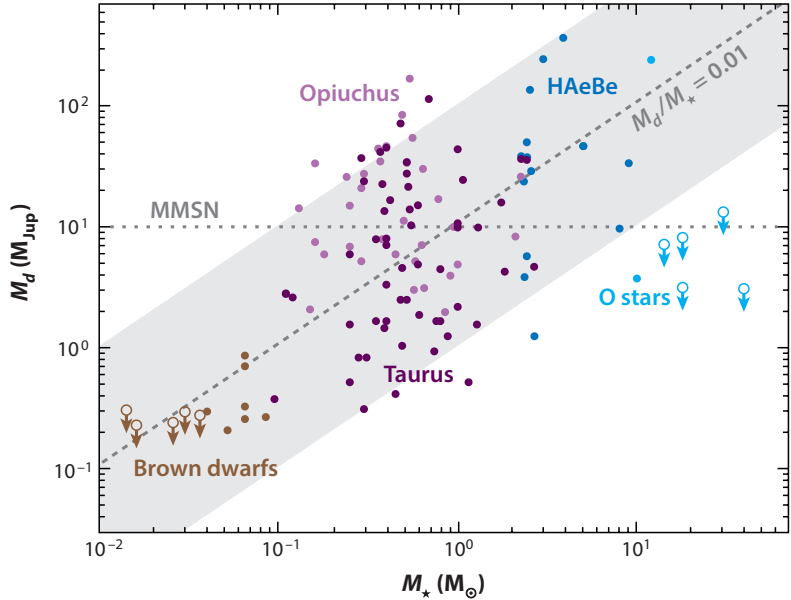

Figure 2.3: Scattering of protoplanetary disk masses according to the mass of the central star. Upper limits are only shown at the extremes of the stellar mass range, where no disks have been detected. The dashed diagonal line delineates where the mass ratio is 1% and is close to the median of the detections. Figure from Williams and Cieza (2011).

where $c_S$ is the sound speed and $\Omega_K$ the Keplerian velocity. The sound speed is involved in the geometry while it is depending on the temperature, which is depending on the strength of the radiation received which in turns depends on the geometry. The solution is thus not trivial but the analytical solution derived by Chiang and Goldreich (1997) consistent with more accurate numerical model (e.g. Dullemond et al., 2002), is a power-law $H \propto r^h$ with an index $h$ between 1.3 and 1.5.

### 2.1.6 Temperature

Dust temperature derived from SED modelling (Andrews and Williams, 2005) and interferometric CO images (Guilloteau and Dutrey, 1998; Piétu et al., 2007) show that the radial temperature gradient follows a power law, $T \propto r^{-q}$ with an index $q$ between 0.4 and 0.7 for T Tauri ($M_* < 1.5 M_\odot$) and Herbig Ae/Be stars ($1.5 M_\odot < M_* < 8 M_\odot$). Gas observations give also the disk vertical structure with a dynamic atmosphere above a cool midplane so that the temperature is increasing with height. The thermal structure is accompanied by phase changes from molecular to atomic and finally ionised. Dust grains undergo sublimation and condensation of icy mantles as they cross those regions from the superficial layers to the midplane.

## 2.2 Processes driving the evolution of an accretion disk

Protoplanetary disks evolution is governed by

**Accretion onto the star** Accretion onto the central host star is relatively well constrained observationally and understood with magnetospheric infall accretion models that are able to reproduce the profiles of emission lines such as Hα, Brγ and CaII (e.g. Muzerolle et al., 1998). Moreover, viscous evolution models are roughly consistent with the observational constraints for disk masses and sizes as well as the decrease in accretion rate over time (from about a value of $10^{-7}$ to $10^{-9}$ with a median accretion rate of $10^{-8}$ $M_\odot$ $yr^{-1}$ for T Tauri stars of age about 1 million years, see Hartmann et al., 1998). To first order, especially in the early phases of the disk, viscous transport drives the disk evolution. However, a disk only subject to viscous evolution will fail to explain the observed variety of spectral energy distribution of the dust as well as the rapid dissipation of the disk in a few million of years.

**Photoevaporation from local or external radiation sources** Photoevaporation is one of the main mechanisms by which primordial disks are believed to lose mass and eventually dissipate. This mechanism is due to the escape of matter, mainly gas, because of an excess of energy/heat brought by energetic radiation (ultraviolet or X-rays) coming from the central host star and, sometimes, by a nearby massive star also. Photoevaporation will be developed in the following section (Sect. 2.3).

**Agglomeration into larger bodies (and eventually planet formation)** Viscous accretion and photoevaporation operate on the gas, the main component of the disk, but other processes act on solid particles, i.e. grain growth and dust settling. Indeed, the smallest grains have the largest surface-to-mass ratio and are swept along with the gas but, as grains collide and stick together, their surface-to-mass ratio decreases and their motions decouple from the gas. Consequently, they suffer a strong drag force and settle toward the midplane. The increase of dust density in the interior of the disk accelerates grain growth and results in even larger grains settling deeper into the disk. Those processes tend to stratify the disk according to the grain size but turbulence add some degree of mixing (Dullemond and Dominik, 2005). Observational evidence of grain growth can be obtained following the shape of the silicates characteristic features in the mid-IR for the sub-micron to micron-sized grains (e.g. Kessler-Silacci et al., 2006), and via the slope evolution of the SED at submillimeter wavelengths (e.g. Mannings and Emerson, 1994). Dust settling implies a scale height and a flaring angle reduction. As a consequence, the surface intercepts less radiation and emits less in the mid-IR so that the slope of the spectral energy distribution of the dust at these wavelengths can be used as a diagnostics (Dullemond and Dominik, 2005).

**Dynamical interactions with stellar or substellar companions** Star formation is a highly dynamic and chaotic process. Therefore, dynamical interactions between the disk and a companion stellar or substellar object may affect it. Dynamical encounters can effectively significantly truncate the disk (Bate et al., 2003) or break the azimuthal symmetry within it. However, disruptive encounters with a passing stellar object have been found to be uncommon (Scally and Clarke, 2001) and thus not significant among the processes governing the disk evolution.

## 2.3 Gas dispersal and photoevaporation

### 2.3.1 Disk lifetime

The easiest way to estimate the lifetime of protoplanetary disks is to follow the disk frequency, i.e. how many stars harbour a disk, with evolving age of star clusters (Fig. 2.4). The inferred disk lifetime varies from 2 – 3 Myr at short wavelengths to about 4 – 6 Myr at longer wavelengths.

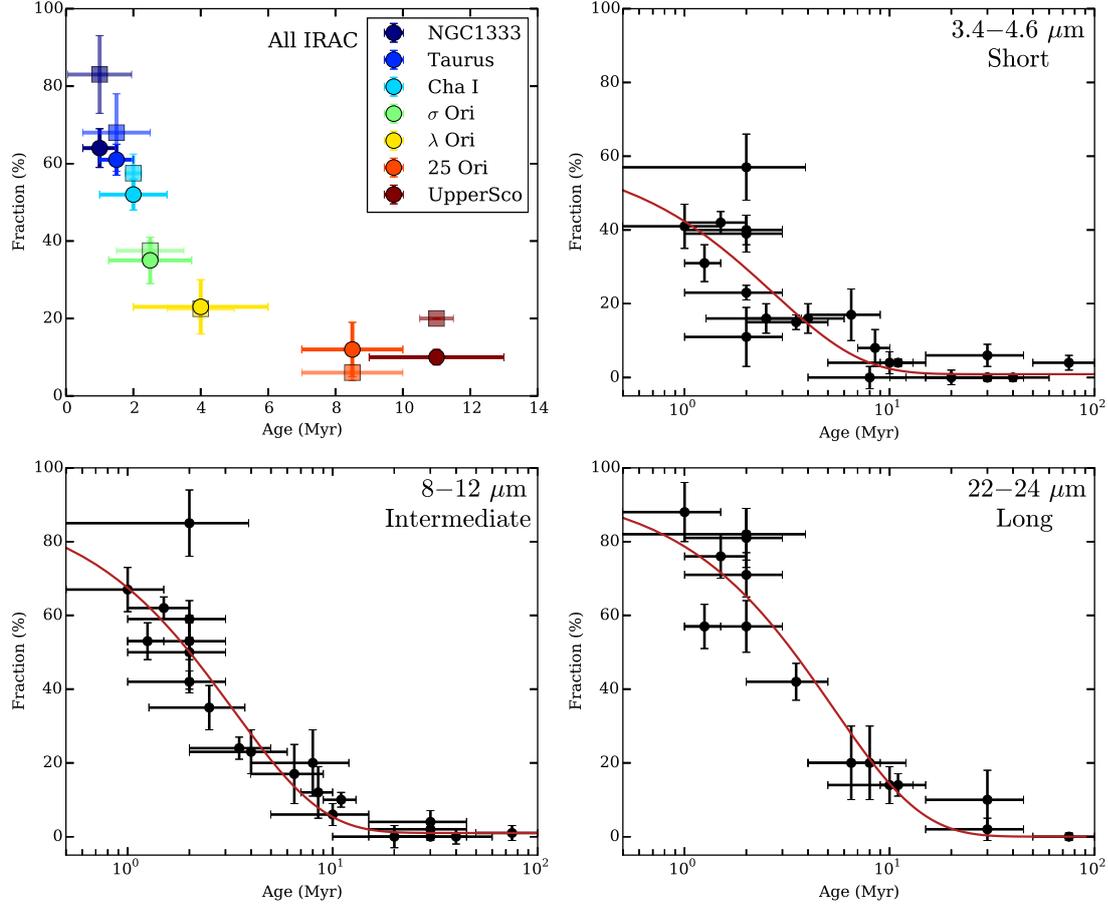

Figure 2.4: Disk frequency as a function of cluster age at different wavelengths as probed by Spitzer photometry. The best-fit exponential law is over-plotted (red line). Figure from Ribas et al. (2015).

This is an evidence than the disk of dust is erode from inside-out since the shortest wavelength are coming from the hot inner disk while the longest ones are coming from the cold outer regions.

Dust evolution may not be representative of the whole disk. Actually, it is possible to deplete the dust by the formation of planet while the disk of gas is not yet affected by dissipation. It is thus necessary to study the gas lifetime too. The lifetime inferred from gas observations is less certain since the emission are very faint but it seems that the disk of gas survive longer in a timescale of tens of million years. However, in the inner disk ($< 1$ AU), gas tends to be remove before the dust (Fedele et al., 2010; Hardy et al., 2015).

Disks are believed to be disperses by a combination of viscous accretion, photoevaporation and, to some extend, planet formation. Viscous accretion is not sufficient to dissipate the disk of gas while the planet formation may deplete the dust and a fraction of the gas (a small fraction of the initial mass is remaining in the form of planets). It is thus very likely that the photo-evaporation becomes a dominant effect of the evolution after the accretion weakens and is the responsible for the disk dissipation. I will focus on this process in the following sections.

### 2.3.2 Internal photoevaporation - energetic radiation from the central star

The impact of energetic radiation on a disk has been first studied for disks around massive stars by Hollenbach et al. (1994), taking into account H-ionizing photons from the central star. Hydrogen can be ionised by photons more energetic than 13.6 eV, defining the extreme-UV (EUV) range. EUV photons, originating at the stellar chromospheres of low-mass stars, ionise and heat the disk surface hydrogen to $10^4$ K. Beyond the so-called "gravitational radius" (about 10 AU for solar-mass star in that case), the thermal velocity of the ionised hydrogen exceeds its escape velocity and the material is lost in the form of a wind. Disk evolution models for low-mass stars, that were then developed, combined viscous evolution with photoevaporation by EUV photons (Clarke et al., 2001; Alexander et al., 2006a,b). The switch-on of the photoevaporation by EUV was able to resolve the two-timescales problem of T-Tauri evolution, for which a sudden dispersion of the disk is observed after a much longer lifetime. According to these models, there is an initial phase where the accretion rate dominates over the photoevaporation rate, and the disk thus undergo standard viscous evolution. In a second phase, where the accretion rate drops sufficiently, the photoevaporation takes the lead, ensuring a sufficient loss of mass to reproduce the observed lifetime. This transition is close to $10^{-10} - 10^{-9}$ $M_\odot$ yr$^{-1}$. At this moment, the outer disk is no longer able to resupply the inner disk ($r \leq 1$ AU) with material so this latter is drained on a viscous timescale of $10^5$ years (after a lifetime of a few million years), and an inner hole of a few AUs in radius is formed. The creation of a hole leads to the direct irradiation of the inner rim and results in a rapid dispersal of the outer disk. Finally this kind of "UV-switch" model accounts for the lifetimes and dissipation timescales of disks, as well as for SEDs of some pre-main sequence stars suggesting the presence of large inner holes.

More recent models include far-UV (FUV) photons (e.g. Gorti et al., 2009; Gorti and Hollenbach, 2009) and/or X-rays (e.g. Owen et al., 2010) in addition, or in competition, with the EUV photons. FUV photons are less energetic than EUV photons, with 6 eV $< h\nu <$ 13.6 eV, and can not ionise hydrogen but can dissociate the H$_2$ molecule, and ionise some other species. Contrary to EUV photons restricted to the inner few AUs of the disk and the surface layers, X-rays and FUV photons are able to penetrate much larger columns of neutral gas and are thus able to heat gas located both deeper in the disk and at larger radii (tens of AUs from the star). According to disk surface density distribution, most of their mass is located at large radii. Photoevaporation of the outer disk can thus have dramatic effects on the disk evolution here: the viscous expansion in the outer disk is restricted and the disk evolves into a shrinking torus of material (Gorti et al., 2009). Photoevaporation rates are of the order of $10^{-9} - 10^{-8}$ $M_\odot$ yr$^{-1}$, i.e. one to two orders of magnitude greater than pure EUV photoevaporation. Those rates are in general sensitive to the density and temperature structure of the disk which, unlike the EUV case, have to be determined involving accurate models on physical and chemical processes.

If there is a reasonable agreement on the qualitative behaviour between different models of viscous evolution under internal photoevaporation, FUV, EUV photons or X-rays act with different ways, and a have relative abundance not yet well understood. The photoevaporation rates depend on a number of parameters, as host star mass, initial disk mass and size, viscosity in the disk, EUV, FUV and X-ray luminosities, and the time-dependent spectrum from X to UV. All of those can vary significantly in young stars, resulting in photoevaporation rates that can vary widely depending on the system, and range from $10^{-11} - 10^{-7}$ $M_\odot$ yr$^{-1}$ for low-mass stars of 0.1 to 3 $M_\odot$. Overall, photoevaporation coupled to viscous evolution can lead to the dispersal of gas on timescales compatible with observations of a few million of years.

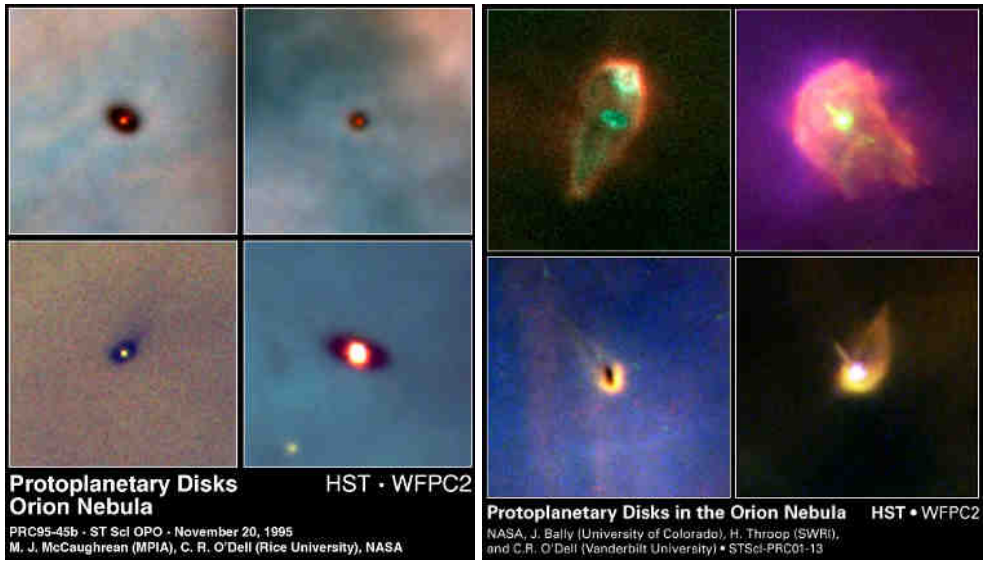

Figure 2.5: Images of protoplanetary disks in the Orion nebula cluster seen by the HST WFPC2 instrument. Left: Images of disks seen only in silhouettes, lying mainly in low-radiation environment (far from the massive stars). Right: Images of disks harbouring a ionised cocoon, laying in high-radiation environment (close to massive stars).

### 2.3.3 External photoevaporation - energetic radiation from nearby massive stars

The intensity of radiation fields generated by massive stars is orders of magnitude larger than those from low-mass stars since the photospheric outputs in terms of EUV and FUV photons are strongly related to the stellar mass (Diaz-Miller et al., 1998). Most stars, including low-mass stars, form in large clusters with hundreds to thousands of stars (Lada and Lada, 2003) including massive O-type stars (very hot and luminous stars, cf. HR diagram in App. A). The disks around low-mass stars, that belong to the cluster, may hence be irradiated by the photons originating from the massive neighbouring star(s).

The effect of external radiation on disks became clear when the *Hubble Space Telescope* observed a high contrast between the large Taurus disks in their low-radiation environment (Padgett et al., 1999) and the photoevaporating disks in the Orion nebula cluster (Bally et al., 2000) (see the two different kinds of objects observed in the Orion nebula cluster in Fig. 2.5 for example). The externally-illuminated photoevaporating disks, also known as "proplyds"[1] (O'dell et al., 1993), generally appear as disks surrounded by a teardrop-shaped ionised envelope with a bright head facing the main UV source (the massive stars) and an elongated tail pointing away. The distribution of disk bearing stars have been studied in massive star clusters where important photoevaporation is expected by comparison of the other star forming region (Fig. 2.6). Results tend to show that the disk frequency decreases more rapidly with time but also that it is lower in the close vicinity of massive stars.

---

[1] The term "ProPlyD" is the diminutive for "PROtoPLanetarY Disk". They were among the first disk imaged. Contrary to what was thought at the time of their discovery, they are not representative of the whole disk population. The name was yet kept for these specific objects.

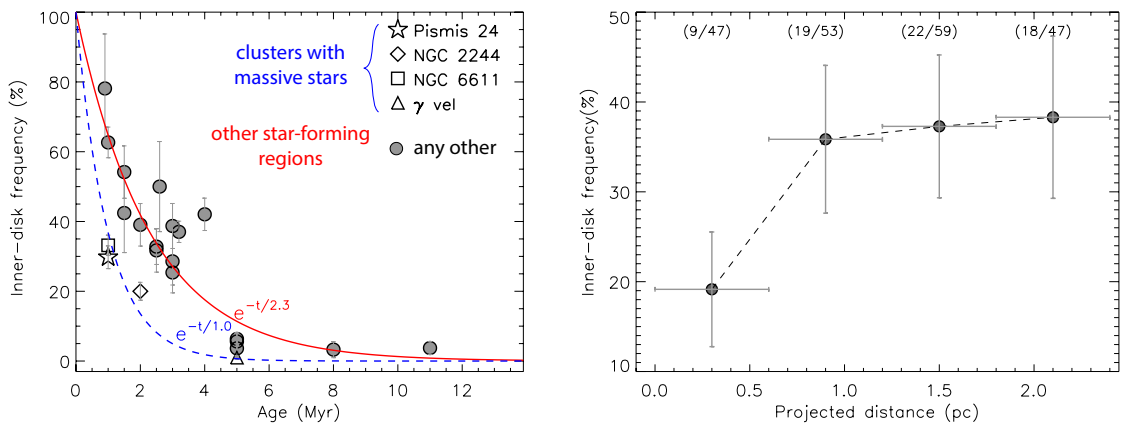

Figure 2.6: Left: The inner disk frequencies in different star formation regions as a function of age. The dashed line is the fit to inner disk frequencies of the four clusters including massive stars, Pismis 24, NGC 2244, NGC 6611, and γ vel, which gives a frequency $e^{-t/1.0}$, whereas the solid line is the fit to all other star formation regions, which gives $e^{-t/2.3}$, where $t$ is the age in Myr. Right: The inner disk frequency as a function of projected distance from Pismis 24-1, the most massive stellar system in the Pismis 24 cluster. Absolute number counts for each bin are given at the top of the panel. Figures adapted from Fang et al. (2012).

The population synthesis exercice of Fatuzzo and Adams (2008) have specified the distributions of radiation environments that forming planetary systems are expected to experience. For typical numbers of stars in a cluster, star forming environments in it are found to span a huge range of many orders of magnitude of radiation field strength, due to an incomplete sampling of the initial mass function (IMF). However, for large clusters (more than 1 or 2 thousands stars), as for the largest in the solar neighbourhood, namely the Orion nebula cluster, the IMF is sufficiently sampled to populate the part of the IMF where most of the UV radiation is emitted and large atypical values of radiation field are expected. For low to moderate radiation fields, photoevaporation may be predominantly in the radial direction from the outer edge of the disk in a process named "sub-critical" photoevaporation (Adams et al., 2004). For strong radiation fields, as in Orion, there is still some missing pieces left to understand what happens precisely and the impact on planet formation. Before discussing those points, I will describe how EUV and FUV photons coming from a nearby massive star may photoevaporate a disk.

The photoevaporating disks in the Orion nebula cluster show cometary morphologies with a bright head facing $\Theta^1$ Ori C, the main ionising star, and a tail pointing away from this star (O'dell et al., 1993). Morphological models assume that the hemisphere of the ionised flow is directly illuminated by the main ionising star whereas the far-side receives a diffuse EUV field derived from recombinations in the nebula (Henney and Arthur, 1998). In the model developed by Johnstone et al. (1998) (cf. Fig. 2.7), EUV radiation impinges on low density gas above the disk, launched by photoevaporation, and sets up an ionisation front in which the integrated number of recombinations per unit area matches the stars ionising flux. The ionisation front is a contact discontinuity where the mass and angular momentum are conserved. Since the gas velocity at the ionisation front is transonic ($v_{\rm II,IF} \approx c_{\rm S}$), this implies that the velocity in the neutral region is subsonic. If the velocity is subsonic throughout the neutral region to the disk surface, there is a causal contact between the ionisation front and the disk surface. Under this assumption, the mass-loss rate at the ionisation front drives the flow. This is the EUV-dominated case which occurs when the EUV mass-loss rate is the most important and defined in this model of Johnstone et al. (1998) by

$$\dot{M}_{\mathrm{EUV}} = 9.5 \times 10^{-9}\ f_r^{0.5}\ \Phi_{49}^{0.5}\ d_{17}^{-1}\ r_{\mathrm{d}14}^{3/2}\ M_\odot\ \mathrm{yr}^{-1}, \qquad (2.5)$$

which is derived by assuming a thin neutral layer, and where $r_{\mathrm{d}14}$ is the disk radius expressed in $10^{14}$ cm, $d$ the distance to the ionizing star expressed in $10^{17}$ cm, $\Phi_{49}$ is the EUV photons luminosity of the UV source expressed in $10^{49}$ s$^{-1}$ ($\Phi_{49} \approx 1$ for $\Theta^1$ Ori C), and $f_r$ the fraction of EUV photons absorbed by recombinations in the ionised portion of the flow. Oppositely to this case, if the neutral layer is larger so that a sonic point and a shock appear in it, the causal contact does not exist anymore and only the FUV photons drive the photoevaporation at the disk surface. This is the FUV-dominated case. In that case, the wind mass flux is set by the maximum density of the flow for which FUV heating is effective, and we have (Johnstone et al., 1998)

$$\dot{M}_{\mathrm{FUV}} = 1.3 \times 10^{-8} \left(\frac{N_{\mathrm{D}}}{10^{21}\ \mathrm{cm}^{-2}}\right)\left(\frac{v_0}{3\ \mathrm{km\ s}^{-1}}\right)\ r_{\mathrm{d}14}\ M_\odot\ \mathrm{yr}^{-1}, \qquad (2.6)$$

where $v_0$ is the velocity at the base of the flow, about 3 km s$^{-1}$ for a temperature of 1000 K, and $N_{\mathrm{D}}$ is the column density of the warm gas between the disk surface and the ionisation front, which is around $10^{21}$ cm$^{-2}$ but depends on the dust properties and the penetration of sufficient FUV photons (more details in Sect. 3.2.2).

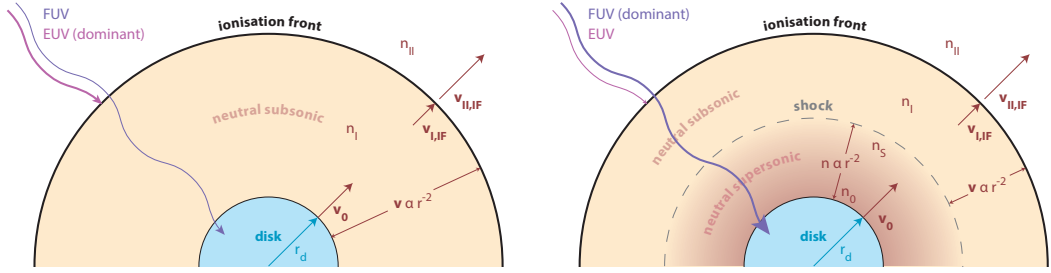

Figure 2.7: Schematic diagrams showing the distinct regions through which EUV-dominated (left panel) or FUV-dominated (right panel) photoevaporation flows pass according to the model of Johnstone et al. (1998).

The here above derived mass-loss rates defined different regions where one or the other regime dominates. Close to the ionizing star (if only one is considered), the loss of mass by EUV photons dominates. By moving farther away, the mass-loss rate by EUV decreases inversely proportional to the distance to the ionising star (see equation 2.5) while the FUV one is almost constant. At some point, the FUV-dominated case takes the lead. From Störzer and Hollenbach (1999), the FUV-dominated region extends from 0.3 – 0.01 pc away from $\Theta^1$ Ori C leading to disks with large neutral region surrounding them. From now, such objects (cf. right panel of the Fig. 2.5) will be the only ones called "proplyds". Farther away, the effect FUV depends on self-shielding by $H_2$ (Sect. 3.2.2) rather than dust absorption, and FUV mass-loss rate drops below the EUV one which becomes dominant again.

**proplyds and their "lifetime problem"**

The spatial distribution of the bulk of the proplyds in the Orion nebula cluster is globally consistent with the FUV-dominated zone of $\Theta^1$ Ori C described above. Some proplyds are seen farther and may be explained by other, more modest, massive stars $\Theta^1$ Ori A et B (Vicente and Alves, 2005), or by a combined effect of internal X-rays and external EUV photons (Clarke and Owen, 2015). The mass-loss rates first deduced from resolved radio free-free observations (Churchwell et al., 1987), and confirmed through esmission line modeling (Henney and O'Dell, 1999), were found consistent with the FUV-dominated case presented above. If the theory globally predicts the spatial distribution, it is also quite correct for the mass-loss rates.

Typical mass-loss rates by external FUV photoevaporation are about $10^{-7}$ $M_\odot$ yr$^{-1}$. If such a rate remains constant with time, a disk with a typical mass of $10^{-2}$ $M_\odot$ will survive only $10^5$ years. Proplyds are thus thought to have a very short lifetime while the majority of the expected disk in this region are currently still observable (between 55 and 90% according to Hillenbrand et al. 1998). This is known as the "proplyd lifetime problem". Many solutions have been proposed:

- a recent sortie out of the protecting core (Störzer and Hollenbach, 1999);

- a photoevaporation in a "subcritical" regime (Adams et al., 2004) supposed less destructive in the long term than the "supercritical" one, because escape flows develop radially at the outer edge instead of being launched from the larger disk surface;

- an unexpectedly high initial disk mass distribution;

- or a recent switch-on (optically revealed) time for $\Theta^1$ Ori C.

None of them is fully conclusive but the current idea is that the sub-critical case dominates in a recent switch-on epoch. Perhaps the main conclusive argument indicating that Orion is being observed at a special epoch is that it is unusual among massive clusters in not showing a deficit of disk bearing stars in the proximity of massive stars (Hillenbrand et al., 1998), while photoevaporation is obviously at play as a deficit of disk mass is clearly visible for the disks close to $\Theta^1$ Ori C (Fig. 2.8).

### 2.3.4 Effect of the photoevaporation on planet formation

Planets form by accreting the solid and gaseous material of the disk, on a timescale of a few million of years (Sect. 1.4). On such timescales, the planet formation may be affected by the dissipation of the disk by photoevaporation. One example is that photoevaporative flows lift preferentially the smallest particle while largest particles, representing the most part of the solid mass, are left behind (e.g. Owen et al., 2011). The initial gas-to-dust mass ratio of 100 thus tends to decrease with time within the remaining disk. Actually, this is visible at the end in the composition of the planets of the Solar System where the gas-to-solid mass ratio is only about 10. This preferential depletion of gas is also visible through statistical analyses of exoplanets: Super-Earths[2] are very frequent while there is a low frequency of gas giants (about 10% for solar-mass stars, see Winn and Fabrycky, 2015) compare to what is expected. This statistic suggest than the gas is considerably present at the time of formation of Super-Earths but is dissipated while forming the giants.

We saw that the presence of gas in the disk is important in many phases of the formation scenario: the early phase of planetesimal formation, the giant planet formation obviously, the effect of migration as well as in the final stage of terrestrial planet formation in case of "planet traps" (where type I migration tend to concentrate bodies in a stable location). The planetesimal formation is supposed to be sufficiently rapid to be unaffected by photoevaporation, at least not critically. If photoevaporation is low, the orbital evolution of the system may be influenced while the planet formation may be not. If it is high, it can possibly, as soon as the earliest stages, decrease the gas-to-dust mass ratio (Gorti et al., 2015) and so potentially trigger the planet formation more rapidly (Throop and Bally, 2005) or even influence the type of planet formed (rocky versus ice or gaseous giant).

---

[2]Super-Earths are exoplanets with a mass higher than Earth's, but below the masses the firsts ice giants, i.e. between 1 and 10 Earth masses approximately. Their compositions are expected to be various: mainly of hydrogen and helium ("mini-Neptune" for low-density super-Earths); water as a major constituent (ocean planets for intermediate density), or have a denser core enshrouded with an extended gaseous envelope (gas dwarf or sub-Neptune). A super-Earth of high density is believed to be rocky and/or metallic, like Earth and the other terrestrial planets of the Solar System.

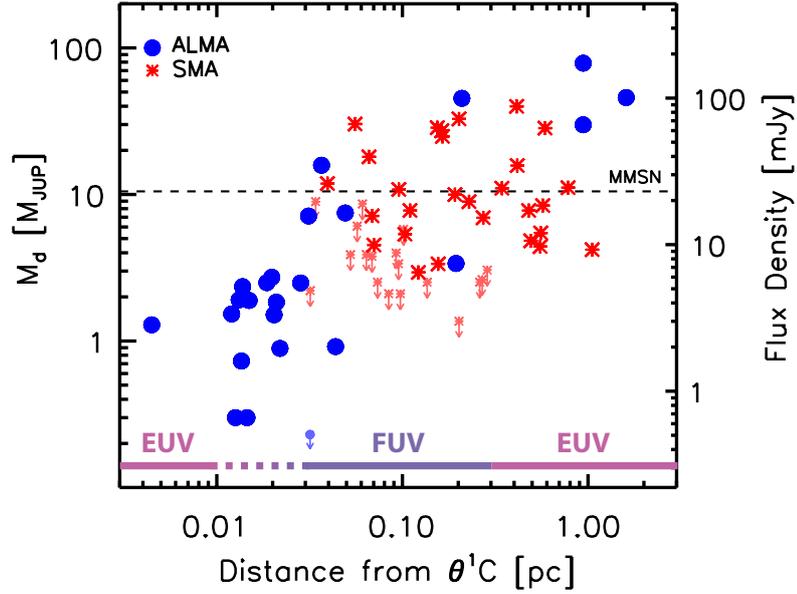

Figure 2.8: Mass of the Orion disks plotted against their projected distances from the massive O-star, $\Theta^1$ Ori C. Large blue dots represent ALMA detections, while red stars represent SMA detections at 880 μm for proplyds not yet observed with ALMA. A clearer dot or stars with an arrow represent the 3σ upper limits for the proplyds not detected in the observed fields. In total, 70 HST-identified proplyds surveyed with both the ALMA and the SMA are plotted here. The dashed line represents the MMSN value of 10 $M_{Jup}$. EUV and FUV-dominated zone are indicated by violet lines where the limit of the inner EUV-dominated region and the FUV-dominated region is at 0.01 pc according to (Störzer and Hollenbach, 1999) or 0.03 pc according to Mann et al. (2014). Figure adapted from Mann et al. (2014).

## 2.4 Objectives of this thesis and thesis plans

The photoevaporation sets the gas disk lifetime and evolution. Consequently, it impacts all stages of planet formation, from planetesimal growth to the formation of giant planets. If we qualitatively understand how disks evolve and disperse, photoevaporation rates are still not measured accurately. In the case of external photoevaporation leading to the formation of objects called "proplyds", there is some missing pieces in the theory with, for instance, the emergence of the "lifetime proplyd problem" and the uncertainty on the possibility to form planets in this kind of environment.

### 2.4.1 How to probe the photoevaporation in proplyds ?

EUV-heated gas is extremely hot and generally supersonic, so easily observable. Gas emission of FUV-heated regions are cooler and slower, so more difficult to detect. By comparison to EUV-dominated photoevaporation, the features expected in the emission line profiles (asymmetries and blue-shifts) for the FUV-dominated photoevaporation are small for detection with current instruments and hard to disentangle from other non-Keplerian sources like turbulence. While it is not easy to measure directly the motion of the flow, I can indirectly estimate the physical conditions at the base of the flow by modelling the emission.

The first part of my PhD thesis was to analyse data from the *Herschel Space observatory* obtained for a few proplyds (Sect. 4) in order to derive their physical conditions and investigate the main processes driving the photoevaporation in that case (Sect. 5). To do that, I used the known physics of photodissociation regions and used a detailed PDR code (Sect. 3).

### 2.4.2 How does a disk evolve under external FUV photoevaporation ?

Determining the decrease in gas mass with time and quantifying disk mass loss rates evolution are essential to develop a comprehensive theory of disk evolution that includes accretion, planet formation and disk dispersal. Based on the results of Sect. 5, I will develop, in Sect. 6 and 7, how ones can consider the viscous evolution of a disk under photoevaporation in a hydrodynamical code. This will permit me to discuss the future of Orion's proplyds, the "lifetime problem" and the possibility to form planets in them (Sect. 8).

# Part II

# Photodissociation regions of protoplanetary disks

# Table of Contents

## Chapter 3

# Photodissociation regions

## 3.1 Their place in the interstellar medium

The interstellar medium (ISM) is defined as the medium filling the space between the stars of our Galaxy. It can be seen as the matter in whatever the form, the cosmic rays, the magnetic and radiation fields. The bulk of the baryonic mass (i.e. common matter) of the Milky Way is mainly located into stars, and the mixture of gas and dust of the ISM is globally so tenuous that it represents only about 10% of the mass. However, those few percents play a crucial role in the cosmic cycle of matter since the ISM gives birth to stars from its densest parts, and recycles matter injected by evolved stars, generation after generation. Today, our local ISM is composed primarily of hydrogen (about 91% in number) followed by helium ($\sim$ 9%), with trace amounts of carbon, oxygen, and heavier elements. That mixture of gas represents 99% of the total mass, and the remaining 1% is located in the solid form, i.e. dust grains. This soup of matter is far from being homogeneous with multiple phases instead, distinguished by whether matter is ionised, atomic, or molecular, plus its temperature and density. The heterogeneity is mainly due to the interaction of the radiation field with matter that impacts on the structure, chemical composition and physical conditions of every phases. Photons that affect the ISM have different origins and altogether lead to the mean interstellar radiation field (ISRF) (cf. Fig. 3.1) :

**Cosmic Microwave Background (CMB)** Photons from the first light of the Universe are present everywhere in a same amount. With an emission corresponding to a black body with a low temperature of around 2.7 K, the CMB emission peaks in the microwave. These low energy photons do not highly interact with matter so that their effect on the ISM is generally negligible compare to other components.

**Dust** Dust grains absorb photons and reemit the energy through infrared radiation. The infrared emission bathing the ISM comes from a very scattered grain population, with galactic or extra-galactic dust. The spectrum is complex because it is a combination of different emission processes from different types of grains (detailed in Sect. 3.2.1).

**Stars (NIR to UV)** From near-IR to UV, the radiation field is dominated by the emission of stars, each one globally corresponding to a black body at a given temperature.

**Stars (FUV and EUV)** The far-UV bump (around 0.15 µm) is due to the emission from the most massive stars. The important drop in the extreme-UV ($\lesssim$ 0.1 µm) corresponds to H-ionising photons that are absorbed by abundant atomic hydrogen, and thus only propagate in ionised region. Contrary to the other components, this one varies significantly from one region to another according to the more or less proximity of a cluster of massive stars.

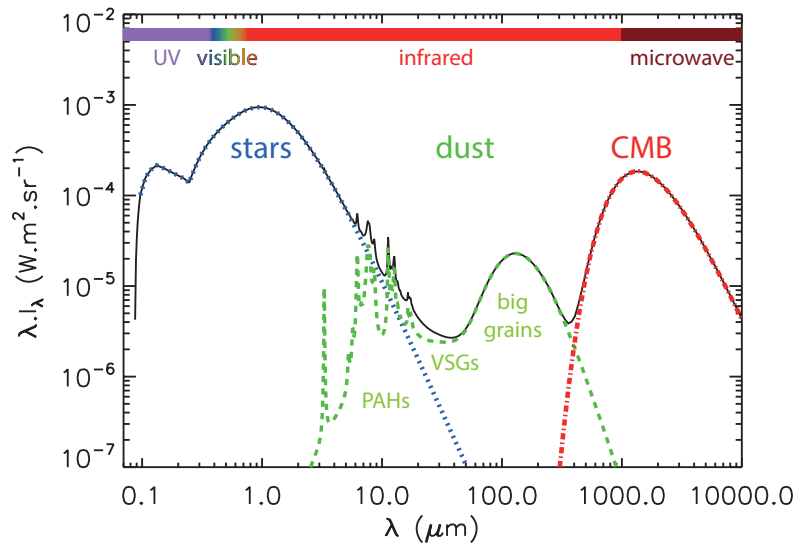

Figure 3.1: Interstellar radiation field (black solid line) with its three main component: the star emission (blue dotted line), dust emission (green dashed line) and the cosmic microwave background (red dash-dotted line). The name of the three types of dust grains are indicated at the wavelength they dominate the emission. Figure inspired by E. Arab based on the model of Mathis et al. (1983).

Regions where the physics and the chemistry of the dust and gas is dominated by the far-UV photons (6 eV < h$\nu$ <13.6 eV or 912 Å< $\lambda$ 2000 Å), arising from nearby massive stars are called photodissociation regions (PDRs). The presence of such regions in the ISM was first observed through bright emissions of atomic oxygen and ionised carbon towards massive star forming regions (e.g. Melnick et al., 1979; Russell et al., 1980). This presence of a neutral layer of gas, between the ionised region and the molecular cloud, was rightly thought to be caused by the radiation field. The first models describing PDRs, i.e. by studying the effect of FUV radiation on matter, were then developed to explain, understand and analyse those observations (Tielens and Hollenbach, 1985b,a). The study of PDRs is now an important field in astronomy since those regions are found everywhere in the ISM. They represent the vast majority of the interstellar matter since all of the atomic gas and most of the molecular gas of the ISM is founds in PDRs. This includes many types of environments such as

- Interstellar clouds as diffuse (neutral) and translucent clouds or reflection nebulae,
- Circumstellar envelopes around evolved stars (e.g. red giants or AGB),
- Neutral envelope and atmospheres of planetary nebulae (glowing shell of gas ejected from old red giant stars late in their lives),
- Surface of molecular clouds (see e.g. Fig. 3.2) and protoplanetary disks,
- and extragalactic-ISM of starburst galaxies (ones with a high star formation rate) and clouds around active galactic nucleus (likely related to a central supermassive black hole).

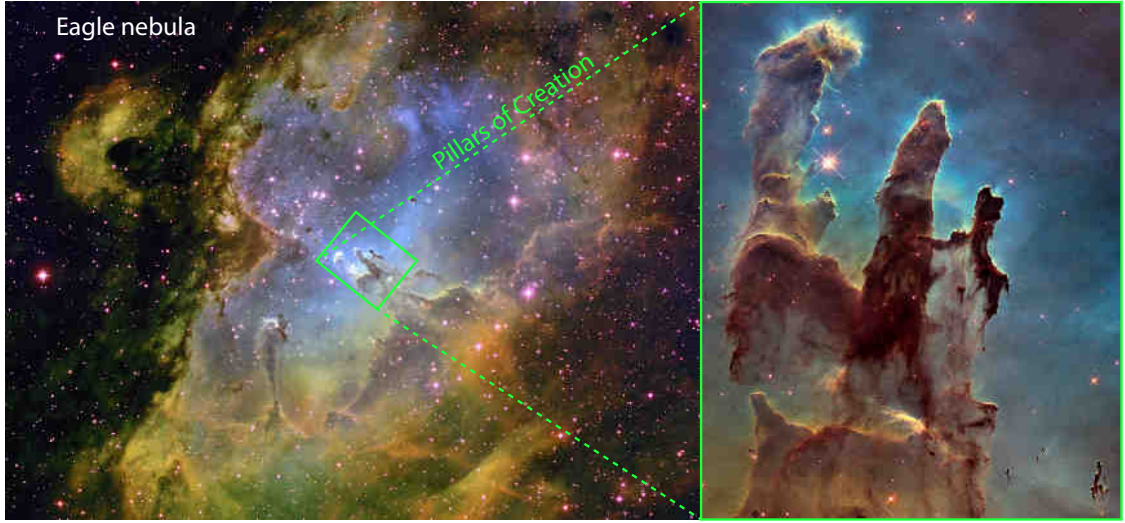

Figure 3.2: Eagle Nebula (M16) and the Pillars of Creation. Left: visible image of the nebula (credit: T. A. Rector & B. A. Wolpa, NOAO, AURA). Right: zoom on the Pillars of Creation with HST (credit: NASA / ESA / Hubble / Hubble Heritage Team).

The critical factor driving the physical structure and composition of the PDR is the UV radiation field. The density structure of the medium and the intensity of the incident UV radiation field will determine how far inside the cloud the UV photons will penetrate. As long as they penetrate sufficiently (defining the extension of the PDR), those photons will regulate the spatial structure, the chemistry and the thermal balance. All of those aspects are interconnected. In the next sections, keeping that interconnection in mind, I will describe them independently starting by the penetration of UV photons and the corresponding spatial structure (Sect. 3.2). I will then describe the chemistry of PDR (Sect. 3.3), the main processes involved in the energy balance (Sect. 3.4) and the resulting temperature profile (Sect. 3.5). Finally, I will present one code that includes that physics and chemistry, the Meudon PDR code (Sect. 3.6).

## 3.2 Spatial structure

### 3.2.1 Far-UV radiation field and penetration of photons

**The local radiation field**

It is common to express the local FUV radiation field according to a reference one, i.e. the interstellar radiation field. Different prescriptions may be used to model it:

**Habing** The Habing's radiation field is the classical reference. Habing (1968) provided estimations of the ISRF for only three wavelengths: 1000, 1400 and 2200 Å. To complete the FUV spectrum, one commonly uses the fit presented in Draine and Bertoldi (1996) where

$$\lambda u_\lambda = \left[ -\frac{25}{6} \left( \frac{\lambda}{1000} \right)^3 + \frac{25}{2} \left( \frac{\lambda}{1000} \right)^2 - \frac{13}{3} \frac{\lambda}{1000} \right] \times 10^{-14} \text{ ergs cm}^{-3}, \quad (3.1)$$

if $\lambda$ is given in Angstroms.

**Draine** Different expressions of Draine's ISRF can be found in the literature. One of them is the expression from Sternberg and Dalgarno (1995) where

$$I_\lambda = \frac{1}{4\pi}\left[\frac{6.3600\ 10^7}{\lambda^4} - \frac{1.0237\ 10^{11}}{\lambda^5} + \frac{4.0812\ 10^{13}}{\lambda^6}\right], \quad \text{for}\lambda \leq 2000\ \text{Å}, \quad (3.2)$$

in ergs cm$^{-2}$ s$^{-1}$ Å$^{-1}$ for $\lambda$ in Å.

**Mathis** The expression of the far-UV radiation field based on Mathis et al. (1983) is

$$I_\lambda = \left[\tanh\left(4.07\ 10^{-3} \times \lambda - 4.5991\right) + 1.0\right] \times \frac{107.182}{\lambda^{2.89}}, \quad \text{for}\lambda \leq 8000\ \text{Å}, \quad (3.3)$$

in ergs cm$^{-2}$ s$^{-1}$ Å$^{-1}$ for $\lambda$ in Å.

The intensity of the local incident FUV radiation field is usually normalised to the value of the integrated ISRF, from one of those, on the far-UV range, i.e. between 912 and 2400 Å or 6 to 13.6 eV. Using those prescriptions and that range, the integrated Habing field is 1.86 10$^{-6}$ W m$^{-2}$. The standard value for the Draine's field is 3.03 10$^{-6}$ W m$^{-2}$, so 1.63 times more than Habing, and the standard value for the field of Mathis is 1.91 10$^{-6}$ W m$^{-2}$, so 1.03 times more[1]. In the following, I will express the FUV flux strength relatively to the Habing field and the multiplicative factor will be called $G_0$.

In PDRs, the far-UV flux is generally a few orders of magnitude above the ISRF. This incident flux may be estimated directly by measuring the related dust infrared emission (reemission following heating), or the observed intensity of H$_2$ ro-vibrational lines. It can also be indirectly estimated according to the spectral type of the massive source star(s) and the distance from the PDR. The incident flux of extreme-UV photons (h$\nu > 13.6$ eV) is generally of lesser importance, since EUV photons do not penetrate significantly in an illuminated cloud of non-ionised mater as they are absorbed rapidly by atomic hydrogen. Far-UV photons are able to go much deeper and are thus the drivers of the PDR. Their penetration is mainly regulated by dust absorption, but dust scattering and absorption lines from the most abundant species, i.e. H$_2$ and CO for the molecules (we will see in Sect. 3.2.2 a such effect), can also be important. I will now focus on the dust dominant absorption.

**The dust of the ISM**

Absorption line spectroscopic studies showed that some elements, such as C, Mg, Si and Fe are under-abundant in the gas-phase of the local ISM, when assuming that the overall composition should be similar to the elemental abundance of the Sun. These observations indicate that those elements could form the bulk composition of the interstellar dust grains with two main families: carbonaceous or silicate grains. Observations also indicate that the dust grains population is dispersed on a large size distribution. To understand and model their emission (see Fig. 3.1), one has to consider different types of dust grains. For instance, Desert et al. (1990) have divided the dust population in three components:

**Polycyclic Aromatic Hydrocarbons (PAHs)** They are a family of planar organic molecules, containing only carbon and hydrogen, i.e. hydrocarbons, organised in multiple aromatic rings. They represent the smallest grains with a typical sub-micron size, corresponding to a few tens or hundreds of carbon atoms.

---
[1] Because of a large variety in the definition of the ISRF, many values of the integrated fields from Mathis, Draine and Habing, or the ratio between them may be found in the literature. Here, I give the ones expressed in the documentation of the Meudon PDR code which i used.

**Very small grains (VSGs)** They are dust grains with a size from 1 nm up to about 15 nm.

**Big grains (BGs)** Dust grains with a size between 15 and 110 nm.

Because of their small size, PAHs and VSGs have a small heat capacity so they may heat up or cool down quickly without reaching an equilibrium temperature. They absorb only a few percents (5% at most) of the far-UV radiation and the cooling reemission make them respectively responsible for the spectral features and continuum in the mid-IR (cf. Fig. 3.1). Oppositely, big grains are responsible for the rest of the dust absorption and are in equilibrium with the radiation field so that, in PDRs, they dominate the emission in the far-IR and submillimetre ranges (in the Sect. 3.4.1, I will discuss the impact of the absorption on the heating).

**Extinction caused by dust**

To discuss their effect on the radiation field, let us define a few notions. First of all, extinction includes absorption and scattering that both can lower the photon flux in a given direction. The optical depth $\tau_\lambda$, for a particular wavelength $\lambda$, measures the extinction and is defined as

$$\tau_\lambda = -\ln\left(\frac{I_\lambda}{I_{\lambda,0}}\right), \tag{3.4}$$

where $I_{\lambda,0}$ is the incident flux (before extinction) and $I_\lambda$ the transmitted flux after crossing the dust cloud, i.e. after extinction. In a similar way[2], the magnitude of the extinction $A_\lambda$ is defined as

$$A_\lambda = -2.5\log\left(\frac{I_\lambda}{I_{\lambda,0}}\right), \tag{3.5}$$

so that each decrease of one order of magnitude in flux is equal to an increase of 2.5 in magnitude (or 1 magnitude point correspond to a factor 2.5 in flux). The apparent magnitude $m_\lambda$ of a star is given by

$$m_\lambda = M_\lambda + A_\lambda + 5\log(d), \tag{3.6}$$

where $M_\lambda$ is the absolute magnitude (only depending on the source star and calibrated from a chosen reference star), $A_\lambda$ the magnitude of extinction because of dust, and the term $5\log(d)$ corresponding to the dilution of flux according to the distance $d$ from the star.

The extinction of light by a grain depends on its composition and its size. Thus, it is hard to compute the extinction curve (the capacity to extinct the light according to the wavelength) without knowing precisely the characteristics of the dust population. One possibility to overcome that is too directly determine the extinction curve by measuring the spectrum of two similar stars, so with same expected absolute magnitude $M_\lambda$, with one subject to extinction on its line of sight while not the other one. The difference in apparent magnitude is thus

$$\Delta m_\lambda = 5\log\left(\frac{d_1}{d_2}\right) + A_\lambda, \tag{3.7}$$

with $d_1$ and $d_2$ the distances (probably unknown) separating us from the two stars. The extinction globally increases with decreasing wavelength. This causes the longest wavelengths, red for visible, to be favoured in an effect called a "reddening". This reddening can be described by a color excess, $E_{\lambda_1-\lambda_2}$, which is defined as the difference of magnitude for two wavelengths $\lambda_1$ and $\lambda_2$ as

$$E_{\lambda_1-\lambda_2} = \Delta m_{\lambda_1} - \Delta m_{\lambda_2} = A_{\lambda_1} - A_{\lambda_2}, \tag{3.8}$$

---

[2] According to the definition of $\tau_\lambda$ and $A_\lambda$, they are almost equal with $A_\lambda = 2.5\log(e)\,\tau_\lambda \simeq 1.086\,\tau_\lambda$

thus leaving out the dependency on distances. Color differences between different stars can readily be compared after normalisation on a common color difference. Often, one compares any wavelength $\lambda$ with the green band V at 540 nm, normalised by the standard color excess taken between the blue band B at 442 nm and the band V, named $E_{B-V}$. Then, we have

$$\frac{E_{\lambda-V}}{E_{B-V}} = \frac{A_\lambda - A_V}{A_B - A_V}, \tag{3.9}$$

where $A_V$ is commonly named the visual extinction. Building the extinction curve consist of measuring this ratio for a large range of wavelengths. If we define the constant term $R_V = A_V/(A_B - A_V)$, this ratio becomes

$$\frac{E_{\lambda-V}}{E_{B-V}} = \frac{A_\lambda}{A_B - A_V} - R_V = R_V \left(\frac{A_\lambda}{A_V} - 1\right), \tag{3.10}$$

where the extinction ratio $A_\lambda/A_V$ can then be expressed in terms of the color excess as

$$\frac{A_\lambda}{A_V} = \frac{1}{R_V}\frac{E_{\lambda-V}}{E_{B-V}} + 1, \tag{3.11}$$

which is another way to represent the extinction curve. The parameter $R_V$, defined as the total-to-selective extinction ratio, represents thus the steepness of the extinction curve.

Numerous observations of the interstellar extinction (e.g. Cardelli et al., 1989; Fitzpatrick and Massa, 2007) have shown that the extinction curve, from the mid-IR through the visible to near- and far-UV, can be characterised by one free parameter, which is then chosen to be $R_V$ (cf. Fig. 3.3). If extinction curves vary from one region to another, and can be separated in four distinct parts (a power-law in the infrared, followed by a less steep optical knee, an "UV" bump centered around 4.6 µm$^{-1}$ and a steep rise to short wavelengths), they can almost all be parametrized by this parameter in the plotted range of interest. The value of $R_V$ depends on the environment traversed by the line of sight and is known to be correlated with the average size of the dust grains causing the extinction. The diffuse ISM of our galaxy is characterized by a value of 3.1 while dense molecular clouds may have higher values up to 5.5.

Since the amount of dust is also related to the total amount of matter, the extinction is related to the neutral hydrogen atoms column density $N_H$ (in cm$^{-2}$). From a large survey, Bohlin et al. (1978) measured the relation between the standard color excess and the column density which was found to be

$$\frac{N_H}{E_{B-V}} = 5.8\ 10^{21}\ \text{atoms cm}^{-2}\ \text{mag}^{-1}. \tag{3.12}$$

In the Milky Way, numerous studies using multiple independent methods (e.g. Reina and Tarenghi, 1973; Gorenstein, 1975; Predehl and Schmitt, 1995; Guver and Ozel, 2009) have found a linear relation between the visual extinction $A_V$ and the column density $N_H$ about

$$\frac{N_H}{A_V} \simeq 2\ 10^{21}\ \text{atoms cm}^{-2}\ \text{mag}^{-1}, \tag{3.13}$$

with single value generally ranging from 1.8 to 2.2 $10^{21}$ atoms cm$^{-2}$ mag$^{-1}$, from one study to another.

### 3.2.2 A layered PDR

**The general pattern**

Since the UV radiation field decreases when crossing matters, the structure of a PDR is stratified in the direction of the incoming UV photons (Fig. 3.4):

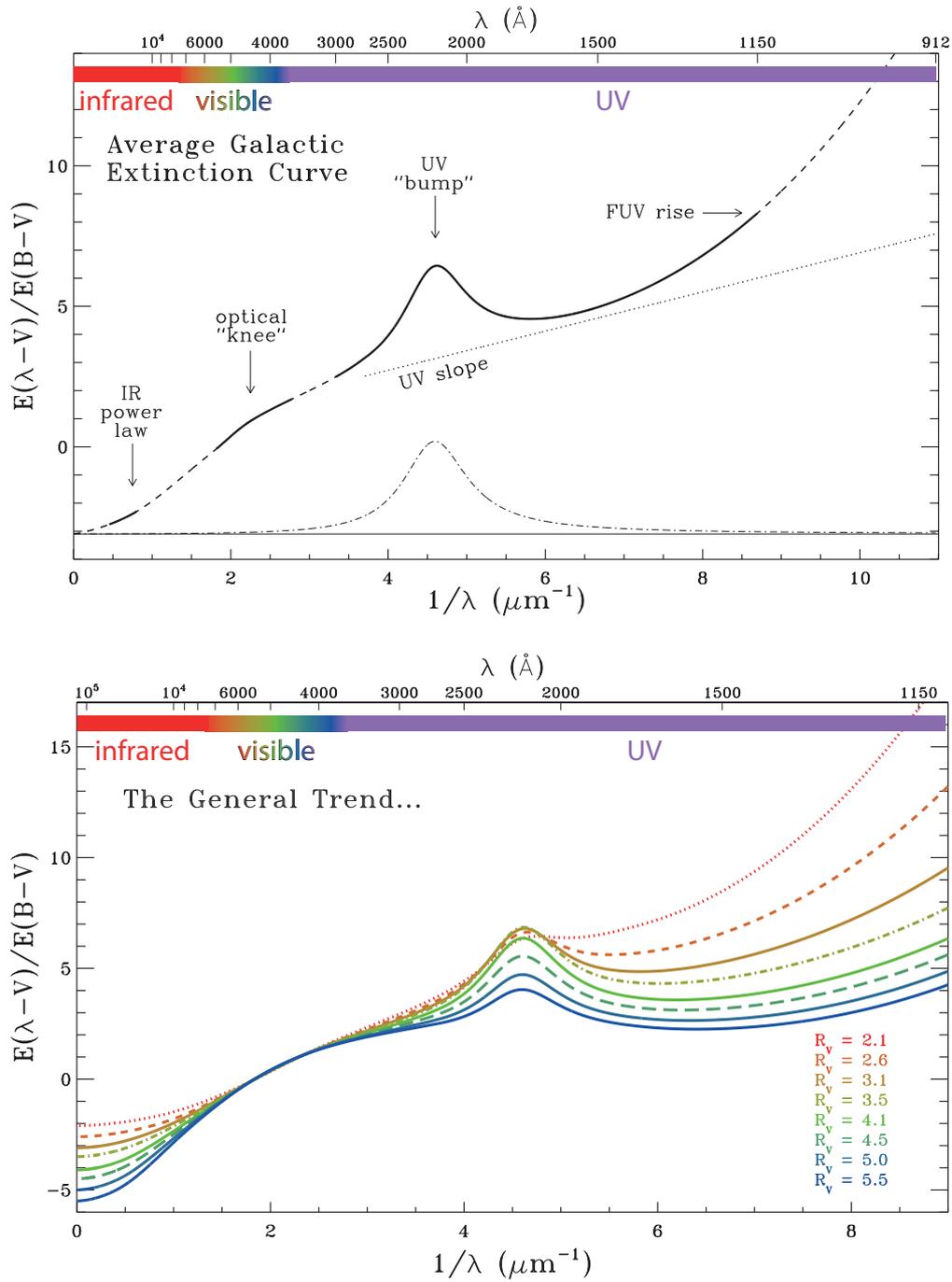

Figure 3.3: Top: average Milky Way extinction curve, corresponding to the case $R_V = 3.1$, as computed by Fitzpatrick (1999). Bottom: a set of normalised extinction curves, with $R_V$ ranging from 2.1 to 5.5, that are able to reproduce the numerous galactic extinction curves. Figures adapted from Fitzpatrick (2004).

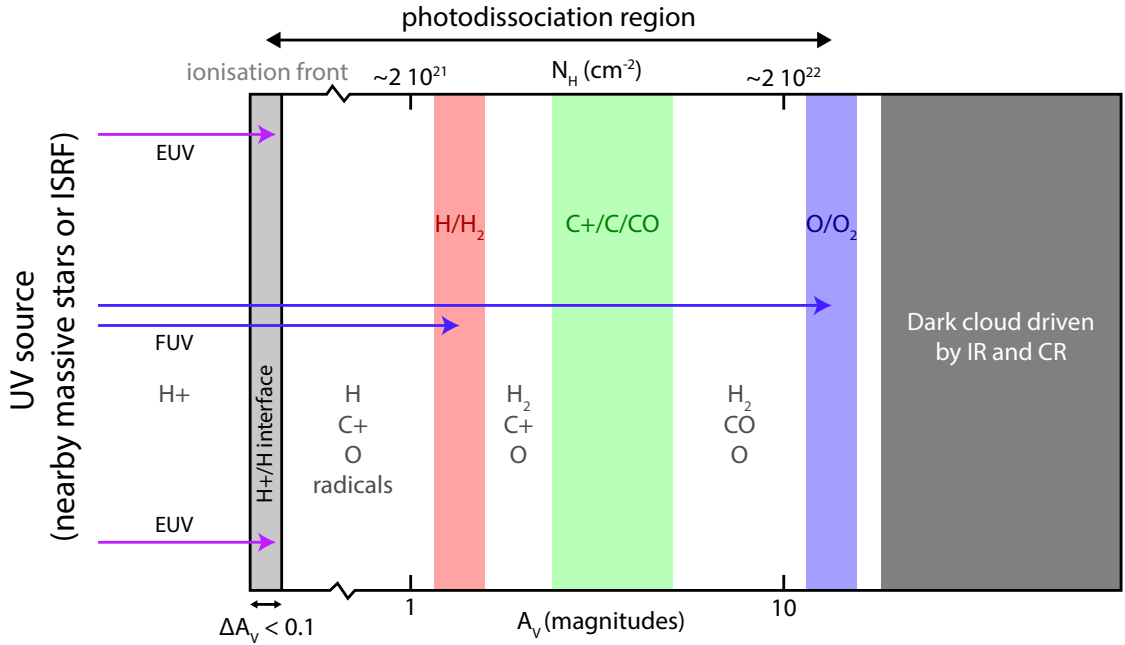

Figure 3.4: Schematic diagram of the structure of a PDR. Figure based on the model of Hollenbach and Tielens (1999).

- EUV photons, i.e. H-ionizing photons, are absorbed in a thin layer ($N_\text{H} \approx 1\ 10^{19}$ cm$^{-2}$ or $A_\text{V} \approx 1\ 10^{-2}$) in which the ionisation structure goes from almost fully ionised in the HII region to almost fully neutral. The limit is called the ionisation front.

- FUV photons penetrate deeper in the cloud, dissociate molecular hydrogen and ionise some species with an ionisation potential below 13.6 eV as carbon forming an HI/CII region.

- Once the flux of FUV photons is substantially attenuated, the composition is dominated by $H_2$. When the attenuation is driven by dust, which is generally the case in dense PDRs, this $H/H_2$ transition takes place for $A_\text{V} \approx 1 – 2$ (depending on $G_0/n$).

- Somewhat deeper in the cloud ($A_\text{V} \approx 2 – 4$, depending on $G_0/n$), the carbon-ionizing flux has dropped sufficiently and ionised carbon recombines to form carbon monoxide.

**The $H/H_2$ transition**

The molecular hydrogen becomes the most abundant species instead of atomic hydrogen when the strength of the photon flux has significantly been attenuated around a visual extinction $A_\text{V}$ about $1 – 2$, for bright UV sources. This is generally true in PDRs if the dust dominates the extinction in the UV but, for some conditions, $H_2$ may protect itself if the first layers absorb all the dissociating photons, so that the transition consequently occurs sooner. When this occurs, the location of the transition between atomic and molecular hydrogen takes place when the photodissociation rate of $H_2$ is equal to the rate of $H_2$ formation. I will now develop rapidly that idea to investigate its location.

The photodissociation rate per volume, $\dot{R}_{\mathrm{pd}}$ in s$^{-1}$ cm$^{-3}$, is

$$\dot{R}_{\mathrm{pd}} = k_{\mathrm{UV}}(N_{\mathrm{H}_2}) \times n_{\mathrm{H}_2} = k_{\mathrm{UV}}(0)\, e^{-\tau_{\mathrm{d}}} \beta_{\mathrm{SS}}(N_{\mathrm{H}_2}) \times n_{\mathrm{H}_2}, \tag{3.14}$$

where $k_{\mathrm{UV}}(N_{\mathrm{H}_2}) = k_{\mathrm{UV}}(0)\, e^{-\tau_{\mathrm{d}}} \beta_{\mathrm{SS}}(N_{\mathrm{H}_2})$ gives the rate of dissociating photons at a given depth, depending on the initial rate $k_{\mathrm{UV}}(0)$ at the ionisation front, then reduced by the optical depth of the dust $\tau_{\mathrm{d}}$ and by the self-shielding of H$_2$. The self-shielding factor $\beta_{\mathrm{SS}}(N_{\mathrm{H}_2})$ is related to the crossed column density of molecular hydrogen $N_{\mathrm{H}_2}$. Draine and Bertoldi (1996) showed, with a model of a semi-infinite static slab of gas irradiated on one surface that, for low-temperatures (a few $10^2$ K) and in an optically-thin case, this factor can be well approximated by a simple power-law that depends only on the H$_2$ column density given by

$$\beta_{\mathrm{SS}}(N_{\mathrm{H}_2}) = \left(\frac{N_{\mathrm{H}_2}}{N_0}\right)^{-0.75}, \tag{3.15}$$

in the column density range $N_0 = 10^{14} \lesssim N_{\mathrm{H}_2} \lesssim 10^{21}$ cm$^{-2}$. Below this range, the column density is so low that the self-shielding is negligible with $\beta_{\mathrm{SS}} = 1$ while, above this range, the dust extinction dominates. These authors also provide a more detailed function taking into account a temperature dependence due to thermal broadening of the lines to improve their fits on observations. I will keep the simplest form for our development.

The H$_2$ formation rate per volume on grains is

$$\dot{R}_{\mathrm{f}} = k_{\mathrm{f}} \times n \times n_{\mathrm{H}}, \tag{3.16}$$

where $k_{\mathrm{f}}$ is the specific temperature-dependent molecule formation rate coefficient expressed in cm$^3$ s$^{-1}$ (e.g. Black and van Dishoeck, 1987; Draine and Bertoldi, 1996), $n_{\mathrm{H}}$ is H-atom density in the gas and n is the total H density (including the atom and molecular form, i.e. $n = n_{\mathrm{H}} + 2n_{\mathrm{H}_2}$).

Let us now assume than the dust extinction is negligible, which is valid only if $N_{\mathrm{H}} \lesssim 10^{21}$ cm$^{-2}$ and then $e^{-\tau_{\mathrm{d}}} \simeq 1$. The transition from atomic to molecular hydrogen occurs when the photodissociation and formation rates are equal, so that

$$\begin{aligned} k_{\mathrm{UV}}(N_{\mathrm{H}_2}) \times n_{\mathrm{H}_2} &= k_{\mathrm{f}} \times n \times n_{\mathrm{H}}, \\ k_{\mathrm{UV}}(0)\left(\frac{N_0}{N_{\mathrm{H}_2}}\right)^{0.75} \times n_{\mathrm{H}_2} &= k_{\mathrm{f}} \times n \times (n - 2n_{\mathrm{H}_2}), \\ \left(\frac{k_{\mathrm{UV}}(0)}{k_{\mathrm{f}}\, n}\left(\frac{N_0}{N_{\mathrm{H}_2}}\right)^{0.75} + 2\right) n_{\mathrm{H}_2} &= n. \end{aligned} \tag{3.17}$$

Noticing that $N = \int_z n\, dz$ along the propagation direction $z$, one can rewrite the previous equation as

$$\begin{aligned} \left(\frac{4\, k_{\mathrm{UV}}(0)}{k_{\mathrm{f}}\, n}\left(\frac{N_0}{N_{\mathrm{H}_2}}\right)^{0.75} + 2\right) N_{\mathrm{H}_2} &= N, \\ \left(\frac{4\, k_{\mathrm{UV}}(0)}{k_{\mathrm{f}}\, n} N_0^{0.75} N_{\mathrm{H}_2}^{0.25} + 2 N_{\mathrm{H}_2}\right) &= N. \end{aligned} \tag{3.18}$$

Ignoring the second term on the left-hand side, we can extract the column density of molecular hydrogen

$$N_{\mathrm{H}_2} \simeq \left(\frac{k_{\mathrm{f}}\, n}{4\, k_{\mathrm{UV}}(0)}\right)^4 \left(\frac{N}{N_0}\right)^4 N_0, \tag{3.19}$$

which, in terms of abundance, is

$$X_{H_2} \simeq \frac{dN_{H_2}}{dN} = \frac{4N_{H_2}}{N} \simeq 4 \left(\frac{k_f\, n}{4\, k_{UV}(0)}\right)^4 \left(\frac{N}{N_0}\right)^3. \tag{3.20}$$

Equating the H/H$_2$ interface, or dissociation front, where the gas is half molecular and half atomic, i.e. with $X_{H_2} = 1/4$, we obtain the needed hydrogen nucleus column density, $N_{DF}$

$$N_{DF} \simeq N\left(X_{H_2} = 1/4\right) = \left[\frac{1}{16}\left(\frac{k_f\, n}{4\, k_{UV}(0)}\right)^{-4}\right]^{1/3} N_0. \tag{3.21}$$

Because (Tielens, 2010)

$$\frac{k_f\, n}{4\, k_{UV}(0)} \approx 1.9\, 10^{-7} \frac{G_0}{n}, \tag{3.22}$$

we have

$$N_{DF} \simeq 3.63\, 10^{22} \left(\frac{n}{G_0}\right)^{4/3} \text{cm}^{-2}, \tag{3.23}$$

so that the structure of the H/H$_2$ transition zone is regulated by the ratio $G_0/n$, where $G_0$ express the strength of the FUV radiation field. This development, and so the estimated location of the transition, is valid only if the self-shielding of H$_2$ dominates the extinction of the photodissociating photons. Since, dust opacity becomes important for a column density about $5\, 10^{20}$ cm$^{-2}$, the self-shielding dominates when the transition occurs before that value, i.e. if

$$\begin{aligned} N_{DF} \simeq 3.63\, 10^{22} \left(\frac{n}{G_0}\right)^{4/3} \text{cm}^{-2} &\lesssim 5\, 10^{20} \text{ cm}^{-2}, \\ \frac{G_0}{n} &\lesssim 4\, 10^{-2} \text{ cm}^3. \end{aligned} \tag{3.24}$$

When this condition is fulfilled, corresponding to a low radiation field or a high density, the self-shielding alone dominates the location of the transition. It can work for dense molecular clouds or diffuse clouds exposed to the relatively low radiation field of the ISRF, but only for dense clumps in PDRs when the radiation field is strong. In such cases, a sharp abundance evolution where $X_{H_2} \propto N^3$ is observed. Oppositely, PDRs associated with bright FUV sources typically have a ratio $G_0/n \simeq 1$ cm$^3$, so that the transition location is dominated by dust absorption and typically occurs at $A_V \simeq 2$ ($N \simeq 4\, 10^{21}$ cm$^{-2}$). At that point, the photodissociation rate of H$_2$ has been reduced sufficiently that an important column of H$_2$ can build up. As a consequence, H$_2$ self-shielding takes over, and the transition will still be very sharp, but deeper.

**The C$^+$/C/CO transition**

Elements with an ionisation potential less than 13.6 eV can still be ionised by FUV photons. These include C, S, and Si. Carbon is the most important one according to its abundance. The transition zone is however different from the hydrogen case: since carbon is less abundant than hydrogen, a higher column density of matter is needed to absorb the FUV ionizing-photons and, while existing, the self-shielding by the corresponding molecule, CO, is not as efficient as in the case of hydrogen. As a consequence, the carbon transition zone from ionised to neutral and molecular occurs later than the H/H$_2$ transition region (Fig. 3.4) and in a smoother way. Transitions zone for other trace species (as oxygen) may takes place even deeper in the cloud.

Table 3.1: Main two-body gas-phase reactions in PDRs.

| Type of reaction | Chemical equation |
|---|---|
| Photodissociation | $AB + h\nu \rightarrow A + B$ |
| Radiative association | $A + B \rightarrow AB + h\nu$ |
| Ionisation | $A + h\nu \rightarrow A^+ + e^-$ |
| Neutral-neutral | $A^+ + B \rightarrow C^+ + D$ |
| Ion-molecule | $AB + h\nu \rightarrow A + B$ |
| Charge-transfer | $A^+ + B \rightarrow A + B^+$ |
| Dissociative recombination | $AB^+ + e^- \rightarrow A + B$ |

## 3.3 Chemistry

### 3.3.1 Gas-phase chemistry

As we have seen previously, PDRs are spatially structured in layers because of the penetration of FUV photons. Consequently, a number of zones can be distinguished and each one is characterized by the chemical reactions that control its molecular composition. Because of penetrating FUV photons, the warm surface layer of PDRs consist largely of H, C+, and O. In this surface layer, the abundance of stable molecules is very limited by the significant flux of photodissociating and ionising photons so that the chemistry is dominated by radicals and their reactions. For this layer, the chemistry is very similar to that of diffuse clouds. However, if the gas is dense, various reactions with molecular hydrogen, which have small activation energies (e.g. with C+, O and OH), can proceed at an appreciable rate. As the molecular hydrogen abundance increases, the chemistry takes off (see e.g. Tielens, 2010). Deep in the cloud, when no more FUV photons can penetrate, the chemical composition will then resemble to the one of dark clouds, with ion-molecule chemistry initiated by cosmic-ray ionisation of $H_2$.

Due to generally low densities in the ISM, as well as in PDRs, the gas-phase chemistry is mainly driven by two-body reactions. The main types of two-body reactions involved in PDRs are presented in Table 3.1. For more details on the chemistry in PDRs, the reader is referred to the paper of Sternberg and Dalgarno (1995).

### 3.3.2 Grain surface chemistry

While dust grains represents only about one percent of the mass in the ISM, they are very important for the physics (for FUV extinction, see Sect. 3.2.1, and energy balance, see Sect. 3.4), but also for the chemistry. Actually, atoms and molecules may be absorbed onto their surfaces, move and encounter other atoms or molecules. They may react altogether to form more complex species and eventually be desorbed back into the gas phase. There exist two types of interactions between the atoms and the surface of dust grains:

**Physisorption** Also called physical adsorption, this is a process involving the fundamental but weak Van der Waals force (binding energies about 0.01 eV). Gas-phase species can easily enter physisorbed sites onto a grain and migrate on its surface even at low temperatures. Physisorbed species are not locked and, if the energy is high enough to overcome the barrier against chemisorption, they can migrate to a chemisorbed site or be evaporated at high temperatures.

**Chemisorption** It is the second kind of adsorption which involves a chemical reaction between

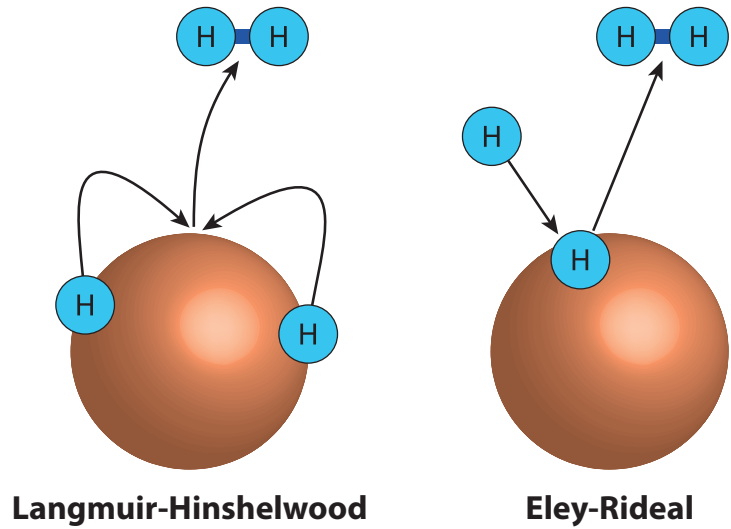

Figure 3.5: Illustration of the two mechanisms to form molecules on the surface of dust grains with the case of the molecular hydrogen.

> the surface and the adsorbate with new strong chemical bonds (about 1 eV). To enter a chemisorbed site, a gas-phase atom has to overcome a barrier that depends on temperature. Once adsorbed, atoms can migrate from one site to another by tunneling effects and thermal hopping. Contrary to physisorbed atoms, chemisorbed atoms are strongly bounded to the grain and, even at high temperatures, may remain onto the surface without being evaporated.

Once atoms are bound with the grain, molecules can be formed on its surface through two mechanisms (Fig. 3.5):

**Langmuir-Hinshelwood mechanism** It involves two physisorbed species (i.e. bound via Van des Waals interactions) that migrate on the surface of a grain and react.

**Eley-Rideal mechanism** It is the reaction of one chemisorbed species (i.e. bound via strong chemical bonds) and an other one coming directly from the gas phase without adsorbing.

## 3.4 Energy balance

Now that we have an idea of the spatial structure and the composition of a PDR, let us take a look at how the energy is absorbed, transformed or released, to understand then the resulting thermal structure. In the energy balance, we are interested in how energy can be transferred to the main component, i.e. the gas, with some processes which also may give energy to the dust. The heating of the gas involves the transfer of energy from the impinging far-UV photons mainly, but also any other form of radiation (X-rays or cosmic rays), to kinetic energy of the gas. On the contrary, the cooling involves the conversion of kinetic energy to photons emitted by atoms and molecules that can escape the cloud.

### 3.4.1 Heating processes

The main source of energy delivered to PDR is the FUV radiation field. The FUV photons absorbed by the dust, or the gas, bring energy to the medium and heat the matter. The coupling process is the ionisation of trace elements, PAHs and dust grains. In this section, I will review the main heating processes that are involved in PDRs, but also more generally in the global ISM.

**Photo-electric effect**

The most important heating process in the neutral ISM, and PDRs, is the photo-electric effect working on large PAH molecules and small dust grains. This process follows the sequence illustrated on Fig. 3.6:

- If a PAH, or a dust grain, absorbs a far-UV photon with an energy higher than its ionisation potential[3], it will extract an electron, called a photo-electron, which contains the excess of energy remaining in the form of kinetic energy.

- In case of a PAH, this photo-electron will be ejected into the gas-phase with that initial excess of energy. In case of a dust grain, the photo-electron will be diffused within the grain where it will lose a part of its energy by collisions. Those collisions heat the dust grain and, if the photo-electron reaches the surface, it will eventually be ejected into the gas-phase with the remaining excess of energy.

- This photo-electron, which carries a fraction of the original photon energy, will then collide with molecules in the gas-phase increasing the global kinetic energy, thus the temperature, of the gas.

The process efficiency differs from one type of grain to another (e.g. PAHs vs. large grains or different grains within a same population). On grains, the effect of charging is important. Actually, if the grain is positively charged, the efficiency of the photo-electric effect will be much less because of the stronger Coulomb interaction (increasing with the charge) and a higher energy needed to ionise (increasing ionisation potential with the charge). Because the charge is a balance between photo-ionisation and electron recombination, which is important for dense regions, this charging effect is more critical for dense regions. Planar molecules as PAHs are less sensitive to that charging effect. The efficiency of the photoelectric heating effect also depends on the size of the grain, favouring the smallest. PAHs and VSGs are thus the most efficient ones at heating the gas through this process. Grains with less than 1500 C atoms (about 1.5 nm in size), including PAHs and VSGs, contribute to half of the photoelectric heating, while grains with 1500 to $4.5\,10^5$ C atoms (a corresponding size of 1.5 to 10 nm) are responsible for the other half (Bakes and Tielens, 1994). Overall, the photo-electric efficiency, corresponding to the energy transferred to the gas normalised by the incoming energy through FUV photons, is at most about 5% and decreases with charging and size. In a similar way, but very much more locally, X-rays may also remove electrons from PAHs or dust grains.

**Photo-pumping of $H_2$**

The most abundant molecule is $H_2$. Through the absorption of an FUV photon, the energy level of the molecule may jump from the ground electronic state to an excited electronic state. As illustrated on Fig. 3.7, this excitation has three possible outcomes:

---

[3]To remove an electron, the energy has to be higher than the work function of the PAH/grain (the energy needed to remove an electron from a solid to a point in the vacuum immediately outside the solid surface) plus the Coulomb potential (the energy to move out the electron while the PAH/grain which may be positively charged tends to attract).

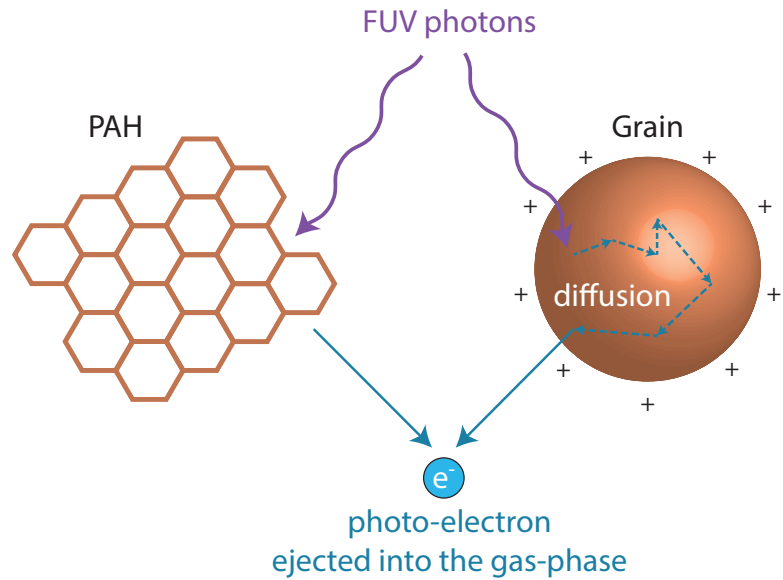

Figure 3.6: Illustration of the photo-electric effect: a far-UV photon gives its energy to a photo-electron that can heat a dust grain through collisions (diffusion) within it and the gas through collisions with molecules after being ejected.

- If the radiative decay, back to the electronic ground state, occurs in the vibrational continuum, the molecule dissociate to form two atoms of hydrogen. This happens between 10 and 15% of the time. After photodissociation, the fragments will carry away some of the photon energy as kinetic energy, thus heating the gas.

- Most of the time, the radiative decay occurs to a bound vibrational excited state of the electronic ground state. In that case, and if the density is below the critical density[4] (around $10^4 - 10^5$ cm$^{-3}$, depending on the temperature), the molecule will then slowly decay radiatively through the emission of infrared photons down to the ground state: this is fluorescence.

- Still considering a radiative decay back to a vibrational excited state but this time with a density higher than the critical density, the vibrationally excited molecule is collisionally de-excited, thereby transferring the vibrational energy to kinetic energy through each collision and heating the gas.

**Formation of H$_2$**

There is no easy way to produce molecular hydrogen through gas-phase reactions. The efficient alternative, generally accepted, is that the formation of H$_2$ proceeds mainly through atomic hy-drogen reactions on the surfaces of grain (as presented in Sect. 3.3.2). The reaction is exothermic so that the heat released by the formation leaves the newly formed species momentarily highly vibrationally and rotationally excited on the grain surface. The energy can be transferred partly to heat the grain and partly to heat the gas if the newly formed molecule is ejected into the gas-phase and collide with other particles.

---
[4]The populations of energy levels for a system depends on the balance between radiative emission and collisions. Considering a system with two levels of energy, the balance is driven by the critical density defined as $n_{\rm cr} = A_{\rm ul}/\gamma_{\rm ul}$, that is the ratio between the Einstein $A_{\rm ul}$ coefficient related to spontaneous radiative de-excitation and $\gamma_{\rm ul}$ the collisional rate coefficient between the upper $u$ and lower $l$ levels. The situation is similar for a multi-levels system. If $n > n_{\rm cr}$, the density is high enough so that the collisions dominate, the level populations are given by the Boltzmann expression at the kinetic temperature of the gas and the gas is in local thermodynamic equilibrium.

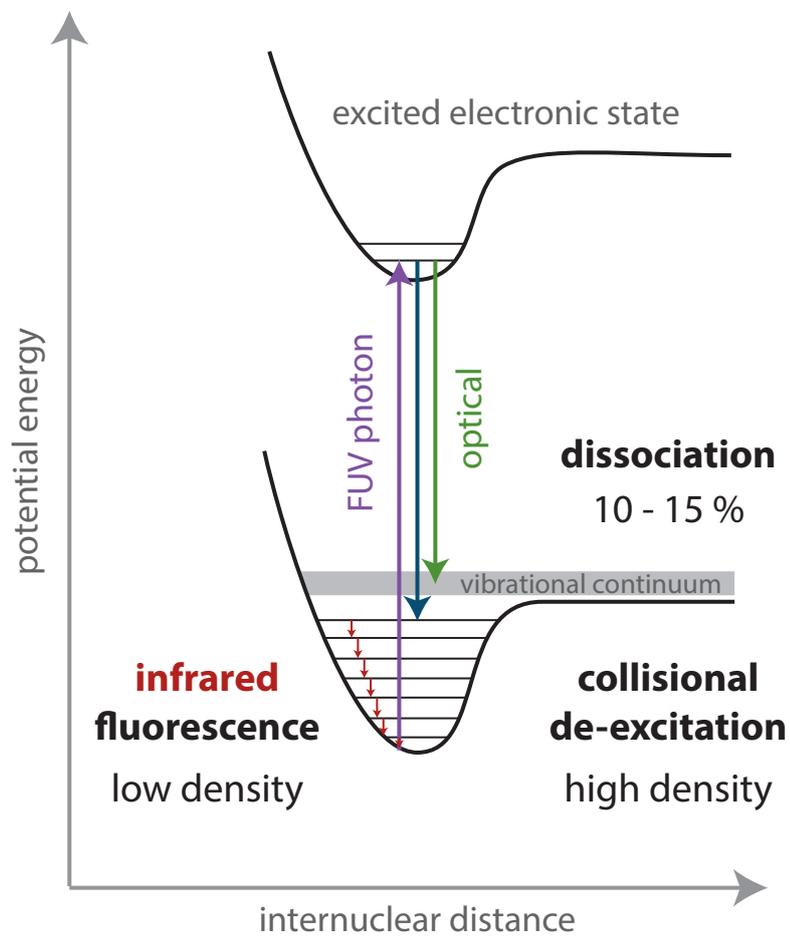

Figure 3.7: Illustration of the FUV-pumping of H$_2$ and its effects. Figure adapted from Hollenbach and Tielens (1999).

**Chemical exothermic reactions**

$H_2$ formation was the most important example, but many other reactions may be exothermic thus transferring energy to the dust or gas-phase.

**Gas-grain collisions**

The kinetic temperature of the gas and the dust grains may be different. In case of hotter dust grains, and if the density is large, collisions between molecules and dust grains can transfer energy to the gas and heat it. This heating mechanism dominates in the densest parts of molecular clouds where no more FUV photons penetrate but where the dust may be heated by infrared emission.

**IR-pumping**

When heated, dust grains strongly emit in IR. In a similar process to FUV-pumping of $H_2$, IR-pumping of gas species followed by collisional de-excitation can also be of importance, mostly deeper in a PDR where no more FUV penetrate.

**Photo-ionisation**

FUV photons can not ionise hydrogen by definition but there are a lot of other species with a lower ionisation potential, including carbon and oxygen. Photo-ionisation of carbon and other trace species is generally of lesser importance but also provide energy to the gas.

**Cosmic rays ionisation and excitation**

Cosmic rays may also ionise, leaving electrons in excess of energy, or directly accelerate free electrons of the medium. Those will eventually collide with molecules and heat the gas in a similar way as the photo-electric effect does. This process occurs everywhere in the ISM but it is negligible by comparison of other processes involved in PDRs.

**Mechanical energy**

Finally, macroscopic motions in PDRs, such as compression through gravitational collapses, winds or shocks, may also lead to local heating. This is also the case with turbulence and ambipolar diffusion (a process related to the magnetic field, see Sect. 6.2.2).

### 3.4.2 Cooling processes

The gas in PDRs is generally cooled by the emission of photons through atomic fine-structure lines or by molecular rotational lines following collisional excitation, but also through opposite mechanisms of heating process described previously. In this section, I will briefly review the main cooling processes.

**Atomic lines**

Some atomic species have fine-structure levels close to their fundamental ground state (with transition energy < 1000 K) so that they can be easily excited by collisions. The cooling of the neutral region (region of atomic neutral hydrogen) of PDRs is mainly dominated by the [CII] 158 µm, [OI] 63 µm and 146 µm lines. Their relative importance is determined by the medium density making them a good probe of this parameter.

The neutral region contains some ionised species (e.g. C, S and Si) and, as a consequence, free electrons. If the temperature is high enough, higher energy levels may be populated by collisions with electrons, thus leading to a cooling through allowed lines (i.e. allowed by the usual electric-dipole approximation), as the Lyman α line of the hydrogen.

**Molecular lines**

Deeper in the PDR, in (or close to) the molecular region (see Sect. 3.2 for the structure), collisional excitation of molecules as $H_2$, CO, $H_2O$ and SiO is important and emission through rotational lines, in particular of CO, provides the dominant cooling process.

**Gas-grain collisions**

In PDRs, the gas is efficiently heated so that it is generally hotter than the dust grains. The collisions between the two phases is an important process to cool the gas and depends directly on the temperature difference.

**Recombination**

In an opposite way to the photo-electric effect may heat the gas, a positively charged PAH or dust grain may recombine with free electrons and absorb energy.

## 3.5 Temperature profile

### 3.5.1 Gas temperature

At steady state, the gas temperature in PDRs is defined as the kinetic temperature obtained from the equilibrium between the heating and the cooling processes. In the first layers of the PDR, mainly heated by the photo-electric effect, the temperature of the gas tends to rise (see Fig. 3.8). Since the FUV photons flux decreases with depth, it could be surprising. Actually, the increase of efficiency of the photo-electric effect, related to the charging effect, dominates over the FUV flux extinction. The charging decreases, and the efficiency increases, from the ionisation front to the almost fully neutral medium. At this point, with a depth of the order of one magnitude in visual extinction, grains are almost fully neutral and the heating rate becomes directly proportionally to the intensity of the penetrating FUV photons. Due to the decreasing UV flux, the gas temperature will then decrease and generally follows an exponential decays if dust grains dominate the extinction.

### 3.5.2 Dust temperature

The temperature of a dust grain is set by the radiative energy balance between the absorption of an incident photon and the emission of another one. Both for silicates or graphite grains in general, smaller the grains are, hotter they are. Because they absorb UV photons more efficiently, the small graphite grains are hotter. However, since they have a lower heat capacity, the small grains are not in a steady state but rather fluctuate significantly around their equilibrium temperature. Let us forget that diversity, and consider a cloud containing a population of grains with a unique size and composition. The energy balance is still at play to set the dust temperature with the FUV photons at the center of interest in PDRs. Actually, the penetrating energetic FUV photons will create a surface layer of warm dust by being gradually absorbed. This surface layer will radiate strongly in infrared wavelengths. The infrared continuum of warm dust will heat up the dust located deeper in the PDR and so on, so that the dust temperature is maximum at the edge of the PDR and decreases with the depth. Microwave background radiation also participate to the heating but is negligible except deep in dense clouds where the other radiation fields are severely attenuated. These dust temperatures are much less than the gas temperature since they much more efficiently cooled by emitting in the continuum rather than through a few lines. The gas-grain collision process is thus mainly a gas cooling effect. However, in depth, no more FUV photons are available to heat the gas, while the dust temperature may still be relatively high because of the infrared emission from the surface layer. As a consequence, the temperature of of the gas is maintained just below the dust temperature by the gas-grains collisions which represent here a gas heating process. This heating, through the dust emission and the gas-grains collision mechanism, explains why the temperature of dust and gas, in regions deeper than the PDR, are still depending on the impinging FUV radiation field at the surface of the PDR.

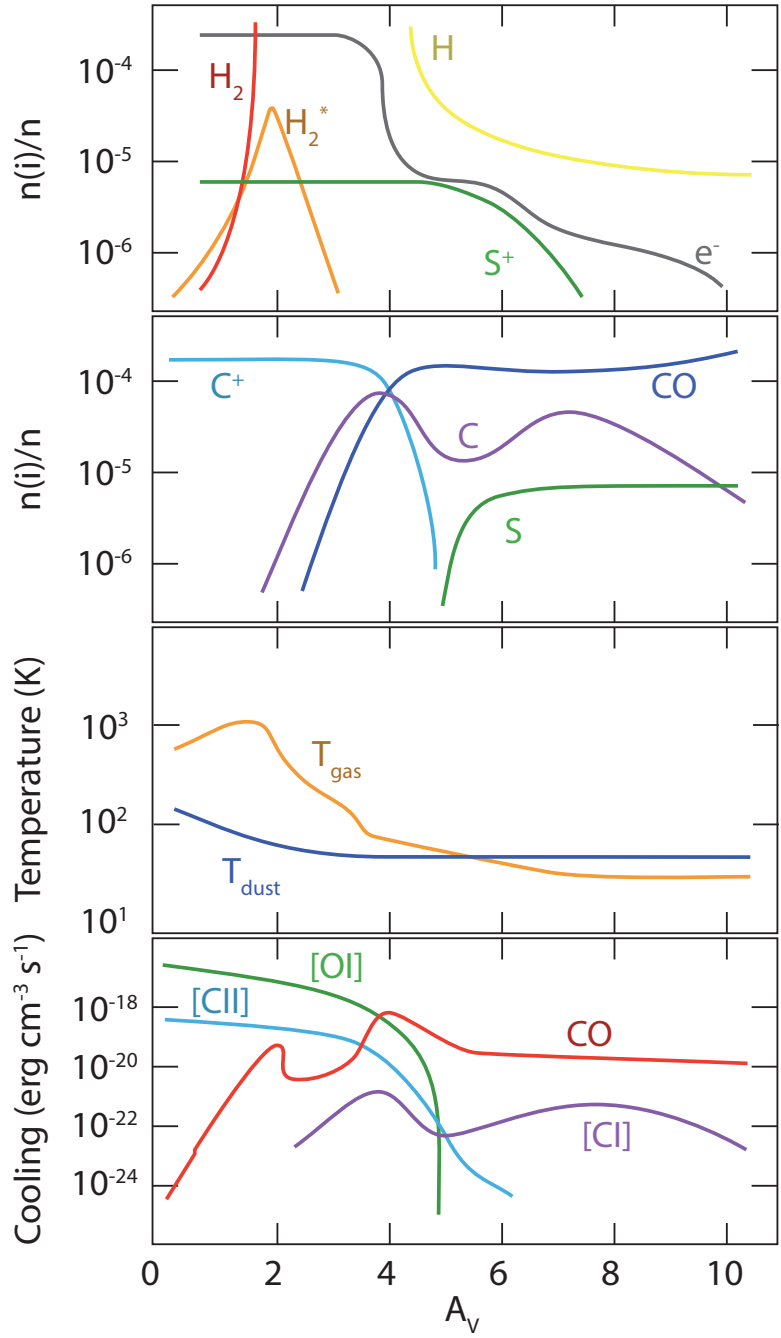

Figure 3.8: Example of a calculated structure of a PDR (here in Orion) as a function of visual extinction $A_V$. The illuminating source is to the left. Top two panels: Abundances relative to total hydrogen. Third panel: Gas and dust temperatures. Bottom panel: Cooling in the various gas lines. Figure adapted from Hollenbach and Tielens (1999).

## 3.6 The Meudon PDR code

The physics and chemistry that I described in the previous chapter may be integrated in models. The Meudon PDR code is one of them. It is a 1D model that considers a stationary plane-parallel slab of gas and dust illuminated by any radiation field coming from one or both sides of the cloud. For each point in the cloud, it solves the radiative transfer in the UV, taking into account the absorption in the continuum by dust and in discrete transitions of H and $H_2$ under assumptions on the abundances and temperatures. Knowing the UV field, it computes the thermal balance, taking into account the heating and cooling processes, and solves the chemistry for many species and reactions.. Once abundances of atoms/molecules and level excitation of the most important species have been computed, at each position in the cloud, line intensities and column densities can be deduced by a post-processor code, and compared with observations.

I used the public version 1.4.4 available during my PhD (downloadable at https://ism.obspm.fr). The Meudon PDR code is constantly updated to include latest improvements on the microphysics (e.g. $H_2$ formation and excitation, photoelectric effect, grain temperature fluctuations etc. (Le Petit et al., 2006; Le Bourlot et al., 2012; Bron et al., 2014, 2016). As the physics of PDRs have already been introduced, I present here a summary of the processes involved in the code and how they are taken into account.

### 3.6.1 Radiation field and radiative transfer

Two radiation fields, that can be combined, can illuminate each side of the cloud: a stellar spectrum (a black body or any given spectrum) or the interstellar standard radiation field (ISRF) eventually scaled by a multiplicative factor. The ISRF is an isotropic radiation field and, in the code, goes from far-UV to the millimetre domain with a sum of four components:

**Cosmic microwave background** Here assumed to be a black body of 2.73 K.

- **Dust emission in infrared** Emission estimated by the DustEM code (code to model the emission of dust, also part of the ISM platform with the PDR code) with a specific intensity resulting from the sum of the emission by PAHs, very small grains and big grains.

- **near-UV to near-IR** "Cold" stars are responsible for the part near-UV, visible and near-IR of the ISRF spectrum. The PDR code uses an updated expression of Mathis et al. (1983) which is a combination of three black bodies at 6184, 6123 and 2539 K respectively.

- **far-UV to near-UV** This component from the emission of massive stars can be modelled with the prescription of Mathis et al. (1983) or Draine (1978), as described in Sect. 3.2.1.

In our case, the FUV radiation field is set up through the multiplicative factor $G_0$ (normalised to the integrated Habing's field). From this impinging FUV field, the code has to treat its absorption in the cloud and consequences. We have seen in Sect. 3.2 that the photons are absorbed by the dust grains and the gas. In the PDR code, absorption by dust is directly given from any observed extinction curve. Parameters for several lines of sight are implemented in the code and are parametrised using the formalism of Fitzpatrick and Massa (1986, 1988, 1990). The default curve is the mean galactic extinction curve, with an $R_V = 3.1$, typical of the diffuse medium. The scattering and emission by dust is included as well as several thousands atomic and molecular lines in the whole spectral domain from the far-UV to the sub-millimeter range. In the UV, the radiative transfer considers only scattering or absorption followed by emission. The radiative transfer of infrared dust emission, and line emission, is done consistently. The resulting wavelength radiation field is used to estimate photo-dissociation and photo-ionisation rates, determine the rate of the photo-electric effect, and determine level excitation of main species as well as their cooling rates. More details on the radiative transfer methods are presented in Le Petit et al. (2006), Goicoechea and Le Bourlot (2007) and Gonzalez Garcia et al. (2008).

### 3.6.2 Chemistry

Elemental abundances and chemical reactions are set up in the parameters of the code. Everything can be modified but, for all my uses, I kept the default values of the ISM. Two-body reactions in the gas phase and on grains are mainly considered but many more reactions on grains and ices can be taken into account in the Meudon PDR code. The aim is to determine the abundance of each species by solving the chemistry, i.e. find the steady-state value that equilibrate the different formation and destruction processes. All of them depend strongly on the temperature, and so the radiative transfer, through the reaction rates, so that an initial temperature is assumed and iterations are done with the thermal balance. The code can finally compute the density at each position of all species taken into account. For some of these species, it can compute their densities in their quantum levels and the corresponding line intensities.

### 3.6.3 Thermal balance

At stationary state, temperature is determined assuming that the sum of the heating rates equals the one of the cooling rates. Heating mechanisms included in the code are the photoelectric effect on grains, cosmic ray ionisations, exothermal chemical reactions, $H_2$ formation on grains. Concerning cooling, the code considers atomic and molecular line emission, and $H_2$ dissociation. $H_2$ vibrational (de-)excitation and gas-grains collisions that can be heating or cooling terms, depending on the physical conditions, are also taken into account. Computations of non-LTE (local thermodynamic equilibrium) level populations, and so of line intensities, are done for the most important species and takes into account collisional and radiative processes (absorption, spontaneous and induced emission), as well as chemical excitation. Radiative excitation includes non local effect as pumping by dust emission.

# Chapter 4

# Observations of dense PDR tracers in proplyds

In this chapter, I present the main observational tracers of PDRs (Sect. 4.1), the *Herschel Space Observatory* (Sect. 4.2) which was suitable to detect them, and our data analysis in the frame of an *Herschel* survey of proplyds (Sect. 4.3).

## 4.1 Observational tracers of PDRs

There exists a great variety of physical conditions inside one PDR or from one region to another. As a consequence, there are many kinds of observable emission that trace different conditions. I briefly describe some of the main "tracers" of PDRs in this section.

### 4.1.1 Gas emission

As we previously saw in Sect. 3.4.2, collisionnally excited gas line emission is one of the main cooling processes of the gas. In the infrared and submillimetre ranges, cooling lines from the ionised, atomic and molecular gas are good probes of the physical conditions and structure of the various gas layers in the PDR.

**Fine-structure lines of [CII] and [OI]**

The ionisation potential of atomic carbon is below that of hydrogen, with 11.26 eV instead of 13.6 eV, which implies that ionised carbon may be found in the ionised region of the ISM and in the neutral phase as well. The [CII] 158 µm line corresponds to the $^2P_3/2 - {}^2P_1/2$ transition of $C^+$. Since carbon is the fourth most abundant element, and since this line is relatively easy to excite, it is one of the most important cooling lines of the ISM and has been observed and studied in a wide variety of objects. The critical density of this line is 50 cm$^{-3}$ for collision with electrons and about $3\ 10^3$ cm$^{-3}$ for collisions with atomic hydrogen. This implies that this line can be seen in emission in diffuse ionised regions, and the neutral gas at the surface layers of PDRs.

    The ionisation potential of atomic oxygen is 13.62 eV, very close but just above the one of hydrogen. As a consequence, atomic oxygen is found in the neutral phase of PDRs. The [OI] 63 µm line, corresponding to the $^3P_1 - {}^3P_2$ transition, has a critical density about $5\ 10^5$ cm$^{-3}$ for collision with atomic hydrogen. The energy of the upper state is 228 K so that this line arises

mostly from the warm dense neutral medium. Along with the [CII] 158 µm line, it is one of the brightest PDR cooling lines.

**H$_2$ spectrum**

Although H$_2$ is the most abundant molecule in the ISM and PDRs, it is very difficult to detect it directly since it is a symmetric molecule, with thus no permanent dipole moment and no possible emission at sub- or millimetre wavelengths related to dipole transitions. Fortunately, rotational quadrupole transitions are allowed and divided into ortho-transitions (antiparallel nuclear spins corresponding to odd quantum $J$ numbers) and para-transitions (parallel nuclear spins, even $J$ numbers). Mid-IR transitions are purely rotational while near-IR transitions are ro-vibrational. They require temperatures about 500 K to be excited and thus probe generally the warm neutral phases rather than the cold molecular cores.

Moreover, we have seen that a FUV-pumped H$_2$ molecule will de-excite most of the time by decaying radiatively through infrared fluorescence, if $n < n_\mathrm{cr}$, or by collisions, if $n > n_\mathrm{cr}$ (Fig. 3.7, Sect. 3.4.1). The observation, or not, of a pure fluorescent spectrum may indicate which regime (emissions or collisions) dominates the de-excitation and thus if the density $n$ is below, or above, the critical density $n_\mathrm{cr}$.

**Rotational lines of CO**

After H$_2$, carbon monoxide, CO, is the most abundant molecule. Molecules of CO can absorb FUV photons but are mainly excited through collisions with H$_2$. The cooling of CO molecules is efficient through rotational transitions. One level of transition is allowed (i.e. $\Delta J = \pm 1$), starting from the transition $J = 1$ to $J = 0$, named $^{12}$CO (1–0), with an energy of 5.5 K. Then, the energy of the upper level and the critical density increase with the level transition. The lower-$J$ ($\lesssim 3$) CO transitions are generally affected by opacity: at low densities ($n \leq 10^4$ cm$^{-3}$) they are optically thin and permit good diagnostics of the density; At high densities, those lines may be optically thick at high densities so they mostly probe the temperature of the surface layer of PDRs instead. Higher-$J$ transition lines, as well as other CO isotopes (e.g. $^{13}$CO or C$^{18}$O), are less affected by optical depth effects and can stay optically thin longer, so they can probe denser medium, located deeper in the molecular region. Note that, if low-$J$ CO lines can be observed from the ground, by single-dish antennas and interferometers, high-$J$ CO lines, which are affected by atmospheric absorption are only visible from space, e.g. with the Herschel Space Observatory.

### 4.1.2 PAH and grain emission

**Aromatic Infrared Bands**

The large aromatic molecules named PAHs are stochastically heated by FUV photons (Sect. 3.2.1 and 3.4.1). The intensities of those aromatic infrared bands depend on the UV radiation field and their abundances so, when observed, those two quantities can be probed. The relative intensities between the bands can be used to extract physical conditions as the ratio between ionised and neutral PAHs (e.g. Berné et al., 2007; Tielens, 2008) which in turn depends on the radiation field, the temperature and electron density (Bakes and Tielens, 1994).

**Dust grains**

Dust grains heated by the radiation field emit most of their energy in the mid-IR (small and warm grains) to submillimetre wavelength range (cold dust). The emission depends on the nature of the grains, their total mass and their temperature, related to the intensity of the radiation field. The dust emission is easily observed in photometry, but several instruments are required to probe the full wavelength range to eventually accurately sample the spectral energy distribution and bring robust constraints on the dust properties.

Table 4.1: Comparison of the capabilities of some airborne or space infrared telescopes.

| Telescope name | Launch date | Wavelength μm | Angular resolution at 100 μm (″) | Spectral resolution |
|---|---|---|---|---|
| KAO | 1974 | 1 – 500 | 30 | $10^4$ |
| ISO | 1995 | 2 – 240 | 60 | $10^2 - 10^4$ |
| Spitzer | 2003 | 3 – 160 | 40 | $10^2 - 10^3$ |
| Herschel | 2009 | 55 – 672 | 10 | $10^2 - 10^7$ |
| SOFIA | 2011 | 0.3 – 240 | 10 | $10^3 - 10^8$ |
| JWST | 2018 | 0.6 – 27 | 0.7 (at 20 μm) | $10^2 - 2.7\,10^3$ |

## 4.2 Herschel Space Observatory

PDR tracers are mostly found in the infrared (Sect. 4.1), a wavelength range unfortunately not accessible from the ground, so that one needs to go above the atmosphere to detect them. Infrared space astronomy was inaugurated in 1983 with the launch of the InfraRed Astronomical Satellite (IRAS). It was followed by several other satellites until the Herschel Space Observatory was launched on May 14th 2009. With a primary mirror of 3.5m in diameter, it was the largest space telescope ever launched. The mission continued until the last drop of Helium, needed to cool the telescope, was consumed on April 29th 2013. During the four years of operation, Herschel has brought the best spatial and spectral resolution in the infrared for more than 25 000 hours of science data. Table 4.1 compares several capabilities of previous infrared airborne (KAO and SOFIA) or space telescopes, with Herschel and the next largest infrared space telescope to come: the James Webb Space Telescope (JWST).

Herschel was designed to observed the cold and obscured Universe by doing photometry and spectroscopy in the range of 55 – 672 μm (Pilbratt et al., 2010). The main scientific goals of the mission were to unveil how the first galaxies formed and how they evolve, to observe star and planet early stage formation, and to understand the physical and chemical processes in the ISM.

The Herschel Space Observatory measured 7.5 m in height and 4 m in width, for a total weight of 3.4 tons. It is composed of a Cassegrain telescope with its primary mirror of 3.5 m in diameter which protected by a sun-shield, a superfluid helium cryostat, a payload and a service module. The optics and scientific instruments are located in the cryostat where the detectors are kept at very low and stable temperatures to make them as sensitive as possible. The service module houses instrument electronics and components responsible for satellite function. Figure 4.1 illustrated the main characteristics of the observatory. The three instruments onboard are the Photodetector Array Camera and Spectrometer (PACS, Poglitsch et al. 2010), the Spectral and Photometric Imaging REceiver (SPIRE, Griffin et al. 2010), and the Heterodyne Instrument for the Far Infrared (HIFI, de Graauw et al. 2010). I will briefly describe them in the following.

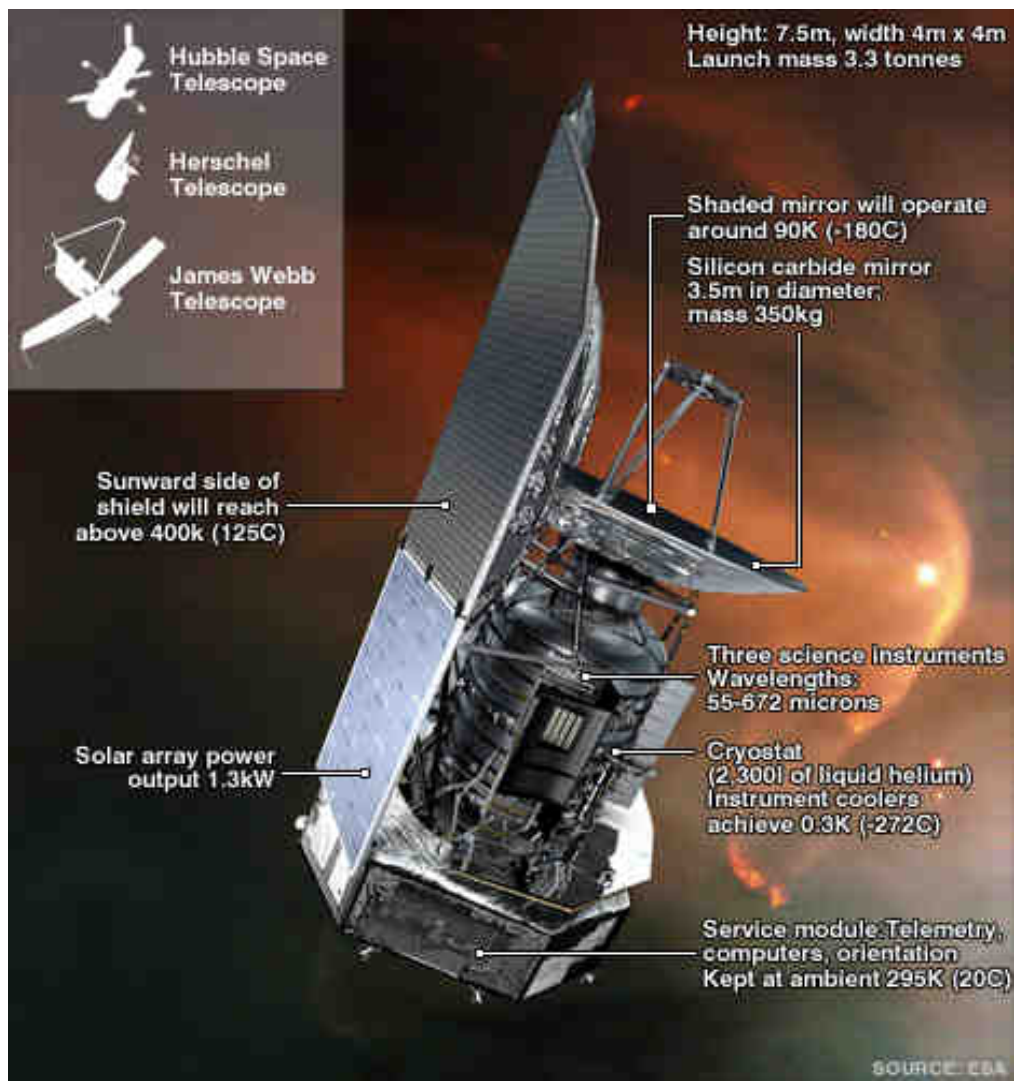

Figure 4.1: Illustration of the Herschel Space Observatory.

### 4.2.1 The Photodetector Array Camera and Spectrometer (PACS)

PACS consisted of two sub-instruments: a color camera to obtain photometric observations and a medium resolution imaging grating spectrometer. It operated at wavelength range of 55 – 210 µm.

The PACS integral-field-unit had a field of view of 47 by 47 arcseconds, resolved into 5 by 5 spatial pixels (or spaxels, with a size of 9.4 by 9.4 arcseconds) with simultaneous imaging. The instantaneous spectral coverage was about 1500 km s$^{-1}$ and the spectral resolution about 75 – 300 km s$^{-1}$. Three observing modes were available with this spectrometer: "chopped line spectroscopy" for single lines, "chopped range spectroscopy" for spectra over larger wavelength ranges, and "wavelength switching" mode for single lines on extended sources without clean background allowing chopping. The PACS spectrometer was optimised to detect faint FIR spectral lines at a high spatial resolution, which are generally several orders of magnitudes lower than the dust continuum over a typical photometric band.

The photometer imaged two bands simultaneously: 60 – 85 µm or 85 – 125 µm and 125 - 210 µm. For all wavelengths, the field of view was 1.75 by 3.5 arcseconds. Three different observing modes were available: "point-source photometry" mode in chopping-nodding technique, "scan map" technique, and "scan map" technique within the PACS/SPIRE parallel mode.

### 4.2.2 The Spectral and Photometric Imaging REceiver (SPIRE)

SPIRE also consisted of two sub-instruments: a three-band imaging photometer (250, 300 and 500 µm) and a low to medium resolution Fourier Transform Spectrometer (FTS) complementing PACS spectrometer for wavelengths in the range 194 – 672 µm.

The FTS covered its wavelength range with two overlapping bands: SSW (194 – 313 µm) and SLW (303 – 671 µm), within a field of view of 2 arcminutes in diameter and with a low spectral resolution up to 0.04 cm$^{-1}$. The beam size, given as Full Width at Half Maximum (FWHM), and depending on the observed wavelength, was 17 – 21 arcseconds for the SSW band and 29 – 42 arcseconds for the SLW band. Spectra were measured in single pointing or by making a raster map by moving the telescope.

The photometer was a three band imaging photometric camera with bands centered on 250, 350 and 500 µm. It had a field of view of 4 by 8 arcminutes. Three observing modes were available: point source photometry, field mapping, and scan mapping.

### 4.2.3 Heterodyne Instrument for the Far Infrared (HIFI)

HIFI was a high-resolution heterodyne receiver spectrometer covering wavelength ranges of 157 – 212 µm and 240 – 625 µm. It observed a single beam on the sky at a time. Once again, this elements consisted of two sub-instruments, two spectrometers: the high resolution spectrometer (HRS) which was a digital auto-correlator that provided a high spectral resolution (from 0.125 to 1.00 MHz, or 0.08 to 0.17 km s$^{-1}$ depending on the observed frequency) over a limited bandwidth, and the HIFI wideband spectrometer (WBS, single resolution of 1.1 MHz, or 0.2 to 0.7 km s$^{-1}$ depending on the observed frequency) provided a wide frequency coverage using an acousto-optical technique. HIFI observations included single-point observations, mapping observations, and spectral scans, with different observing modes: position switch, dual beam switch, frequency switch, and load chop. When position switch mode was used, the beam alternates between the target position ("ON" position) and a reference position ("OFF" position). The reference position should preferably be without emission in the band in question. Dual beam switch mode was similar to the position switch mode, but used internal chopper mirror to steer the beam between target and reference positions. In the frequency switch mode, the local oscillator frequency was changed by a few tens of MHz small enough that the lines of interest remained observable in both frequencies, which made this mode very efficient, since the lines can be observed both in ON and OFF positions. In the load chop mode an internal cold source was used as a reference. A chopping mirror was used to alternate between the target in the sky and the internal load. This method was generally used if there were no emission-free regions near the target. One of the main science goals of HIFI was to observe the [CII] 158 µm line, a dominant cooling line in the ISM, including PDRs.

## 4.3 Herschel survey of proplyds

It is well established that FUV photons play a key role in the gas heating in PDRs (Sect. 3.4.1) and that, externally illuminated photoevaporating protoplanetary disks with large envelope, a.k.a. proplyds (Sect. 2.3.3), should contain one region of this kind. However, there is no quantitative study of classical PDR tracers in proplyds so far, mostly due to their small sizes and strong background emission. Since these tracers are mostly found at far-IR wavelengths, and hence have to be observed from space, the ESA *Herschel Space Observatory* (Pilbratt et al., 2010) was the first to provide sufficient sensitivity and resolution to observe them. I will now describe the observational study that I have conducted.

### 4.3.1 Targets

To maximise the signal-to-noise ratio for the first survey of this kind, we have carefully selected three proplyds (see Table 4.2 but note that 203-506 will be presented later) according to their relatively large size (several arcseconds in diameter) and their strong incoming FUV radiation field ($G_0 > 10^4$). These targets are briefly described below.

**105-600** One of the giant proplyd candidates of Smith et al. (2003) in the Carina nebula (Fig. 4.2(a)), namely 104632.9-600354 in their paper and hereafter called 105-600 following Mesa-Delgado et al. (2016). It is located amongst the southern pillars at a projected distance of 25.4 arcminutes south-east from η Carinae and at 1 arcminute from the Bochum 11 cluster to the west. It was first studied by Vicente (2009) and has been observed with HST/ACS in Hα emission (Smith et al., 2010a), revealing a large envelope pointing towards η Carinae and a collimated bipolar jet that confirms the pre-stellar nature of this object. The large size of this object as well as its high molecular content ($M \approx 0.35$ M$_\odot$, after Sahai et al. 2012) could suggest that the young stellar object is still embedded in the remnant molecular core while the morphology is similar to a proplyd. From this point forward, we will consider that the molecular core inside the object is a disk since the true nature is not important for our study.

**HST10** Also known as 182-413, HST10 (Fig. 4.2(b)) is the most studied proplyd and was one of the first bright objects in which an embedded disk was seen in silhouette (e.g. O'dell et al., 1993; O'dell and Wong, 1996) and in emission through the [OI] 6300 Å and H$_2$ (1-0) S(1) lines (Bally et al., 1998, 2000; Chen et al., 1998). It is located at a projected distance of 56″ from the star $\Theta^1$ Ori C to the SE, and at 32.6″ from $\Theta^2$ Ori A to the NW. It is a teardrop-shaped proplyd containing a prominent nearly edge-on disk.

**244-440** The largest proplyd observed in the Orion nebula cluster (Fig. 4.2(c)). It exhibits a faint nearly edge-on silhouette disk (Vicente and Alves, 2005) and a one-sided microjet drived by the central star (Bally et al., 2000). It is located at a projected distance of 142″ from the star $\Theta^1$ Ori C and at 29″ from $\Theta^2$ Ori A.

### 4.3.2 Observing strategy and data reduction

The main concern about observations of proplyds is that they are small objects, so their emission is likely to be diluted in the *Herschel* beam (Fig. 4.2), and their emission can also be confused with emission from the nebula where they lie. Since the PDR density of proplyds is expected to be high at some locations, especially at the disk surface (about $10^6$ cm$^{-3}$), some of the main PDR tracers such as the [OI] 63 µm and the high-$J$ CO lines ($J = 15$-14, 17-16 and 19-18) are expected to be bright compared to background nebular emission (which has a lower density) and we thus proposed to observe them with PACS (Poglitsch et al., 2010). With HIFI (de Graauw et al., 2010), the [CII] 158 µm line and fainter lower $J$ CO lines ($J = 7$-6 and 10-9) were observed assuming that the proplyd emission can be spectrally separated from background/foreground nebular emission with a velocity resolution of 190 m s$^{-1}$ or less. Thanks to these two instruments, we were able to observe for the first time in proplyds selected excited CO lines spanning a wide range of densities (critical densities from $10^5$ to $10^7$ cm$^{-3}$, see e.g. Yang et al. 2010), and the [OI] 63 µm and [CII] 158 µm fine-structure lines (critical densities from about $10^3$ to $10^5$ cm$^{-3}$) which are generally the two major cooling lines in PDRs (see Tielens and Hollenbach, 1985b, and Sect. 3.4.2).

From one object or one line to another, different observing modes have been used to avoid contamination by the background emission (see Table 4.3). For HIFI, the mode Dual Beam Switch uses an OFF position (i.e. away from the source) on the sky if a nearby emission-free zone is expected, as for 105-600 mainly. If there is no such zone, the Load Chop mode allows to use internal loads as references and the mode Frequency Switch make its reference by switching between two frequencies in the local oscillator. In a similar way with PACS, we use the Standard Chopping-nodding mode for 105-600 assuming a vicinity free of emission while the Unchopped grating scan mode was used for the Orion proplyds that enables to go farther away from the nebula to find a clean OFF position.

The data reduction and parts of post-processing were done using the *Herschel Interactive Processing Environment* (HIPE version 13, see Ott, 2010) in order to obtain level-2 calibrated spectra (Fig. 4.3). Line profiles were then fitted using gaussian functions to extract the line integrated intensities and other characteristics such as the full width at half maximum (FWHM) and the velocity with respect to the local standard-of-rest, $v_{\mathrm{LSR}}$ (see Table 4.4). Uncertainties have been calculated by quadratically summing the instrumental (see user manuals of the instruments for more details) and fitting uncertainties.

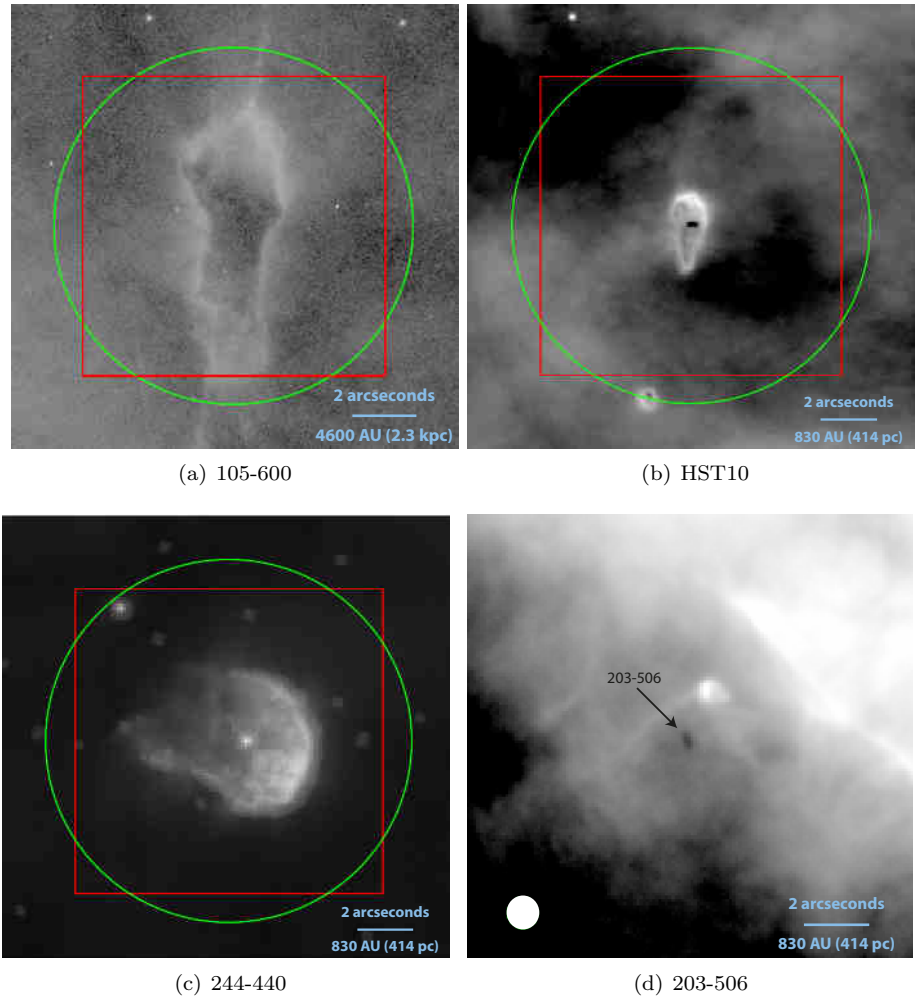

Figure 4.2: $14'' \times 14''$ Hα images of the observed proplyds from HST/WFPC2/F656N or WFC/F658N. Overlaid are the dimensions of one spaxel of PACS (red square of $9.4'' \times 9.4''$) and the minimal FWHM of HIFI (green circle with a diameter of $11.3''$) for the ones observed with *Herschel*. The ALMA beam of $0.99'' \times 0.97''$ is represented for 203-506.

Table 4.2: General properties of the studied proplyds.

| Proplyd name (parent nebula) | RA (J2000.0) | DEC (J2000.0) | Apparent envelope size | Apparent disk size | FUV field $G_0$ (Habing field) | Heliocentric distance (parsec) |
|---|---|---|---|---|---|---|
| 105-600 (Carina) | 10 46 32.97[a] | − 60 03 53.50[a] | $9.5'' \times 3.7''$[b] | ... | $2.2 \times 10^{4b}$ | 2300[c] |
| HST10 (Orion) | 05 35 18.22[d] | − 05 24 13.45[d] | $2.6'' \times 1.0''$[e] | $0.4'' \times 0.1''$[e] | $2.4 \times 10^{5f}$ | 414[g] |
| 244-440 (Orion) | 05 35 24.38[d] | − 05 24 39.74[d] | $5.6''$[h] | $0.86'' \times 0.69''$[h] | $1 \times 10^{5b}$ | 414[g] |
| 203-506 (Orion) | 05 35 20.32[d] | − 05 25 05.55[d] | ... | $0.75'' \times 0.61''$[j] | $2 \times 10^{4i}$ | 414[g] |

**Notes.** Units of right ascension (RA) are hours, minutes, seconds and units in declination (DEC) are degrees, arcminutes and arcseconds (J2000.0). Sizes are given as the major axis diameter times the minor axis diameter of an elliptical object (only one value if circular). FUV fields are expressed in units of the Habing field which represents the estimated average FUV flux in the local interstellar medium (Habing, 1968).
[a] Sahai et al. (2012), [b] Vicente (2009), [c] Smith et al. (2003), [d] Ricci et al. (2008), [e] Chen et al. (1998), [f] Störzer and Hollenbach (1998), [g] Menten et al. (2007), [h] Vicente and Alves (2005), [i] mean value of Marconi et al. (1998) and Walmsley et al. (2000), [j] Noel (2003).

Table 4.3: Observing strategy for *Herschel* spectroscopy.

| Line | Target | Instrument | Observing mode | Band | Beam size (FWHM or spaxel witdth) |
|---|---|---|---|---|---|
| [CII] 158 μm | 105-600 HST10 244-440 | HIFI - HRS | Dual beam switch Load chop Load chop | 7b (1795.2 - 1902.8 GHz or 158 - 177 μm) | 11.3 '' |
| CO (7-6) | 105-600 HST10 244-440 | HIFI - HRS | Dual beam switch Frequency switch Frequency switch | 3a (807.1 - 851.9 GHz or 352 - 372 μm) | 28.4 '' |
| CO (10-9) | 105-600 HST10 244-440 | HIFI - HRS | Frequency switch | 5a (1116.2 - 1240.8 GHz or 242 - 269 μm) | 18.6 '' |
| CO (15-14), (17-16), (19-18) | 105-600 HST10 244-440 | PACS | Chopping nodding Unchopped grating scan Unchopped grating scan | 51-73 μm and 103-220 μm | 9.4 '' |
| [OI] 63 μm | 105-600 HST10 244-440 | PACS | Chopping nodding Unchopped grating scan Unchopped grating scan | 51-73 μm and 103-220 μm | 9.4 '' |

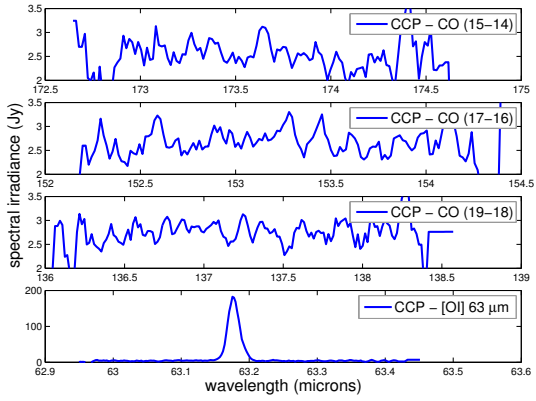
(a) PACS spectroscopy - 105-600

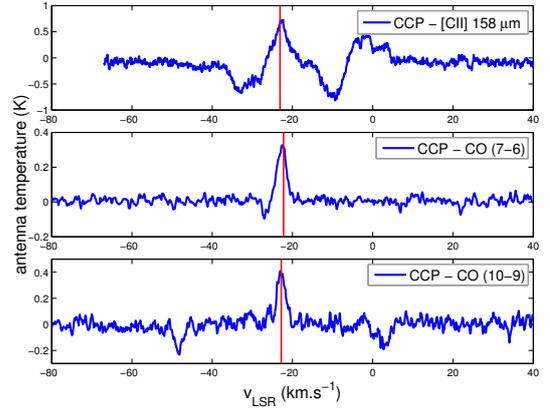
(b) HIFI spectroscopy - 105-600

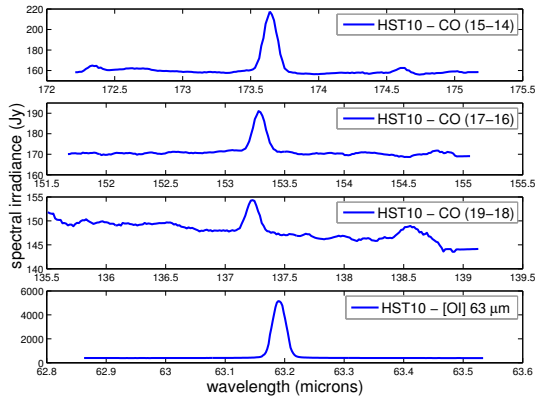
(c) PACS spectroscopy - HST10

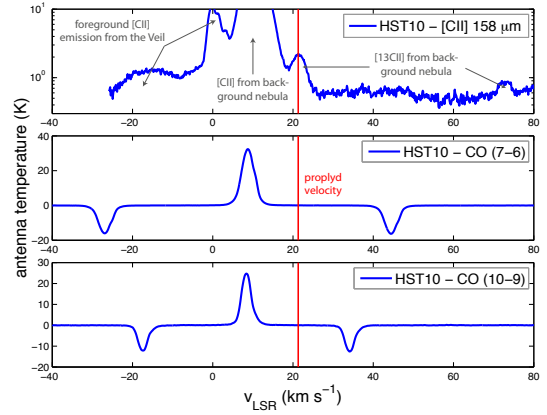
(d) HIFI spectroscopy - HST10

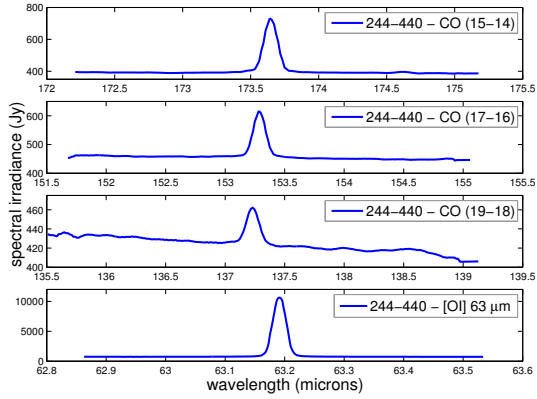
(e) PACS spectroscopy - 244-440

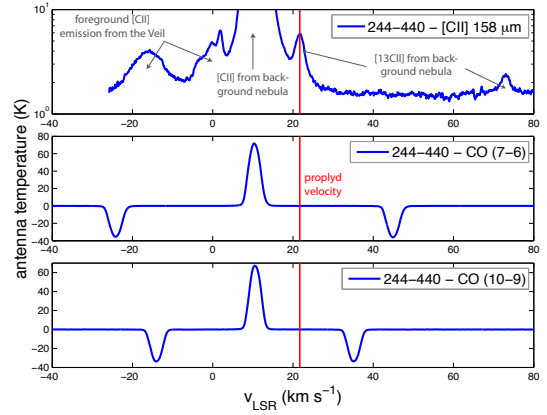
(f) HIFI spectroscopy - 244-440

Figure 4.3: Level-2 calibrated spectra for the three targets: 105-600, HST10 and 244-440 respectively from top to bottom. Spectrally unresolved lines observed with PACS are presented on the left and resolved lines observed with HIFI lied on the right. The red vertical bars on HIFI spectra indicate the $v_{LSR}$ of the objects.

Table 4.4: Features of the observed lines
(*Herschel* plus published/archival data available).

| Target | Line | Wavelength (μm) | Instrument | Integrated intensity (W m$^{-2}$) | | Integrated intensity (W m$^{-2}$ sr$^{-1}$) | FWHM (km s$^{-1}$) | $v_\text{LSR}$ (km s$^{-1}$) |
|---|---|---|---|---|---|---|---|---|
| | [CII] 158 μm | 157.74 | HIFI | 2.39e-16 | (12%) | 3.68e-07$^c$ | 6.4 | -23.1 |
| | CO (7-6) | 371.65 | HIFI | 1.18e-17 | (11%) | ... | 2.7 | -22.7 |
| | CO (10-9) | 260.24 | HIFI | 2.26e-17 | (14%) | ... | 2.5 | -22.8 |
| | CO (15-14) | 173.63 | PACS | 5.50e-18 | (24%) | ... | ... | ... |
| | CO (17-16) | 153.27 | PACS | 7.59e-18 | (24%) | ... | ... | ... |
| | CO (19-18) | 137.20 | PACS | < 2.65e-17$^a$ | (20%) | ... | ... | ... |
| 105-600 | [OI] 63 μm | 63.18 | PACS | 2.90e-15 | (12%) | 4.47e-06$^c$ | ... | ... |
| | CO (3-2) | 866.96 | APEX/FLASH+ | 2.98e-18 | (20%) | ... | 3.2 | -22.7 |
| | CO (4-3) | 650.25 | APEX/FLASH+ | 6.67e-18 | (20%) | ... | 3.1 | -22.7 |
| | CO (6-5) | 433.56 | APEX/CHAMP+ | 1.26e-17 | (20%) | ... | 3.5 | -22.6 |
| | CO (7-6) | 371.65 | APEX/CHAMP+ | 1.40e-17 | (20%) | ... | 4.1 | -22.3 |
| | $^{13}$CO (3-2) | 906.83 | APEX/FLASH+ | 7.12e-19 | (20%) | ... | 2.2 | -22.6 |
| | HCO+ (4-3) | 840.41 | APEX/FLASH+ | 2.97e-19 | (20%) | ... | 2.0 | -22.6 |
| | HCN (4-3) | 845.66 | APEX/FLASH+ | 1.72e-19 | (20%) | ... | 1.8 | -22.8 |
| | [CII] 158 μm | 157.74 | HIFI | < 2.27e-16$^b$ | (12%) | 4.72e-06$^{b,c}$ | 3.9 | 21.3 |
| | CO (7-6) | 371.65 | HIFI | < 4.32e-18$^a$ | (11%) | 5.84e-06$^{a,c}$ | ... | ... |
| | CO (10-9) | 260.24 | HIFI | < 2.63e-17$^a$ | (14%) | 3.56e-05$^{a,c}$ | ... | ... |
| | CO (15-14) | 173.63 | PACS | < 7.15e-16$^b$ | (13%) | 9.68e-04$^{b,c}$ | ... | ... |
| HST10 | CO (17-16) | 153.27 | PACS | < 3.47e-16$^b$ | (13%) | 4.70e-04$^{b,c}$ | ... | ... |
| | CO (19-18) | 137.20 | PACS | < 1.44e-16$^b$ | (13%) | 1.95e-04$^{b,c}$ | ... | ... |
| | [OI] 63 μm | 63.18 | PACS | < 1.01e-13$^b$ | (12%) | 2.10e-03$^{b,c}$ | ... | ... |
| | H2 1-0 S(1) | 2.12 | HST/NICMOS | 1.85e-18$^c$ | (20%) | 2.50e-06 | ... | ... |
| | [OI] 6300Å | 0.63 | HST/WFPC2 | 1.11e-17$^c$ | (34%) | 1.50e-05 | ... | ... |

Table 4.4 – Continued

| Target | Line | Wavelength (μm) | Instrument | Integrated intensity (W m$^{-2}$) | | Integrated intensity (W m$^{-2}$ sr$^{-1}$) | FWHM (km s$^{-1}$) | $v_{\mathrm{LSR}}$ (km s$^{-1}$) |
|---|---|---|---|---|---|---|---|---|
| | [CII] 158 μm | 157.74 | HIFI | < 4.68e-16$^b$ | (12%) | 8.09e-07$^{b,c}$ | 2.2 | 21.7 |
| | CO (7-6) | 371.65 | HIFI | < 2.69e-18$^a$ | (11%) | 2.46e-07$^{a,c}$ | ... | ... |
| | CO (10-9) | 260.24 | HIFI | < 1.53e-17$^a$ | (14%) | 1.40e-06$^{a,c}$ | ... | ... |
| 244-440 | CO (15-14) | 173.63 | PACS | < 4.19e-15$^b$ | (13%) | 3.83e-04$^{b,c}$ | ... | ... |
| | CO (17-16) | 153.27 | PACS | < 2.72e-15$^b$ | (13%) | 2.49e-04$^{b,c}$ | ... | ... |
| | CO (19-18) | 137.20 | PACS | < 8.16e-16$^b$ | (13%) | 7.45e-05$^{b,c}$ | ... | ... |
| | [OI] 63 μm | 63.18 | PACS | < 2.06e-13$^b$ | (12%) | 3.55e-04$^{b,c}$ | ... | ... |
| | OH 84 μm | 84.45 | PACS | 6.83e-18 | (13%) | 6.24e-07$^c$ | ... | ... |
| | HCO+ (4-3) | 840.41 | ALMA | 3.04e-20$^c$ | (10%) | 3.60e-09 | 4.0 | 9.5 |
| 203-506 | H2 1-0 S(1) | 2.12 | CFHT | 7.01e-18$^c$ | (11%) | 8.30e-07 | ... | ... |
| | [OI] 6300Å | 0.63 | HST/WFPC2 - MUSE | 1.01e-16$^c$ | ... | 1.20e-05 | ... | ... |

**Notes.** 1σ-relative uncertainties of integrated intensities are given in parentheses after each value and take into account instrumental accuracy and fitting uncertainties. Other uncertainties are not indicated, they are below 5% for the full width at half maximum (FWHM) and are estimated to be smaller than a percent for the velocity with respect to the local standard of rest ($v_{\mathrm{LSR}}$).
$^{(a)}$ Upper limit determined by the noise (peak of the line supposed to be smaller than three times the noise level).
$^{(b)}$ Upper limit determined by the observed background emission.
$^{(c)}$ Values converted using the apparent size given in Table 4.2 and assuming that all the lines are coming from the disk except for [CII] 158 μm and [OI] 63 μm lines that are supposed to originate in the envelope.

### 4.3.3 Observing results

Fig. 4.3 shows the spectra extracted from the *Herschel* data. A summary of all detected lines and upper limits is given in Table 4.4.

For the Carina candidate proplyd, 105-600, PACS spectra are not contaminated by the surrounding nebula so that the target emission is directly observed (Fig. 4.3(a)). The [OI] 63 µm line is clearly visible while the sensitivity of the instrument allows the detection, at the limit, of the CO (15-14) and CO (17-16) lines (with peaks about 3 - 4$\sigma$) but is insufficient to detect the CO (19-18) line. Estimation of the integrated intensities, or their upper limits, can be obtained with reasonable uncertainties (about 24%, see Table 4.4) since the theoretical FWHM of the lines, related to instrumental characteristics, are known. Emission lines observed by HIFI are well spectrally resolved and with high signal-to-noise ratios (Fig. 4.3(b)). However, the observing mode chosen for the [CII] 158 µm line has caused a contamination from the background emission of the HII region. The strategy to remove this contamination and extract the [CII] 158 µm line flux is detailed in Appendix B.1.

In the case of the two Orion proplyds observed with *Herschel* (HST10 and 244-440), the low-resolution spectroscopy obtained by PACS (Figs. 4.3(c) and 4.3(e)) is highly contaminated by the nebula so that the observed emissions mainly come from the background and not from the targets. Only upper limits of the target emission can thus be extracted for the CO (15-14), CO (17-16), CO (19-18) and [OI] 63 µm lines. The high-resolution spectroscopy provided by HIFI enables to spectrally resolve the emission of the Orion proplyds and the emission from the nebula. This is clearly visible with the [CII] 158 µm fine-structure line, on Figs. 4.3(d) and 4.3(f), where the targets emission lines are detected at a $v_{\text{LSR}}$ about 21 - 22 km s$^{-1}$ and located on the wing of the nebula emission line that peaks close to a $v_{\text{LSR}}$ of 10 km s$^{-1}$. Unfortunately, the velocities of the Orion proplyds shift the [CII] lines at the same frequency of the strongest [$^{13}$CII] line from the nebula (Goicoechea et al., 2015). Here again, only upper limits can thus be extracted. For HST10, the maximum brightness of the [CII] 158 µm line, after correcting the beam dilution, is 4.72 × 10$^{-6}$ W m$^{-2}$ sr$^{-1}$ which is the same order of magnitude as the one predicted by Störzer and Hollenbach (1999) who probably overestimated it. At low velocities, several emission lines are observed: three components close to a $v_{\text{LSR}}$ of 0 km s$^{-1}$ and one located about -15 km s$^{-1}$. Those lines likely correspond to the emission of the $^{12}$C$^+$ from the Veil in front of the nebula seen in HI emission and absorption at those velocities (van der Werf et al., 2013). The line at a $v_{\text{LSR}}$ about 70-75 km s$^{-1}$ is another $^{13}$C$^+$ line from the nebula (Ossenkopf et al., 2013). Even with the high spectral resolution of HIFI, no emission from HST10 and 244-440 is visible for the CO (7-6) and CO (10-9) transitions. The peak in the corresponding spectra is the emission from the nebula and the negative signatures are ghosts related to the observing mode (frequency switch, see Table 4.3). Constraining upper limits of the emission from the targets can still be derived (see Table 4.4) assuming that the width of the lines should be smaller than the one detected from the $^{12}$C$^+$ emission and that the peak of the lines is below the detection threshold, i.e. three times the noise level.

Level-2 calibrated spectra from the PACS line spectroscopy also allow to extract the continuum emission which is underlying the line (see Fig. 4.3). The continuum extraction was done for the Carina candidate proplyd (105-600) only because the far-infrared signal from the two Orion proplyds is drowned in the emission of the nebula. In a five-by-five spaxels image of PACS, the emission from the target is estimated by extracting the flux in the central spaxel, correcting it for losses to the neighbouring spaxels because of the point spread function, and subtracting the mean observed flux in the 16 external spaxels. The obtained values are presented in Table 4.5.

Table 4.5: Continuum emission of the Carina candidate proplyd 105-600.

| Wavelength (μm) | Instruments | Continuum emission (mJy) | AOR[a] or reference |
|---|---|---|---|
| 3.55 | IRAC | 3.5 (11 %) | 12914432 |
| 4.49 | IRAC | 3.5 (11 %) | 12914432 |
| 5.73 | IRAC | 13.9 (11 %) | 12914432 |
| 7.87 | IRAC | 28.3 (11 %) | 12914432 |
| 24.00 | MIPS | < 1144.2[b] | 23788800 |
| 57.95 | PACS | 1153.5 (36 %) | OT2_oberne_4 |
| 63.22 | PACS | 839.4 (45 %) | OT2_oberne_4 |
| 68.70 | PACS | 1013.1 (25 %) | OT2_oberne_4 |
| 70.00 | PACS | 1483.0 (11 %) | 1342255062 |
| 137.25 | PACS | 1086.4 (15 %) | OT2_oberne_4 |
| 153.19 | PACS | 1144.2 (16 %) | OT2_oberne_4 |
| 160.00 | PACS | 1035.0 (11 %) | 1342255062 |
| 173.75 | PACS | 1122.7 (21 %) | OT2_oberne_4 |
| 189.50 | PACS | 1007.7 (43 %) | OT2_oberne_4 |
| 250.00 | SPIRE | 50.3[b] (14 %) | 1342255061 |
| 350.00 | SPIRE | 10.9[b] (36 %) | 1342255061 |
| 350.00 | LABOCA | 196.0 (34 %) | Sahai et al. (2012) |
| 870.00 | SABOCA | < 40.0 | Sahai et al. (2012) |

**Notes.** 1σ-relative uncertainties of the continuum emission are given inside parentheses after each value, in case of detection, and take into account instrumental accuracy and post-processing uncertainties. For the MIPS data, upper limit corresponds to the maximal signal detected plus three times the noise level.
[a] Astronomical observation request.
[b] Values in MJy/sr.

### 4.3.4 Complementary data

**Other infrared to submillimetre data for the proplyds observed by Herschel**

For the Carina candidate proplyd 105-600, the dataset is complemented with optical to submillimeter data from other instruments. From the online archives, we retrieved the available data of the *Spitzer*/IRAC and MIPS (PI: N. Smith, program-ID: 3420, 30848, Smith et al. 2010b) instruments, plus the photometry obtained by *Herschel*/PACS and SPIRE (PI: S. Molinari, program-ID: OT2_smolinar_7). From this data, we have extracted the continuum emission by direct measurement of the spectral radiance or by aperture photometry (Table 4.5). We also retrieved the data of the Atacama pathfinder experiment (APEX) 12-m telescope from Sahai et al. (2012) who used the LABOCA and SABOCA bolometers (beam size of respectively 7.8″ and 19″) to measure continuum emission and CHAMP+ and FLASH+ spectrometers (beam sizes from 7.7″ to 18.6″) to observe molecular lines. Finally, we used the optical HST/ACS Hα image (PI: N. Smith, program-ID: 10475, Smith et al. 2010a). Performing photometry at the ionisation front, we obtained a brightness (including Hα and [NII] 6583 Å lines) of $4 \times 10^{-7}$ W m$^{-2}$ sr$^{-1}$ which has been used to estimate the electron density at the ionisation front of $n_e \leq 680$ cm$^{-3}$.

For the proplyd 244-440, we also included the OH 84 µm line detected in a PACS survey of the Orion Bar and described in Parikka et al. (2016). HST10 has been observed in the H$_2$ 1-0 S(1) line by Chen et al. (1998), and we use their value for the intensity of this line. The [OI] 6300 Å line was observed with HST and we use the value given in the table 4 of Bally et al. (1998).

**ALMA detection of the 203-506 proplyd**

During my analysis of these data, coworkers have serendipitously detected intense HCO$^+$ (4-3) emission towards the 203-506 proplyd with ALMA. This detection was quite surprising since only a few, and generally very massive, proplyd silhouette disks have been detected with ALMA in HCO$^+$ and reported in the literature (Williams et al., 2014). Since HCO$^+$ is a PDR tracer, and given that other PDR tracers have been detected towards 203-506 (see below), we have decided to include it in our study. It also complements our sample quite well since it is a silhouette proplyd (Bally et al., 2000), i.e. seen against the bright background in visible. The disk is edge-on and situated just south of the Orion Bar. Its surface is bright in [OI] 6300 Å and a faint [OI] jet is also visible (Bally et al., 2000). It does not exhibit any visible envelope as the others probably because it is lying in the neutral PDR resulting from the photodissociation of the Orion bar. The fact that the ionisation front of the M42 HII region has not yet reached 203-506 suggests that this disk has emerged out of the molecular cloud only very recently and hence that it has been exposed to UV irradiation for a relatively short time compared to classical proplyds. The 203-506 proplyd was detected as part of the ALMA cycle 1 project dedicated to the observation of the Orion bar (PI: J. Goicoechea). The observations were conducted in Band 7 with an angular resolution of 1″ insufficient to spatially resolve the disk. The detailed analysis and data reduction are presented in Goicoechea et al. (2016). 203-506 was clearly detected in the HCO$^+$ (4-3) line. For other tracers with lower critical densities such as low-$J$ CO lines in Band 7 of ALMA, the emission is dominated by the Orion molecular cloud and hence the disk cannot be detected. The derived integrated intensity for the HCO$^+$ (4-3) line is presented in Table 4.4. Emission from the H$_2$ 1-0 S(1) line was also observed with the BEAR instrument (Noel et al., 2005) at the Canada France Hawaii Telescope. This data is presented in Noel (2003) and we use their intensity. The [OI] 6300 Å line was also observed by HST. The line integrated intensity was extracted from the HST image and corrected from extinction ($A_V = 1.26$) using the extinction map of the Orion nebula derived from the MUSE data by Weilbacher et al. (2015). The intensity

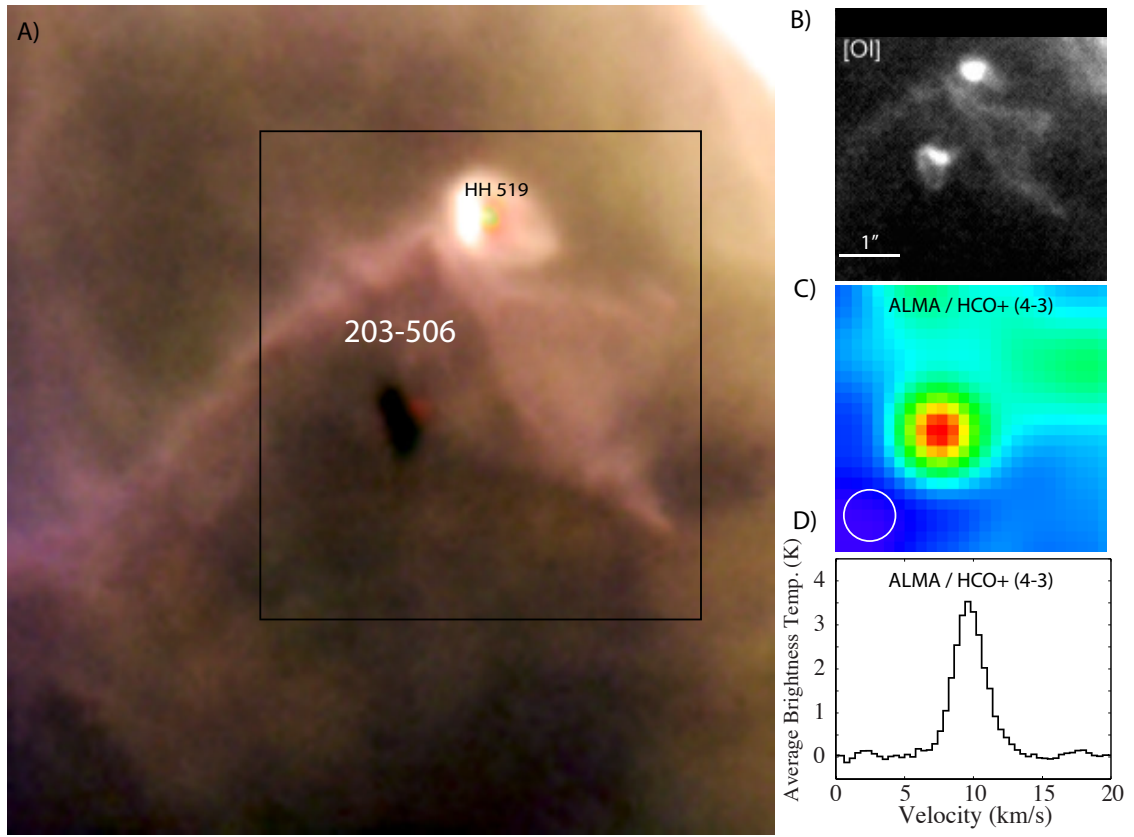

Figure 4.4: Overview of the 203-506 proplyd. A) Hubble Space Telescope color image (NASA, ESA and L. Ricci (ESO)), 203-506 is seen in silhouette against the bright background. B) [OI] 6300 Å image (Bally et al., 2000), tracing emission from the warm disk surface where OH is photodissociated (Störzer and Hollenbach, 1998). C) ALMA HCO+ (4-3) image of the same field as B), the ALMA beam (1") is the large circle in the corner. D) HCO+ (4-3) spectral line towards 203-506 extracted over a region 1.5 × 1.5 "

is reported in Table 4.4.

## Chapter 5

# Modelling the PDRs of proplyds

Since their discovery, proplyds have been extensively studied with the help of PDR models in order to determine their physical properties (e.g. Johnstone et al., 1998; Störzer and Hollenbach, 1999; Richling and Yorke, 2000; Adams et al., 2004; Clarke, 2007; Walsh et al., 2013). This approach is well-suited when trying to explain the FIR lines tracing the warm gas present at the surface layer of the disk and in the envelope. In this section I describe our methodology, and results, about the study of PDRs in proplyds.

## 5.1 1D-model of the PDR in an externally-illuminated protoplanetary disk

Observations (Sect. 4.3.3), and preliminary modelling we performed, showed that two different components, with different densities, are necessary to describe a proplyd and correspond very likely to the two distinct regions of the PDR, respectively the disk surface layer and the envelope. Given that, the geometry and parameters that we finally used are summarised in Fig. 5.1 and Table 5.1. We consider that a proplyd consists of an envelope (fed by the photoevaporation of the disk) with the apparent shape of an ellipse with semiaxes $a_{\rm env}$, $b_{\rm env}$, and defined by a mean density $n_{\rm env}$, a mean dust temperature $T_{\rm D,env}$ and $f_{\rm C}^{\rm PAH}$ the fraction of elemental carbon in the gas phase locked in Polycyclic Aromatic Hydrocarbons (PAHs). At the center of the envelope lies the disk with an apparent surface $S_{\rm disk}$, a mean dust temperature $T_{\rm D,disk} < T_{\rm D,env}$ and a mean density $n_{\rm disk}$ in the surface layer. Note that we are not modelling the properties of the bulk of the disk since we limit ourselves to the PDR which corresponds to the external layers of the proplyd (envelope and disk surface). Finally, the object is irradiated by FUV photons along the major-axis. The corresponding FUV radiation field $G_0$ is expressed in units of the Habing field. These parameters are used to predict both the dust spectral energy distribution (Sect. 5.2), and the gas emission (Sect. 5.3.1).

Table 5.1: Parameters of the models for the four studied proplyds.

| Symbol | Description | 105-600 | HST10 | 244-440 | 203-506 |
|---|---|---|---|---|---|
| $a_{env}$ | Envelope semi-major axis | $4.75''$ | $1.3''$ | $2.8''$ | ... |
| $b_{env}$ | Envelope semi-minor axis | $1.85''$ | $0.5''$ | $2.8''$ | ... |
| $n_{env}$ | Envelope density | $8.5 \times 10^3$ cm$^{-3\,a}$ | $2.5 \times 10^5$ cm$^{-3\,a}$ | $1 \times 10^5$ cm$^{-3\,a}$ | $4.5 \times 10^4$ cm$^{-3\,a}$ |
| $T_{D,env}$ | Mean dust temperature in the envelope | $39.8$ K$^a$ | ... | ... | ... |
| $f_C^{PAH}$ | Fraction of gas phase C locked in PAHs | $0.24\,\%^a$ | $0.08\,\%^b$ | ... | ... |
| $S_{disk}$ | Apparent surface of the disk | $3.82$ arcsec$^{2\,a}$ | $0.0314$ arcsec$^2$ | $0.47$ arcsec$^2$ | $0.36$ arcsec$^2$ |
| $n_{disk}$ | Mean disk surface layer density | $7 \times 10^5$ cm$^{-3\,a}$ | $4 \times 10^6$ cm$^{-3\,a}$ | $2 \times 10^6$ cm$^{-3\,a}$ | $1 \times 10^6$ cm$^{-3\,a}$ |
| $T_{D,disk}(<T_{D,env})$ | Mean dust temperature in the disk | $19.5$ K$^a$ | ... | ... | ... |
| $G_0$ | Radiation field | $2.2 \times 10^4$ | $2.4 \times 10^5$ | $1.0 \times 10^5$ | $2.0 \times 10^4$ |

($a$) Values obtained from the best-fit model.
($b$) Value derived by Vicente et al. (2013).

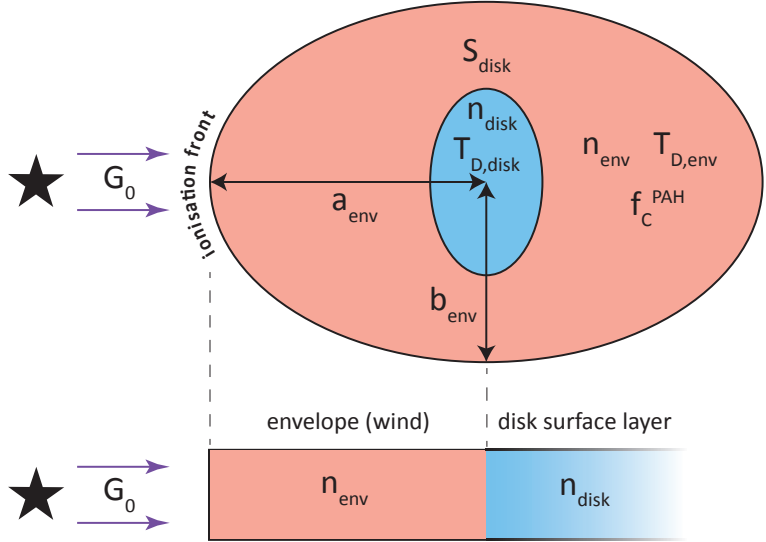

Figure 5.1: Illustration of the geometrical model used to represent proplyds. Top: schematics of the apparent 2D geometry with parameters used in our adapted model of an externally-illuminated protoplanetary disk composed of a disk (blue) surrounded by an envelope (red). Bottom: 1D corresponding slab used to model the PDR of the proplyd.

## 5.2 Dust spectral energy distribution of a proplyd

Interstellar dust is mainly composed of small grains (10 nm to 0.1 µm, see e.g. Compiègne et al., 2011) of silicates and carbonaceous compounds. Disks form from this medium and evolve with time in a way that their grains get larger and their dust-to-gas mass ratio increases (see e.g. the review of Williams and Cieza, 2011). The dust properties in the photoevaporation flow (envelope) are expected to differ from that of the disk because of a physical selection in size (e.g. Owen et al., 2011). However, we will initially assume that their dust populations are quite primitive and close to the ones of the ISM. We will then discuss the effects of grain evolution (Sect. 5.3.1 and App. B.2).

Dust grains participate strongly to the extinction of light (Sect. 3.2.1) but also emit in the infrared. The far-IR spectrum of proplyds is dominated by the emission of the largest grains, which are at thermal equilibrium with the radiation field. The thermal component of their spectral energy distribution can be modelled knowing the opacity $\tau_\nu$ and the mean dust temperature ($T_{D,env}$, $T_{D,disk}$[1]). Indeed, for a given dust temperature $T_D$, the emitted spectral radiance follows the relation

$$L_\nu(T_D) = \tau_\nu \, B_\nu(T_D), \qquad (5.1)$$

where $B_\nu$ is Planck's law. The dust opacity $\tau_\nu$ at the frequency $\nu$, is defined by

$$\tau_\nu = \tau_{\nu_0} \left(\frac{\nu}{\nu_0}\right)^\beta, \qquad (5.2)$$

---

[1] Contrary to the PDR approach which is limited to the surface layer of the disk, the mean dust temperature in the disk $T_{D,disk}$ is valid for the bulk of the disk.

where $\tau_{\nu_0}$ is the dust opacity at the reference frequency $\nu_0$ and $\beta$ is the spectral index. Following Planck Collaboration et al. (2011), we used a spectral index of 1.8 and a reference wavelength of 250 µm (1200 GHz) for which they found that, in the molecular phase of the ISM, the total column density of hydrogen, $N_H$, is related to the dust opacity by the empirical relation

$$\sigma = \frac{\tau_{250}}{N_H} = 2.32 \pm 0.3 \times 10^{-25} \text{ cm}^2. \quad (5.3)$$

The mean column density of hydrogen is related to the total mass of gas and dust by

$$M = N_H \, \Omega \, \mu \, m_H, \quad (5.4)$$

where $\Omega$ is the observed surface in the instrumental beam knowing the distance to the source, $\mu \, m_H$ the mean particle mass per hydrogen atom with $m_H$ the mass of one hydrogen atom and $\mu = 1.4$. Finally, for each component (disk and envelope), the emission can be written as a function of the total mass $M$ and the dust temperature $T_D$,

$$L_\nu(T_D) = \sigma \left( \frac{M}{\Omega \, \mu \, m_H} \right) \left( \frac{\nu}{\nu_{250}} \right)^\beta B_\nu(T_D). \quad (5.5)$$

The mid-infrared dust emission is dominated by PAHs which are found in the envelope and disk surface of proplyds (Vicente et al., 2013). This component is included in the model using the PAH emission modelled with the DustEM code (Compiègne et al., 2011). We use the standard PAH population given by them, i.e. a log-normal size distribution of neutral molecules and one more for singly-charged molecules. The intensity of PAH emission then only depends on the incident FUV radiation field $G_0$, the abundance of PAHs given as $f_C^{PAH}$ and the total mass of the envelope $M_{env}$.

## 5.3 Modelling strategy

### 5.3.1 Input parameters of the model

To predict emission lines from a PDR, one needs a model that correctly describes geometrical and micro-physical elements. Doing both with accuracy is not currently possible for computational time reasons. In this study, we choose to favor a proper description of the physics and chemistry rather than geometry, also taking into account that we do not have enough angular resolution to perform a careful geometrical analysis. From this perspective, the 1D Meudon code (Sect. 3.6) is well suited since it is constantly updated to include latest improvements on the microphysics. In our case, the integration of the local emissivities of lines, an output of the code, gives us the brightness of the disk surface layer and the brightness of the envelope. Weighted those quantities by the size of the corresponding emitting region enables us to derive the line integrated intensity that can be compared with the spatially unresolved observations that we have.

In our modelling, we use the default values for the dust-to-gas mass ratio, dust properties, cosmic ray ionisation rate and turbulent velocity (Table 5.2), as well as elemental abundances appropriate for the ISM (see Le Petit et al., 2006). However, grain properties in proplyds are likely to differ from those found in the ISM. Indeed, the smallest grains or PAHs seem to be underabundant in proplyds compared to the ISM: this is observed for HST10 by Vicente et al. (2013) and for 105-600 in this study (see Sect. 5.4.1). This observational fact is in agreement with the low detection rate of PAHs in isolated disks around young low-mass stars (Geers et al., 2006; Oliveira et al., 2010). We set the minimum grain radius to 3 nm in the model, i.e. excluding the smallest particles such as PAHs, for consistency with that observational fact and assuming that this is a general trend for proplyds. In the PDR model, the main impact of this modification is to remove the contribution of these small grains to the photo-electric heating and to the $H_2$ formation rate. This parameter has not a significant impact on the modelled line fluxes because proplyds are dense PDRs where the grain photoelectric effect may not always dominate the heating (this is discussed in App. B.2). Grain growth inside an evolving object tends to modify the grain distribution by increasing the maximum grain radius $a_{max}$, lowering the absolute value of the power-law index and finally it reduces the FUV extinction curve. Since these parameters can not be constrained accurately based on current observations of proplyds, we have explored their effects by changing independently:

- the maximal radius $a_{max}$ up to 3 μm, consistently with observations of Orion proplyds (Shuping et al., 2003; Vicente, 2009);

- the power-law index down to 3.0 (e.g. Testi et al., 2014);

- the FUV extinction using various curves of Fitzpatrick and Massa (1988) including a case with a dust FUV extinction cross section per H nucleus $\sigma_{ext} \simeq 8 \times 10^{-22}$ cm$^2$ as used for a sample of proplyds in the Orion nebula by Störzer and Hollenbach (1999).

The effects of grain parameters on the model results are discussed in App. B.2 and the general conclusion is that the changes do not affect the model results significantly. Hence, we keep the default ISM properties throughout this paper.

The input parameters of the Meudon code are the density profile as a function of visual extinction $A_V$, and the incoming FUV radiation field. These parameters are related to those of the proplyd model presented in Sect. 5.1 in the following way. The impinging radiation field is set by the value $G_0$ arriving at the surface of the envelope. The density profile between the envelope surface, or ionisation front, at $A_V = 0$ and the disk surface at $A_V^{jump}$ is constant with a value $n = n_{env}$. At the envelope - disk interface (i.e. at at $A_V^{jump}$), the density jumps to $n = n_{disk}$ and is kept constant up to $A_V^{max}$. Since the parameter $A_V^{jump}$ gives the position of the envelope - disk interface, it is also the visual extinction of the envelope and is set by

$$A_V^{jump} = \frac{N_{env}}{\kappa} = \frac{a_{env} \times n_{env}}{\kappa}, \quad (5.6)$$

where $\kappa$ is the parameter which relates the envelope column density $N_{env}$ to the extinction at the density jump, and can be calculated from Table 5.2, $\kappa = N_H/A_V = 1.87 \times 10^{21}$ cm$^{-2}$ (assuming ISM properties). The envelope column density is defined by $N_{env} = n_{env} \times a_{env}$. Finally, the value of $A_V^{max}$ is set to 10. The exact value is not critical but we have carefully checked that it is high enough. Indeed, most of the UV radiation is rapidly absorbed and most of the emission arises from regions of low extinction so increasing this parameter does not impact the results.

### 5.3.2 Fitting strategy

In order to estimate the unknown parameters of the model (see Table 5.1), we are looking for the set of values that best fit the observations based on a chi-square minimisation. For the spectral energy distribution of the dust, the model numerically converges to the best solution in the space of parameters using a Nelder-Meadwith non-linear minimisation script. For the line emission, we run the model hereabove on a grid for $n_{env}$ and $n_{disk}$ in a range of realistic values. The apparent size of the disk $S_{disk}$ is adjusted in the case of 105-600 while is known for the other proplyds. The best model is the one which minimizes the chi-square between observed and predicted line integrated intensities.

Table 5.2: Default values of some parameters used in the Meudon PDR code.

| Parameter | Value |
|---|---|
| Cosmic ray ionisation rate | $5 \times 10^{-17}$ s$^{-1}$ |
| Turbulent velocity (Doppler broadening) | 2 km s$^{-1}$ |
| $R_V = A_V/E(B-V)$ | 3.10 |
| $N_H/E(B-V)$ | $5.8 \times 10^{21}$ cm$^{-2}$ |
| Dust-to-gas mass ratio $\delta$ | 0.01 |
| Power-law index of the grain size distribution | 3.50 |
| Minimum grain radius $a_{\min}$ | 3 nm |
| Maximum grain radius $a_{\max}$ | 300 nm |

Table 5.3: Parameters derived from the fit of the spectral energy distribution for 105-600.

| Parameters | Best value | Interval at 1$\sigma$ |
|---|---|---|
| $M_{\text{disk}}$ (M$_\odot$) | 0.67 | 0.46 - 0.97 |
| $T_{\text{D,disk}}$ (K) | 19.5 | 17.9 - 21.4 |
| $M_{\text{env}}$ (M$_\odot$) | 0.048 | 0.036 - 0.062 |
| $T_{\text{D,env}}$ (K) | 39.8 | 37.7 - 42.4 |
| $f_C^{\text{PAH}}$ (%) | 0.24 | 0.22 - 0.25 |

**Notes.** Confidence intervals at 1$\sigma$ are calculated based on a method with constant $\chi^2$ boundaries as confidence limits (see Press, 2007, chap. 15.6).

## 5.4 Physical conditions in the PDRs

### 5.4.1 Proplyd 105-600

**Dust spectral energy distribution**

The spectral energy distribution of 105-600 is shown in Fig. 5.2. Parameters extracted from the fit to the observed SED are given in Table 5.3. Under the assumptions described in Sect. 5.2, the first component, the disk, is composed of a cold dust population ($\sim$ 20 K) and a total mass (dust and gas, assuming a dust-to-gas mass ratio of 1%) of 0.67 M$_\odot$. The envelope corresponds to a hotter dust population ($\sim$ 40 K) and contains a total mass of 0.048 M$_\odot$. This mass estimate is compared to other methods in Sect. B.4.2 which yield similar values both for the disk and envelope. The estimated PAH abundance in the envelope is $f_C^{\text{PAH}} = 0.0024$, about 30 times less than in the PDR of the NGC 7023 reflection nebula (Berné and Tielens, 2012). The depletion of PAHs within a proplyd-like object has already been observed in HST10 (Vicente et al., 2013) with a relative abundance three times lower than in 105-600, i.e. about 90 times less than in NGC7023. These low abundances in proplyd-like object could result from PAH destruction by the strong UV radiation field or sedimentation of PAHs in the disk but this is currently still unclear.

**Line emission**

To compare our models with observations, we have ran a grid of models with density ranges $n_{\text{env}} = 1 \times 10^3 - 2 \times 10^4$ cm$^{-3}$ and $n_{\text{disk}} = 1 \times 10^5 - 1 \times 10^7$ cm$^{-3}$, with constant grain properties as defined previously in Table 5.2 ($a_{\min} = 3$ nm, $\delta = 0.01$, see App. B.2 for the impact of those parameters). Fig. 5.3 presents the comparison between all the models and the observations.

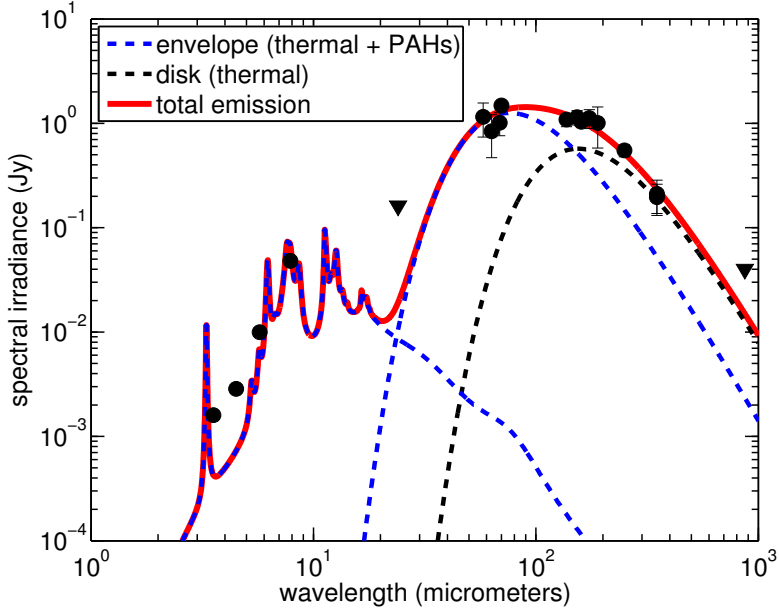

Figure 5.2: Best fit to the observed spectral energy distribution of the dust in 105-600 with a model including two components for the thermal emission and one for the emission of PAHs. Dots are detections while triangle represents upper limits.

From the $\chi^2$ analysis (Fig. 5.3(b)), a minimum is clearly found where the best-fit model is the one with $n_{\rm env} = 8.5 \times 10^3$ cm$^{-3}$, $n_{\rm disk} = 7 \times 10^5$ cm$^{-3}$ and a value $S_{\rm disk} = 2 \times 10^7$ AU$^2$. Fig. 5.4 presents the comparison between the observed line integrated intensities and the best-fit model. All lines are reproduced to a factor of a few or better, with the notable exception of HCN (4-3) which is under-predicted by more than an order of magnitude. HCN emission is a tracer of dense gas, whereas our model with a constant density is adopted only for the surface layers of the disk. To go further and improve the modelling of the low-$J$ CO and HCN lines would require a detailed disk model, not limited to the PDR as here.

The physical structure of the PDR (see the following section), extracted form the model, gives the gas temperature profile. The gas temperature at the edge of the envelope, i.e. at the ionisation front, increases with the envelope density chosen as input parameter, but in a relative small range ($\sim$ 150 to 400 K, see Fig. 5.3(c)) compared to the variations associated to the grains properties (Fig. B.2 and B.3). The gas temperature at the surface of the disk depends on its density but mainly on the envelope density. Higher the envelope density, lower the temperature at the disk surface (see Fig. 5.3(d), and Sect. 5.5 for more details). The best-fit model corresponds to a value of about 1100 K. Models with slightly lower envelope densities, and thus slightly higher disk surface temperatures, are still close to the observations while the decrease in disk surface temperature caused by an increase in the envelope density makes models diverge significantly from observations. This is strong evidence that the disk surface is hot with temperature around 1000 K or more.

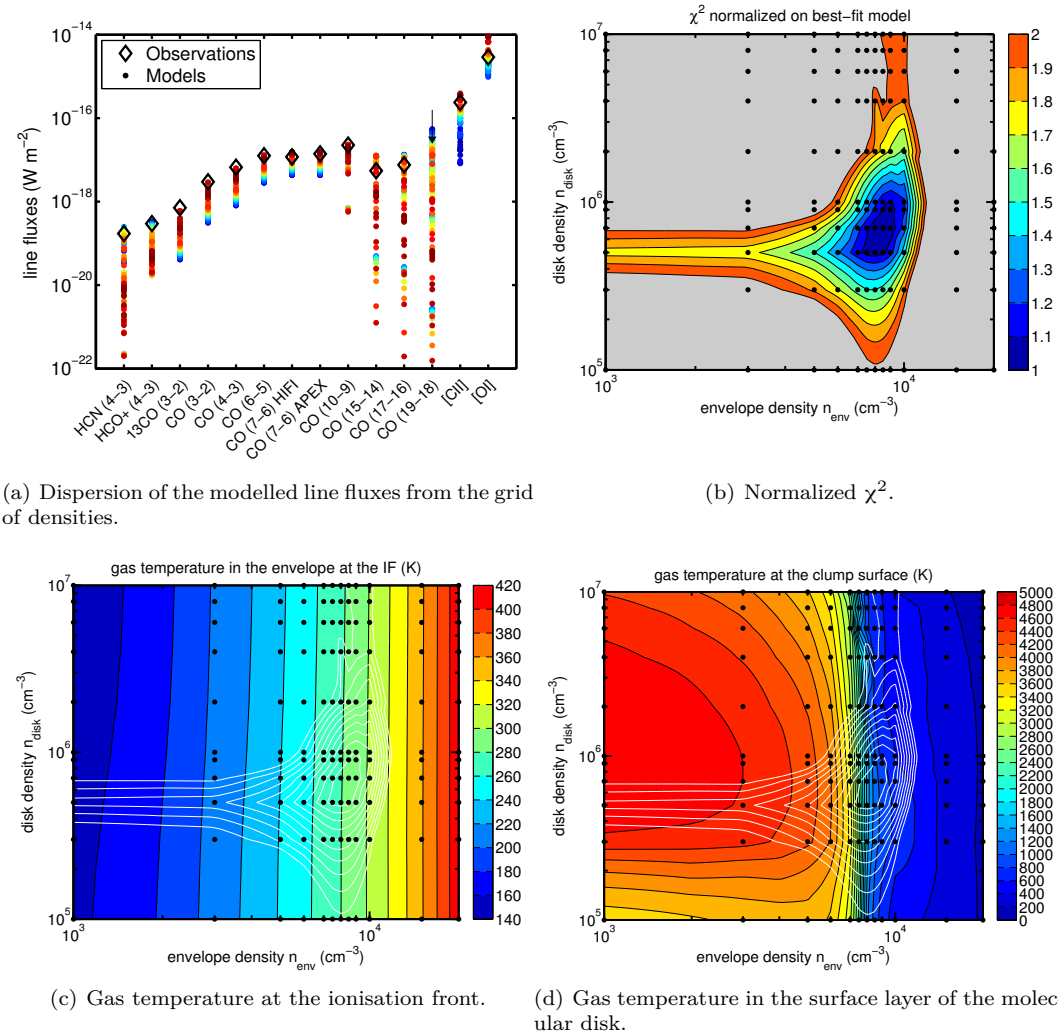

(a) Dispersion of the modelled line fluxes from the grid of densities.

(b) Normalized $\chi^2$.

(c) Gas temperature at the ionisation front.

(d) Gas temperature in the surface layer of the molecular disk.

Figure 5.3: Model outcomes for 105-600 and comparison with observations: (a) Modelled lines fluxes (105-600) for the considered density ranges (colored points) compared to observations (black diamonds). Uncertainties of observations are not plotted because they are smaller than the marker. (b) $\chi^2$ comparing observed line integrated intensities of 105-600 and modeled values for the considered envelope and disk densities. (c) temperature at the ionisation front, overlaid are the $\chi^2$ contours. (d) temperature at the disk surface. In panels (b,c,d), points are the positions where the models are calculated and the colors and contours result from interpolating the model results over these points.

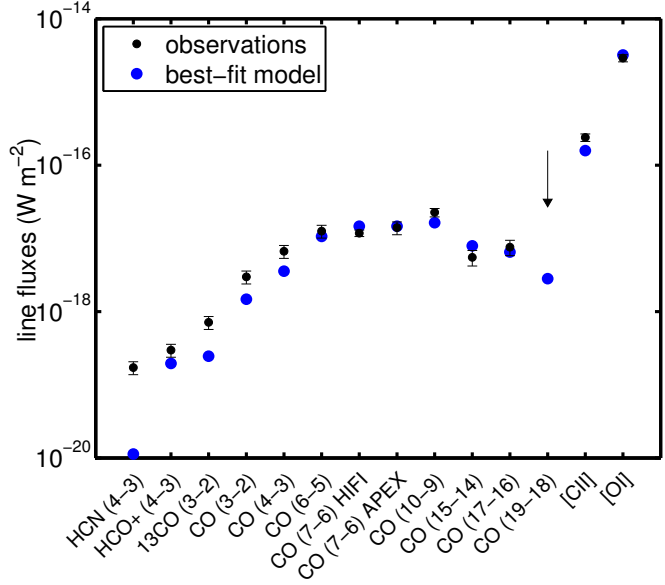

Figure 5.4: Comparison between best-fit model line integrated intensities and observed line integrated intensities for 105-600.

**Physical structure of the PDR**

As an output of the model it is possible to obtain the 1D physical structure of the PDR in the studied proplyds. Note that this structure is only limited to the PDR, i.e. the envelope and the surface layer of the disk, and so that the extracted physical properties are not valid deeper in the disk where much larger densities are expected and UV photons are strongly attenuated. The 1D profiles for the best model of 105-600 are shown in Fig. 5.5. If there are similarities a with classical PDR structure (see e.g. Fig. 3.8), the change of two density medium (with but also without a jump) have some impacts. It can be seen that the $H/H_2$ transition, defined here as the region where 10 to 90% of H is in $H_2$, encompasses the envelope - disk interface ($A_V^{\text{jump}} = 0.742$, see eq. (5.6)) ranging from $A_V = 0.729$ (inside the envelope, right before the interface) to $A_V = 0.748$ (inside the disk). The transition is sharp and is here located early ($A_V < 2$), so that the self-shielding of $H_2$ should play an important role, as expected for such conditions (see Sect. 3.2.2). The $C^+$/CO transition starts inside the molecular disk, at $A_V = 0.742$ and stops at $A_V = 2.49$ and is thus thicker as also expected from classical PDR studies.

The gas temperature is 285 K at the ionisation front and increases up to 650 K at $A_V = 0.63$ (Fig. 5.5, second pannel), because of the charging effect on photo-electric efficiency. It then decreases to about 320 K before jumping to 1100 K at the disk surface and then slowly decreases. Note that a smoother density profile (which we tested), rather than a jump, does not significantly change the temperature profile except that it is slightly smoothed too. We thus choose to keep a jump for simplicity. The initial increase inside the envelope can be attributed to an increase of the photoelectric-effect efficiency related to the increasing concentration of neutral grains (Fig. 5.5, third pannel). The trend is inverted just before reaching the disk (and the $H/H_2$ transition) where the efficiency drops. The photoelectric-effect is by far the main heating mechanism inside the envelope. However, at the disk surface, several heating mechanisms contribute significantly

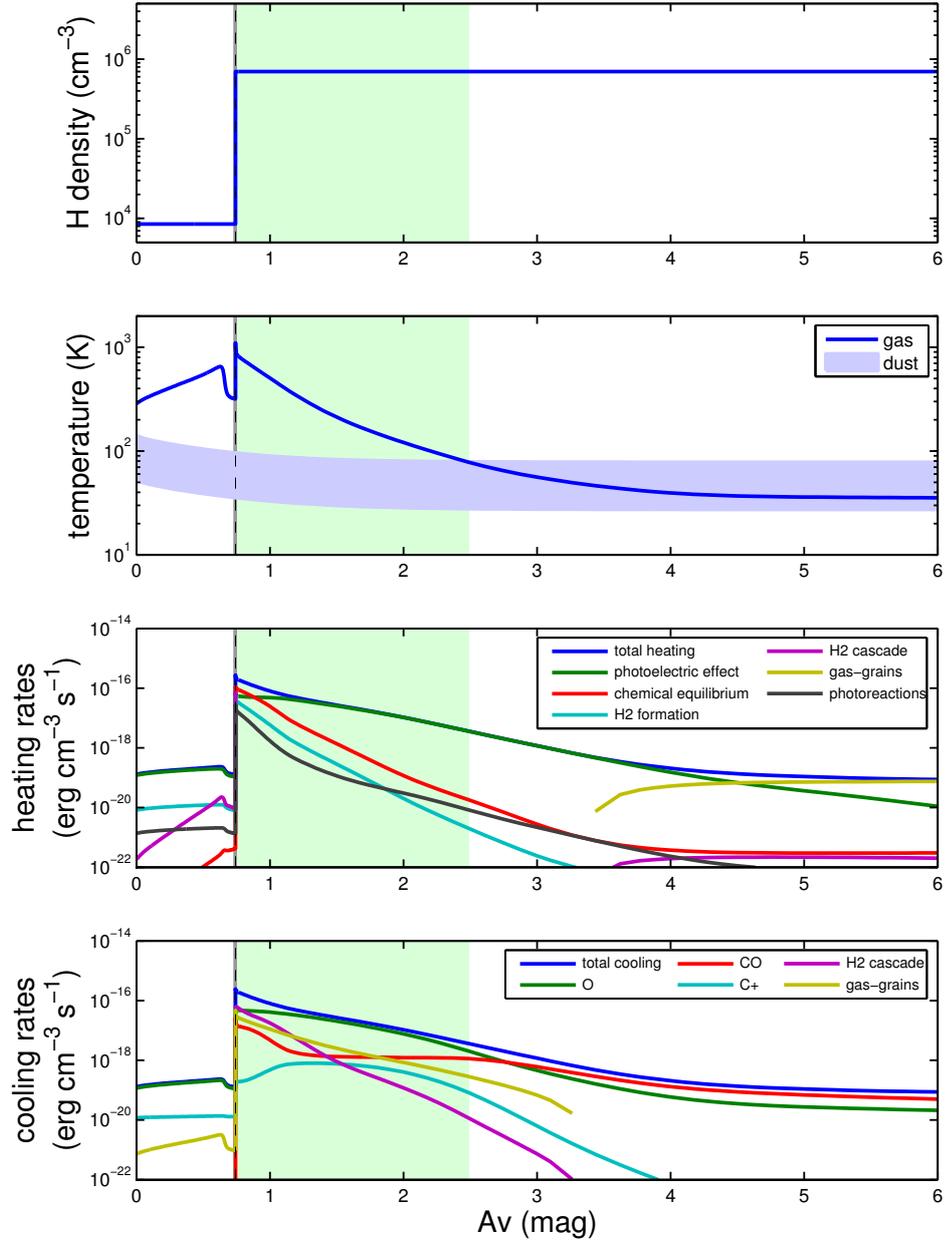

Figure 5.5: PDR structure of the best-fit model for the proplyd 105-600. FUV radiation field is incoming from the left where the ionisation front is located at $A_V = 0$. The disk surface, represented by a vertical dashed black line, is located at $A_V = 0.742$. The small dark grey patch close to the disk surface illustrates the H/H$_2$ transition (10 to 90% of H in H$_2$) and the bright green patch illustrates the C$^+$/CO transition (10 to 90% of C in CO). The dust temperature is given as a range corresponding to the temperature of the biggest to the lowest studied grains (radius of 3 to 300 nm). Note that the x-axis is in magnitude of visual extinction. In distance, the envelope represents about 11000 AU while the green patch in the disk represents 300 AU.

Table 5.4: Physical parameters estimated with the model for the studied prolyds.

| Object | $P_{trans}$ | $T_{surf}$ (K) | $N_{env}$ (cm$^{-2}$) | $r_{disk}$ (AU) | $\dot{M}$ supercritical (M$_\odot$/yr) | $\dot{M}$ subcritical (M$_\odot$/yr) | $\dot{M}$ observed (M$_\odot$/yr) |
|---|---|---|---|---|---|---|---|
| 105-600 | 1.00 | 1100 | 1.4e+21 | 2500$^a$ | 7.5e-05 | ... | 1e-06 |
| HST10 | 1.05 | 1150 | 2.0e+21 | 85$^b$ | 5.0e-07 | 1.7e-08 | $(7 \pm 5)$e-07 |
| 244-440 | 1.01 | 1270 | 1.9e+21 | 193$^b$ | 1.4e-06 | 8.2e-09 | $(15 \pm 7)$e-07 |
| 203-506 | 1.12 | 1080 | 1.0e+21 | 150$^b$ | 9.4e-08 | 9.9e-09 | ... |

**Notes.** Densities and temperature are derived form the best fit model for each object. The observed mass-loss rates derived at the ionisation front are from Henney and O'Dell (1999) for HST10 and 244-440, and derived by us from H$\alpha$ images for 105-600 (see text for details). A central star of 1 M$_\odot$ is used in the calculation of the mass-loss rate for 105-600 while, for the Orion proplyds, we used a 0.1 and 1 M$_\odot$ for the *supercritical* and *subcritical* cases respectively.
$^{(a)}$ Disk radius roughly approximated by this study (see Sect. 5.4.1).
$^{(b)}$ Disk radius estimated by spatially resolved observation: HST10 from Chen et al. (1998), 244-440 and 203-506 from Vicente and Alves (2005).

(Fig. 5.5, third pannel): chemical reactions (37%), photoelectric-effect (24%), formation of H$_2$ on grains (17%,) collisional de-excitation of UV-pumped H$_2$ (15%) and direct ionisation/dissociation by FUV photons (7%). Beyond $A_V = 1$, the photoelectric-effect becomes again the main heating mechanism up to $A_V = 4.5$, where the gas-grains collisions take the lead and keep the gas temperature close to the dust temperature. However, at this location, our simple constant density PDR model is not appropriate anymore and proper modelling will require specific disk models.

Inside the envelope, owing to the relatively high gas density, the main cooling mechanism is the fine-structure line emission of [OI] 63 μm, which is one order of magnitude stronger than the [CII] 158 μm line. At the surface of the disk, the cooling is shared between the cascade of H$_2$, the gas-grains collisions, the emission of O and CO. Deeper in the disk, CO and O dominate the cooling.

### 5.4.2 Results for Orion proplyds

For the three Orion proplyds, few detections of gas lines are available. For these sources the far-infrared dust emission was not detected since it is drowned in the emission from the nebula. We therefore only attempt to reproduce the observed line emission with our model. Intense emission of the [OI] 6300 Å line is observed towards the disks of HST10 and 203-506 as reported by Bally et al. (1998). This line arises from the hot neutral layers of the PDR. Its excitation is the result of the rapid formation of OH through the reaction of O with H$_2$, followed by photodissociation leaving a large fraction of O in an electronically excited state which radiatively decays through emission in the 6300 Å line. Since this process was not included in the public version of the Meudon PDR code (it will be in the next one), we therefore calculate the intensity of the [OI] 6300 Å line using the prescription of Störzer and Hollenbach (1998), using the column density of OH predicted by our model. In order to fit the observations, we follow the same strategy as for 105-600, i.e. we vary the envelope and disk density. Comparison between observations and the best-fit model are shown in Fig. 5.6, while parameters and outputs of the best-fit models are given in Tables 5.1 and 5.4.

For HST10, models cannot produce more than $10^{-6}$ W m$^{-2}$ sr$^{-1}$ of [CII] emission – including changing the envelope density, the small grains abundance or dust extinction properties – which is well below the upper limit extracted from observation. This is consistent with the idea that the observed emission should be dominated by the [$^{13}$CII] emission from the HII region expected at the same frequency. Overall this suggests that the envelope properties are not well constrained here. The [OI] 6300 Å line intensity is initially strongly underestimated but is better reproduced if the mechanism of formation for the excited oxygen described by Störzer and Hollenbach (1998) is included in the model and if an $n_{env} \leq 4 \times 10^5$ cm$^{-3}$ is used. Above this value, the envelope column density is large enough that OH starts to form inside the envelope, which is incompatible with observations showing that the emission occurs close to the disk surface (Bally et al., 1998). An envelope density of $n_{env} \leq 4 \times 10^5$ cm$^{-3}$ corresponds to an envelope column density of about $N_{env} = 3 \times 10^{21}$ cm$^{-2}$ which is slightly below the estimated value of $5.5 \times 10^{21}$ cm$^{-2}$ from Störzer and Hollenbach (1998). The H$_2$ and [OI] 6300 Å lines are best explained with a disk density of the order of $10^6$ cm$^{-3}$. The best-fit model gives $n_{env} = 2.5 \times 10^5$ cm$^{-3}$ and $n_{disk} = 4 \times 10^6$ cm$^{-3}$. The disk density is consistent with the estimate of Störzer and Hollenbach (1998) who found the same value. The temperature at the disk surface predicted by the best fit model is 1150 K.

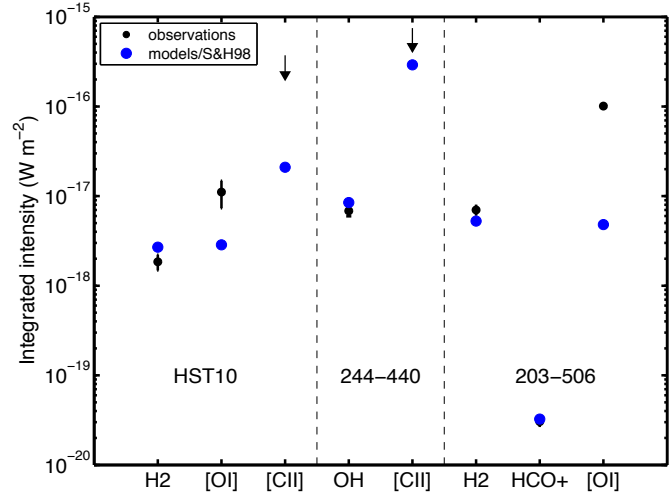

Figure 5.6: Comparison between observed line integrated intensities and best-fit models for the Orion proplyds. Observations are shown in black while estimated line integrated intensities using the outputs of the models (including the method of Störzer and Hollenbach (1998) for the [OI] 6300 Å line) are shown in blue. Heads of black arrows give the upper limits for [CII] emission in the case of HST10 and 244-440.

For 244-440, [CII] arising mainly from the envelope is probably detected but, as HST10, drowned in the nebula emission. The OH line is well detected and should arise from the the disk (under the assumptions that this line is being emitted by the proplyd and no other object of the Orion Bar in the line of sight). With two lines, whose one is an upper limit, the two densities will be poorly constrained, but we nevertheless found a clear best-fit for a model with $n_{env} = 1 \times 10^5$ cm$^{-3}$ and $n_{disk} = 2 \times 10^6$ cm$^{-3}$. The temperature at the disk surface predicted by the best fit model is 1270 K. These values are consistent with those found for the other Orion proplyds.

For 203-506, the disk density is well constrained by the observation of the H$_2$ and HCO$^+$ lines and the best-fit value is $n_{disk} = 1 \times 10^6$ cm$^{-3}$. The "envelope" density that corresponds to the ambiant PDR resulting from the photoevaporation of the Orion bar (Goicoechea et al., 2016) is estimated to be $n_{env} = 4.5 \times 10^4$ cm$^{-3}$. The temperature at the disk surface (which we consider being at the H/H$_2$ transition) predicted by the best fit model is 1080 K. Here again, we need to include the mechanism of Störzer and Hollenbach (1998) in the model in order to get closer to the observed [OI] 6300 Å line intensity but is still highly underestimated. This discrepancy could be explained by the emission from shocked regions or that another mechanism to form OH is missing, e.g. by reactions with vibrationally excited H$_2$.

The physical structures extracted for the Orion proplyds are not shown here because they are less well constrained than 105-600. They are found similar to the one given for 105-600 in Fig. 5.5, with a rapid increase of the gas temperature at the disk surface and the H/H$_2$ transition layer located close to it.

### 5.4.3 Comparison to other models

It is interesting to compare the results obtained here with those of earlier studies of proplyds using PDR models. When doing this comparison it is important to keep in mind that these studies usually do not compare their predictions to actual observations of PDR tracers towards these objects.

Störzer and Hollenbach (1999) and Adams et al. (2004) use a PDR model based on the classical Tielens and Hollenbach (1985b) model. Contrary to what is presented here, they consider a single constant density structure. Störzer and Hollenbach (1999) compare their models to the emission of vibrationally excited H$_2$ and [OI] 6300 Å emission observed in Orion proplyds (including HST10) and find good agreement. They also predicted the far infrared line intensities of the [CII] line, for which they give values ranging between 2.4 and $3.1 \times 10^{-6}$ W m$^{-2}$ sr$^{-1}$, which assuming an emitting size equal to the size of the envelope projected on the plane of the sky ($2.6'' \times 1''$, see Table 4.2) corresponds to a flux of 1.15 - 1.50 $\times 10^{-16}$ W.m$^{-2}$, just below the upper limit of $2.3 \times 10^{-16}$ W.m$^{-2}$ that we obtained. Their value is probably overestimated since the observation is supposed to be highly contaminated by the nebula emission. They have also predicted strong [OI] 63 µm emission close to our high upper limit of detection leading to the same conclusion. Detailed observation are needed to conclude clearly. The study of Adams et al. (2004) is essentially theoretical. While focusing on different FUV field ranges, the temperature profile is characterized in both studies by a region of constant and high temperature ($> 4000$ K) at low column densities followed by a drop of temperature at the location of the H/H$_2$ transition, which is situated, in their models, just above the disk. The temperature profile which we find here (Fig. 5.5) differs greatly from that of Störzer and Hollenbach (1999) and Adams et al. (2004) in that the temperature we find at low column densities is much lower (typically a few 100 K). This is mainly due to two fundamental differences which reduce the heating in the envelope in our models: 1) the lower density does not allow collisional de-excitation of H$_2$ to be important, 2) the low abundance of small grains that we have considered (in agreement with the observation that PAHs are underabundant in HST10 and in 105-600 (Vicente et al., 2013, and Sect. 5.4.1), reduces photoelectric heating drastically. Nevertheless, an important aspect in the frame of photoevaporation is that the temperatures at the disk surface derived by Störzer and Hollenbach (1999) and Adams et al. (2004) are typically of the order of a few 100 K, i.e. lower than ours. Our prediction for the H$_2$ (1-0) S(1) line for HST10 is in good agreement with theirs and with our observations.

## 5.5 Effect of the H/H$_2$ transition layer location on the gas thermal balance at the disk surface

In this section, we explore the effect of the location of the H/H$_2$ transition layer on the gas thermal balance, and the gas temperature, at the disk surface, i.e. the density jump in our model. As we have seen in Sect. 3.2.2, the location of this layer, i.e. the column density of hydrogen nucleus, when dominated by the H$_2$ self-shielding, is proportional to the FUV flux G$_0$ and inversely proportional to the density. Since the column density between the ionisation front and this layer is mainly set by the envelope density in the case of proplyds, we run our model for different values of $n_{\text{env}}$ and G$_0$ from $2 \times 10^3$ to $2 \times 10^5$. We also considered different envelope sizes: a large one corresponding to the size of the biggest studied proplyd, 105-600, and a small one corresponding to the size of HST10.

### 5.5.1 Temperature

First of all, we extracted the gas temperature at the disk surface for each model of the grid to see its evolution with the position of the H/H$_2$ transition layer. This is illustrated in Fig. 5.7 which presents the gas temperature at the disk surface as a function of the position of the H/H$_2$ transition which is captured by the ratio $P_{\text{trans}} = A_{\text{V}}^{\text{jump}}/A_{\text{V}}^{\text{H}-\text{H}_2} = N_{\text{env}}/N_{\text{H}-\text{H}_2}$, where $A_{\text{V}}^{\text{H}-\text{H}_2}$ is the visual extinction, or position, at which the transition layer starts, i.e. where 10% of the hydrogen is in molecular form, and $A_{\text{V}}^{\text{jump}}$ is the visual extinction setting the position of the density jump corresponding to the disk surface in our model. The H/H$_2$ transition is thus totally located inside the disk when $P_{\text{trans}} < 1$, and is located or starts in the envelope when $P_{\text{trans}} > 1$. A similar behaviour is observed for all the studied cases: a *hot* ($T = 1500 - 5000$ K) regime prevails when the transition is located inside the disk ($P_{\text{trans}} < 1$) while a *warm* regime ($T < 500$ K) is observed when the transition is located inside the envelope ($P_{\text{trans}} > 1$). A *transitional* regime ($T \sim 500 - 1500$ K) lies between the *hot* and *warm* regimes at $P_{\text{trans}} \sim 1$.

### 5.5.2 Heating and cooling mechanisms

To gain insight into the mechanisms driving this temperature behaviour, we also studied the evolution of the heating and cooling mechanisms as a function of the H/H$_2$ transition layer position. Results are given in Fig. 5.8 and Fig. 5.9 respectively. The following observations can be made: when the envelope density (and visual extinction) is small and the transition is inside the disk ($P_{\text{trans}} < 1$, $T \sim 1500 - 5000$ K), the collisional de-excitation of UV-pumped H$_2$ is very efficient and dominates the heating, except for very intense FUV fields ($G_0 > 10^5$) where the heating by the photo-electric effect can be even higher, while cooling is dominated by gas-grain collisions. This corresponds to the *hot* regime. When the transition is located inside the envelope ($P_{\text{trans}} > 1$, $T < 500$ K), the photoelectric effect on grains dominates the heating, and cooling is dominated by emission in fine structure lines of oxygen. This corresponds to the *warm* regime which is the classical solution for thermal balance usually observed in PDRs. Finally, in the *transitional* regime, i.e. when the transition is close to the surface but starts inside the envelope ($P_{\text{trans}} \sim 1$, $T = 500 - 1500$ K), the contribution of a collection of heating processes (exothermal reactions, H$_2$ formation and direct absorption of ionizing/dissociation photons, exothermal chemical reactions) control the heating and cooling also occurs through multiple processes (e.g. collisional excitation of H$_2$, OH and CO emission lines)

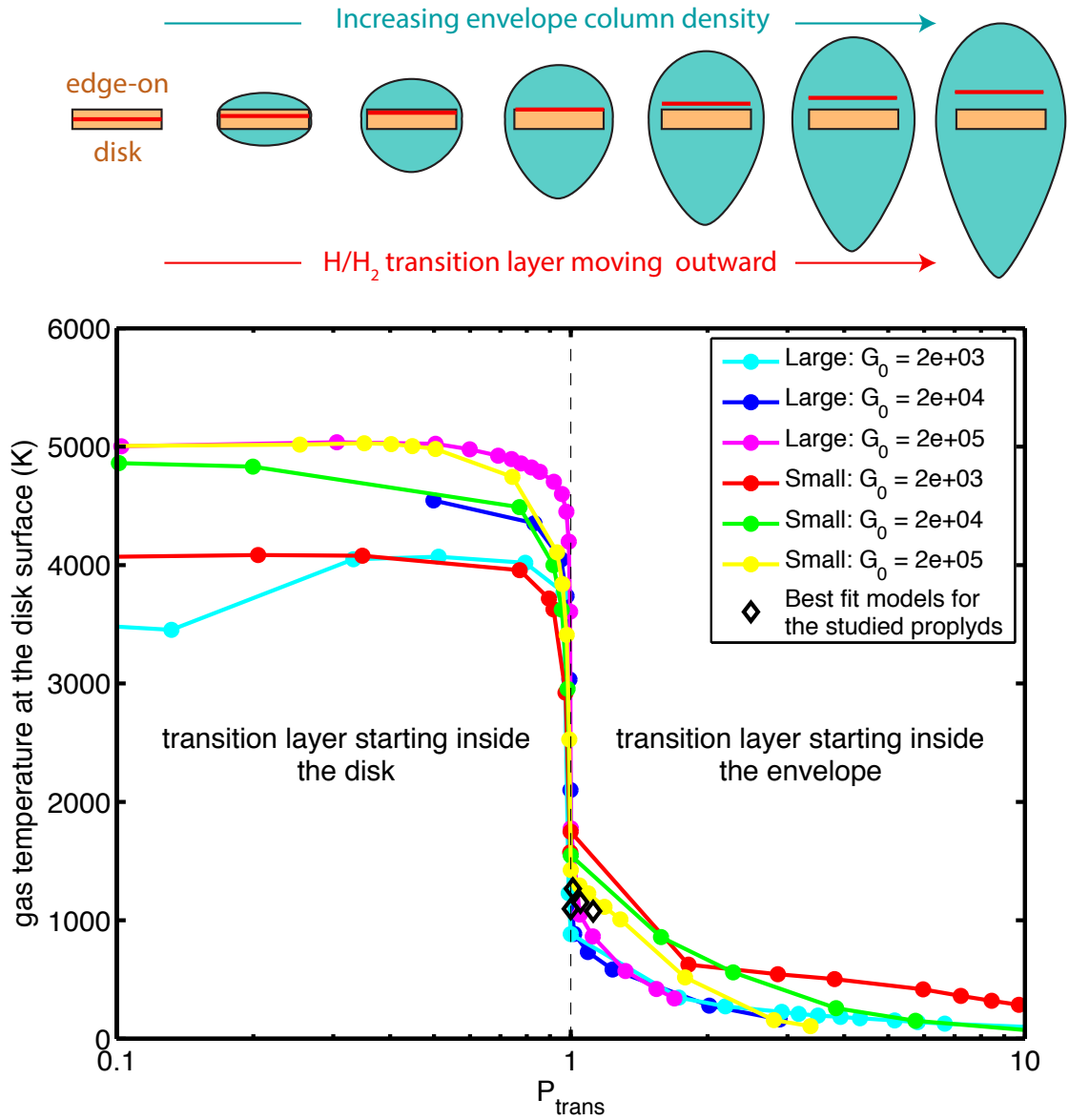

Figure 5.7: Modelled temperature at the disk surface as a function of the position of the H/H$_2$ transition for various radiation fields and adopting the geometries of 105-600 as a large proplyd and HST10 as a small proplyd. The points are the positions where the models are calculated and corresponding solid lines are cubic interpolations. The dotted vertical line corresponds to $P_{\mathrm{trans}} = 1$. On the left side of this line, the H/H$_2$ transition is inside the disk, while on the right side of this line the H/H$_2$ transition is inside the envelope. The evolution from left to right when the envelope is growing is illustrated by the top panel.

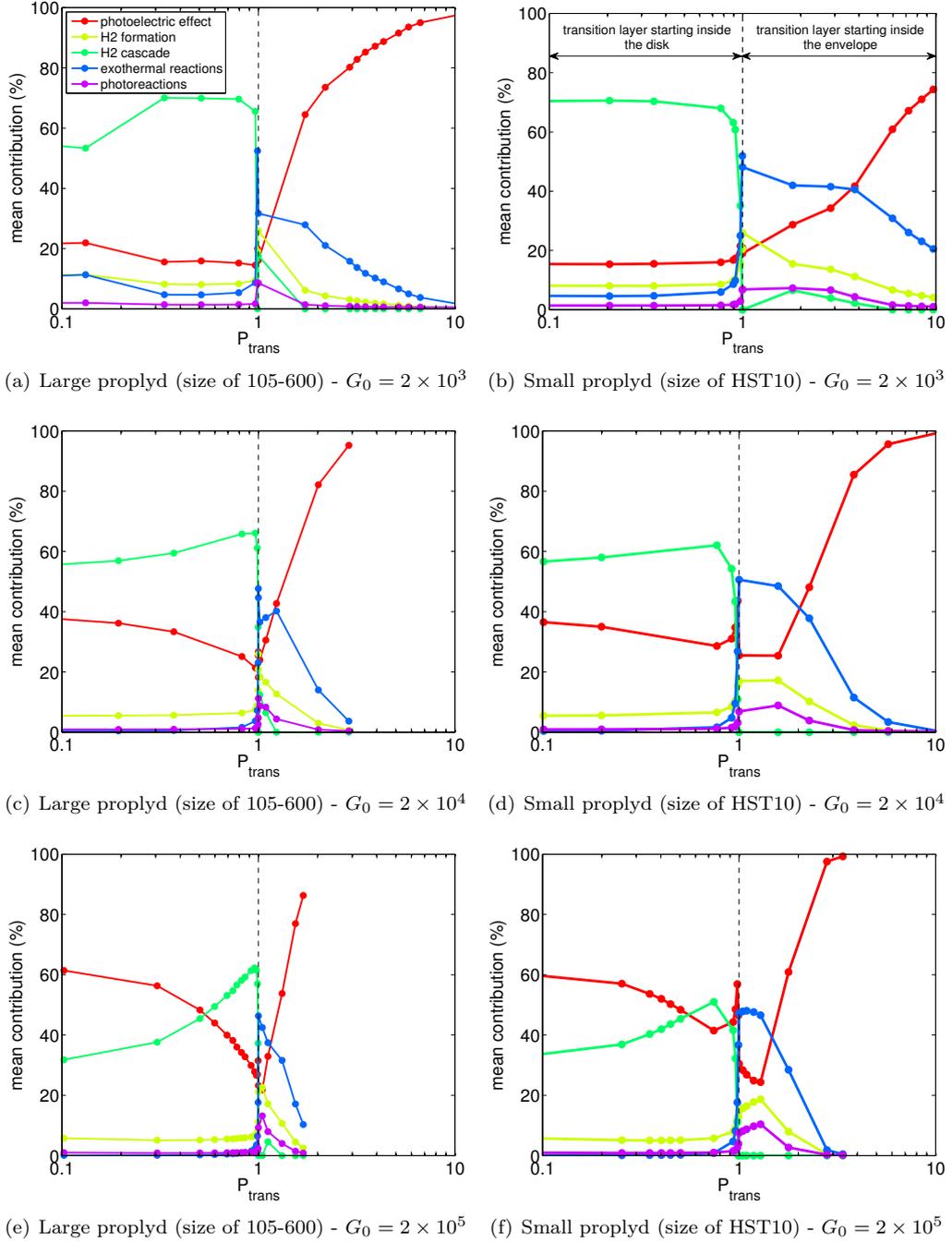

Figure 5.8: Relative contribution of the main heating processes at the surface of the disk for different sizes of proplyd (size of 105-600 or HST10) and FUV fields ($2 \times 10^3$, $2 \times 10^4$ and $2 \times 10^5$) as a function of the location of the H/H$_2$ transition.

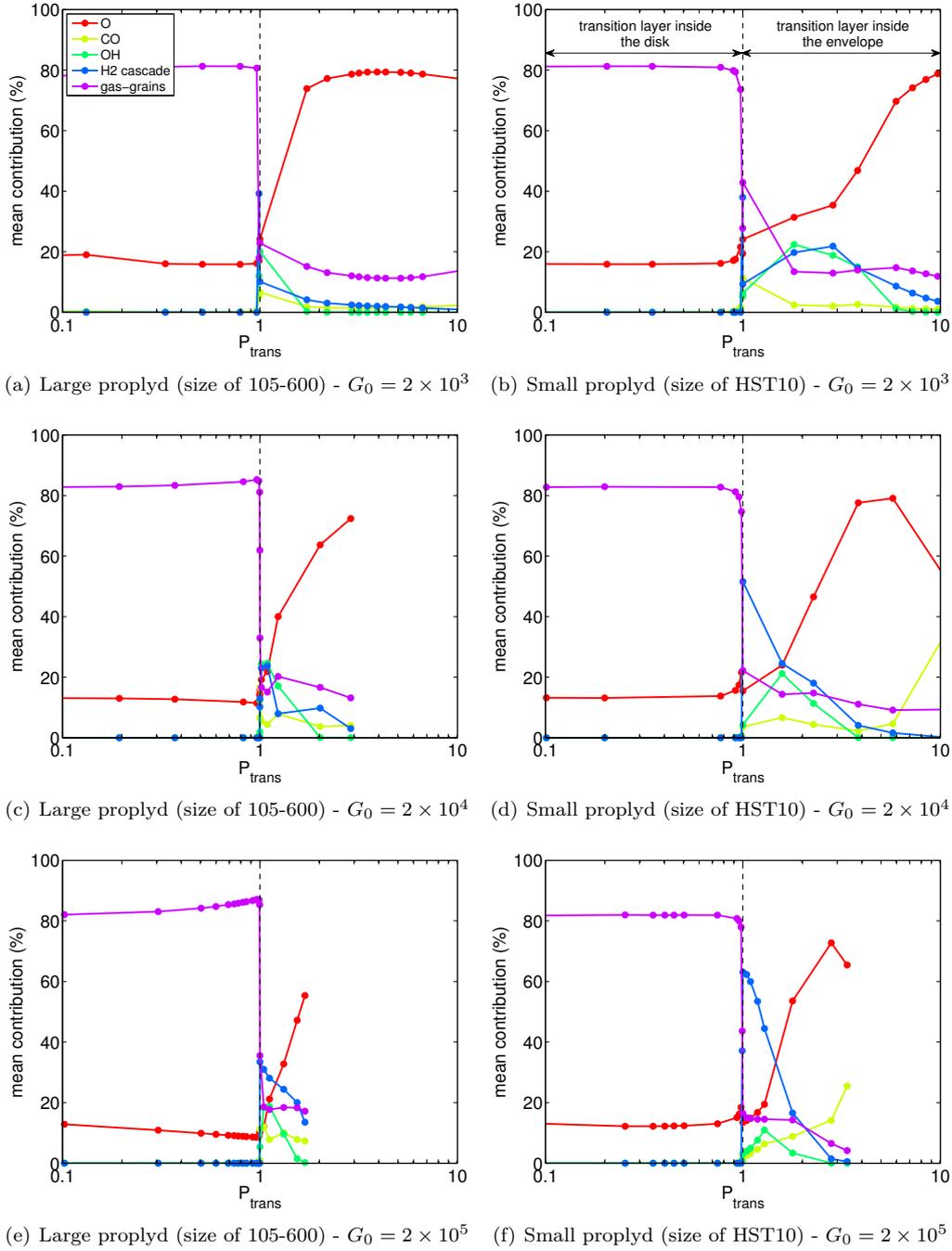

Figure 5.9: Relative contribution of the main cooling processes at the surface of the disk for different sizes of proplyd (size of 105-600 or HST10) and FUV fields ($2 \times 10^3$, $2 \times 10^4$ and $2 \times 10^5$) as a function of the location of the H/H$_2$ transition.

### 5.5.3 Evidence for a specific energetic regime

At this point it is interesting to note that for each object, the H/H$_2$ transition in the best-fit model is located close to the disk surface ($P_{\text{trans}} \gtrsim 1$, cf. Table 5.4 and Fig. 5.7). This is in agreement with spatially resolved observations which for instance show that the emission of vibrationally excited H$_2$ (for HST10) and [OI] 6300 Å (for HST10 and 203-506) arises from the disk surface or just above (Chen et al., 1998; Bally et al., 1998). This implies that proplyds appear to be in the *transitional regime* with temperatures at the disk surface close to 1000 K. Guided by Fig. 5.7, we build the following reasoning:

- When the isolated disk (before the formation of the envelope, corresponding to the left edge of the Fig. 5.7) is irradiated by the external FUV photons, the H/H$_2$ transition forms inside the disk, and the temperature rises to some several 1000 K, driving an intense photoevaporation flow.

- This increases the column density of the envelope $N_{\text{env}}$, dragging the H/H$_2$ transition out of the disk ($P_{\text{trans}} = N_{\text{env}}/N_{\text{H}-\text{H}_2} > 1$) which lowers the temperature at the disk surface (corresponding to a move towards the right on the figure till the middle or a bit more).

- Consequently, the evaporating mass flux will decrease and so will the envelope column density due to spherical divergence (corresponding to a move towards the left). At one point the H/H$_2$ transition will go back inside the disk ($P_{\text{trans}} < 1$), temperature will rise, mass-loss increase as well as envelope column density.

- This continues iteratively so that a stable equilibrium may be found around $P_{\text{trans}} \simeq 1$ (middle of the figure).

Overall, this suggests that proplyds are in an equilibrium where the temperature of the disk surface is set so that mass loss is sufficient to keep the envelope column density to a value that maintains the H/H$_2$ transition close to the disk surface. In this specific case, the disk surface temperature is set by a combination of numerous heating and cooling processes and reaches values of the order of 1000 K (for radiation fields between $G_0$=2000 and $G_0$=2 × 10$^5$).

## 5.6 Photoevaporation and mass-loss rates

At this point it is interesting to discuss photoevaporation and mass loss of proplyds in the context of our results. Namely, knowing the size of the disks and a possible range of masses for the central stars from independent studies, we can use the surface temperature and densities from our models (Tables 5.1 and 5.4) to estimate photoevaporation rates.

### 5.6.1 Definitions

The gravitational radius where the kinetic energy of gas particles is equal to the gravitational potential is defined as (Hollenbach et al., 1994)

$$r_{\text{g}} = \frac{G M_*}{c_{\text{S}}^2}, \tag{5.7}$$

where $c_{\text{S}}$ is the sound speed, i.e.

$$c_{\text{S}} = \sqrt{\frac{\gamma \, k_{\text{B}} \, T}{\mu \, m_{\text{H}}}}. \tag{5.8}$$

The term $\mu\, m_H$ is the mean particle weight, $\gamma$ is the adiabatic index taken as 7/5 here - the value for a perfect diatomic gas - and $k_B$ the Boltzmann constant. The mass loss rate is defined for two cases, which depend on the value of the disk radius $r_{\text{disk}}$:

**Supercritical case** when $r_{\text{disk}} > r_{\text{g}}$.

The mass loss rate occurs mostly from the disk surface (vertical mass-loss) and is given by (Johnstone et al., 1998),

$$\dot{M} = \mu m_H n_{\text{disk}} c_S \times 2\pi(r_{\text{disk}}^2 - r_{\text{g}}^2). \tag{5.9}$$

**Subcritical case** when $r_{\text{disk}} < r_{\text{g}}$.

The mass loss rate occurs from the disk edge (radial mass-loss) and is given by (Adams et al., 2004),

$$\dot{M} = C_0\, N_C\, \mu\, m_H\, c_S\, r_g \left(\frac{r_g}{r_d}\right) \exp\left[-r_g/2r_d\right], \tag{5.10}$$

where $C_0$ is a dimensionless constant of order unity and $N_C$ is defined as the atomic region of the PDR in their models which must be of the order of $N_{\text{env}}$ in ours.

### 5.6.2 Carina candidate proplyd 105-600

For 105-600, the disk (or candidate disk) size is unknown. The value of $S_{\text{disk}}$ derived from the model can be converted in an equivalent radius considering that the disk is in the plane of the sky. This yields a value of about $1.1''$ or 2500 AU. The temperature at the surface of the disk is estimated to be about 1100 K. For such a large radius, $r_{\text{disk}} > r_{\text{g}}$ for masses of central star up to 30 $M_\odot$, and the *supercritical* regime prevails everywhere. We derive for this case a mass-loss rate $\dot{M} = 7.5 \times 10^{-5}$ $M_\odot$/year. The mass-loss at the ionisation front can be derived from the H$\alpha$ emission, and we find a value of the order of $2 \times 10^{-6}$ $M_\odot$/year, about 40 times smaller. This discrepancy could be explained if the envelope is in a phase where it is gaining mass (see Sect. 5.5.1 ) while depleting the disk. We note that the Orion proplyds, which seem to be at equilibrium (mass-loss at the ionisation is roughly equal to the "disk" mass loss computed here, see Sect. 5.6.3), generally exhibit a tear-drop shaped envelope as HST10 (Fig. 4.2(b)) as the consequence of this equilibrium. It is thus possible that the sharper shape of 105-600 (Fig. 4.2(a)) results from an envelope that is still growing in mass. Since the growth phase is expected to be quick, e.g. forming a 0.05 $M_\odot$ envelope with such a rate would take few hundred years, 105-600 would thus be a very young object. Alternatively, it is possible that the disk radius we consider here is too large by about an order of magnitude. This would however pose a serious problem to our model when trying to reproduce the high-$J$ CO lines. Since a protoplanetary disk with a radius of 2500 AU appears unrealistic, what we call disk in this case could most likely refer to a young forming disk from the original cloud material (similar to other large "pseudo-disks" observed, e.g. as in Quanz et al., 2010). Further observations at (sub)millimeter wavelengths to resolve the structure and kinematics of the disk will be required to settle this issue.

### 5.6.3 Orion Proplyds

For HST10 and 244-440, Henney and O'Dell (1999) derived the mass-loss rates at the ionisation front from spectroscopic observations of ionised gas tracers and found values of $(7 \pm 5) \times 10^{-7}$ and $(15 \pm 7) \times 10^{-7}$ $M_\odot\,\text{yr}^{-1}$ respectively. For 203-506, since the envelope ionisation front is not observed because the disk is embedded in neutral gas, the mass-loss cannot be derived from ionised gas tracers.

For the Orion proplyds whose disk outer radii are known (Vicente and Alves, 2005), we can derive the mass-loss rates from our model using the analytical prescription presented above. The masses of their central stars are not precisely known but taking into account the stellar mass function of the cluster (e.g. Hillenbrand and Carpenter, 2000), they are very likely about 0.1 $M_\odot$ or less. A surface temperature of the order of 1000 K, as obtained from our study, would result in a gravitational radius of 20 AU, or less, and implies the *supercritical* case ($r_{\text{disk}} > r_{\text{g}}$) for all disks. The *subcritical* case is still possible, but unlikely because we need to assume that central stars have a mass of about 1 $M_\odot$, resulting in a gravitational radius of the order of 200 AU. We nevertheless consider both cases in our calculations. The mass-loss rates for each case are given in Table 5.4. In the *subcritical* regime, we find values of the order a few $10^{-8}$ $M_\odot$/yr, in agreement with the results of Adams et al. (2004), but which are too low compared to the observed values. On the other hand, the *supercritical* values of mass-loss are in better agreement with the mass loss derived at the ionisation front. Hence, to reconcile the photoevaporation rates derived from our best fit models with those derived from the observations by Henney and O'Dell (1999), one must assume that the flows are in a *supercritical* regime. This is compatible with the idea that central stars have masses well below one solar mass. If this is indeed the case, given that disk masses do not significantly exceed 0.01 $M_\odot$ in mass, this implies short lifetimes for these disks (of the order of several $10^4$ years) which suggests that we should not see so many of them, or that the ionizing star $\Theta^1$ Ori C is particularly young, a conundrum that is known as the "proplyd lifetime problem" (Henney and O'Dell, 1999). Clarke (2007) studied in details the dynamical evolution of proplyds assuming they are in the *subcritical* regime and found that their lifetime could in this case be relatively large, solving this issue. In the light of the results presented here it appears that the *subcritical* assumption may not be correct. Instead our results are in line with the idea that the Orion Nebula is young, i.e. that $\Theta^1$ Ori C is less than a few $10^4$ years old, and that proplyds are indeed rare objects which we happen to see in Orion by chance. Another possibility is that the *supercritical* regime is not sustained at the same rate during the whole proplyd life and declines with time so that the lifetimes could be not so short.

## 5.7 Conclusions on photoevaporation within PDRs of proplyds

In this part, I have presented the first far-infrared observations of dense PDR tracers emitted by proplyds, obtained with the *Herschel space observatory* and ALMA. Based on the detailed Meudon PDR code, I have developed a 1D-model for the PDR of a proplyd to predict the emission arising from it. This model successfully reproduces most observed lines, with the exception of HCN (4-3) in 105-600, and is somewhat lower than the observed intensities of [OI] 6300 Å (and possibly [CII] 158 µm) in HST10 and 244-440. The [OI] 6300 Å line is most sensitive to the densest gas (while the most diffuse for [CII] 158 µm) in proplyds which are not well described in our simplified model.

For all the sources, at the disk surface, we found densities of 0.7 to $4 \times 10^6$ cm$^{-3}$ and temperatures of 1000-1300 K. We found that the position of the H/H$_2$ transition layer is the critical parameter determining the disk surface temperature. Related to that, our results suggest that proplyds are in a self-regulated regime where the temperature of the disk surface is set in a way that the resulting mass loss keeps the envelope column density sufficient to bring the H/H$_2$ transition out of the disk, yet close to its surface. We found that gas energetics in this specific case results from a complex combination of several heating and cooling mechanisms, hence it can not be captured by the classical solution where photoelectric effect on grains dominates the heating and [OI] and [CII] far infrared emission dominates the cooling.

We derived mass-loss rates that are large (of the order of a few $10^{-7}$ M$_\odot$/yr for the Orion proplyds and a few $10^{-5}$ M$_\odot$/yr for 105-600). Most of them are consistent with spectroscopic studies of the ionisation front in theses objects and suggest that proplyds undergo supercritical photoevaporation, and indeed may have a short lifetime.

To understand proplyd evolution, in a next step, it is necessary to couple hydrodynamic simulations of a supercritical evaporation flow with a correct treatment of PDR physics and chemistry (this will be developed in the Part III). Concomitant with the development of models, spatially resolved observations of molecular lines close to the surface of the disks, or PDRs in general, would help to constrain the physical and dynamical properties of the photoevaporative flows, and to understand the evolution of a proplyd. In particular, the high angular resolution provided by ALMA and the future JWST will enable us to resolve photoevaporative flows and study their properties and dynamics (the gain in spatial resolution in the IR brought by the JWST is illustrated on Fig. 5.10).

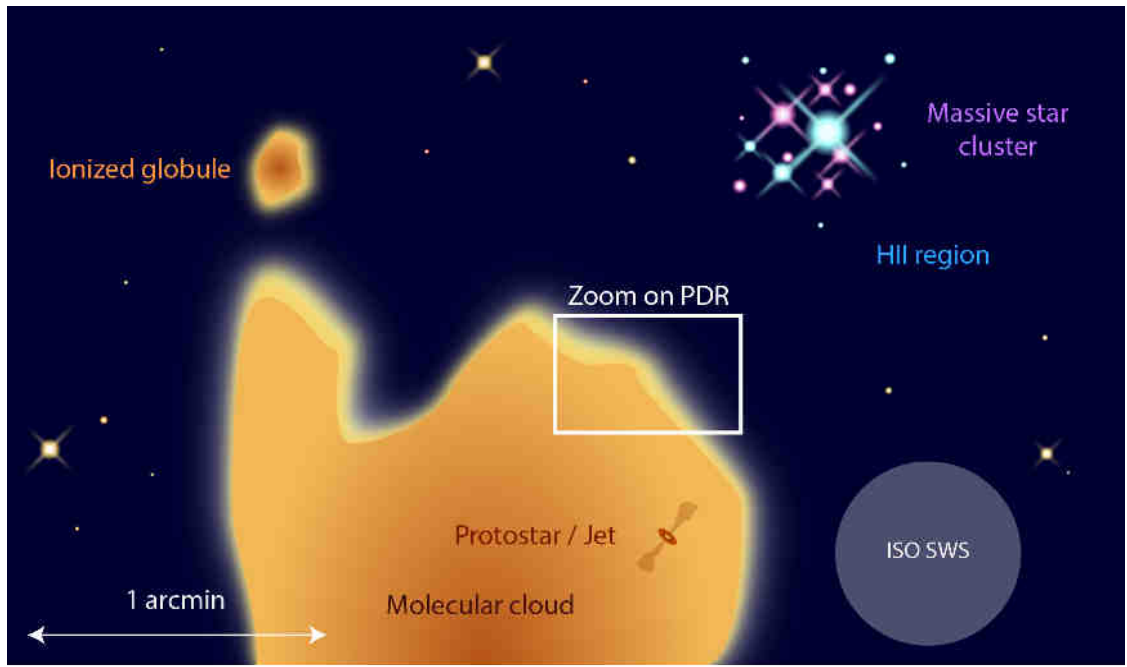

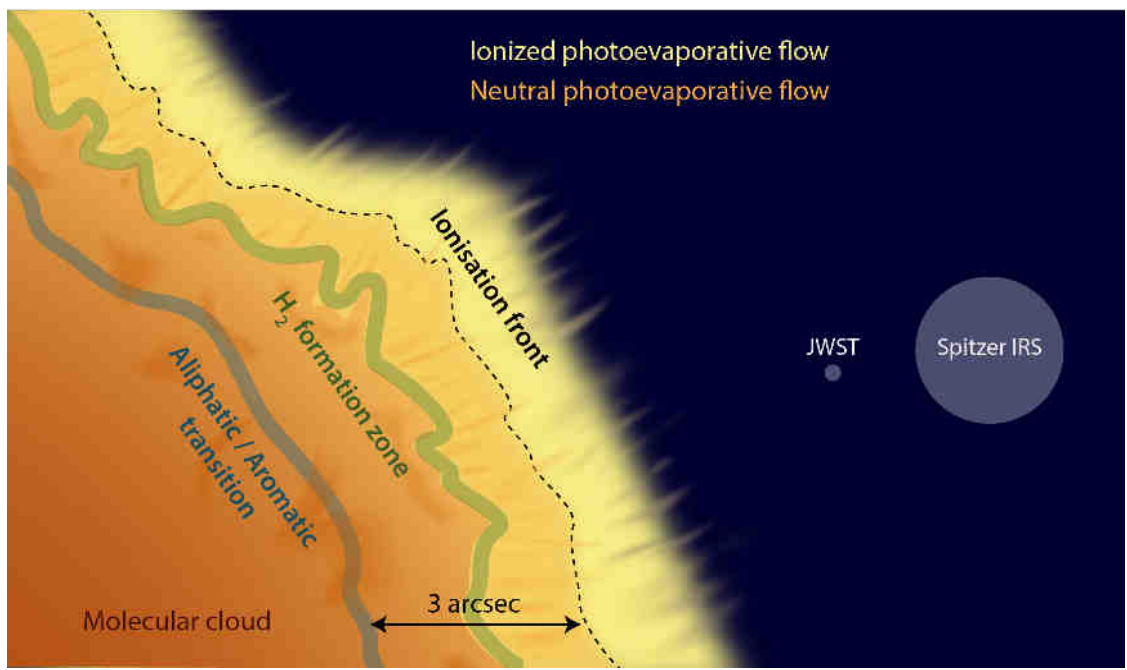

Figure 5.10: Illustration of the PDR at the edge of a molecular cloud close to a massive star cluster, and the gain in spatial resolution that should be able to bring the *James Webb Spatial Telescope*.

# Part III

# Modelling the dynamical evolution of externally illuminated photoevaporating protoplanetary disks

# Table of Contents

## Chapter 6

# Viscous motion of disks

## 6.1 Effective viscosity

### 6.1.1 Need for a process to transport angular momentum

Protoplanetary disks have a limited lifetime (Sect 2.3.1). One reason is that matter is accreted onto the central star. Within the protoplanetary phase, this input of mass is not significant for the star anymore, as it is already almost formed, but this represents a substantial fraction of the disk mass that is lost. Accretion onto the central star is possible only if the equilibrium between the pressure, the gravitational and the centrifugal forces is broken. More precisely, to produce an inward motion of matter, and so accretion, a process that is able to extract angular momentum is needed.

A natural solution would be that molecular viscosity of an hydrodynamical flow could play that role. Indeed, since the angular velocity decreases with the radius, according to the law of gravitation, there is a differential motion between two neighbouring annuli of matter. The friction between the two makes the inner annulus, which is the fastest, to be slowed down. Reciprocally, the outer annulus is accelerated. The side effect of this outward transfer of angular momentum is a spreading of the disk: the inner annulus, by losing energy and angular momentum, moves inwards and eventually is accreted, while the energy and angular momentum transferred to the outer annulus makes it to move outwards. However, the molecular viscosity of protoplanetary disks, that can be roughly approximated by the product of the velocity and the mean free path of a molecule (e.g. Lynden-Bell and Pringle, 1974), is by many orders of magnitude too weak to be responsible for the observed accretion.

Other processes have been proposed (see the review of Papaloizou and Lin, 1995), such as non axisymmetric propagating global waves, or magneto-hydrodynamical winds or jets[1]. They are however not convincing to explain the omnipresence of accretion in disks. Indeed, global waves are only suitable for cases with a source of assymmetry, for example when there is a companion stellar object within the disk, or a different inclination of rotational axis between the star and the disk. Winds and jets (e.g. Blandford and Payne, 1982; Shu et al., 1994) may both represent a loss of angular momentum via the loss of mass, but generally require an important magnetic field. This process is thus not believed to be dominant in the general case. Photoevaporation winds also remove mass, but the loss of angular momentum that it represents primary affects the ejected particles, so that the accretion in the remaining disk is not significantly enhanced.

---

[1] supersonic collimated flow ejected perpendicularly from the disk plane at the star location

Finally, the solution generally accepted today is that the turbulence is the main source of local angular momentum transfer leading to global scale motions.

### 6.1.2 An effective viscosity driven by turbulence

Turbulence is characterised by a large distribution of local speeds and directions for the particles of a flow. The relative motion between two fluid particles, resulting from turbulent fluctuations, leads to a local exchange of energy. Including turbulence, any process introducing a local exchange of energy, acts in a similar way than a molecular viscous constrain does. Consequently, the turbulence of a disk can be treated as being an effective viscosity.

There is two different approaches to describe the effective viscosity when its driven by turbulence. The classical and easiest approach is to postulate a simple scaling relation between the effective viscosity, $\nu$, and some locally defined properties of the flow. Shakura and Sunyaev (1973) and Lynden-Bell and Pringle (1974) were the first to use an ad-hoc effective viscosity, without any assumption on its origin at this time, but based on observational constraints and theoretical considerations (see Sect. 6.1.3). In their description, known as the "α-model", the effective viscosity is given by

$$\nu = \alpha \frac{c_S^2}{\Omega_K} = \alpha c_S H, \tag{6.1}$$

with $c_S$ is the sound speed, $\Omega_K$ the Keplerian angular velocity, and $H = c_S/\Omega_K$ the pressure scale height derived from vertical hydrostatic equilibrium in the disk. In this model, $\alpha$ is a dimensionless parameter that captures the efficiency of the underlying process, believed to be the turbulent diffusion, and here assumed constant within the disk. Assuming its turbulent origin, another way to treat the effective viscosity is to theoretically compute its value. This requires to estimate the structure of the turbulence and evaluate the resulting stresses, described as the Reynolds (hydrodynamical) and Maxwell (magnetic) stresses (Balbus and Hawley, 1998). In our case, and as it is generally done in the literature, we will focus on the ad-hoc α-model.

### 6.1.3 Quantifying the effective viscosity

The ad-hoc value of $\alpha$, that parametrise the efficiency viscosity, can be estimated thanks to theoretical considerations and observational constraints. Studies of the evolution of disk populations with time can constrain angular momentum transport, via the time dependence derived in the theoretical model, and thus the value of $\alpha$ (e.g. Hartmann et al., 1998, who studied the mean values of mass-loss rates for clusters of different ages). Such studies lead to typical values around $10^{-2}$ (Hartmann et al., 1998; Andrews and Williams, 2007a). The value of α may also be derived from observations by measuring the magnetic fields and/or turbulent velocity fields (e.g. Hughes et al., 2011). The determination of the surface density profile, along with the mass-loss rate, yield the recovery of the viscosity profile under some assumptions (Andrews et al., 2009). More recently, Rafikov (2017) used a sample of protoplanetary disks resolved by ALMA with measured masses and stellar accretion rates to derive the value of $\alpha$ for each object and the general distribution (Fig. 6.1). They found that the inferred values do not cluster around a single value, but instead have a broad distribution centered around $10^{-3}$ and extending from $10^{-4}$ to 0.04. They found no significant correlation between those values and the global disk parameters (mass, size, surface density) nor the stellar characteristics (mass, luminosity, radius) or irradiation pattern, thus confirming the difficulty to estimate the effective viscosity a priori without specific observations.

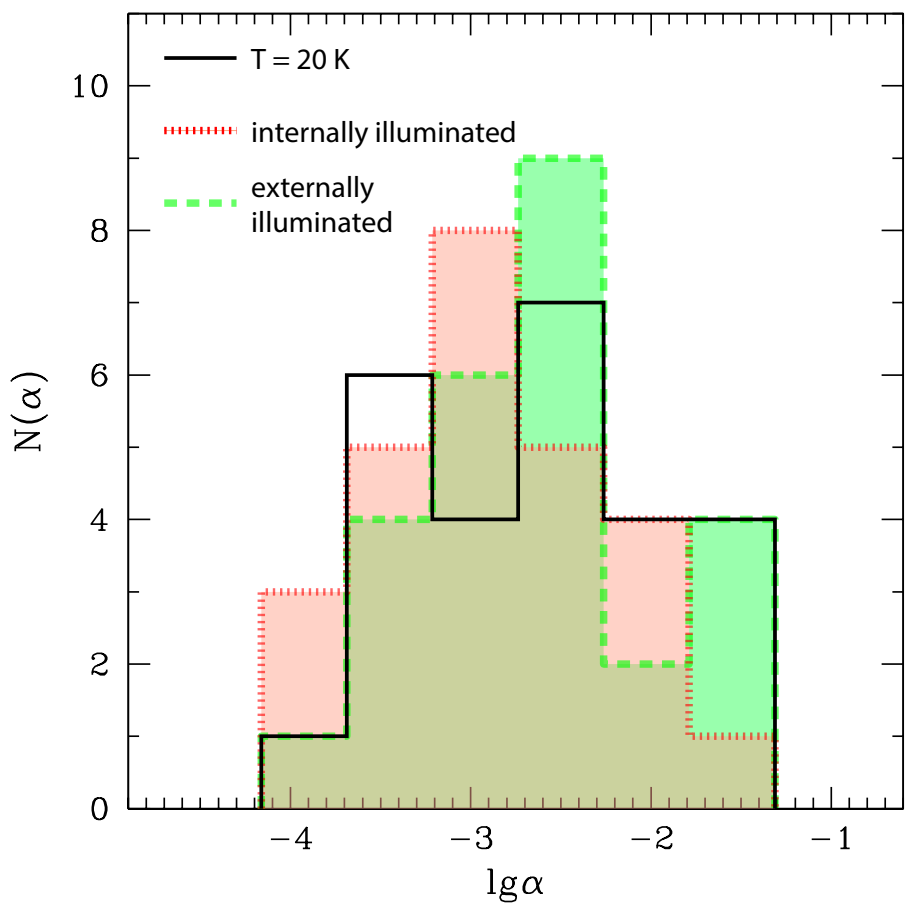

Figure 6.1: Distribution of inferred values of α for disks observed by ALMA, computed using different prescriptions of the outer disk temperature: the solid black histogram corresponds to a uniform value of 20 K, the dotted red histogram correspond to a temperature obtained by assuming optically thin dust directly illuminated by its central star, and the dashed green corresponds to an optically thick and externally irradiated passive protoplanetary disk. Figure adapted from Rafikov (2017).

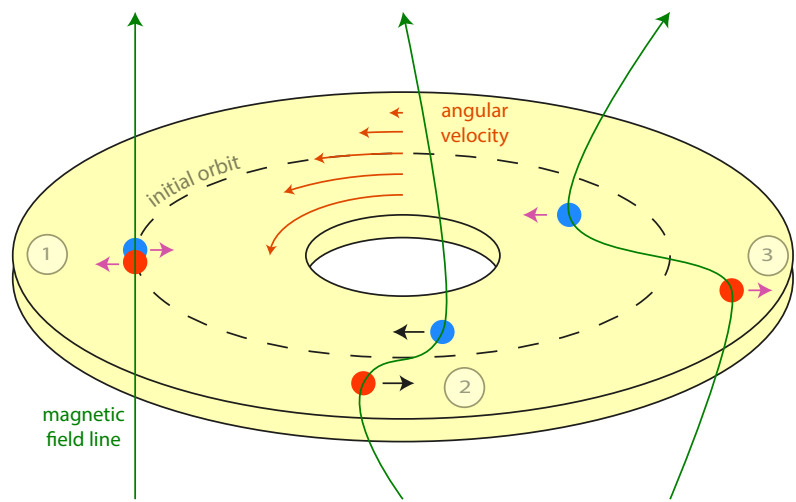

Figure 6.2: Illustration of the magneto-rotational instability initiated by the coupling of the magnetic field (green) and the differential rotation of the disk (orange). We follow the motion of two fluid parcels (red and blue dots) with time (time given by steps 1 to 3) where the black arrows indicate the direction of the magnetic force along the orbit while the pink arrows indicate the radial motions.

## 6.2 Turbulence in disks

### 6.2.1 Magneto-rotational instability as the source

When the α-model was introduced, the origin of the effective viscosity was not related to turbulence since, at this time, theoretical studies were indicating a stable flow in the disk. Since the highlighting of the turbulence, many sources have been proposed: self-gravity effects, global waves, hydrodynamic instabilities driven by the local shearing of the disk, or vertical convective instabilities in the case of heating sources in the midplane of the disk for example. None of them was fully concluent, but if one considers a disk coupled to a magnetic field, this leads to a strong linear instability (Balbus and Hawley, 1991), which is today supposed to be the main source of turbulence in disks, and can be consistent with the α-model.

To depict what happens, let us consider a case where the gas of the disk is perfectly coupled to the magnetic field, i.e. where the gas is fully ionised. In such a perfect plasma, the magnetic field lines are frozen in the fluid and follow its motion. The action of the magnetic field is then to link neighbouring fluid parcels that lie along a common field line. The fluid parcels can be imagined of being tied together on an elastic string. If, for any reason, the fluid parcels start to diverge, the tension in the magnetic "string" acts to bring the connected fluid elements back together. At first glance, this looks like a stable situation, but if we consider the differential rotation of a disk, the situation is inverted. Let us follow, step by step, the motion of two parcels initially on the same orbit and magnetically connected by a field line (see Fig. 6.2):

**Step 1** A initial displacement separates the two parcels radially, one moving inwards and the other moving outwards.

**Step 2** As the angular velocity decreases with radius, the differential rotation makes the inner parcel to rotate more rapidly than the outer one. The parcels thus separate on the azimuthal direction also. The magnetic tension opposes this displacement and, by trying to bring the two parcels back together again, causes the inner parcel to slow down while the outer parcel is sped up.

**Step 3** This outwards transfer of angular momentum causes the inner parcel to lose energy and migrate even more inwards while the outer parcel is pushed outwards.

**Repetition** The initial displacement has been enhanced and the cycle can continue.

In the presence of a differential rotation and a magnetic field, any displacement may thus be enhanced. This process is called the magneto-rotational instability, or MRI. Although there is no observational proof, the ubiquity of the conditions necessary for the MRI strongly suggest that Keplerian disks are almost always subject to it, and that it is the main source of angular momentum transport in non self-gravitating disks. However, the MRI does not set in if the magnetic field is too strong since it is able to bring back the two parcels together. In that case, the magnetic tension will instead cause the parcels to oscillate around their initial position. The magnetic field on Fig. 6.2 is illustrated perpendicular to the disk plane but a radial or orthoradial field can also destabilise the flow. Numerical simulations of the MRI show that the instability initially increases exponentially until it saturates, setting the given turbulence of the disk (e.g. Balbus and Hawley, 1991; Hawley et al., 1995). For more details on this process, the reader is refer to Balbus and Hawley (1991).

### 6.2.2 Heterogeneity

The MRI, in the way we have dealt with previously, is an ideal magneto-hydrodynamical process, since we considered a perfect plasma. In reality, this is not the case: the disk is not fully ionised nor fully coupled to the magnetic field. When taking into account a partially ionised gas, some effects are added:

**Ambipolar diffusion** Effect of the neutral gas that collides the ions and perturb the magnetic field.

**Ohmic resistivity** Effect of the neutral gas that collides the electrons and perturb the magnetic field.

**Hall effect** Any magnetic field modification will create an electric field, here called "Hall field", that will also result in a perturbation of the magnetic field. This effect may perturb or enhance the MRI depending on the field polarity quench.

By setting a good coupling between the plasma and the magnetic field, favourable to the MRI, or by creating the above dissipative non-ideal MHD effects, the ionisation fraction is a crucial parameter. The ionisation may be caused by the absorption of energetic photons and energetic particles, or by collisions with other gas particles (a process called thermal ionisation). Because these processes do not act with the same efficiency everywhere, the ionisation is not homogeneous. In regions of weakly ionised gas, taking into account the consequent perturbations (e.g. Blaes and Balbus, 1994) led to the introduction of the dead zone concept (Gammie, 1996). The dead zone is a region deep in the disk, extended on $10 - 20$ AU, where the thermal excitation and ionisation by absorption are so low that recombination dominates over the ionisation and the MRI can not develop turbulence (Fig. 6.3). Within this region that acts like a wall against the accretion, angular momentum transport is then probably very weak. While laminar transport within the dead zone is still possible, the accretion is then supposed to be governed by the turbulence driven within the disk surface layers (or by mass-loss via an MHD wind). This heterogeneity is critical when one wants to investigate the chemical mixing in a 2D model for example. In the case of the global 1D dynamical evolution, as in the α-model, this can be neglected.

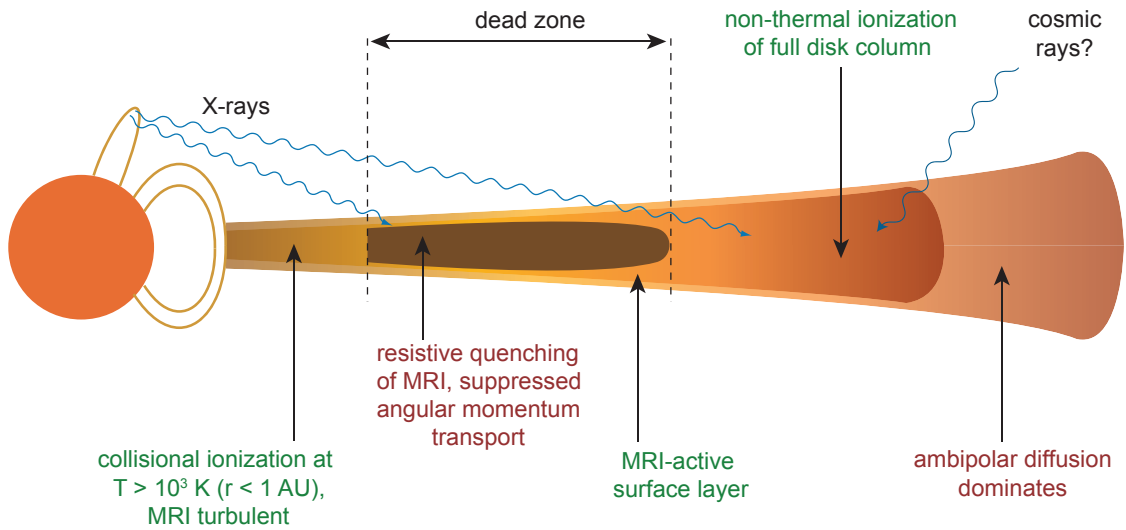

Figure 6.3: Schematic structure of the protoplanetary disk if the low ionisation fraction at radii about 1 AU quenches angular momentum transport due to the MRI, forming a "dead zone". X-rays, produced from the cooling of plasma confined within magnetic field loops in the stellar corona, ionize the disk surface, but fail to penetrate to the mid-plane. Green text indicate ragions where the MRI develops while red text indicate regions where perturbation effects dominate. Figure adapted from Armitage (2011).

# Chapter 7

# A 1D hydrodynamical model for the viscous evolution of a disk under external photoevaporation

## 7.1 The viscous evolution equation for accretion disks

To derive the general equation of the disks viscous evolution, we follow the motion of an annulus of matter by considering the mass and angular momentum conservations. The continuity equation, or mass conservation, expresses that, in any steady state process, the variation of the density $\rho$ is related to the mass flux. With a velocity $\vec{v}$, it gives

$$\frac{\partial \rho}{\partial t} + \vec{\nabla} \cdot (\rho \vec{v}) = 0. \tag{7.1}$$

For cylindrical coordinates, the equation becomes

$$\frac{\partial \rho}{\partial t} + \frac{1}{r}\frac{\partial}{\partial r}(r\rho v_\mathrm{r}) + \frac{1}{r}\frac{\partial}{\partial \theta}(\rho v_\theta) + \frac{\partial}{\partial z}(\rho v_z) = 0. \tag{7.2}$$

Assuming that the disk is axisymmetric, i.e. with no azimuthal variation or $\frac{\partial}{\partial \theta} = 0$, and that the vertical velocity $v_z = 0$, the equation is reduced to

$$\frac{\partial \rho}{\partial t} + \frac{1}{r}\frac{\partial}{\partial r}(r\rho v_\mathrm{r}) = 0. \tag{7.3}$$

Neglecting any variation of motion with height $z$, one can integrate the equation through the depth of the disk. If we introduce the surface density $\Sigma(r) = \int \rho(r,z)\,dz$, the continuity equation can now be written as

$$\frac{\partial \Sigma}{\partial t} + \frac{1}{r}\frac{\partial}{\partial r}(r\Sigma v_\mathrm{r}) = 0. \tag{7.4}$$

An accretion disk is not rotating as a solid body but can be seen as a system of concentric annuli with their own Keplerian angular velocity (Fig. 7.1) defined by

$$\Omega_K = \sqrt{\frac{GM_*}{r^3}}, \tag{7.5}$$

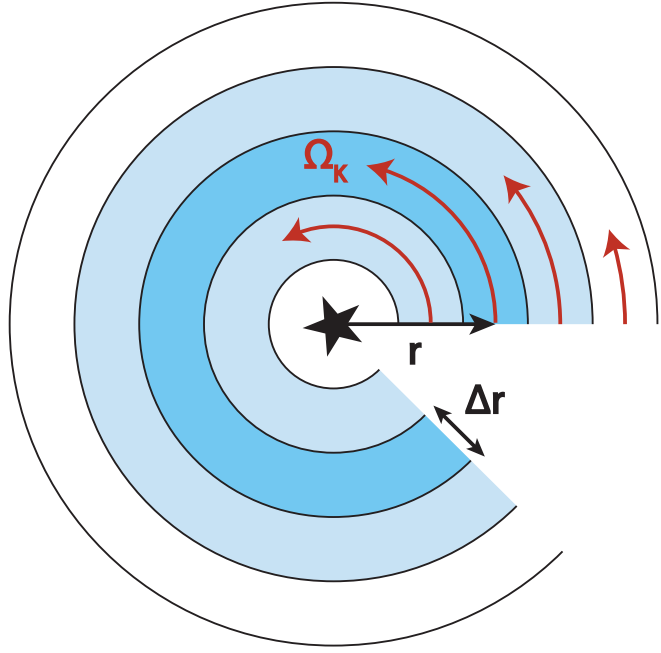

Figure 7.1: Illustration of the differential rotation between an annulus of matter (dark blue) compare with its direct neighbouring annuli (light blue).

where $M_*$ is the mass of the central host star, and $G$ the gravitational constant. Let us consider an annulus with radius $r$ and thickness $\Delta r$ so that its angular momentum is $L = M_{\text{ann}} \times r^2 \Omega_K = 2\pi r \Delta r \Sigma \times r^2 \Omega_K$, where $M_{\text{ann}}$ is its mass. The variation of its angular momentum is given by the advection of mass with different angular momentum from both outer and inner annuli, plus the viscous torque $\tau$ exerted by the shearing of this annulus with its neighbours. It reads

$$\begin{aligned}\frac{\partial}{\partial t}\left(2\pi r \Delta r \Sigma r^2 \Omega_K\right) =& + v_{\text{r}}(r) \cdot 2\pi r \Sigma(r) \cdot r^2 \Omega_K(r) \\ & - v_{\text{r}}(r+\Delta r) \cdot 2\pi(r+\Delta r)\Sigma(r+\Delta r) \\ & \quad (r+\Delta r)^2 \Omega_K(r+\Delta r) \\ & + \tau(r+\Delta r) \\ & - \tau(r).\end{aligned} \quad (7.6)$$

Taking the limit $\Delta r \to 0$ brings to

$$\frac{\partial}{\partial t}\left(\Sigma r^2 \Omega_K\right) = -\frac{1}{r}\frac{\partial}{\partial r}\left(v_{\text{r}} r \Sigma r^2 \Omega_K\right) + \frac{1}{2\pi r}\frac{\partial \tau}{\partial r}. \quad (7.7)$$

The torque exerted by a neighbouring annulus is proportional to the difference in orbital velocities with a rate of shearing $A = r\partial \Omega_K/\partial r$. The torque $\tau$ is the product of the annulus circumference ($2\pi r$), the viscous force per unit length ($\nu \Sigma A$), and the lever ($r$), so that the momentum angular conservation is then written as

$$\frac{\partial}{\partial t}\left(\Sigma r^2 \Omega_K\right) + \frac{1}{r}\frac{\partial}{\partial r}\left(v_{\text{r}} r \Sigma r^2 \Omega_K\right) = \frac{1}{r}\frac{\partial}{\partial r}\left(r^3 \nu \Sigma \frac{\partial \Omega_K}{\partial r}\right). \quad (7.8)$$

The term on the left hand side of previous equation can be developed and simplified as

$$\begin{aligned}
&\frac{\partial}{\partial t}\left(\Sigma r^2 \Omega_K\right) + \frac{1}{r}\frac{\partial}{\partial r}\left(v_{\rm r} r \Sigma r^2 \Omega_K\right) \\
&= \Sigma \frac{\partial}{\partial t}\left(r^2 \Omega_K\right) + r^2 \Omega_K \frac{\partial}{\partial t}\left(\Sigma\right) \\
&\quad + v_{\rm r}\Sigma \frac{\partial}{\partial r}\left(r^2 \Omega_K\right) + r\Omega_K \frac{\partial}{\partial r}\left(v_{\rm r} r \Sigma\right) \\
&= \Sigma \frac{\partial}{\partial t}\left(r^2 \Omega_K\right) \\
&\quad + r^2 \Omega_K \left(\frac{\partial \Sigma}{\partial t} + \frac{\partial}{\partial r}\left(r\Sigma v_{\rm r}\right)\right) \\
&\quad + v_{\rm r}\Sigma \frac{\partial}{\partial r}\left(r^2 \Omega_K\right) \\
&= v_{\rm r}\Sigma \frac{\partial}{\partial r}\left(r^2 \Omega_K\right),
\end{aligned} \qquad (7.9)$$

where, on the second line, the first term is obviously 0 because there is no time dependency while the second term is also 0 according to the continuity equation (7.4). The angular momentum conservation can be finally written as

$$v_{\rm r}\Sigma \frac{\partial}{\partial r}\left(r^2 \Omega_K\right) = \frac{1}{r}\frac{\partial}{\partial r}\left(r^3 \nu \Sigma \frac{\partial \Omega_K}{\partial r}\right). \qquad (7.10)$$

Noting that $\partial \Omega_K / \partial r = -3/(2r) \cdot \Omega_K$ and developing this last equation yields the expression of the radial velocity

$$v_{\rm r} = \frac{-3}{\Sigma\sqrt{r}} \frac{\partial}{\partial r}\left(\nu \Sigma \sqrt{r}\right). \qquad (7.11)$$

Combining the equations (7.4) and (7.10), one can derive the temporal evolution of the surface density $\Sigma$ as

$$\begin{aligned}
\frac{\partial \Sigma}{\partial t} &= -\frac{1}{r}\frac{\partial}{\partial r}\left(r\Sigma v_{\rm r}\right) \\
&= -\frac{1}{r}\frac{\partial}{\partial r}\left[\frac{\partial}{\partial r}\left(r^3 \nu \Sigma \frac{\partial \Omega_K}{\partial r}\right)\left(\frac{\partial}{\partial r}\left(r^2 \Omega_K\right)\right)^{-1}\right] \\
&= \frac{3}{r}\frac{\partial}{\partial r}\left[\sqrt{r}\frac{\partial}{\partial r}\left(\nu \Sigma \sqrt{r}\right)\right],
\end{aligned} \qquad (7.12)$$

which is the partial differential equation that we have to solve.

## 7.2 Numerical model to solve the surface density equation

### 7.2.1 Generalities

Numerical approaches are used to estimate a solution when analytical solutions do not exist. This is the case with the equation that we are interested in. Keeping a general form, the equation describing the time evolution of the disk surface density is

$$\frac{\partial \Sigma(r,t)}{\partial t} = G\bigl(\Sigma(r,t)\bigr), \qquad (7.13)$$

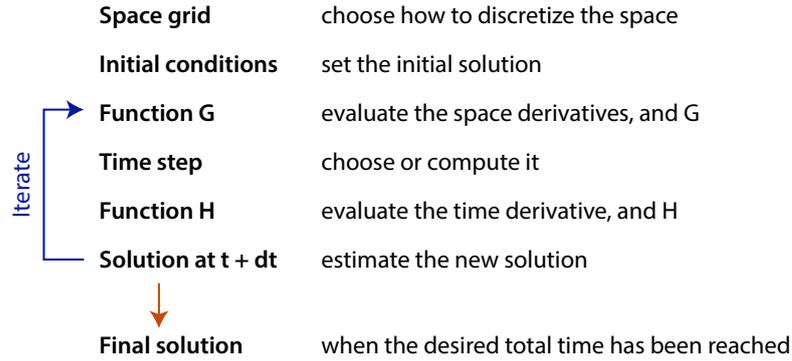

| | | |
|---|---|---|
| | **Space grid** | choose how to discretize the space |
| | **Initial conditions** | set the initial solution |
| | **Function G** | evaluate the space derivatives, and G |
| Iterate | **Time step** | choose or compute it |
| | **Function H** | evaluate the time derivative, and H |
| | **Solution at t + dt** | estimate the new solution |
| | **Final solution** | when the desired total time has been reached |

Figure 7.2: Illustration of the iteration process to numerically solve a partial differential equation at a given time of evolution.

where $G$ is a function depending on $\Sigma(r,t)$, its spatial derivatives and other terms (cf. eq. 7.12). We have there a partial differential equation, i.e. a mathematical equation that relates a function, here $\Sigma(r,t)$, with its partial derivatives[1]. Since this equation describes a temporal evolution, the problem is completely set when an initial condition, for a time $t = t_0$, is given: $\Sigma(r,t_0) = \Sigma_0$.

When the initial condition is set, the next stage is to discretise the time with a given step $dt$ and to estimate how the solution evolves with time. We have

$$\Sigma(r, t+dt) = \Sigma(r,t) + H\big(G(\Sigma, dt)\big), \qquad (7.14)$$

with $H$ a function to assess the temporal derivative, and depending on the method chosen (according to the simplicity, accuracy and stability).

To end our development, it is then necessary to discretise space on a grid to evaluate numerically the spatial derivatives on the corresponding control points. Once every methods have been defined, the numerical approach evaluates the solution $\Sigma(r,t)$ at a given time starting from the initial condition and estimating iteratively the solution step by step to move forward in time. Illustrated in Fig. 7.2, each step of the process will be developed in the following sections.

### 7.2.2 Space grid

A disk is generally extended over a few tens or hundreds of astronomical units in radius (cf Sect. 2.1.4) and starts from a small separation from the central star, generally a small fraction of an AU. These limits are respectively named outer radius and inner radius. There is no reason for a disk to keep its initial extension with time since it can spread or lose matter. The inner radius will not get closer to the star because it is supposed to be set by the magnetic field of the central star. However, it can move further away because of internal photoevaporation. The outer radius may move inwards or outwards. To take into account a possible expansion of the disk, control points on the space grid have to be placed outside of the initial extent, at least one or two orders of magnitude further to be safe. Since the inner radius and the outer disk, or the numerical outer limit if different, may span many orders of magnitude, it is often useful to use a logarithmic radial spacing. Parameters related to the space grid in the model are given in Table 7.1 and illustrated on Fig. 7.3.

---
[1] a partial derivative is the derivative of a function of several variables with respect to only one of them while the others are held constant.

Table 7.1: Parameters setting the space grid in the model.

| Parameter | Possible values/choices | Typical values |
|---|---|---|
| number $N$ of control points | $N > 0$ | 500 |
| inner radius $r_{in}$ (AU) | $0 < r_{in} < r_{out}$ | 0.01 |
| outer radius $r_{out}$ (AU) | $0 < r_{in} < r_{out}$ | 100 |
| scale | linear / logarithmic | logarithmic |
| expansion | enable / disable | enable |
| numerical outer limit $r_{out,num}$ (AU) | $r_{out,num} \geq r_{out}$ | $10^4$ |

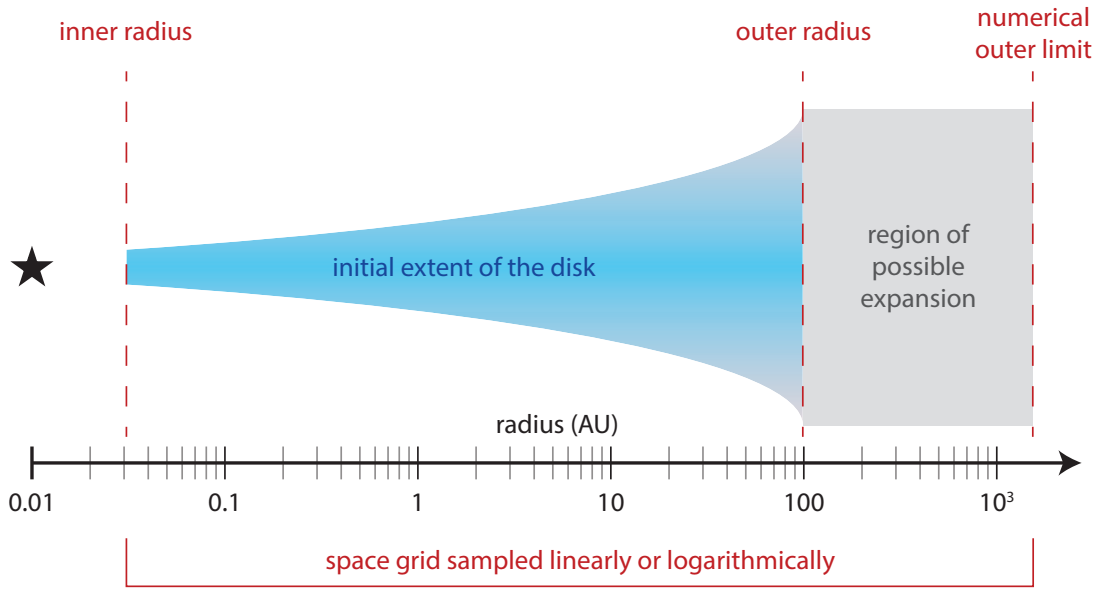

Figure 7.3: Illustration of the key parameters involved in the setting of the space grid (values are chosen arbitrarily as an example).

Table 7.2: Parameters setting the initial mass distribution in the model.

| Parameter | Possible values/choices | Typical values |
|---|---|---|
| mass of the disk $M_{\text{disk}}$ | $M_{\text{disk}} < 0.1 M_*$ | $10^{-3} \, M_*$ |
| type of power-law | pure or exponentially tapered | exponentially tapered |
| power-law index $\beta_{\text{m}}$ | $\beta_{\text{m}} < 0$ | -1 |

**Notes.** $M_*$ is the mass of the central host star.

### 7.2.3 Initial conditions

The initial conditions set the properties of the disk at the beginning of the evolution, generally taken close to conditions observed in real disks. The mass distribution, or surface density profile, is one of them. The other initial conditions set the processes acting on it: the viscosity and the photoevaporation. These processes depend on the temperature at different locations which, as any physical properties that is not computed in the simulation, has to be set initially and supposed constant with time.

**Mass distribution**

The mass distribution is following one of the function derived from observations (see Sect. 2.1.3). For the study, I preferentially adopted the exponentially tapered profile of a power-law of index $-1$, also used in Anderson et al. (2013),

$$\Sigma_0(r) = \Sigma(r, t=0) = C \left(\frac{r}{r_{\text{out}}}\right)^{-1} e^{-r/r_{\text{out}}}, \tag{7.15}$$

where $C$ is a constant depending on the total initial mass of the disk $M_{\text{disk}}$, which, according to the distribution function, is

$$\begin{aligned} M_{\text{disk}} &= \int_{r_{\text{in}}}^{r_{\text{out}}} 2\pi r \Sigma_0(r) dr, \\ &= 2\pi C r_{\text{out}} \left[-r_{\text{out}} e^{-r/r_{\text{out}}}\right]_{r_{\text{in}}}^{r_{\text{out}}}, \\ &= 2\pi C r_{\text{out}}^2 \left(e^{-r_{\text{in}}/r_{\text{out}}} - e^{-1}\right). \end{aligned} \tag{7.16}$$

Equations (7.15) and (7.16) lead to

$$\Sigma_0(r) = \frac{M_{\text{disk}}}{2\pi r_{\text{out}} \left(e^{-r_{\text{in}}/r_{\text{out}}} - e^{-1}\right)} \frac{e^{-r/r_{\text{out}}}}{r}. \tag{7.17}$$

Parameters related to the mass initial distribution are given in Table 7.2.

**Viscosity**

The effective viscosity $\nu(r)$, modelling the turbulent transport of angular momentum, is parametrised by the α-model (Sect. 6.1.2) with a constant α parameter:

$$\nu(r) = \alpha \frac{c_{\text{S}}^2(r)}{\Omega_{\text{K}}(r)}. \tag{7.18}$$

Table 7.3: Parameters setting the effective viscosity in the model.

| Parameter | Possible values/choices | Typical values |
|---|---|---|
| mass of the central star $M_*$ (M$_\odot$) | $M_* > 0$ | $10^{-1}$ |
| efficiency $\alpha$ | $10^{-4} < \alpha < 10^{-2}$ recommended | $10^{-3}$ |
| mid-plane ref. temperature radius $r_{0,T}$ (AU) | $r_{0,T} > 0$ | 1 |
| mid-plane ref. temperature $T_{\text{mid},0}$ (K) | $T_{\text{mid},0} > 0$ | 300 |
| mid-plane temperature power-law index $\beta_T$ | $\beta_T < 0$ | -0.5 |

The speed of sound,

$$c_S(r) = \sqrt{\frac{\gamma k_B T_{\text{mid}}(r)}{\mu m_H}}, \quad (7.19)$$

is defined with $\gamma$ the adiabatic index, $k_B$ the Boltzmann constant, $T_{\text{mid}}(r)$ the mid-plane temperature, and $\mu m_H$ the mean particle mass. We assume here that the mid-plane temperature follows a pure power-law

$$T_{\text{mid}} = T_{\text{mid},0} \left(\frac{r}{r_{0,T}}\right)^{\beta_T}, \quad (7.20)$$

where $T_{\text{mid},0}$ is the reference temperature at radius $r_{0,T}$. The viscosity is finally given by

$$\nu = \frac{\alpha \gamma k_B T_{\text{mid},0} r^{\beta_T + 3/2}}{\mu m_H r_{0,T}^{\beta_T} \sqrt{GM_*}}. \quad (7.21)$$

The parameters related to the viscosity of the disk are given in Table 7.3.

**Photoevaporation**

A term $\dot{\Sigma}_{\text{evap}}(r)$, describing the loss of mass by photoevaporation, can be added to the viscous evolution equation. It becomes

$$\frac{\partial \Sigma}{\partial t} = \frac{3}{r}\frac{\partial}{\partial r}\left[\sqrt{r}\frac{\partial}{\partial r}\left(\nu \Sigma \sqrt{r}\right)\right] - \dot{\Sigma}_{\text{evap}}(r). \quad (7.22)$$

The supercritical photoevaporation (cf. Sect. 5.6) occurs when the kinetic energy of the gas particles is higher than the gravitational potential energy so that the particles can efficiently escape from the disk. For heating dominated by an external source of FUV photons, the kinetic energy at the disk surface can be considered constant with the distance to the central star. However, the gravitational potential energy decreases with radius. Therefore, photoevaporation flows develop in the outer parts of the disk beyond what is called the gravitational radius, for which the kinetic energy of gas particles is equal to the gravitational potential energy. The gravitational radius is defined as (Hollenbach et al., 1994)

$$r_g = \frac{GM_*}{c_{S,\text{surf}}^2}, \quad (7.23)$$

where $c_{S,\text{surf}}$ is the sound speed at the disk surface which depends on the local gas temperature $T_{\text{surf}}$. If $r_g < r_{\text{out}}$ (supercritical case), the outer disk loses mass from the surface at a rate per unit area given by (Johnstone et al., 1998)

$$\dot{\Sigma}_{\text{sup}}(r) = 2\,\mu\,m_H\,n_{\text{surf}}(r)\,c_{S,\text{surf}}, \quad (7.24)$$

with $n_{\text{surf}}$ the particle density at the disk surface. Assuming this parameter constant along the disk and integrating over the photoevaporation region gives the following total mass loss rate

$$\dot{M}_{\text{sup}} = \int_{r_{\text{g}}}^{r_{\text{disk}}} \dot{\Sigma}_{\text{sup}} \times 2\pi r dr \qquad (7.25)$$
$$= \mu \, m_{\text{H}} \, n_{\text{surf}} \, c_{\text{S,surf}} \times 2\pi(r_{\text{out}}^2 - r_{\text{g}}^2).$$

In this study, the external supercritical photoevaporation is thus described by the two input parameters $n_{\text{surf}}$ and $T_{\text{surf}}$ which can be estimated from observations. In Part II, we have found that $n_{\text{surf}} \approx 10^6$ cm$^{-3}$ and $T_{\text{surf}} \approx 1000$ K are representative of the proplyds in the Orion cluster and the Carina nebula. Moreover, those values are likely to be constant in time when photoevaporation is at play, because of a dynamical equilibrium (see Sect. 5.5.3). We do not state that those values are universal nor constant with time but, as they are our only estimates and are consistent with observations for the Orion proplyds (e.g. line emission and mass-loss rates), they will be our fiducial values.

Other prescriptions for photoevaporation in published papers were used for tests or comparison with other processes. Anderson et al. (2013) proposed the following expression for the external sub-critical case ($r_{\text{out}} < r_{\text{g}}$),

$$\dot{\Sigma}_{\text{sub}} = \frac{A}{4\pi} \left(\frac{r_{\text{g}}}{r}\right)^{3/2} \left(1 + \frac{r_{\text{g}}}{r}\right) e^{-r_{\text{g}}/2r}, \qquad (7.26)$$

where the first parameter is $A \approx n_{\text{d}} c_{\text{S}} \mu m_{\text{H}}$, with $n_{\text{d}}$ the gas particle density at the disk edge, $c_{\text{S}}$ is the sound speed at the sonic radius (radius outside of the disk where the flow becomes supersonic and mass can freely escape) and $\mu m_{\text{H}}$ the mean particle mass. The gravitational radius $r_{\text{g}}$ is here given as the second input parameter which is computed from surface temperature based on PDR modelling.

The total mass-loss rate variation with X-ray luminosity, $L_X$, of the central star, described by Owen et al. (2012) from numerical simulations, is

$$\dot{M}_X = 6.25 \times 10^{-9} \left(\frac{M_*}{1 M_\odot}\right)^{-0.068} \left(\frac{L_X}{10^{30} \text{erg s}^{-1}}\right)^{1.14} M_\odot \text{ yr}^{-1}. \qquad (7.27)$$

It is then spread on the mass-loss rate profile of the form

$$\begin{aligned}
\dot{\Sigma}_X(x > 0.7) &= 10^{(a_1 \log(x)^6 + b_1 \log(x)^5 + c_1 \log(x)^4)} \\
&\quad \times 10^{(d_1 \log(x)^3 + e_1 \log(x)^2 + f_1 \log(x) + g_1)} \\
&\quad \times \left(\frac{6a_1 \ln(x)^5}{x^2 \ln(10)^7} + \frac{5b_1 \ln(x)^4}{x^2 \ln(10)^6} + \frac{4c_1 \ln(x)^3}{x^2 \ln(10)^5} \right.\\
&\quad + \left.\frac{3d_1 \ln(x)^2}{x^2 \ln(10)^4} + \frac{2e_1 \ln(x)}{x^2 \ln(10)^3} + \frac{f_1}{x^2 \ln(10)^2}\right) \\
&\quad \times \exp\left[-\left(\frac{x}{100}\right)^{10}\right]
\end{aligned} \qquad (7.28)$$

where $a_1 = 0.15138$, $b_1 = -1.2182$, $c_1 = 3.4046$, $d_1 = -3.5717$, $e_1 = -0.32762$, $f_1 = 3.6064$, $g_1 = -2.4918$ and

$$x = 0.85 \left(\frac{R}{\text{AU}}\right) \left(\frac{M_*}{1 M_\odot}\right)^{-1}, \qquad (7.29)$$

where $\dot{\Sigma}_X(x < 0.7) = 0$. In addition to the central star mass, the input parameter defining this mass-loss is the internal X-ray luminosity.

Table 7.4: Parameters setting the photoevaporation in the model.

| Photoevaporation | Parameter | Possible values | Typical values |
|---|---|---|---|
| Ext. supercritical | density at the surface $n_{\text{surf}}$ (cm$^{-3}$) | $n_{\text{surf}} > 0$ | $10^6$ |
| Ext. supercritical | surface temperature $T_{\text{surf}}$ (K) | $T_{\text{surf}} > 0$ | $10^3$ |
| Ext. subcritical | parameter $A$ (M$_\odot$ AU$^{-2}$ yr$^{-1}$) | $A > 0$ | – |
| Ext. subcritical | gravitational radius $r_{\text{g}}$ (AU) | $r_{\text{g}} > r_{\text{out}}$ | – |
| Internal X | X-ray luminosity $L_X$ (erg s$^{-1}$) | $10^{29} < L_X < 10^{31}$ | $10^{30}$ |
| Internal UV | profiles extracted from Gorti and Hollenbach (2009) | | |

We also used mass-loss rate profiles due to internal UV photons, extracted from the study of Gorti and Hollenbach (2009). Parameters related to all of those prescriptions are given in Table 7.4.

### 7.2.4 Spatial derivatives

Once physical conditions are set on the control points, we can estimate the spatial derivatives involved in the equation (7.12) that gives the temporal evolution of the surface density profile. The more intuitive way to do that is to use the finite differences locally on each control point in replacement of the spatial derivatives. The major advantage of the finite difference method is its simplicity and its almost universal application. A drawback of this method is that it does not take into account conservative laws (mass, energy, etc.) a priori, because of numerical integration errors, but this effect may still be controlled and corrected if needed.

**Finite differences**

Let us consider an unidimensional function $f(x)$ discretised on a regular grid with coordinates noted $x_i$ and separated by a space step $\delta x$. The method of the finite differences consists in expressing the space derivative of $f$ of any order $n$, noted $f^{(n)}$, based on the values of the function $f_i = f(x_i)$.

The Taylor series is a way to link the value of a function at a point $x$ to its derivatives since it is a representation of a function as an infinite sum of terms that are calculated from the values of the function's derivatives at a single point. If the coordinate of the point in question is $x_i$ then we have

$$f(x) = \sum_{n=0}^{\infty} \frac{f^{(n)}(x_i)}{n!} (x - x_i)^n. \tag{7.30}$$

Since we can not evaluate the infinite sum, this expression is generally evaluated up to an order $p$, and we have

$$f(x) = \sum_{n=0}^{p} \frac{f^{(n)}(x_i)}{n!} (x - x_i)^n + o\left((x - x_i)^p\right), \tag{7.31}$$

where $o\left((x - x_i)^p\right)$ represents the terms of orders higher than $p$ and expect to be negligible. The expression of $f(x_{i+1}) = f(x_i + \delta x)$ developed at order 1 is thus

$$f(x_{i+1}) = f(x_i) + f'(x_i)\delta x + o\left(\delta x^1\right), \tag{7.32}$$

from which we can extract the 1st order derivative

$$f'(x_i) \simeq \frac{f(x_{i+1}) - f(x_i)}{\delta x}. \tag{7.33}$$

This relation is called the 1st order forward since it needs the value of the function $f$ at $x_i$ and $x_{i+1}$. Now using the expression of $f(x_{i-1}) = f(x_i - \delta x)$ which is

$$f(x_{i-1}) = f(x_i) - f'(x_i)\delta x + o\left(\delta x^1\right). \tag{7.34}$$

we can develop, in a similar way to previously, the 1st order backward as

$$f'(x_i) \simeq \frac{f(x_i) - f(x_{i-1})}{\delta x}. \tag{7.35}$$

The 1st order central is obtained by the difference between the expressions (7.32) and (7.34),

$$\begin{aligned} f(x_{i+1}) - f(x_{i-1}) &\simeq 2f'(x_i)\delta x, \\ f'(x_i) &\simeq \frac{f(x_{i+1}) - f(x_{i-1})}{2\delta x}, \end{aligned} \tag{7.36}$$

which is the one that I used.

I will not detailed the finite differences method more here but, in a very similar way, I used the development of the order 2 of the Taylor series to extract the 2nd order derivative needed for the tests (cf. Sect. 7.3). This development is still valid when the space grid is not regular by assuming two different steps instead: $\delta x_+$ forward and $\delta x_-$ backward. This leads to other formulas, but using the same development. Finite differences are not truly local since they need values of the neighbouring points to be estimated (the number of neighbors to take into account is depending on the order of the derivative which is estimated and the order of the Taylor development). At the edges of the space grid, there is thus the need to use ghost points that set what are called the boundary conditions.

**Boundary conditions**

Boundary conditions are in the core of the problem when solving a differential equation since they are involved in the estimation of the space derivatives and strongly impact the final solution.

One common method is the 0-torque, which simply supposes that ghost points (points out of the space grid used to compute the spatial derivatives) have a surface density equal to zero. Another possibility is to force the surface density profile to follow a power-law of a given index, by setting the value of the ghost points according to this law.

The final possibility that I tested was to impose the steady-state when photoevaporation is not active. According to the equation (7.12), the steady-state is reached when

$$\frac{\partial}{\partial r}\left[\sqrt{r}\frac{\partial}{\partial r}\left(\nu\Sigma\sqrt{r}\right)\right] = 0, \tag{7.37}$$

which is effectively the case when

$$\frac{\partial}{\partial r}\left(\sqrt{r}\ \nu\Sigma\right) = 0, \tag{7.38}$$

or when

$$\sqrt{r}\frac{\partial}{\partial r}\left(\nu\Sigma\sqrt{r}\right) \neq f\left(r\right), \tag{7.39}$$

meaning that the left-hand side term is uniform. The first solution represents the static case since the radial velocity, proportional to this derivative (see eq. 7.11), is then zero. Since the viscosity is proportional to $r^{\beta_T+3/2}$ in the alpha model developed here (eq. 7.21), the condition is fulfilled only if $\Sigma \propto r^{-\beta_T-2}$. The uniform case occurs when

$$\frac{\partial}{\partial r}\left(\nu \Sigma \sqrt{r}\right) = \frac{A}{\sqrt{r}}, \tag{7.40}$$

where $A$ is a constant. At this time we can note that this implies that the radial velocity is non zero and is $v_r = -\frac{3}{\sqrt{r}\Sigma}\frac{A}{\sqrt{r}}$, so that the term $\Sigma r v_r$ is uniform too. The mass-loss rate (by radial transport) given by

$$\dot{M}(r) = -2\pi r v_r(r) \Sigma(r), \tag{7.41}$$

is thus also uniform in that case[2]. The term $\sqrt{r}\frac{\partial}{\partial r}(\nu\Sigma\sqrt{r})$ is uniform if $\frac{\partial}{\partial r}(\nu\Sigma\sqrt{r}) \propto r^{1/2}$. If we note $\Sigma \propto r^q$, we have

$$\frac{\partial}{\partial r}\left(\nu\Sigma\sqrt{r}\right) \propto \frac{\partial}{\partial r}\left(r^{\beta_T+3/2}r^q r^{1/2}\right) \propto \frac{\partial}{\partial r}\left(r^{\beta_T+2+q}\right) \propto r^{\beta_T+1+q}. \tag{7.42}$$

So, this uniform case is obtained when $q = \beta_T - 3/2$.

Figure 7.4 shows the different evolutions of the surface density profile according to the boundary conditions chosen. The 0-torque method implies a drop at the edges, which also increases the mass-loss. This is quite correct at the inner edge with accretion but should be treated carefully at the outer edges since the loss of mass there could be important and not realistic. This effect is avoided when the disk is allowed to expand as we generally do. The power-law method tends to make the entire profile to follow the power-law as shown here (the initial profile is already a power-law of index -1 exponentially tapered at the outer edge so that the profile tends to be a pure power-law with time). The uniform mass-loss rate in the steady state case here is also a power-law, but less steep so that we see the slope evolving. Finally, from what we can see there, the boundary conditions affects initially the edges but propagate then to the entire profile. The choice of the method is thus crucial.

In the end, there are two important choices in setting the boundary conditions: using the 0-torque method or a power-law, which can be arbitrary or carefully chosen to ensure the convergence to a static or uniform-transport steady state. In the case of our study, we used the 0-torque method coupled to an allowed expansion, so that the accretion is the only source of mass-loss related to viscosity and impacted by the boundary conditions. More generally, this method is the most consistent since it is implicitly produced by the photoevaporation that removes mass efficiently after a given radius (we will check this assertion in Sect. 8.1.1).

### 7.2.5 Time step

Once the spatial derivatives have been computed, the temporal derivative is obtained. It is then necessary to find the time step on which the model will progress while ensuring a small numerical error. One arbitrary method could be to use a constant time step corresponding to a small fraction of the smallest orbit in the disk. It is hard to find a good fraction to avoid too much computing time and to avoid too large numerical errors. A better way is to use a method to compute iteratively the optimal time step. The Courant–Friedrichs–Lewy (Courant et al., 1928), or CFL, condition is one of them. This is a necessary condition for stability while solving numerical problem using finite differences. It stipulates that the evolution of a quantity only depends on the value of its nearest neighbouring control points. As a consequence, the step of time must be shorter than the time needed to cross a cell by the most rapid physical process[3].

---

[2]We retrieve more directly the steady-state in the uniform case when writing the viscous equation using the mass-loss rate : $\frac{\partial \Sigma}{\partial t} = \frac{-3}{2\pi}\frac{\partial}{\partial r}\dot{M}(r)$.

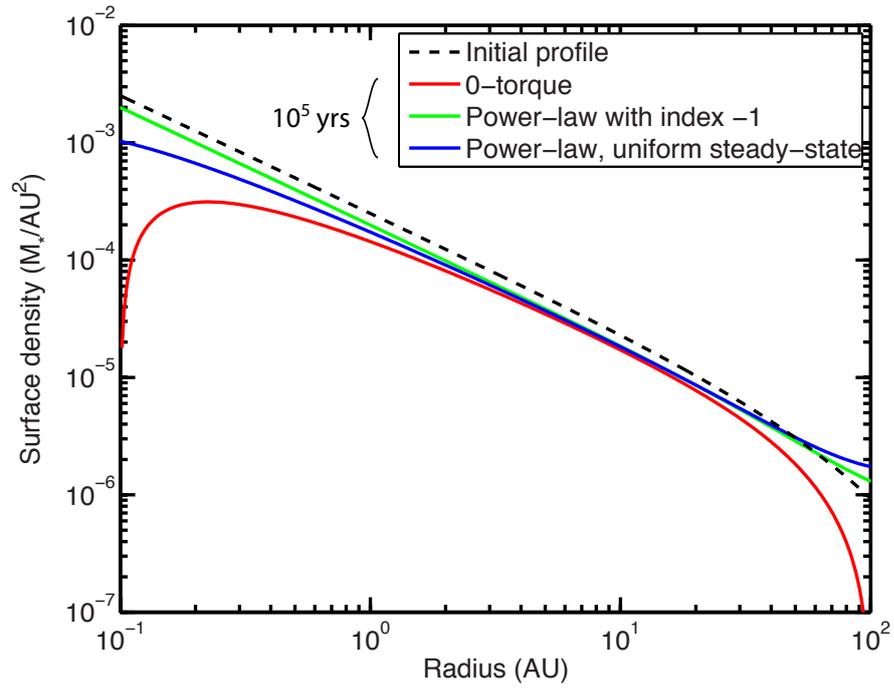

Figure 7.4: Viscous evolution of a disk of mass $M_{\rm disk} = 0.1 M_*$ extended between $r_{\rm in} = 0.1$ AU and $r_{\rm out} = 100$ AU and applying different boundary conditions at the disk edges. The initial profile is represented with a black dashed line while the evolution at $10^5$ years is given in colours for the different methods.

In our model, two radial transports have to be taken into account: the advection by radial velocity and the viscous diffusion. For a radial step $\delta r$, the characteristic time for advection is

$$\delta t_{v_r} = \frac{\delta r}{v_r}, \tag{7.43}$$

where $v_r$ is the radial velocity expressed in equation (7.11), which depends on the local viscosities and surface densities. The characteristic viscous time is

$$\delta t_\nu = \frac{(\delta r)^2}{4\nu}. \tag{7.44}$$

The time of the CFL condition is a combination of these two processes according to

$$\delta t_{\text{CFL}} = C_0 \times \frac{1}{\sqrt{\left(\frac{1}{\delta t_\nu}\right)^2 + \left(\frac{1}{\delta t_{v_r}}\right)^2}}, \tag{7.45}$$

where $C_0 < 1$ is an arbitrary number, known as the Courant number describing the fraction of the cell that is crossed during the time step. In our case, $C_0 = 0.5$. This CFL time is computed at every cell and the smallest one is chosen.

### 7.2.6 Applying the temporal evolution

The most basic explicit method[4] for numerical integration of differential equations is the Euler method which simply states that,

$$\Sigma(r, t + \delta t) = \Sigma(r, t) + k_1 \times \delta t, \tag{7.46}$$

where solutions are computed iteratively from the initial condition at $t_0$ to the final time. This first-order numerical procedure uses the slope at the beginning of the interval as the increment, so that $k_1 = \frac{\partial \Sigma}{\partial t}$.

This Euler method represents the first order of the Runge-Kutta development, that may be refined when using higher orders. The Runge-Kutta method is iterative since a first estimate of the solution is made before a second one, more precise, is computed and so on and so forth. Any order can be used but the classical method is the one of order four, named RK4, and for which the solution is given by

$$\Sigma(r, t + \delta t) = \Sigma(r, t) + \frac{\delta t}{6}(k_1 + 2k_2 + 2k_3 + k_4), \tag{7.47}$$

where $k_i$ are functions of the time $t$, time step $\delta t$ and lowest-order $k_{i-1}$ functions. For this RK4 method, $k_1$ is, as previously, the increment based on the slope at the beginning of the interval, $k_2$ is the increment based on the slope at the midpoint of the interval, $k_3$ is again the increment based on the slope at the midpoint but using the previous estimation of $k_2$ and $k_4$ is the increment based on the slope at the end of the interval. Contrary to the Euler method that only uses the slope of $k_1$ at the beginning of the interval, here the four increments are averaged in equation (7.47) with greater weights given to the increments at the midpoint. This method reduces significantly the local truncation error but needs to add an iterative process again and increases significantly the time of computing.

---
[3]With that condition, it is clear that reducing the step in the space grid or using a logarithmically scaled grid with very small steps on the lowest scales will have a critical impact on the time step and then on the computing time. Adding more control points is very expensive in terms of time because of that indirect impact.

[4]An explicit method determines the solution at a time $t + \delta t$ based only on time $t$, contrary to implicit methods that involve both solutions in the calculation.

In our case, we kept the simple Euler method which gives good enough results (see Sect. 7.3 for tests). The iterative process is done until the desired total time, given as an input parameter of the model, has been reached.

## 7.3 Tests

In order to test our numerical scheme, we can compare its outcomes with analytical solutions in some simple cases, or with previous numerical studies.

### 7.3.1 Analytical solution in a simple case

If one considers the viscosity $\nu$ constant along the disk, the viscous equation simplifies to

$$\frac{\partial \Sigma}{\partial t} = \frac{3\nu}{r} \frac{\partial}{\partial r} \left[ \sqrt{r} \frac{\partial}{\partial r} \left( \Sigma \sqrt{r} \right) \right], \tag{7.48}$$

This equation can be solved analytically by separations of variables. It may not corresponds to any real case but is useful, first to understand what is going on inside a viscous disk, and second to test our code. If we write $s = 2\sqrt{r}$, the equation becomes

$$\begin{aligned}
\frac{1}{\sqrt{r}} \frac{\partial}{\partial t} \left( \Sigma \sqrt{r} \right) &= \frac{3\nu}{r} \frac{\partial s}{\partial r} \frac{\partial}{\partial s} \left[ \sqrt{r} \frac{\partial s}{\partial r} \frac{\partial}{\partial s} \left( \Sigma \sqrt{r} \right) \right], \\
\frac{1}{\sqrt{r}} \frac{\partial}{\partial t} \left( \Sigma \sqrt{r} \right) &= \frac{3\nu}{r} \frac{1}{\sqrt{r}} \frac{\partial}{\partial s} \left[ \sqrt{r} \frac{1}{\sqrt{r}} \frac{\partial}{\partial s} \left( \Sigma \sqrt{r} \right) \right], \\
\frac{\partial}{\partial t} \left( \Sigma \sqrt{r} \right) &= \frac{12\nu}{s^2} \frac{\partial^2}{\partial s^2} \left( \Sigma \sqrt{r} \right).
\end{aligned} \tag{7.49}$$

Writing $\Sigma \sqrt{r} = T(t) S(s)$, this becomes

$$\frac{\frac{\partial T}{\partial t}}{T} = \frac{12\nu}{s^2} \frac{\frac{\partial^2 S}{\partial s^2}}{S} = \text{constant}, \tag{7.50}$$

where one can recognise that the integration will lead to an exponential temporal dependency while the space dependency is a Bessel function.

Let us look at a mass initially located within an infinitesimal annulus of matter at radius $r_0$. The initial condition is thus

$$\Sigma(r, 0) = \frac{M_{\text{disk}}}{2\pi r_0} \delta(r - r_0). \tag{7.51}$$

Using the dimensionless variables $x = r/r_0$ and $\tau = 12\nu t r_0^{-2}$, the analytical solution derived from the equation (7.50) is

$$\Sigma(x, \tau) = \frac{M_{\text{disk}}}{\pi r_0^2 \tau x^{\frac{1}{4}}} \exp\left(-\frac{1 + x^2}{\tau}\right) I_{\frac{1}{4}}\left(\frac{2x}{\tau}\right), \tag{7.52}$$

where $I_{\frac{1}{4}}\left(\frac{2x}{\tau}\right)$ is a modified Bessel function of the first kind of order 1/4.

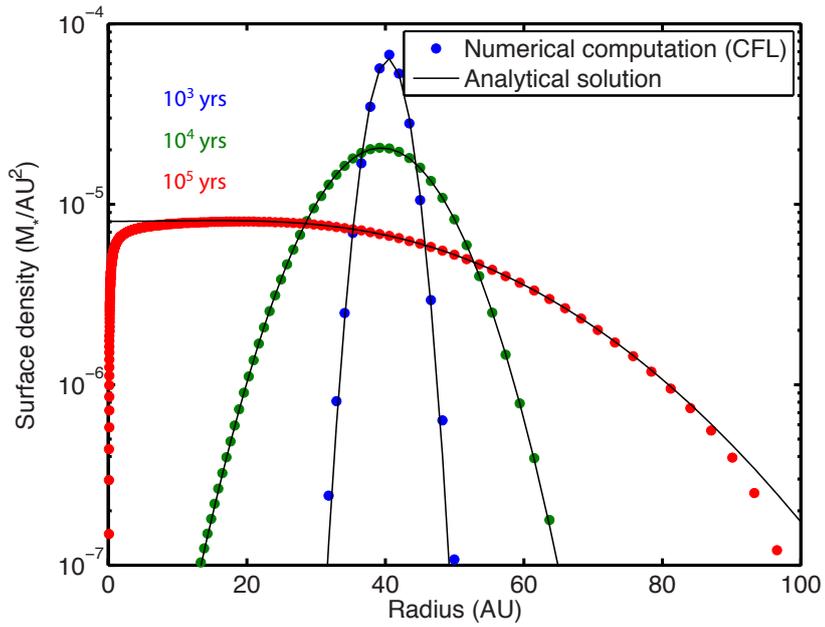

Figure 7.5: Comparison between the analytical and numerical viscous evolution of an initial infinitesimal ring of matter of mass 0.01 M$_\odot$, located at $r_0 = 40$ AU, and evolving under a constant viscosity of $10^{-3}$ AU$^2$ yr$^{-1}$. The numerical computation is made on a logarithmic space grid between 0.1 and 100 AU with the 0-torque boundary conditions and using a time step satisfying the CFL condition.

Figure 7.5 illustrates the spreading of a viscous ring according to this analytical solution. Most of the mass moves inwards by losing energy and angular momentum, but a small amount of matter moves outwards taking up the angular momentum lost by the inner rings. Analysing the asymptotic behaviours of the analytical solution shows that the outer parts, where $2x > \tau$, move outwards while the inner parts move inwards and tend to be accreted onto the star. The radius of this change of direction moves thus steadily outwards with time (as $\tau$ increases). At large times, where any $x << \tau$, almost all the mass drifts inwards and is accreted onto the star with the exception of a very small fraction of the mass that carries the angular momentum outwards. This trend to move inwards preferentially is visible in Fig. 7.5, where the center of mass of the "ring", or the maximum of the profile, is moving to smaller radii with time.

This analytical evolution allows us to test the numerical solution of the code. In Fig. 7.5, it is clear that the code with a time step given by the CFL condition gives a very good estimation of the analytical solution. However, when the mass has reached the edges of the grid, numerical boundary conditions impact the result that derives from the analytical solution which have no limit because the zero-torque boundary condition does not correspond to the equation that we solve. This behaviour illustrates the crucial effect of boundary conditions and the necessity to choose them carefully or to use a larger grid to avoid them to have a significant impact, as in our case when we consider free expansion and a numerical outer edge located further away.

### 7.3.2 Comparison with a previous study

Another way to test the code is to check if it is able to reproduce results of previous studies of this kind. To do so, we add the prescription of the sub-critical photoevaporation development by Anderson et al. (2013) who parametrised the mass-loss rate using a given gravitational radius (initially estimated with a temperature extracted through a PDR model for a given FUV field strength) and a parameter "A", through gas particle density. It is similar to our case for supercritical photoevaporation since this prescription only depends on the temperature and density. The input parameter $A$ (Sect. 7.2.3) has been adjusted (not accurately known) to retrieve roughly the same conditions as input in their paper. The comparison between their results and ours is presented in Fig. 7.6. The evolution is the same and the quantitative results, taking into account the rough estimation of one of the two parameters, are very similar. This gives us confidence that our code is robust.

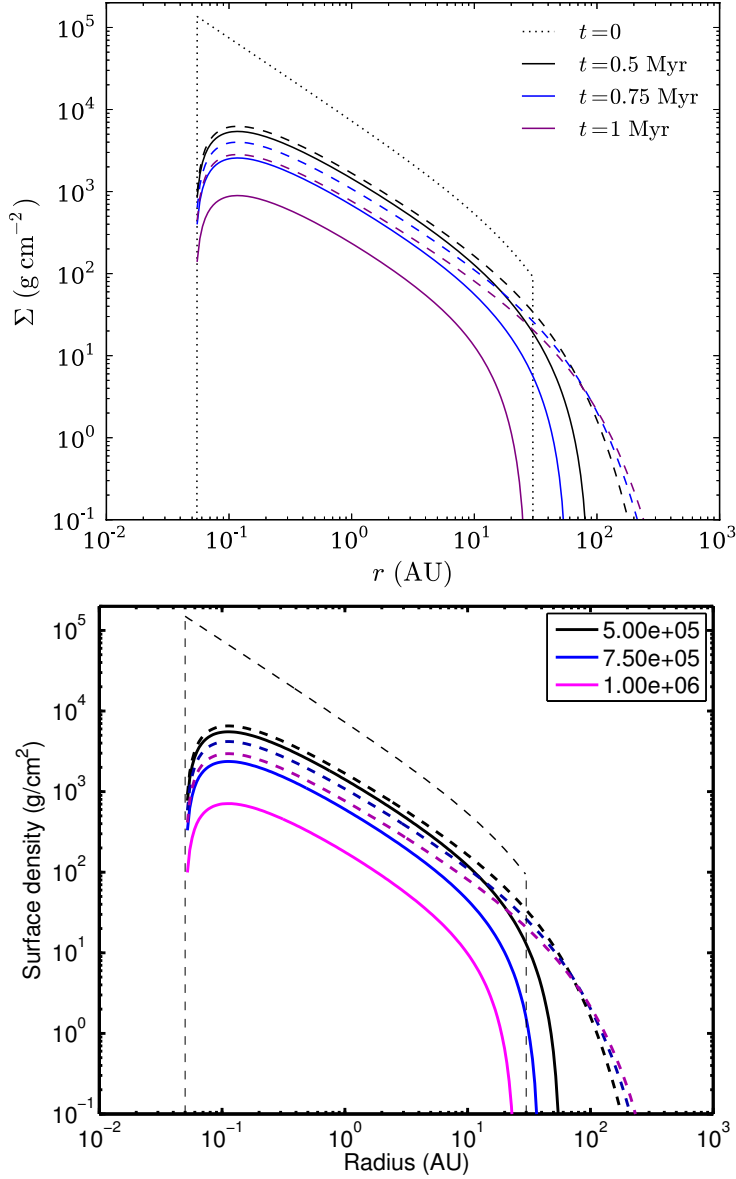

Figure 7.6: Example of the viscous evolution of the surface density. The surface density profile is given three distinct times, as labeled, with viscosity parameter $\alpha = 10^{-3}$. The solid curves show the disk immersed in an external FUV radiation field with dimensionless strength $G_0 = 3000$, and the dashed curves show pure viscous evolution (for comparison). The initial surface density profile is also shown in tiny black dashed line. The initial disk mass and radius are $M_{\rm disk} = 0.1$ $M_\odot$ and $r_{\rm out} = 30$ AU respectively. The initial surface density profile has discontinuities at both the inner boundary ($r_{\rm in} = 0.05$ AU) and at the initial outer radius ($r_{\rm out} = 30$ AU). The disk is free to expand outwards while the inner boundary method is the 0-torque. Top: original figure from Anderson et al. (2013). Bottom: reproduction using our code with their prescription for photoevaporation and parameters as close as theirs.

## Chapter 8

# Results: evolution of a disk under photoevaporation

## 8.1 Disk evolution under supercritical photoevaporation

### 8.1.1 General pattern: truncation of the disk extension and lifetime

As we have seen, a viscous disk tends to spread. If we add the external supercritical photoevaporation, this remains true but, intuitively, a substantial part of the disk is expected to be lost. Since photoevaporation does not impact the whole disk in a same way, two regions have to be defined:

**Inner disk** Inner part of the disk where the gravitational field of the central host star prevents gas from escaping. By definition, it is extended from the inner radius, $r_{\text{in}}$, to the gravitational radius, $r_{\text{g}}$.

**Outer disk** Outer part of the disk where the gravitational field is no longer strong enough to avoid the development of photoevaporative flows. It is extended from the gravitational radius, $r_{\text{g}}$, to the outer radius of the disk, initially defined by the outer radius, $r_{\text{out}}$.

As seen in equation (7.23), the gravitational radius depends on the central star mass and the disk surface temperature, due to the external FUV radiation field, by

$$r_{\text{g}} \approx 107.5 \left(\frac{M_*}{1\text{M}_\odot}\right)\left(\frac{1000 \text{ K}}{T_{\text{surf}}}\right) \text{ AU}. \tag{8.1}$$

The outer disk, which experiences photoevaporation flows, is supposed to be removed efficiently, and the whole disk should be eroded from outside in. This is clear indeed in the temporal evolution of the case presented in Fig. 8.1. This example case corresponds to a disk with a mass of $M_{\text{disk}} = 1.7 \times 10^{-3}$ M$_\odot$, initially extended to a radius $r_{\text{out}} = 100$ AU, and with a value of viscosity set by $\alpha = 10^{-3}$. For a star of mass $M_* = 0.13$ M$_\odot$ and conditions at the disk surface defined by $n_{\text{surf}} = 10^{-6}$ cm$^{-3}$ and $T_{\text{surf}} = 1000$ K, the gravitational radius is located at about 14 AU. When subject to supercritical photoevaporation at time 0, the outer disk (from $r_{\text{g}} \approx 14$ AU to $r_{\text{out}} = 100$ AU), that represents a large fraction of the disk mass, will thus lose mass directly. It takes a few $10^3$ years before the effect of photoevaporation becomes visible. Its first visible effect is to continuously decrease the disk size to finally truncate it down to the gravitational radius. According to Fig. 8.1(b), it takes about $10^5$ years to achieve truncation and to completely remove the outer disk. While the photoevaporation is dissipating the outer disk, the disk surface density in the "inner disk" ($r \leq r_g \approx 14$ AU) seems less affected by photoevaporation at first glance, but still decreases because of viscous evolution.

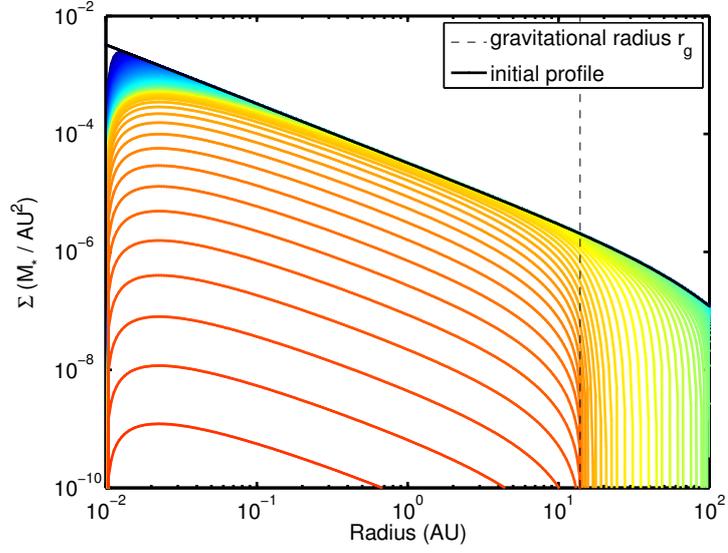

(a) Surface density profile

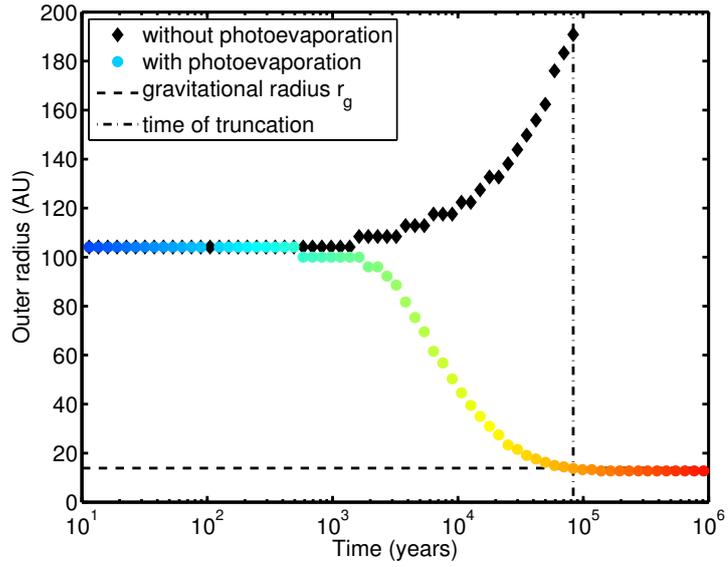

(b) Outer radius

Figure 8.1: Time evolution of the surface density $\Sigma(r)$ profile (a) and corresponding outer radius (b) for a disk with parameter $M_* = 0.13$ $M_\odot$, $M_{\text{disk}} = 1.7 \times 10^{-3}$ $M_\odot$, $r_{\text{out}} = 100$ AU and $\alpha = 10^{-3}$, under a supercritical photoevaporation defined by $n_{\text{surf}} = 10^{-6}$ cm$^{-3}$ and $T_{\text{surf}} = 1000$ K. 100 output iterations are plotted, logarithmically spaced in time from the initial state (blue) to 10 Myrs (red). Its evolution for the same set of parameters but with no photoevaporation is also given for comparison (black).

Actually, the viscosity drives the radial motion within the disk and is always responsible for part of the mass loss. While spreading, a disk not subjected to photoevaporation loses mass by accretion onto the central star, with a rate around $10^{-8}$ M$_\odot$ yr$^{-1}$ (Hartmann et al., 1998). A disk under supercritical external photoevaporation could lose mass much more rapidly and be dissipated prematurely. Fig. 8.2 gives the evolution of the mass, and components of the mass-loss rate (detailed at Sect. 8.1.2), for our example case. Here, the mass-loss rate by accretion starts slightly below $10^{-8}$ M$_\odot$ yr$^{-1}$, while the photoevaporation leads to an initial value slightly above $10^{-7}$ M$_\odot$ yr$^{-1}$. If the conditions for photoevaporation do not change in time, i.e. constant density and temperature at the disk surface, the mass-loss rate is maximum when the disk is the most extended (see equation 7.25). This happens generally at the beginning of photoevaporation, except if the viscosity and the reservoir of mass are sufficiently high to still allow some expansion while photoevaporation is on. Since the main trend for the disk size is to be reduced down to the gravitational radius with time, the mass-loss rate then tends to decrease as well. This is what happens for our example case (cf. Fig. 8.2), where the mass-loss rate continuously decreases. However, the decrease is not regular and a few components for the mass-loss rate can be distinguished. I will describe those in the next section.

### 8.1.2 Different mass-loss processes

The mass loss rate of the disk may be decomposed in components according to the location of the mass-loss, or the processes involved (cf. Fig. 8.3). Two processes may effectively lead to a definitive loss of mass: the accretion mass-loss rate $\dot{M}_\mathrm{acc}$ (in the inner disk), and photoevaporation $\dot{M}_\mathrm{evap}$ (in the outer disk). If one wants to follow the mass of the inner or outer disk independently, the mass transfer $\dot{M}_\mathrm{in \to out}$ between the two parts, i.e. at the gravitational radius, have to be taken into account. Taking mass-loss, $\dot{M}$ as positive for a loss and choosing $\dot{M}_\mathrm{in \to out}$ to be positive for a mass motion from the inner to the outer disk, we have

$$\dot{M}_\mathrm{in} = \dot{M}_\mathrm{acc} + \dot{M}_\mathrm{in \to out}, \qquad (8.2)$$

for the inner disk mass-loss, and

$$\dot{M}_\mathrm{out} = \dot{M}_\mathrm{evap} - \dot{M}_\mathrm{in \to out}, \qquad (8.3)$$

for the outer disk mass-loss. The total mass-loss rate thus equal to

$$\dot{M}_\mathrm{total} = \dot{M}_\mathrm{in} + \dot{M}_\mathrm{out} = \dot{M}_\mathrm{acc} + \dot{M}_\mathrm{evap}. \qquad (8.4)$$

When looking at the disk evolution with time (Fig. 8.2), we can see that the loss of mass may be split into two main phases:

**Phase I** The mass loss is dominated by the photoevaporation flows that develop in the outer disk, $\dot{M}_\mathrm{evap}$.

**Phase II** The mass loss is dominated by viscous motion in the inner disk, inwards through accretion (matter lost to the central star, $\dot{M}_\mathrm{acc}$) or outwards to the outer disk (matter that will be then lost by photoevaporation, $\dot{M}_\mathrm{in \to out}$).

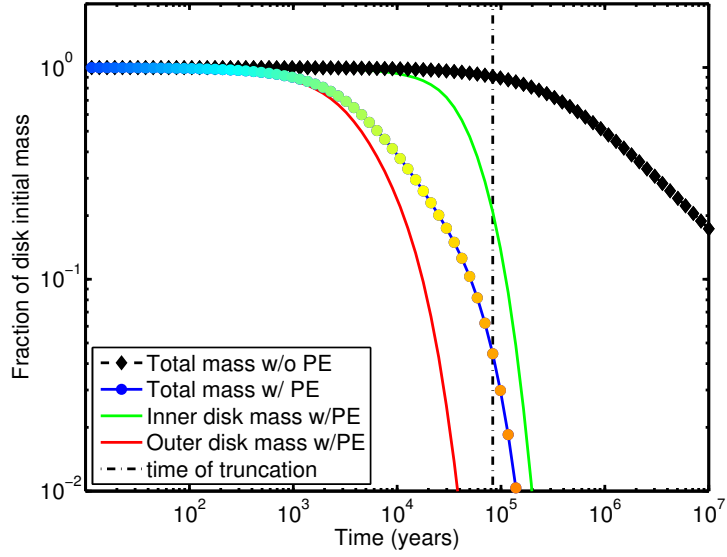

(a) Disk mass to initial mass ratio

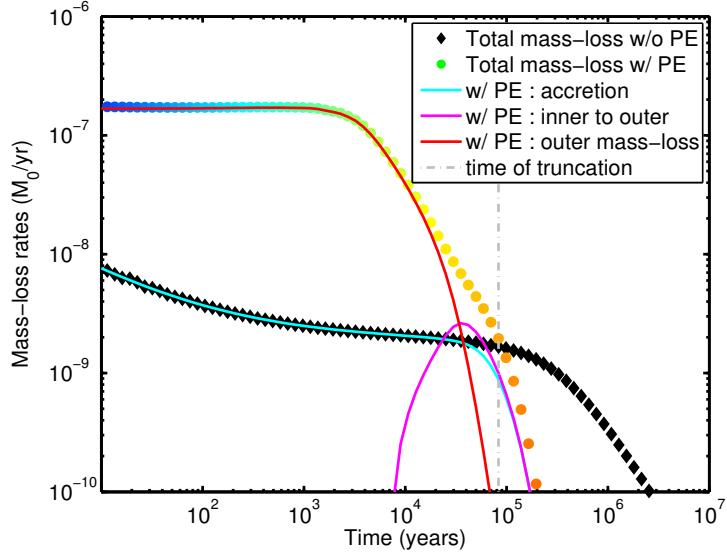

(b) Mass-loss rate

Figure 8.2: Time evolution of the disk mass to initial mass ratio (a) and mass-loss rate (b) for a disk with parameter $M_* = 0.13$ $M_\odot$, $M_{\text{disk}} = 1.7 \times 10^{-3}$ $M_\odot$, $r_{\text{out}} = 100$ AU and $\alpha = 10^{-3}$, under a supercritical photoevaporation defined by $n_{\text{surf}} = 10^{-6}$ cm$^{-3}$ and $T_{\text{surf}} = 1000$ K. 100 output iterations are plotted, logarithmically spaced in time from the initial state (blue) to 10 Myrs (red). The inner disk ($r \leq r_g$) and outer disk ($r > r_g$) components of the mass-loss rate are also depicted in cyan and magenta lines. Results for the same set of parameters but with no photoevaporation is also shown for comparison (black and dotted lines).

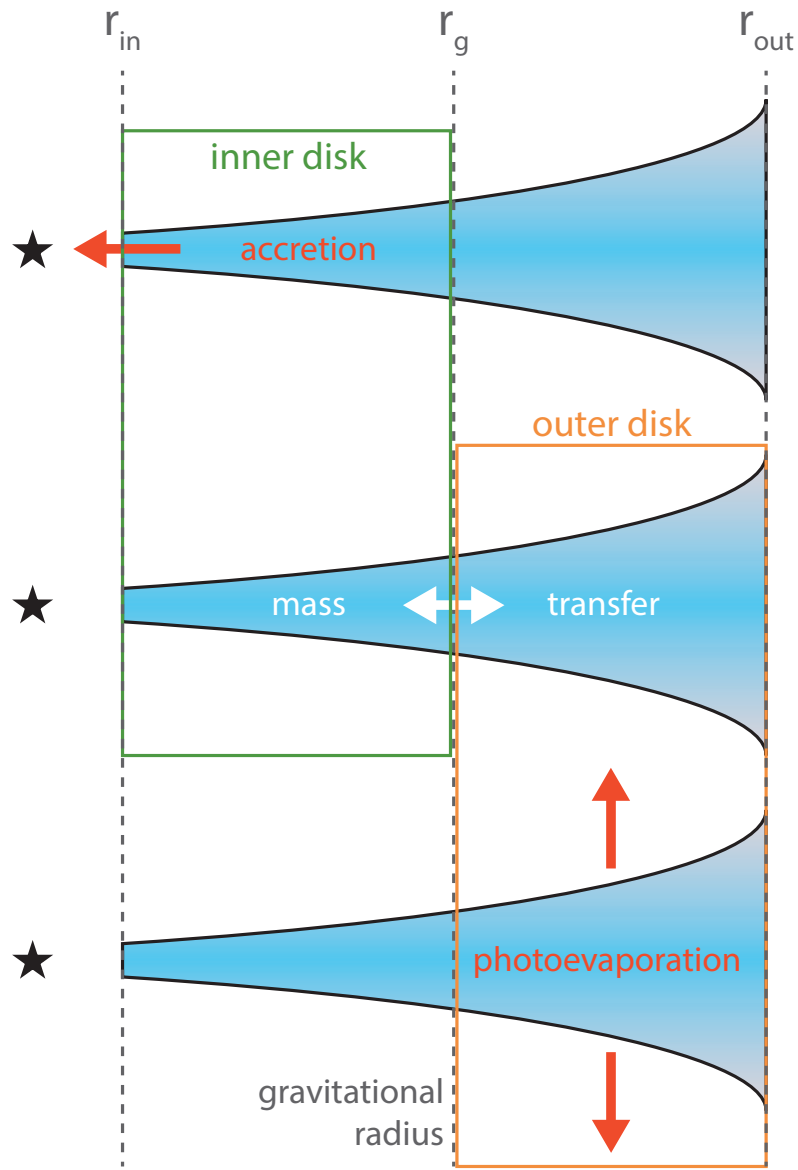

Figure 8.3: Illustration of the mass-loss components involved in a disk. The two processes that effectively lead to an absolute loss of mass within the disk are represented in red: accretion (for the inner disk) and photoevaporation (for the outer disk).

Table 8.1: Parameters of the models used to study the impact of viscosity.

| Category | Parameter | Symbol | Value |
|---|---|---|---|
| Masses | Central star mass | $M_*$ | 0.3 $M_\odot$ |
| | Disk initial mass | $M_\text{disk}$ | 0.01 $M_\odot$ |
| | Disk surface density profile | $\Sigma_0(r)$ | $\propto r^{-1}e^{-r}$ |
| Disk extension | Inner radius | $r_\text{in}$ | 0.01 AU |
| | Outer radius | $r_\text{out}$ | 100 AU |
| Viscosity | Mid-plane ref. temperature | $T_\text{mid,0}$ | 300 K |
| | Mid-plane ref. temperature position | $r_{0,\text{T}}$ | 1 AU |
| | Mid-plane temperature power-law index | $\beta_\text{T}$ | -0.5 |
| | Efficiency parameter | $\alpha$ | $10^{-4}$ to $10^{-2}$ |
| Photoevaporation | Density at the surface | $n_\text{surf}$ | $10^6$ cm$^{-3}$ |
| | Temperature at the surface | $T_\text{surf}$ | 1000 K |

The phase I starts as soon as the photoevaporation is on, here at the beginning of the simulation. During this phase, the outer disk, which may represent a significant fraction of the total disk mass, is removed by photoevaporation. The loss of the outer disk is apparently quite quick, with a duration of a few $10^4$ years in the example case. The phase II starts just before the truncation, when the outer disk is almost completely cleared. From this time on, there is no more substantial mass to lose in the initial outer disk and the loss of mass becomes driven by the viscous motion in the inner disk: accretion, which is present since the start of the simulation, and an outward motion at the gravitational radius, followed by photoevaporation, which settles when the outer disk is almost dissipated. The accretion onto the central star follows the same evolution as in a case without photoevaporation except that it drops when the inner disk starts significantly losing through its outer edge (cf. Fig. 8.2). Indeed, the outward motion at the gravitational radius transfers the inner disk mass to the outer disk region, where the matter will inevitably be lost by photoevaporation. This loss generally dominates over accretion at the start of the phase II. The final evolution of the inner disk is thus indirectly affected by photoevaporation, but mainly driven by its viscosity since it rules the motion of mass at both edges.

### 8.1.3 Model sensitivity to parameters

In this section, I describe the effect of varying the viscosity, as well as the mass of the star and disk, on out results.

**Effect of viscosity**

To study the effect of viscosity on the disk dispersal, one can compare the evolution of disks that differ only because of their viscosity. Here, I ran models with parameters given in Table 8.1, all fixed with the exception of the value of α, which describes the efficiency of viscosity, and which ranges between $10^{-4}$ and $10^{-2}$.

The impact of varying viscosity on the different components of the mass loss is presented in App. C.1. Table 8.2 summarises the impact of iscosity for the dispersal of the disk. The dispersal of the outer disk by photoevaporation, during the first phase, seems very slightly enhanced by a stronger viscosity, as illustrated with the anti-correlation between the value of α and the time of truncation. Indeed, a more efficient spreading of the disk leads to a larger surface subjected to photoevaporation. Consequently, the loss of mass is enhanced, and the dispersal of the outer disk occurs a bit faster. Here, an increase of two orders of magnitude in the viscosity leads to a truncation only two times faster.

Table 8.2: Effect of the viscosity on the disk evolution.

| Viscosity ($\alpha$ value) | Truncation time $t_{\text{tr}}$ ($10^6$ years) | Initial mass fraction at $t_{\text{tr}}$ (%) | Lifetime ($10^6$ years) |
|---|---|---|---|
| $1 \times 10^{-4}$ | 0.21 | 34.8 | 4.96 |
| $3 \times 10^{-4}$ | 0.21 | 27.4 | 1.73 |
| $1 \times 10^{-3}$ | 0.21 | 12.4 | 0.61 |
| $3 \times 10^{-3}$ | 0.15 | 4.2 | 0.25 |
| $1 \times 10^{-2}$ | 0.09 | 0.7 | 0.09 |

The processes leading to loss of mass in the inner disk, i.e. accretion and outward mass transfer, are driven by the viscosity and thus strongly depend on its efficiency. During the phase I, only the accretion is effective. Since the accretion is roughly proportional to the efficiency of the viscosity, the mass remaining in the disk at the end of the phase I is significantly reduced if the viscosity is high. The outward motion at the gravitational radius, which initiates at the start of the phase II, is also depending on the viscosity. Related to the late evolution of the inner disk, the lifetime of the disk[1] is thus strongly anti-correlated with the strength of viscosity. This is something well known for a disk not subjected to photoevaporation, but here the loss at both edges strengthen the relation.

**Effect of masses: $M_*$, $M_{\text{disk}}$ and $M_{\text{disk}}/M_*$**

Now that we have found that viscosity is highly involved in the evolution of the inner disk, and late evolution of the disk, let us study the impact of the masses in the system, i.e. the disk and its central star. The impacts on each component of the mass-loss rate, and their evolution, are detailed in App. C.2. We focus here on the final effect on the lifetime.

A simple way to estimate the duration of the phase I is to assume that the removed mass is the initial outer disk mass $M_{\text{out}}$ (mass located between $r_g$ and $r_{\text{out}}$), and that the mass-loss rate is constant, and given by equation (7.25). The lifetime of the outer disk is thus estimated to be $\tau_{\text{out}} = M_{\text{out}}/\dot{M}_{\text{evap}}$. However, stating that only the outer disk is removed is not completely true since it neglects the input from the inner disk, which can be important depending on the viscosity (cf. Sect. 8.1.3). Instead, the outer disk mass represents a lower limit of the available mass. A second wrong statement is to assume that the mass-loss rate is constant. Indeed, if there may be an initial enhancement, the trend is that the mass-loss rate will decrease with time. These two approximations both lead to an underestimation of the outer disk lifetime. The expression that can be derived from this simple calculation is thus a lower-limit, and is given by

$$\tau_{\text{out}} \approx 6.663 \text{Myr} \times \left(\frac{M_{\text{disk}}}{M_\odot}\right) \times \frac{\frac{e^{-r_g/r_{\text{out}}} - e^{-1}}{e^{-r_{\text{in}}/r_{\text{out}}} - e^{-1}}}{\left(\frac{r_{\text{out}}}{100 \text{ AU}}\right)^2 \left(1 - \left(\frac{r_g}{r_{\text{out}}}\right)^2\right)}, \quad (8.5)$$

An important outcome is that the lifetime is reduced if the initial disk mass $M_{\text{disk}}$ is lowered or the ratio $r_g/r_{\text{out}} \propto M_*$ is increased. To study more rigorously the effect of the stellar and disk masses, we have explored a grid of star masses from $10^{-2}$ to $1$ $M_\odot$, disk initial masses from $10^{-5}$ to $10^{-1}$ $M_\odot$, and assuming that the initial disk-to-star mass ratio remains be between $10^{-3}$ and 0.1 (a constraint from observations, see Fig. 2.3). The other parameters, give in Table 8.3 are kept fixed. Fig. 8.4(a) gives the theoretical lower limit of the outer disk lifetime, from equation (8.5), for the range studied. It shows that the lifetime is mainly depending on the disk initial mass and may reach about $10^5$ years at most. Fig. 8.4 gives the same information but from our model outputs. The same dependency is visible, confirming the correct idea behind the development of equation (8.5), but the true lifetime is obviously higher than the theoretical lower-limit, with here a factor from 1 to about 23 here. If the development is interesting, this shows that one can not estimate a lifetime based on one measurement of the mass-loss rate at one time, but only a lower-limit than can be very rough. However, numerical results confirm that the outer disks do not live more than a few $10^5$ years.

---
[1] The lifetime is here defined as the time for which only 1% of the total disk, or the region concerned, is remaining

Table 8.3: Parameters of the models used to study the impact of the stellar and disk masses.

| Category | Parameter | Symbol | Value |
|---|---|---|---|
| Masses | Central star mass | $M_*$ | 0.01 to 1 $M_\odot$ |
| | Disk initial mass | $M_{\text{disk}}$ | $10^{-5}$ to $10^{-1}$ $M_\odot$ |
| | Disk surface density profile | $\Sigma_0(r)$ | $\propto r^{-1} e^{-r}$ |
| Disk extension | Inner radius | $r_{\text{in}}$ | 0.01 AU |
| | Outer radius | $r_{\text{out}}$ | 100 AU |
| Viscosity | Mid-plane ref. temperature | $T_{\text{mid},0}$ | 300 K |
| | Mid-plane ref. temperature position | $r_{0,\text{T}}$ | 1 AU |
| | Mid-plane temperature power-law index | $\beta_{\text{T}}$ | -0.5 |
| | Efficiency parameter | $\alpha$ | $10^{-3}$ |
| Photoevaporation | Density at the surface | $n_{\text{surf}}$ | $10^6$ cm$^{-3}$ |
| | Temperature at the surface | $T_{\text{surf}}$ | 1000 K |

The dynamical evolution of the inner disk is more difficult to estimate a priori. However, it is mainly driven by viscosity (Sect. 8.1.3), where $\nu \propto \alpha/\sqrt{M_*}$ (see 7.21), so that its lifetime is supposed to be anti-correlated with the central star mass. This is effectively what can be seen Fig. 8.5. The lifetime of the inner disk can be much longer than for the outer disk, lasting here for more than a million of years for the disks with a central mass star more than $10^{-0.5} \approx 0.3$ $M_\odot$. Those disks have still a few tenth of percents of their initial inner disk mass after a million of years (up to about 45% here). Fig. 8.5(b) illustrates the part of the mass loss due to the photoevaporation by mass transfer from the inner to the outer disk. It appears that the accretion is generally the dominant source of mass loss while the indirect loss due to photoevaporation can still be responsible for around 60% at most, for the lowest disk masses around the most massive stars.

### 8.1.4 Comparison with other photoevaporation mechanisms

We also ran models including prescriptions for internal photoevaporation caused by UV photons (Gorti and Hollenbach, 2009), or X-rays (Owen et al., 2012), originating from the central star. This is illustrated in Fig 8.6 with the example of a rapidly evolving disk under those processes. UV photons and X-rays act preferentially at some locations of the disk, creating firstly a hole and further eroding the disk from each side of it. The maximum of their induced mass-loss rate is generally located at a few astronomical units, so within the inner disk in general. Logically, the typical supercritical external photoevaporation when the external FUV field leads to an expected surface temperature of 1000 K (as for the studied value of $G_0$ from $10^3$ to $10^5$), dominates the mass-loss of the outer disk. This is true even when taking high FUV, EUV or X-ray activities of the central host star. When the surface temperature is about 100 K, as it could be expected for a mean interstellar radiation field, external photoevaporation is not so efficient but still significant. Compare to this case, photoevaporation by UV photons from the central star may also cause significant mass-loss in the outer disk. X-rays, however, have no real impact there except at the inner disk for strong X-ray luminosity.

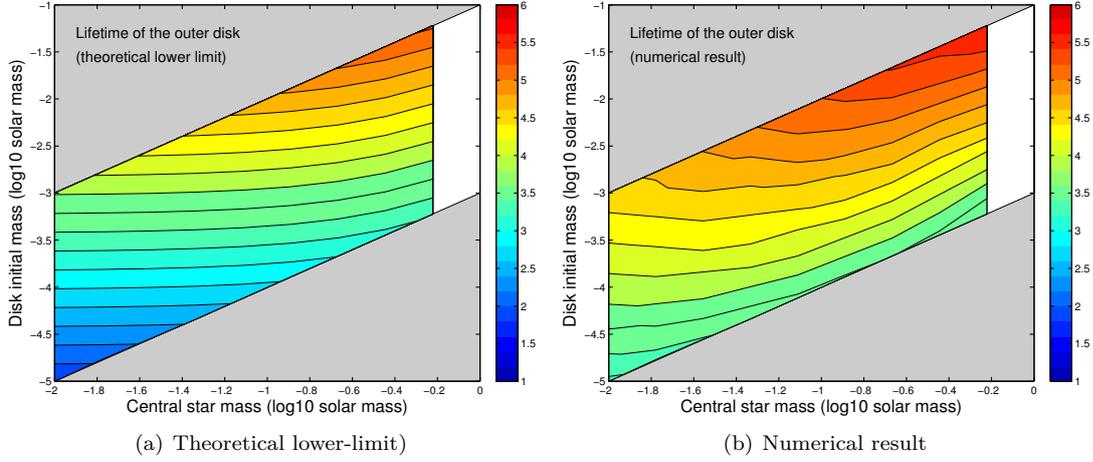

(a) Theoretical lower-limit)　　　　　　(b) Numerical result

Figure 8.4: Lower-limit theoretical estimate and numerical result of the lifetime of the outer disk, given as the log10 of years, for a grid of star and disk masses. The color filled contour and white area show the range modeled, i.e. star mass between $10^{-2}$ and 1 $M_\odot$ with a disk mass between $10^{-3}$ and $10^{-1}$ $M_\odot$. Each color indicates a given range of values while white means any value below the range of the color bar. The grey regions are regions not modeled here.

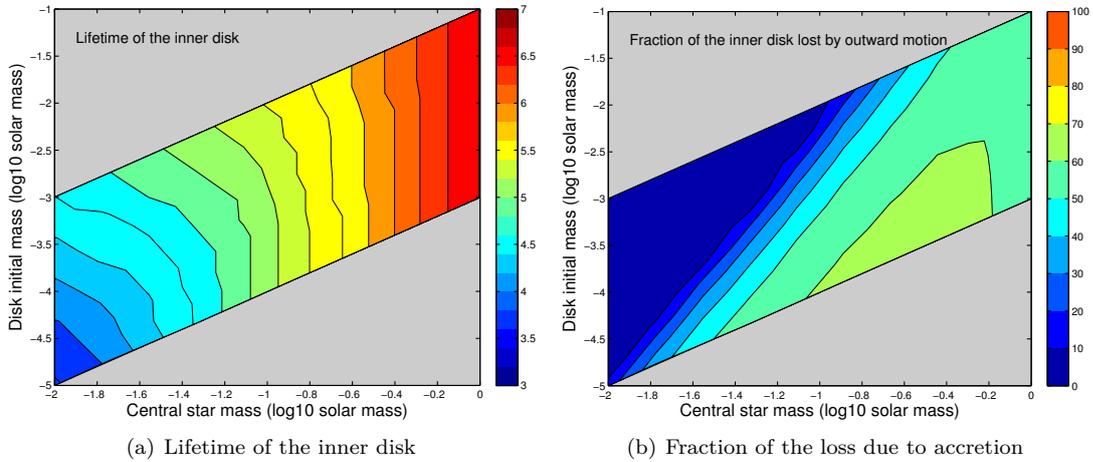

(a) Lifetime of the inner disk　　　　　　(b) Fraction of the loss due to accretion

Figure 8.5: Lifetime of the inner disk (a), given as the log10 of years, and fraction of the mass loss due to outward motion followed by photoevaporation (b), for a grid of star and disk masses. The color filled contour and white area show the range modeled, i.e. star mass between $10^{-2}$ and 1 $M_\odot$ with a disk mass between $10^{-3}$ and $10^{-1}$ $M_\odot$. Each color indicates a given range of values while white means any value below the range of the color bar. The grey regions are regions not modeled here.

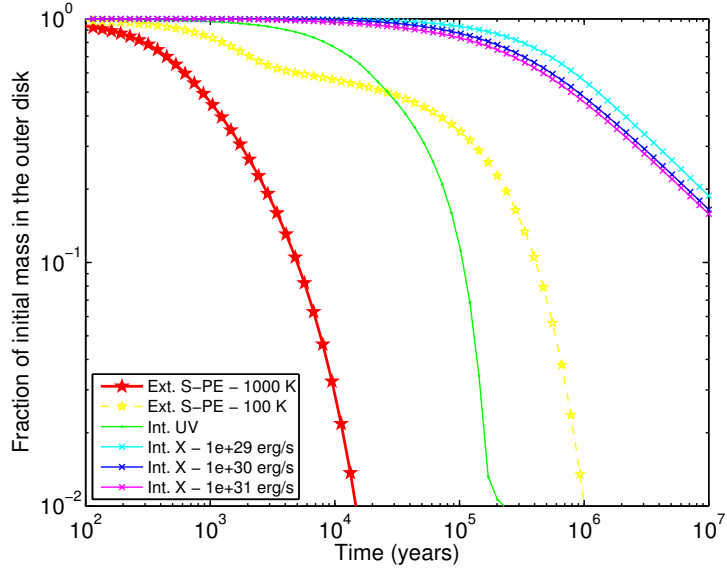

(a) Evolution of the mass in the outer disk

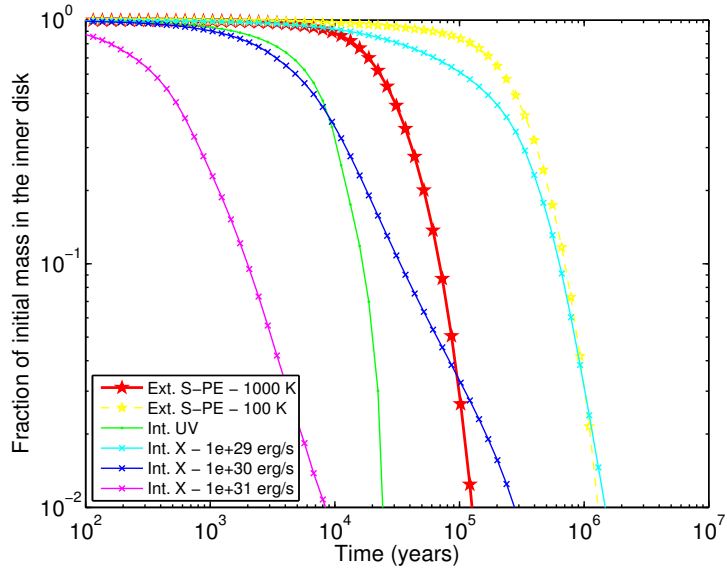

(b) Evolution of the mass in the inner disk

Figure 8.6: Time evolution of the mass for a rapidly evolving disk subject to different photoevaporation mechanisms. The top panel gives evolution for the outer disk while the bottom panel gives the one for the inner disk. The typical external supercritical photoevaporation, with a surface temperature of 1000 K (red curve), is compared to: external photoevaporation in case of a mean interstellar radiation field, with a temperature around 100 K (yellow); internal photoevaporation by UV photons (green); internal photoevaporation by X-rays with different central star luminosities from a weak value of $10^{29}$ erg s$^{-1}$ (light blue) to a typical value of $10^{30}$ erg s$^{-1}$ (dark blue) and then a very high value of $10^{31}$ erg s$^{-1}$ (purple).

As the effect of external supercritical photoevaporation is not direct in the inner disk, this process does not dominate the loss of mass. The internal UV photons and potentially X-rays, if the luminosity is relatively high, are indeed more efficient there. This confirms that the evolution of the inner disk is not driven by the external photoevaporation.

## 8.2 Statistical analysis of the Orion Cluster

It has been proposed that the lack of low-mass disks in the Orion nebula compared to the Taurus nebula is due to external photoevaporation (see e.g. Adams et al., 2004; Mann and Williams, 2010). In case of supercritical external photoevaporation, our modelling confirms this idea through the low outer disk lifetimes of low-mass disks (cf. Fig. 8.4). Since disks evolve differently according to their mass parameters (Sect. 8.1.3), one can wonder how a population of disks will evolve with time under supercritical external photoevaporation. We explored if general patterns observed in the disk population of the Orion nebula, which likely experiences significant external photoevaporation, can be reproduced by our disk evolution model. We simulated this by taking an initial population within the range of our studied grid of masses (see Table 8.3), scattered to be consistent with the observed distributions: the initial mass function of the Orion Nebula Cluster (ONC) for the central stars (Hillenbrand and Carpenter, 2000); and a flat distribution of disk masses as observed in the Taurus nebula (Williams and Cieza, 2011), taken as the initial state of the ONC before photoevaporation is switched on. The evolution of the disk mass are then interpolated between the exact values obtained for the models on the grid.

Fig. 8.7 gives the modeled evolution of the disk mass distribution while comparing to observations. Actually, starting with a flat distribution of disk masses similar to the one in the Taurus, it is clear that photoevaporation makes the distribution evolve in a way that low-mass disks become underabundant with a peak value around a mass of $10^{-2}$ M$_\odot$, as observed today in Orion, which is a few $10^5$ years likely.

Another prediction of our model is that all disks tend to reach their gravitational radius at some time, either by photoevaporation of the outer disk or by expansion of the inner disk. Since the gravitational radius is defined by the central star mass, the size distribution of the disks is supposed to evolve towards the one of the IMF, i.e. a power-law of index about $-1$. Only few observations of disk sizes in the ONC have been carried (Vicente and Alves, 2005) but the current distribution seems to looks like a power-law of index $-1.5$ rather. If we believe our asymptotic behaviour to reach an index of -1, this means either that the disks in the ONC are still too young to have reached this state, or that other effects drive significantly the disk size so that there is no reason to reach this predicted state. The observed index around $-1.5$, potentially transient, is consistent with our simulations but is not conclusive since a large population of transitional distribution may be found according to the initial distribution of size chosen. Unfortunately, we have no real constraint on the initial one.

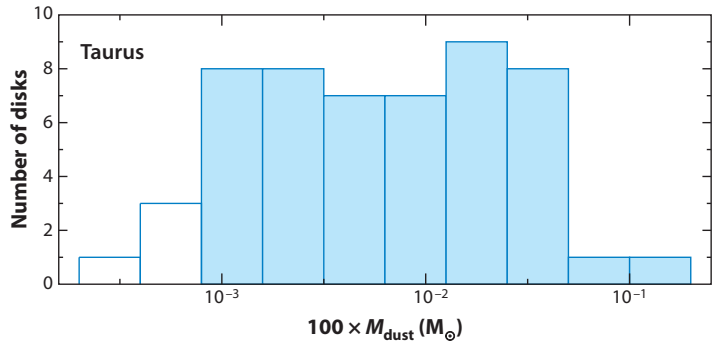

(a) Mass distribution of the disk in the Taurus

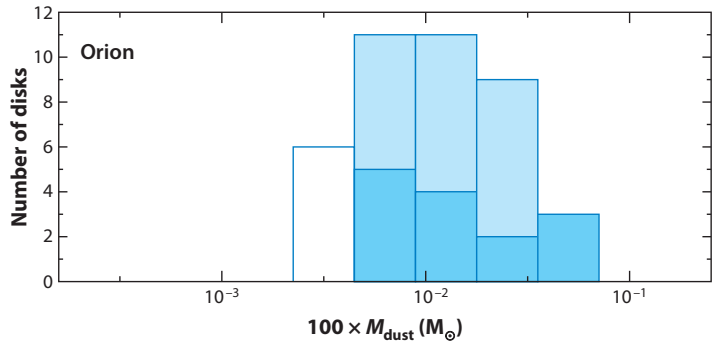

(b) Mass distribution of the disk in the Orion nebula cluster

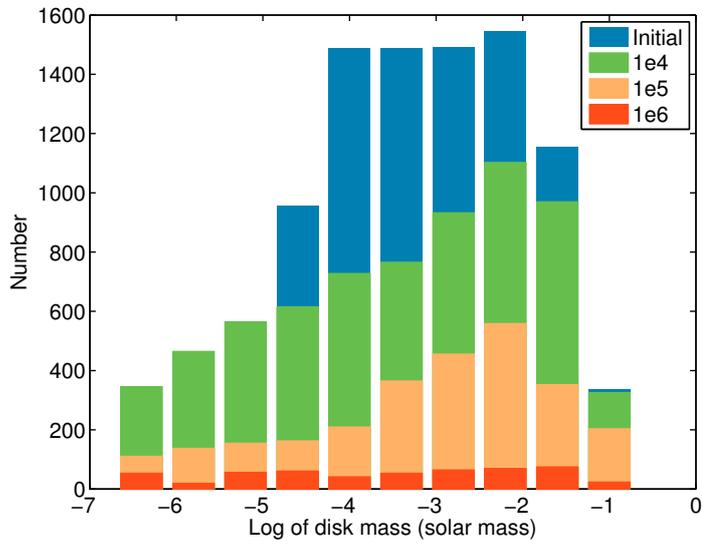

(c) Model of the evolution of the mass distribution in the Orion nebula cluster for 10 000 disks

Figure 8.7: Comparison between the observed and modeled disk mass distribution: observations in the Taurus (a), observations (b) and modelling after 0, $10^4$, $10^5$ and $10^6$ of evolution (c) in the Orion nebula cluster. Observed mass distributions are taken from Williams and Cieza (2011).

## 8.3 Discussion

### 8.3.1 Limits and caveats of the model

**Prescription for the supercritical photoevaporation**

Our prescription for the external supercritical photoevaporation has the advantage to be derived from physical parameters determined owing to PDR models constrained by far infrared observations (Sect. 5). However, the estimates of the physical conditions are only valid for environmental conditions close to the observations, that it to say the harsh environments of Orion and Carina nebula clusters ($G_0$ between $10^3$ and $10^5$) and one should be careful when trying to extrapolate to other conditions where the supercritical regime may be less efficient or even marginal.

An ideal solution would be to be able to run an accurate PDR model in order to extract, at each iteration of the viscous disk model, or for a few of them, the physical conditions at the disk surface to iteratively estimate the mass-loss rate. This would imply to know also precisely how the atomic envelope surrounding the disk evolves since it has been found to be a crucial parameter to set the disk surface temperature (Sect. 5.5). Currently, this modelling would be very time consuming and not necessarily more precise without new observational constraints. The future JWST, with its high spatial resolution (about 0.1″) in the infrared, should bring new observational constraints on the structure and physical conditions of photoevaporating disks.

**Viscosity via the α-model**

The α-model, used here to describe turbulent transport of angular momentum, is not completely realistic when considering a single value representative of the whole disk. Indeed, the value of α is highly dispersed from one disk to another but can also be different from one region of the disk to another, and may as well depend on time. However, this model has been extensively used in the literature and provides parsimonious description of a substantial body of data.

**Robustness**

The trends of our model outcomes are robust since different values and sets of parameters have been tested and give the same conclusions: a short lifetime for the outer disk and late viscous evolution for the inner disk. The parameters describing the supercritical photoevaporation may evolve with time or be very different from one photoevaporating disk to another. However, our study in Sect. 5.5 indicates that the surface temperature is rather stable because of the adjustment of the proplyd envelope, and the location of the H/$H_2$ transition (Sect. 5.5.3). By using a single value to set the density at the surface of the disk, we postulate that it is supposed to be of the same order of magnitude along the disk. This assumption was previously checked by Gorti and Hollenbach (2009) using PDR models. Unless there is a very critical variation in density or temperature, at the disk surface, that would lead to a large drop of efficiency in the photoevaporation process, the outer disk is expected to be removed quickly and the disk to be truncated. The radius of truncation, which is here precisely the gravitational radius, may be slightly different in reality if one considers a more detailed treatment of the natural velocity dispersion within a disk (related to the natural thermal dispersion) that may lead to a broad transitional region between the inner and outer disk rather than a particular radius. This may affect the surface density profile and potentially the mass transfer between the inner and outer disk. However, it remains true that the inner disk is protected from direct photoevaporation and that the viscosity will drive the final evolution of it. This late evolution is thus highly dependent on the viscous strength of the disk. We will thus discuss the consequences of our results in the following sections.

### 8.3.2 Is there a proplyd lifetime problem?

A quick estimate of the disk lifetime with equation (8.5) based on observational constraints at one time leads to values of the order of $10^4$ years for the population of disks studied here. Hence proplyds should be extremely rare in the ONC, assuming an age of a few $10^5$ years, unless they are not subject to photoevaporation since the cluster birth but much more recently instead. In our modelling, we expect high mass-loss rates during the life of the outer disk, which is about $10^5$ years, while the inner disk may survive a few $10^6$ years depending mainly on the viscosity and the central star mass. Our estimates are thus consistent with the fact that we still see disks, and some in the form of proplyds, in the ONC even if disk photoevaporation happens very shortly in the cluster history and without any need to use non-typical disk properties.

### 8.3.3 Consequences on planet formation

If protoplanetary disks dissipate far more rapidly with external photoevaporation, one may wonder what the consequences are on planet formation. On a timescale of $10^5$ years, it seems difficult to form any gas giant planet except in the case of gravitational instabilities. Following the usual core accretion scenario, we do not therefore expect to form any giant planets in the outer disk on such short timescales. However, the inner disk can survive longer. This is especially true for the disks around the most massive of the low-mass stars, which have a more extended inner disk that survive longer, and for disks with low viscosities. In such a harsh environment, one could thus still expect terrestrial and giant planet formation to happen in the inner disk. This is consistent with recent observations of exoplanets which indicate that planets, of any kind, form around most of the stars. Actually, to date, the vast majority of exoplanets have been found closer than 10 AU from their host star. While there are unambiguously a lot of observational biases in exoplanets statistics, this is probably also reflecting that planets form preferentially close to their star, or migrate inward. A very small number of exoplanets have been detected, mostly by direct imaging, at a few tens or hundreds astronomical units. Those planets exist but more statistical data are needed to state about their occurence or not around a low-mass star located in the vicinity of a massive star.

The formation of a gas giant planet in the outer disk, at several tens or hundreds of astronomical units to the central star, is possible if one considers the gravitational instabilities scenario since it is very fast and may occur before the outer disk is completely dissipated. Besides, Throop and Bally (2005) found significant photoevaporative enhancement of the dust-to-gas ratio that can trigger the gravitational instability. If this happens, such a planet will be subject to the high FUV radiation when the outer disk is dissipated, leading to the existence of an interesting "hot jupiter" far from its host star and heated by external massive stars. With a surface temperature about 1000 K, this kind of planet could be easily detectable in NIR direct imaging with a peak emission close to 3 µm and a high angular separation from its host star that may, moreover, be hidden by the possibly still existing inner disk if observed edge on. The technology to image such a planet already exists (see e.g. HD 100546 b studied by (Quanz et al., 2015) but the arrival of the *James Webb Space Telescope* and numerous observations to come in star forming regions will confirm if such objects may exist.

## 8.4 Conclusions on disk evolution under external supercritical photoevaporation

The external photoevaporation, in the supercritical regime, tends to truncate disks to their gravitational radius (a few tens of AUs for $0.1 - 1$ $M_\odot$ stars), in a time less than or about $10^5$ years. During this truncation, which sets a first phase in the mass loss evolution, the outer disk, that correspond generally to the most part of the mass, is removed. When the disk is almost truncated, the viscous motion dominates the loss of mass of the remaining inner disk (mass transport from the inner to the outer disk and accretion). This second phase drives the lifetime and the late evolution of the remaining mass in the planet formation region. As for disks not subjected to photoevaporation, the dispersal of the inner disk is mainly depending on the efficiency of the viscosity, while the indirect effect of the photoevaporation slightly enhances the loss.

Because of a short lifetime (below or about $10^5$ years), giant planet formation, as depicted in the classical core accretion scenario, might be difficult in the outer parts of a disk subjected to external photoevaporation. However, it is still theoretically possible to form a giant via gravitational instabilities, and could lead to the formation of a "hot distant jupiter" (if the planet does not migrate significantly) heated by external FUV photons. The inner disk, whose extension depends on the central star mass, can be only slightly affected by photoevaporation, especially if the host star is relatively massive and if the viscosity of the disk is low. Even under harsh conditions, planet formation via the classical core accretion scenario is still possible there.

Disk and star masses also affect the evolution of the disk in such a way that disks with a low mass will be dissipated earlier so that the disk population evolves naturally towards a distribution favouring high-mass disks. If only external supercritical photoevaporation drives the disk size, the disk population tends to converge to their gravitational radius so that the disks size distribution finally reflects the initial mass function. Unfortunately, not enough statistical data are available yet to confirm this prediction observationally. For the case of the Orion nebula cluster, our prescription for the external photoevaporation, which is consistent with observations and mass-loss rate estimates, is also able to reproduce the disk mass distribution currently observed. In such a region, disks can survive long enough to be still observed as proplyds today, solving partly the "proplyd lifetime problem".

# Conclusions and perspectives

# Conclusions et perspectives (français)

Les objectifs de cette thèse étaient d'étudier la photoévaporation dans le cas particulier où elle est due à des photons FUV, d'identifier les principaux paramètres physiques (densité, température) et processus (chauffage et refroidissement) impliqués, puis d'en estimer l'impact sur l'évolution dynamique des disques.

La première partie du travail de recherche consistait à sonder les régions de photodissociation (PDR) de quatre disques protoplanétaires subissant de la photoévaporation externe. Appelés "proplyds", ces objets se distinguent par une large enveloppe de matière qui entoure leur disque, et qui témoigne de la photoévaporation intense qui s'y produit. Grâce au caractéristiques de l'observatoire spatial Herschel, des objets de ce type ont pu être observés pour la première fois dans l'infrarouge lointain, domaine spectral privilégié pour l'émission des PDRs denses. A l'aide d'un modèle 1D d'une PDR, j'ai développé un modèle pour l'émission des proplyds. Ce modèle a été utilisé pour interpréter ces observations, ainsi que d'autres données du visible, avec Hubble, au submillimétrique, avec ALMA. Bien que les objets étudiés soient différents sur de nombreux points, il apparait que les conditions physiques en surface de leur disque sont similaires : une densité de l'ordre de $10^6$ cm$^{-3}$ et une température d'environ 1000 K. Le régime thermique particulier de ces objets, resultant de la compétition de nombreux processus de chauffage et refroidissement, apparait lorsque la transition H/H$_2$ est proche de la surface du disque. La température de surface du disque, qui découle de ce régime, est maintenue par un équilibre dynamique : si la surface se refroidit, la perte de masse diminue et l'enveloppe se réduit. L'atténuation UV produite par l'enveloppe entre le front d'ionisation et la surface du disque diminue alors, et le disque, recevant plus de photons UV, chauffe. Dans ces conditions, la majorité du disque peut s'échapper sous forme de flots de photoévaporation. On parle alors de photoévaporation en régime *super-critique*. Laissé de côté dans les recherches récentes, ce régime semble pourtant dominer dans le cas des proplyds, et les taux de perte de masse qui en découlent, de quelques $10^{-7}$ M$_\odot$ yr$^{-1}$ ou plus, sont en accord avec les observations précédentes des traceurs du gaz ionisé.

A la suite de ce travail, j'ai développé un modèle hydrodynamique 1D pour étudier l'évolution dynamique d'un disque en photoévaporation. Le code inclut l'évolution visqueuse due à la turbulence et des prescriptions pour différents types de photoévaporation, interne ou externe, incluant la prescription dans le cas *super-critique* défini par les observations et modèles décrits précédemment. Dans ce dernier cas, deux régions du disque évoluent différemment. L'évolution dynamique du disque externe, où les flots de photoévaporation se développent, est dominée par la photoévaporation. Comme peuvent le prédire de simples calculs analytiques, le disque externe se dissipe dans un temps de l'ordre de $10^5$ ans au plus, laissant un disque rapidement tronqué. Le disque interne, où la gravitation retient les flots, est aussi affecté, mais indirectement. Une fois le disque externe quasiment dissipé, le disque interne perd de la masse via un transfert de matière en direction du disque externe. Cette matière est ensuite perdue par photoévaporation. La perte de masse induite par ce dernier processus est du même ordre que celle provoquée par l'accretion. L'impact de la photoévaporation est donc limité et le disque interne, dont l'évolution dynamique est plutôt dominée par la viscosité, peut survivre quelques $10^6$ années. Finalement, j'ai effectué une étude statistique en utilisant ce modèle et montré qu'il était capable de reproduire la fonction de masse des disques d'Orion, où la photoévaporation externe est supposée importante. D'après mes résultats, la photoévaporation externe en régime *super-critique* est très efficace dans le disque externe. La formation de planètes y est alors très difficile, mais la possibilité de former des géantes par instabilité gravitationnelle reste existante, ce qui donnerait lieu à un type de planète théorique : des "Jupiters chaudes lointaines". Malgré cet environnement défavorable, la survie suffisamment longue du disque interne (pouvant être de quelques millions d'années) permettrait également la formation de planètes par le scénario classique d'accrétion par un coeur.

Dans le futur proche, l'arrivée du *James Webb Space Telescope* et sa haute résolution spatiale dans l'infrarouge, en synergie avec ALMA dans le submillimétrique, vont permettre de nous apporter une vision plus locale des propriétés des flots de photoévaporation et de leurs effets. A partir des observations, la dérivation des conditions physiques devrait en être améliorée, donnant ainsi la possibilité de commencer à distinguer la structure d'un disque en photoévaporation. Une proposition d'observation du proplyd 203-506 avec ALMA, dans le cadre du Cycle 5, a été soumise dans ce but. L'observation directe de la dynamique locale des flots de photoévaporation permettrait également d'étudier plus précisément l'effet direct, et indirect, de la perte de masse sur le disque externe, et interne, respectivement. Les effets sur le disque interne pourraient être suffisamment important pour influer sur les caractéristiques du futur système planétaire. Ces conséquences se cachent probablement dans les exosystèmes que nous observons actuellement. Les modèles de synthèse de populations planétaires, étudiant les impacts à l'échelle d'une grande population de planètes, sont une perspective prometteuse de ce travail. Ils pourraient offrir la possibilité d'étudier la divergence de la trajectoire évolutive des systèmes planétaires en formation et irradiés par des étoiles massives voisines, par rapport à ceux qui se forment de manière plus isolée.

# Conclusions and perspectives (english)

The objectives of this thesis were to study the photoevaporation, in the particular case where it is due to FUV photons, to identify the main physical parameters (density, temperature) and process (heating and cooling) involved, and then to estimate the impact on the dynamical evolution of disks.

The first part of the research work was to probe the photodissociation regions (PDR) of four protoplanetary disks undergoing external photoevaporation. Called "proplyds", these objects are distinguished by large envelopes of material that surround their disks, and which highlight the intense photoevaporation that occur there. Thanks to the characteristics of the Herschel space observatory, objects of this type could be observed for the first time in the far infrared region, the main spectral domain for the emission of dense PDRs. Using a 1D model of a PDR, I developed a model for the emission of proplyds. This model was used to interpret these observations, as well as other data from the visible, with Hubble, to the submillimeter range, with ALMA. Although the studied objects are different on many aspects, it appears that the physical surface conditions of their disk are similar: a density of the order of $10^6$ cm$^{-3}$ and a temperature about 1000 K. The particular thermal regime of these objects, resulting from the competition of many heating and cooling processes, arises when the transition H/H$_2$ is located close to the surface of the disk. The surface temperature of the disk, resulting from this regime, is maintained by a dynamic equilibrium: if the surface cools, the loss of mass decreases and the envelope is reduced. The UV attenuation produced by the envelope, between the ionisation front and the surface of the disk, then decreases and the disk, receiving more UV photons, heats up. Under these conditions, most of the disk can escape through photoevaporation flows. This corresponds to the *super-critical* regime of photoevaporation. Put aside in recent researches, this regime seems yet to dominate in the case of proplyds, and the resulting mass-loss rates of a few $10^{-7}$ M$_\odot$ yr$^{-1}$ or more, are in agreement with earlier spectroscopic observations of ionised gas tracers.

Following this work, I developed a 1D hydrodynamic model to study the dynamical evolution of a disk under photoevaporation. The code includes the viscous evolution driven by turbulence and prescriptions for different types of photoevaporation, internal or external, including the prescription in the *supercritical* case defined by the observations and models described above. In the latter case, two regions of the disk evolve differently. The dynamical evolution of the outer disk, where photoevaporation flows develop, is dominated by photoevaporation. As it can be predicted from simple analytical calculations, the outer disk dissipates in a time of the order of $10^5$ at most, leaving a disk quickly truncated. The inner disk, where gravity holds the flows, is also affected, but indirectly. Once the outer disk has been almost dissipated, the inner disk loses mass through mass transfer towards the outer disk. This matter is then lost by photoevaporation. The loss of mass induced by this mass transfer is of the same order as the one caused by accretion. The impact of photoevaporation is therefore limited and the inner disk, whose dynamical evolution is rather dominated by viscosity, can survive for a few $10^6$ years. Finally, I carried out a statistical study using this model and showed that it was able to reproduce the mass function of the Orion disks, where external photoevaporation is supposed to be of importance. According to my results, external photoevaporation in *supercritical* regime is very effective in the outer disk. The formation of planets is then very difficult there, but the possibility to form giants by gravitational instability remains, which would give rise to a theoretical planet type: "distant hot Jupiters". Despite this harsh environment, the long survival of the inner disk (which could be of a few million years) could also allow the formation of planets by the classic core accretion scenario.

In the near future, the upcoming *James Webb Space Telescope* and its high spatial resolution in the infrared, in synergy with ALMA in the submillimeter, will permit to probe the properties of the photoevaporation flows more directly and to get a local view of its effects. From the observations, the derivation of the physical conditions should be improved, thus giving the possibility to start distinguishing the structure of a disk in photoevaporation. An ALMA Cycle 5 proposal to observe the proplyd 203-506 was submitted for this purpose. The direct observation of the local dynamics of the photoevaporation flows would also permit to study more precisely the direct and indirect effects of the mass-loss on the outer, and inner, regions of the disk respectively. The effects on the inner disk could be sufficiently important to affect the characteristics of the future planetary system. These consequences are probably hidden in the exosystems we are currently observing. The planetary population synthesis models, studying the impacts on a scale of a large population of planets, are promising perspectives of this work. They could offer the possibility to study the divergence in the evolutionary track of planetary systems in formation and irradiated by neighbouring massive stars, in comparison with systems in formation that are relatively isolated.

# Appendices

# Appendix A

# Hertzsprung-Russel diagram

The Hertzsprung-Russel diagram (or HR diagram) links the absolute luminosity of a star to its actual temperature (or associated magnitude). The specificity of such a diagram is that it allows, at the same time, to visualise the different morphologies of the stars and their evolution.

90% of the stars that we can see are at a stage that we could call "adult", which correspond to the main sequence of the HR diagram (Fig. A.1). The particularity of these stars is that there is a well-defined relationship between their actual temperature, their absolute luminosity, their mass and their lifetime. For example, the average lifetime $\tau_{\text{life}}$ of a star can be obtained from its mass, $M_*$, or its luminosity, $L_*$, by

$$\begin{aligned}\tau_{\text{life}} &\approx 10^{10} \left(\frac{\text{M}_\odot}{M_*}\right)^2 \quad \text{years,} \\ &\approx 10^{10} \left(\frac{\text{L}_\odot}{L_*}\right)^{2/3} \quad \text{years,}\end{aligned} \tag{A.1}$$

where $\text{M}_\odot \approx 1.989 \times 10^{30}$ kg and $\text{L}_\odot \approx 3.828 \times 10^{26}$ W are the mass and luminosity of the Sun respectively. This diagram can also be used to illustrate the spectral classification used to distinguish stars according to spectroscopic criteria. The standard classification is the one of Harvard, based on the temperature of the stars (see Table A.1). It contains 9 categories of which the first 7 (O, B, A, F, G, K, and M) covers about 99% of the stars. Each class is divided into subclasses numbered from 0 to 9, from the hottest to the coldest. Spectral classes are used to describe all stars, regardless of their type. To differentiate the types, we then use a Roman numeral, following the class to describe the luminosity and nature of the object according to the following convention (e.g. G0II, for a one of the hottest G star which is a luminous giant):

**I** Supergiant,
**II** Luminous giant,
**III** Giant,
**IV** Sub-giants,
**V** Dwarfs,
**VI** Sub-dwarfs,
**VII** White dwarfs.

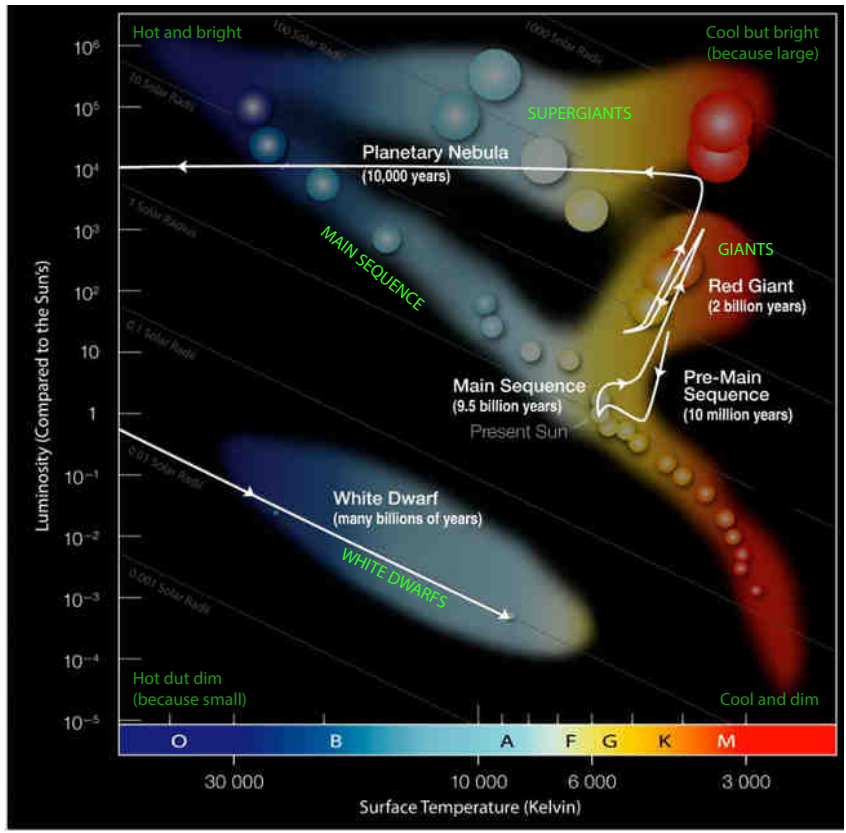

Figure A.1: Evolutionary track of the Sun across the H-R diagram. Light green words, in capital, indicate the position of different classes of objects in the diagram. Dark green words in the corners the trends in terms of temperature and luminosity. The white line illustrates the evolutionary track of the Sun, associated with indications on the stage denomination and duration. Figure adapted from the educational website of the Chandra X-ray observatory (http://www.chandra.harvard.edu).

Table A.1: Harvard spectral classification

| Spectral type | Effective temperature (K) | Color | Spectral signatures |
|---|---|---|---|
| O | $> 25000$ | blue | He II |
| B | $11000 - 25000$ | blue – white | He I |
| A | $7500 - 11000$ | white | H |
| F | $6000 - 7500$ | yellow – white | H, Ca II |
| G | $5000 - 6000$ | yellow | Ca II, H, K, CH |
| K | $3500 - 5000$ | orange – yellow | metals |
| M | $< 3500$ | red | Ti, TiO |
| C (rare) | $< 3500$ | red | C2, CN, CH, ZrO |
| S (rare) | $< 3500$ | red | Zr0 |

# Appendix B

# More details about *Herschel* observations and PDR modelling

## B.1   105-600: case of the [CII] emission

As can be seen in Fig. 4.3(b), the fine-structure line of [CII] from the target does not seem well defined. Indeed, since the observing mode chosen was dual beam switch (see Table 4.3), the final spectrum is the difference between a spectrum obtained on the target (ON position) and others off the target (OFF position). While expected to be flat, the OFF-position spectrum is apparently not free of emission for this line so that we sum a positive line with a negative one. Fortunately, we have some information about the target as well as for the nebula which is seen in the OFF position. It is known from Sahai et al. (2012) and from our other resolved lines (see Table 4.4) for this object that it has a $v_{\text{LSR}}$ about $-23$ km s$^{-1}$. This confirms that the positive peak seen on the resulting spectrum is the target emission. Concerning the OFF emission from the nebula, it is known that it comes from a HII region where the temperature is about $10^4$ K. At this high temperature, the width of the line is dominated by the thermal doppler broadening. With the constraints on the position of the target line and the width of the nebular line, we can fit the spectrum (Fig. B.1).

Our model, along with these two gaussian lines, fits quite well the part of the spectrum with $v_{\text{LSR}} < -10$ km s$^{-1}$ while there are still some features on the remaining part which do not overlap the target emission. The constraints are sufficient to retrieved the target profile with a good uncertainty (see Table 4.4).

## B.2   Effect of grain parameters (test on 105-600)

While grains represent only a small fraction of the total mass, a correct treatment of their role is crucial since the photoelectric effect on dust grains (Bakes and Tielens, 1994) is supposed to be one of the major sources of heating and because they are responsible for the UV extinction in the PDR. The grain population is generally not precisely constrained in an evolving protoplanetary disk. It should slightly differs from the one of the ISM, so we investigated the effect of changing the main grain parameters in the model: the grain size distribution given by the minimum radius ($a_{\text{min}}$), the maximum radius ($a_{\text{max}}$) and the power-law index, and the dust-to-gas mass ratio $\delta$.

If, contrary to the models presented in the main text, we include the smallest particles, i.e. we set $a_{\text{min}}$ to 0.4 nm instead of 3 nm, we observe very minor changes in the modelled line intensities:

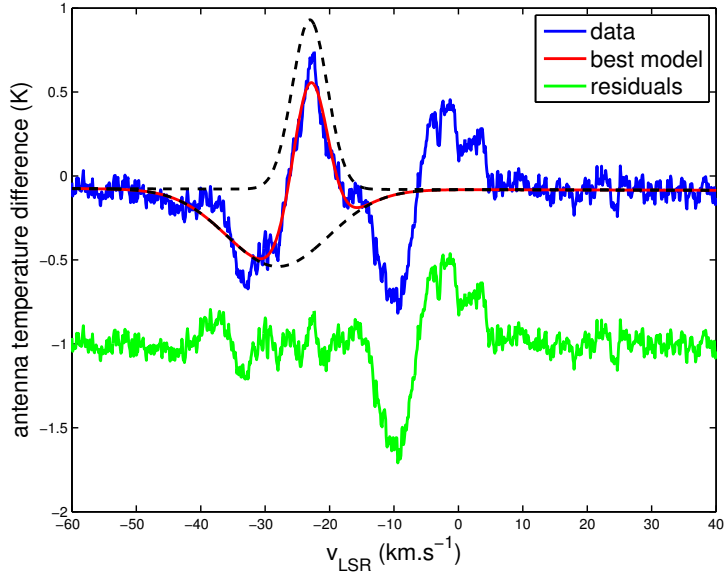

Figure B.1: Fit of the [CII] fine-structure line of 105-600. The blue curve is the level-2 calibrated spectrum which is best fitted by the red curve. The positive dashed black line is the best model without the OFF line corresponding to the observation that we could have done if there was no contamination from the nebula. The negative one is OFF contamination model. Both gaussians are fitted together and the residuals of the fit are given by the green curve.

low-$J$ CO lines are slightly lower and the [OI] 63 µm is slightly higher (Fig. B.2). In the absence of strong dependency, those lines cannot significantly constrain with confidence this grain parameter. The temperature at the surface of the disk is almost constant but the thermal profile is significantly changed in the envelope with strong increase of the temperature when small grains are included because they improve heating by photoelectric effect. According to the electron density estimated at the ionisation front, $n_e \leq 680$ cm$^{-3}$, and assuming a pressure equilibrium (constant $P = nT$) between the neutral and ionised gas, we can derive the temperature on the side of the envelope. Note that, rigorously, the pressure in the envelope is likely somewhat larger than in the ionised gas in order to maintain the pressure gradient of the photoevaporation flow. However, the temperature on the side of the envelope should be

$$T_{\rm env,IF} \approx \frac{n_e T_{\rm HII}}{n_{\rm env}}, \tag{B.1}$$

where $T_{\rm HII} = 10^4$ K is the temperature in the ionised region. In the case of the best-fit model for 105-600 where $n_{\rm env} = 8.5 \times 10^3$ cm$^{-3}$, the temperature is thus estimated to be $T_{\rm env,IF} \leq 800$ K. Models are consistent with this result only if small grains are removed by setting the minimal radius to 3 nm or a bit less, i.e. without PAHs. This is consistent with the observed low abundance of PAHs (Sect. 5.4.1).

We investigate the effect of changing the dust-to-gas mass ratio by varying it up to an order of magnitude (Fig. B.3). In that case, line fluxes are very sensitive and the best model is the one with the standard value of 100 which is, moreover, compatible with the mass estimates (see Sect. B.4.2). Assuming a slightly higher dust-to-gas mass ratio is still possible but it will increase the temperature at the ionisation front and will not be consistent with the 800 K limit for the envelope side. We therefore adopt gas-to-dust mass ratio of 100 ($\delta = 0.01$) and $a_{\rm min} = 3$ nm.

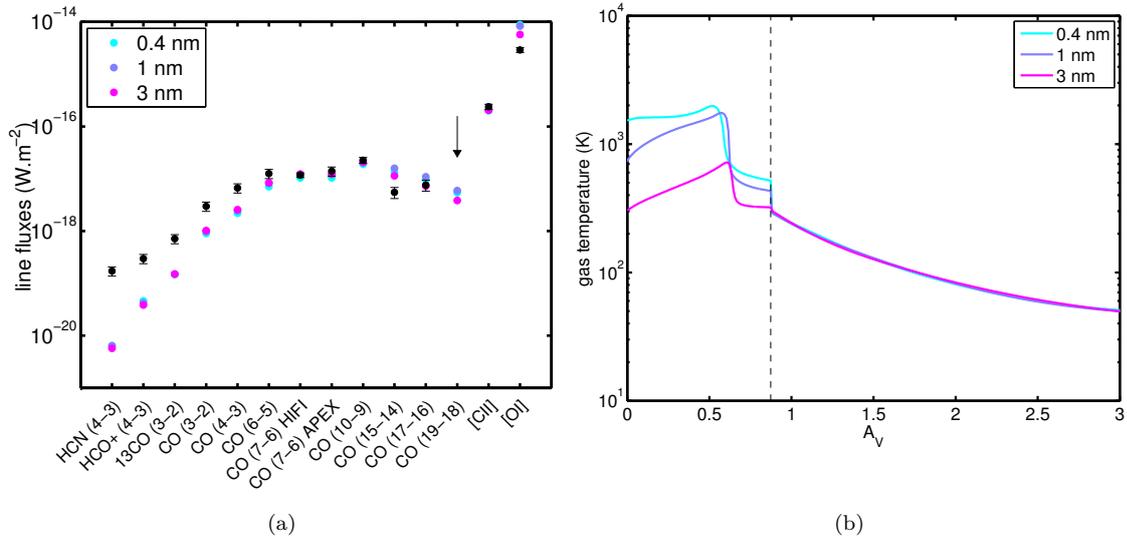

Figure B.2: Variations of the modelled line fluxes (a) and temperature profile (b) for various minimal grain size in the PDR model given in the legend. Here, densities are fixed to $n_{\rm env} = 1 \times 10^4$ cm$^{-3}$, $n_{\rm disk} = 1 \times 10^6$ cm$^{-3}$, the FUV field is $G_0 = 2 \times 10^4$ and a gas-to-dust mass ratio of 100 is used.

We also investigated the effect of the grain growth by increasing the maximum radius and decreasing the power-law index. Adding large grains does not significantly impact the photo-electric effect but the FUV extinction curve is changed. To better describe this effect we ran tests using a standard ISM size distribution ($a_{\rm max} = 0.3$ µm and a power-law index of 3.50), but with lower extinctions with the curves measured towards HD38087, HD36982, and HD37023. These span a wide range of dust FUV extinction cross section per H nucleus $\sigma_{\rm ext}$ including a value as low as $\simeq 8 \times 10^{-22}$ cm$^{-2}$ which is the value observed towards $\Theta^1$ Ori C and appropriate for Orion proplyds (Störzer and Hollenbach, 1999). The main effect of lowering the FUV extinction is to shift the H/H$_2$ transition deeper in the PDR. This does not change the main heating and cooling processes involved nor the general results of our study. However, this could result in a slightly higher mean density and subsequent total column density for the envelope (as a comparison, for 105-600, the best-fitted density is $n_{\rm env} = 8.5 \times 10^3$ cm$^{-3}$ for the Galaxy curve and $n_{\rm env} = 9.0 \times 10^3$ cm$^{-3}$ for the one of HD37023). Since, we did not obtain better fits to our observations by changing the extinction curve, we kept a standard Milky Way extinction curve.

## B.3 105-600: the possibility of emissions from shocks

The Carina candidate proplyd exhibits a bipolar collimated jet that is visible in HST/ACS Hα image (Smith et al., 2010a). Shocks along jets around protostars at comparable evolutionary stages are known to be sources of optical and FIR emission lines (see e.g. Podio et al., 2012). We should thus verify if some of the observed lines could originate in shocks rather than in the PDR.

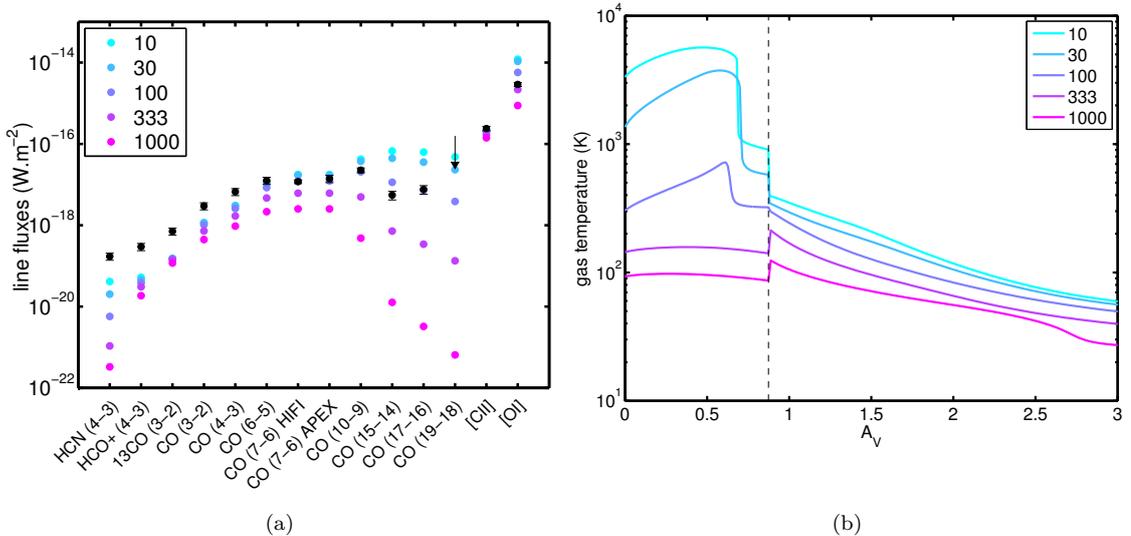

Figure B.3: Variations of the modelled line fluxes (a) and temperature profile (b) for various gas-to-dust mass ratio in the PDR model given in the legend. Here, densities are fixed to $n_\mathrm{env} = 1 \times 10^4$ cm$^{-3}$, $n_\mathrm{disk} = 1 \times 10^6$ cm$^{-3}$, , the FUV field is $G_0 = 2 \times 10^4$ and a minimal grain radius of 3 nm is used.

Fast and dissociative $J$-type shock models (with velocity between 30 and 150 km s$^{-1}$) of Hollenbach and McKee (1989) could explain a fraction of the observed atomic and molecular lines but predict, for a pre-shock density of $10^4$ cm$^{-3}$ or more, that the shocked surface should be bright in the H$\alpha$ image which is not the case (see Fig. 4.2(a)). For lower densities, the emissions in the FIR become negligible compared to the observations. We conclude that J-shocks are unlikely to be responsible for the observed infrared lines. Moreover, except if the jet is almost perfectly in the plane of the sky, the relatively low FWHM of the [CII] line (6.4 km.$^{-1}$) suggests that this line is not coming from a high-velocity shock.

For low-velocity shocks ($\leq 30$ km s$^{-1}$) associated with a jet, we looked at the estimated line fluxes from C and J-shock models of Flower and Pineau Des Forêts (2010). C-shocks produce negligeable [OI] emission compared to the observed one while J-shocks are not able to produce the observed [OI] emission without being, at least, two orders of magnitude brighter in CO compared to the observed emission. Atomic lines are thus very unlikely produced by such shocks. Finally, we conclude that atomic lines arise from the PDR of 105-600.

The very low FWHM of CO lines observed with *Herschel*/HIFI or APEX rule out the origin of this emission from relatively high-velocity shocks associated with the jet. The presence of the jet, however, suggest an accretion of matter onto a young star (see e.g. Konigl and Pudritz, 2000). Different low-velocity shocks can be expected in forming disk such as an accretion shock at the surface of the molecular disk, shock waves launched deeper in the disk or residual shocks from the interaction of the jet surface with the disk (Boss and Durisen, 2005). We used a C-shock code assuming a shock velocity of 4 - 6 km s$^{-1}$, a magnetic field of 1 - 2 mG and a pre-shock density of $10^6$ cm$^{-3}$ to estimate the radiated energy to compare it with the UV energy absorbed. In the most favorable case, the radiated energy flux, or radiated power per unit of shocked surface, reaches $8.65 \times 10^{-5}$ J m$^{-2}$ s$^{-1}$. The photo-electric effect injects a fraction $\epsilon_\mathrm{pe} \approx 1\%$ of the UV flux ($G_{0,\text{mol}}$ times the Habing's field) into the gas. Hence, the power per unit of disk surface is $\epsilon_{\text{pe}}\, G_{0,\text{mol}}\, 1.6 \times 10^{-6}$ J m$^{-2}$ s$^{-1}$ where $G_{0,\text{mol}}$ is the UV radiation field incoming at the surface of the disk which is about 10% of the initial flux incoming on the envelope for the range of studied envelope densities. With these values and $G_{0,\text{mol}} = 2.2 \times 10^3$, the energy flux is $3.52 \times 10^{-5}$ J m$^{-2}$ s$^{-1}$. To have an important contribution from the shock, the comparison of the two sources of energy implies that the shocked surface has to be about 45% of the disk surface. Moreover, if it happens, some of the emitted lines can be of the order of the observations only if the shock is not FUV-illuminated. Otherwise the emission drops significantly. Hence we cannot fully exclude that shocks could have a contribution to the CO emission, however this requires that the shocked surface is large and that it is protected from UV radiation. Both constraints seem difficult to concile. Altogether, we conclude that shocks are unlikely to play a major role in the observed emission.

## B.4  105-600: mass and mass-loss rate

### B.4.1  Molecular mass from $^{12}$CO and $^{13}$CO (3-2) lines

Column density of hydrogen in the molecular disk, and so its mass, can be estimated from the $^{12}$CO (3-2) and $^{13}$CO (3-2) transition lines. Under the assumption that the $^{13}$CO line to be optically thin, the column density of the upper energy level (here 3), noted $N_\text{u}$, is linked to the integrated intensity of the line,

$$N_\text{u} = \frac{8\pi k_\text{B} \nu^2}{hc^3 A_\text{ul}} \int T_\text{B}\left(^{13}\text{CO}\right) d\nu, \tag{B.2}$$

where $\nu$ is the frequency of the transition, $A_\text{ul}$ the Einstein emission-coefficient between the upper and the lower levels of the transition. $h$ is the Planck's constant, $c$ the vacuum speed of light, $k_\text{B}$ the Boltzmann's constant and $\int T_\text{B}\left(^{13}\text{CO}\right) d\nu$ the integrated intensity of the line. Additionally assuming that Local Thermodynamical Equilibrium (LTE) conditions are verified for $^{13}$CO, the column density of the upper level, $N_\text{u}$, is linked to the total one, $N$, by

$$N_\text{u} = \frac{N}{Z} g_\text{u} \exp\left(-\frac{E_\text{u}}{k_\text{B} T_\text{ex}}\right), \tag{B.3}$$

where $g_\text{u}$ and $E_\text{u}$ are respectively the statistical weight and energy of the upper level. $Z$ is the partition function and $T_\text{ex}$ is the excitation temperature of the transition. This latter can be calculated thanks to the optically thick emission of $^{12}$CO since we have

$$T_\text{B}\left(^{12}\text{CO}\right) = \frac{h\nu}{k_\text{B}} \left[ \frac{1}{\exp\left(\frac{h\nu}{k_\text{B} T_\text{ex}}\right) - 1} - \frac{1}{\exp\left(\frac{h\nu}{k_\text{B} T_\text{bb}}\right) - 1} \right], \tag{B.4}$$

where $T_\text{bb} = 2.7$ K is the background temperature.

From Sahai et al. (2012), we have $T_\text{B}\left(^{12}\text{CO}\right) = 33.0$ K so we get $T_\text{ex} = 40.8$ K. With a line integrated intensity of $\int T_\text{B}\left(^{13}\text{CO}\right) d\nu = 0.96$ K.km s$^{-1}$, we find $N\left(^{13}\text{CO}\right) = 3.53 \times 10^{14}$ cm$^{-2}$. Supposing an abundance ratio $[\text{H}_2]/\left[^{13}\text{CO}\right] = 7 \times 10^5$, we obtain the column density of $H2$, $N_{\text{H}_2} = 9.84 \times 10^{22}$ cm$^{-2}$. The molecular mass, $M_\text{mol}$, is finally calculated by

$$M_\text{mol} = \mu\, m_\text{H}\, S\, N_{\text{H}_2}, \tag{B.5}$$

where $\mu = 2.8$ is the mean molecular weigh, $m_\text{H}$ is the mass of a hydrogen atom and $S$ is the apparent surface visible is the beam of the instrument at the distance of the observed object. In our case, the molecular mass is $M_\text{mol} \approx 0.187$ M$_\odot$.

### B.4.2  Consistency of mass estimates

We can roughly estimate of the disk mass assuming spherical geometry and making the strong hypothesis that the density is constant throughout the disk. In reality, the density deep in the disk could be much higher, so this estimate can be seen as a lower limit. From our PDR best-fit model, this mass is $M_\mathrm{disk} \gtrsim 0.19$ M$_\odot$. In the same way, we can evaluate the mass of the envelope assuming that the 3D shape is the one of an ellipsoid with axis of $9.5'' \times 3.7'' \times 3.7''$ (from Table 4.2) which gives $M_\mathrm{env} \approx 0.028$ M$_\odot$ for the best-fit model.

We have used different methods to estimate the mass of each component in the Carina candidate proplyd. The mass of the atomic envelope is estimated to be about 0.05 M$_\odot$ from the SED (Sect. 5.4.1) and about 0.03 M$_\odot$ from the PDR models (Sect. 5.4.1). For the mass of the disk, the SED gives a value of about 0.7 M$_\odot$ (Sect. 5.4.1), consistent with the lower limit of 0.19 M$_\odot$ from the PDR model (Sect. 5.4.1). The fit on the thermal components of the dust emission using modified Planck's functions is based on empirical parameters such as the spectral index $\beta$ and the efficiency section $\sigma$. The latter is used to convert the dust opacity obtained from the fit in a total mass. Since the empirical value comes from observations in the galactic plane, it is valid only in environments relatively similar to the ISM, specially in terms of dust-to-gass mass ratio which typical value is 0.01. Our models with the Meudon PDR code also take into account this standard value of the dust-to-gas mass ratio and the comparison with the observations have shown that this value is most likely correct (see Appendix B.2).

Another estimation of the molecular mass, independent of the dust-to-gas mass ratio can be obtained using emission of molecular isotopologues. Sahai et al. (2012) have estimated, using their measurement of $^{12}$CO and $^{13}$CO (3-2) line fluxes, a mass of 0.35 M$_\odot$. With their observations, we use a slightly different method (see details in Appendix B.4.1) and obtained a value of 0.187 M$_\odot$ under the assumption that the $^{13}$CO line is optically thin while the $^{12}$CO is optically thick. Calculations with the online non-LTE molecular radiative transfer code RADEX (van der Tak et al., 2007) confirms that the $^{12}$CO line is optically thick but indicates that the $^{13}$CO starts to be thick considering the size and the density obtained from our models. Hence, this value of 0.187 M$_\odot$ is a lower limit.

Overall, estimates of masses are consistent but it is possible that values based on the dust emission are slightly overestimated. Using a dust-to-gas mass ratio a little higher (a few percent) than the typical value of 0.01, in agreement with what is often found for protoplanetary disks (Williams and Best, 2014), could correct the small differences.

### B.4.3  Mass-loss rate at ionisation front

Recombinations of electrons and protons in the HII region give rise to the H$\alpha$ emission at the ionisation front with an intensity depending on the electron density $n_\mathrm{e}$ (see Tielens, 2010, chap. 8). From the observed emission of $4 \times 10^{-7}$ W m$^{-2}$ sr$^{-1}$ in the HST/ACS/F658N filter (including H$\alpha$ and [NII] 6583 Å lines), we have derived an upper limit for the electron density of $n_\mathrm{e} \leq 680$ cm$^{-3}$ when considering that all the flux is coming from the H$\alpha$ line, which is probably close to the reality. Assuming a constant electronic density at the ionisation front all around the envelope, a shape of an ellipsoid of revolution with a surface $S$, and that the gas is escaping at the sound speed $c_\mathrm{S}$ calculated here for a temperature of $10^4$ K, the mass-loss rate is

$$\dot{M} = \mu \, m_\mathrm{H} \, n_\mathrm{e} \, S \, c_\mathrm{S}, \tag{B.6}$$

where $\mu m_\mathrm{H}$ is the mean particle weight. With our values, this gives a mass-loss rate of the order of $2 \times 10^{-6}$ M$_\odot$/year for 105-600.

# Appendix C

# More details about the hydrodynamical code

## C.1 Model sensitivity to parameters: effect of viscosity

This sections details the results discussed in Sect. 8.1.3. To study the effect of viscosity on disk dispersal, I present how the components of the mass-loss rate evolve when changing viscosity.

First, let us look at the mass-loss rate by photoevaporation that dominates the phase I of mass loss. Fig. C.1(a) gives the time evolution of this mass-loss rate for the different models. We can see that the mass-loss rates, which logically start initially at the same value, slowly diverge with time from one model to another. The higher the viscosity, the more important the increase in mass-loss. At about $5\ 10^3$ years, the difference reaches a maximum which is not significant. A variation of two orders of magnitude in the α value leads to an increase about 80% in the mass-loss rate. Photoevaporation is set in the same way for all the models here so that the differences are only related to the different dynamical evolution of the disks, and especially their sizes. Fig. C.1(b) gives the same mass-loss rates but this time according to the extension of the disks at any time. Here, it is clear that the rates increase with the extension of the disks, and more precisely by an expected proportional relation with the surface of the outer disk that is not depending on the viscosity at all. However, starting from an extent of 100 AU initially, a model with a higher viscosity will spread out more easily and will potentially increase the disk size leading to a higher surface to erode. This is the case here with, for example, 135 AU reached by the model with $\alpha = 1\ 10^{-2}$ at the same time as the model with $\alpha = 1\ 10^{-4}$ is still at about 100 AU. This increase by 35% of the radius leads to an increase of about 80% in the disk surface. The mass-loss rate by photoevaporation is proportional to the disk surface of the outer disk so that this increase is retrieved in the difference of the mass-loss rates. The outer disk is then being truncated and the mass-loss rates by photoevaporation decrease continuously to 0.

Photoevaporation is responsible for the direct loss of almost the entire outer disk and, lead to a loss of mass of the inner disk by outwards mass transfer. When the disk is much more extended than its gravitational radius, the mass transport because of the spreading is initially inward. Actually, this motion represents only a very small and negligible fraction of the outer mass (that is yet more or less proportional to α and for those examples is a fraction of or about 1%) so I will not go in detail of this process here. However, when the disk is initially less extended than the gravitational radius or when it is almost truncated, the motion is from the inner to the outer disk. From there, the mass is lost by photoevaporation. Fig. C.2 gives the time evolution

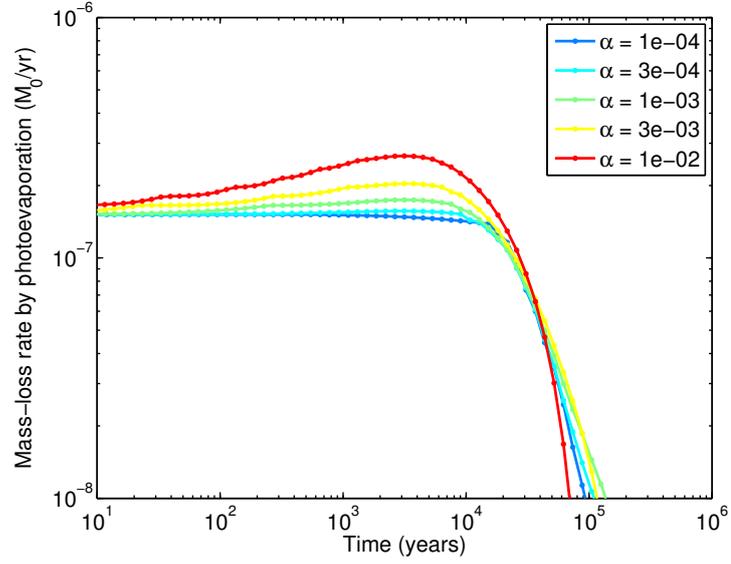

(a) Time evolution of the mass-loss rate related to photoevaporation

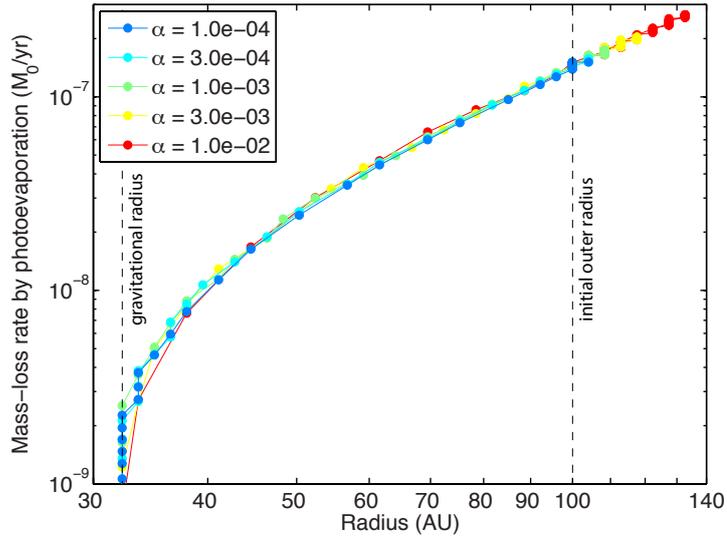

(b) Link between the disk size and the mass-loss rate related to photoevaporation

Figure C.1: Illustration of the impact of the viscosity on the mass-loss rate by photoevaporation (including the mass motion from the inner disk, i.e. $\dot{M}_{\mathrm{evap}} + \dot{M}_{\mathrm{in} \to \mathrm{out}}$). Top: time evolution of the mass-loss rates. Bottom: mass-loss rates according to the disk outer radius.

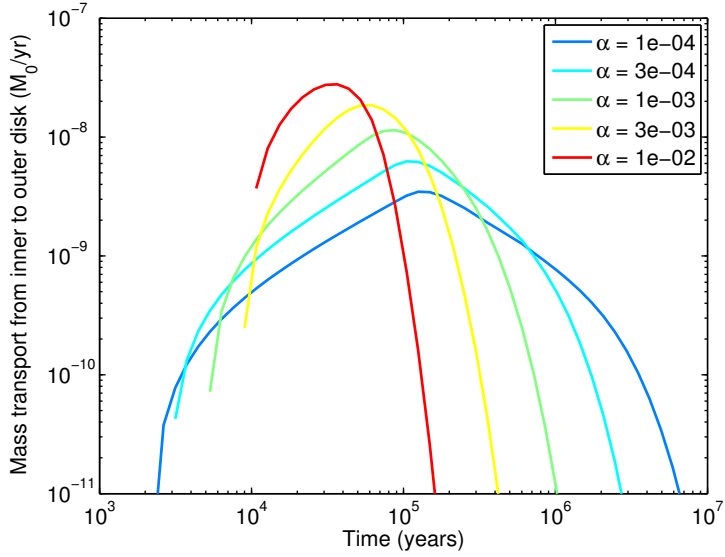

Figure C.2: Illustration of the impact of the viscosity on the mass-loss rate by mass transport from inner to the outer disk (followed by photoevaporation).

of the mass-loss rate by outward mass transfer, $\dot{M}_{\text{in}\rightarrow\text{out}}$, for the same set of models. Here, we can see that, when the viscosity is higher, the beginning of this motion happens later. This is because of the extension of the disk at high viscosities that delays when the direction of motion changes. It happens later but stronger and lasts less time since it is very efficient and actively participate in the loss of the inner disk. For those examples, at the end, this process is responsible for the loss of about the half of the inner disk.

For the inner disk, the other part is thus removed by the accretion. Fig. C.3 gives the time evolution of the mass-loss rate by accretion, $\dot{M}_{\text{acc}}$, for the set of models studied here. Once again, as a direct viscous related process, the mass-loss rates are more or less proportional to the α value. The accretion is maximal at the start of the simulation and then decreases with the spreading of the disk. It rapidly drops when the inner to outer disk motion starts to erode the inner disk.

## C.2 Model sensitivity to parameters: effect of masses

This sections details the results discussed in Sect. 8.1.3. To study the effect of the star and disk masses on the disk dispersal, I now present how the components of the mass-loss rate evolve when varying those parameters, and how their impacts evolve with time. To separate more clearly the effect of the masses, I will plot the mass-loss rate given as the fraction of mass lost per year, i.e. $\dot{M}/M_{\text{disk}}$.

For the same conditions of density and temperature at the disk surface, the mass-loss rate by photoevaporation is only depending on the disk extension (checked on Fig. C.1(b)). Consequently, $\dot{M}_{\text{evap}}/M_{\text{disk}} \propto 1/M_{\text{disk}}$. This relation means that less massive disks will be dissipated by photoevaporation much more rapidly than more massive ones. This is particularly clear at the beginning of the photoevaporation in Fig. C.4(a) and this is retrieved in the outer disk lifetime (Fig. 8.4). This relation is true if the disk is far more extended than its gravitational radius. But if the gravitational radius tends to be close to the outer radius of the disk or even farther, then the mass-loss rates drop, and become zero. This is what is seen in the right-hand side of the figure 8.4 for the most massive stars that have a large gravitational radius. This dependency remains over time, but the outer parts of the least massive disks are rapidly eroded, so that their fraction of mass lost by photoevaporation decreases rapidly over time and the maximum of relative loss shifts to higher disk mass (Fig. C.4(b) to C.4(d)).

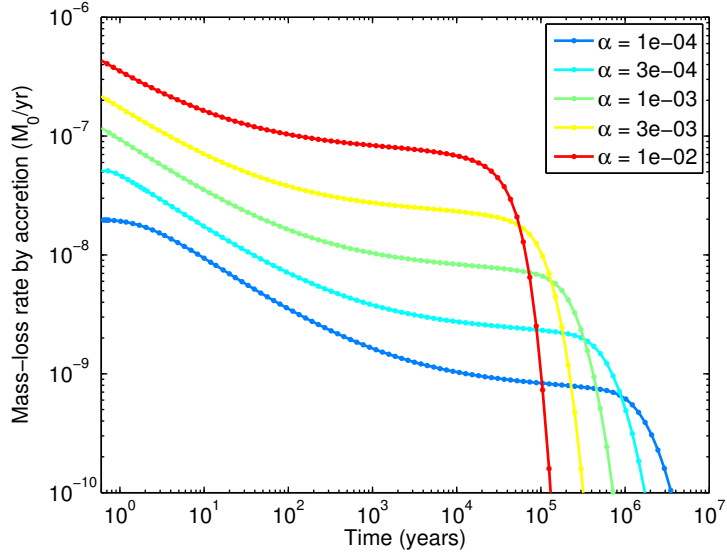

Figure C.3: Illustration of the impact of the viscosity on the mass-loss rate by accretion.

The more extended the disk is compared to its gravitational radius, the higher the inward velocity at the gravitational radius. The lowest gravitational radii are found around the least massive stars. It is thus logical to find that the inward motion from the outer to the inner disk is inversely proportional to the central star mass (Fig. C.5(a)). This configuration at the start of the simulation will not evolve much without photoevaporation, except with a very small but continuous decrease with time as the disk is spreading. With photoevaporation, this small decrease with time is still present for the whole grid but here the disks tend to be truncated rather than spreading. Consequently, the inward motion will rapidly decrease with the truncation and then become 0 before the motion reverts and becomes outward. From then, the inner disk will lose mass. As the truncation happens more rapidly for the disks around the most massive stars, because the gravitational radius is farther away and easier to reach, and for low-mass disks, easier to erode (as we saw in Fig. C.4), those are the ones for which the motion is slowed down and reverted more rapidly (cf. Fig. C.5(b) to C.5(d)). As in the example case of Sect. 8.1.2, the rate of motion to the inner disk is a few orders of magnitude below the photoevaporation so that, only a very small fraction of the initial outer disk may be protected for a while by moving into the inner disk.

Some of the disks in the grid have no "outer disk", meaning that they are initially less extended than the gravitational radius. This happens for the most massive stars. If this is a bit out of range of the classical "supercritical" photoevaporation, where a substantial part of the disk is directly subject to escape, this case can be treated as the others. For those disks, the motion at the gravitational radius may only be outwards. Those disks are the ones that rapidly lose mass by this mean (Fig. C.6(a)). As for the inward motion, the outward motion is enhanced when the gravitational radius is separated from the disk outer radius. Consequently, the more massive the central star, the farther away the gravitational radius and the higher the outward motion when the gravitational radius has been reached. Without photoevaporation, this does not change except a small decrease in the rate with time while the disk is spreading. However, we recover here that, with photoevaporation, the disks around the most massive central stars and the disks with the lowest masses are the first to revert their motion (when almost truncated) to become outward at the gravitational radius. Finally, the less massive disks around the most massive stars are thus those that are more subject to this mass-loss.

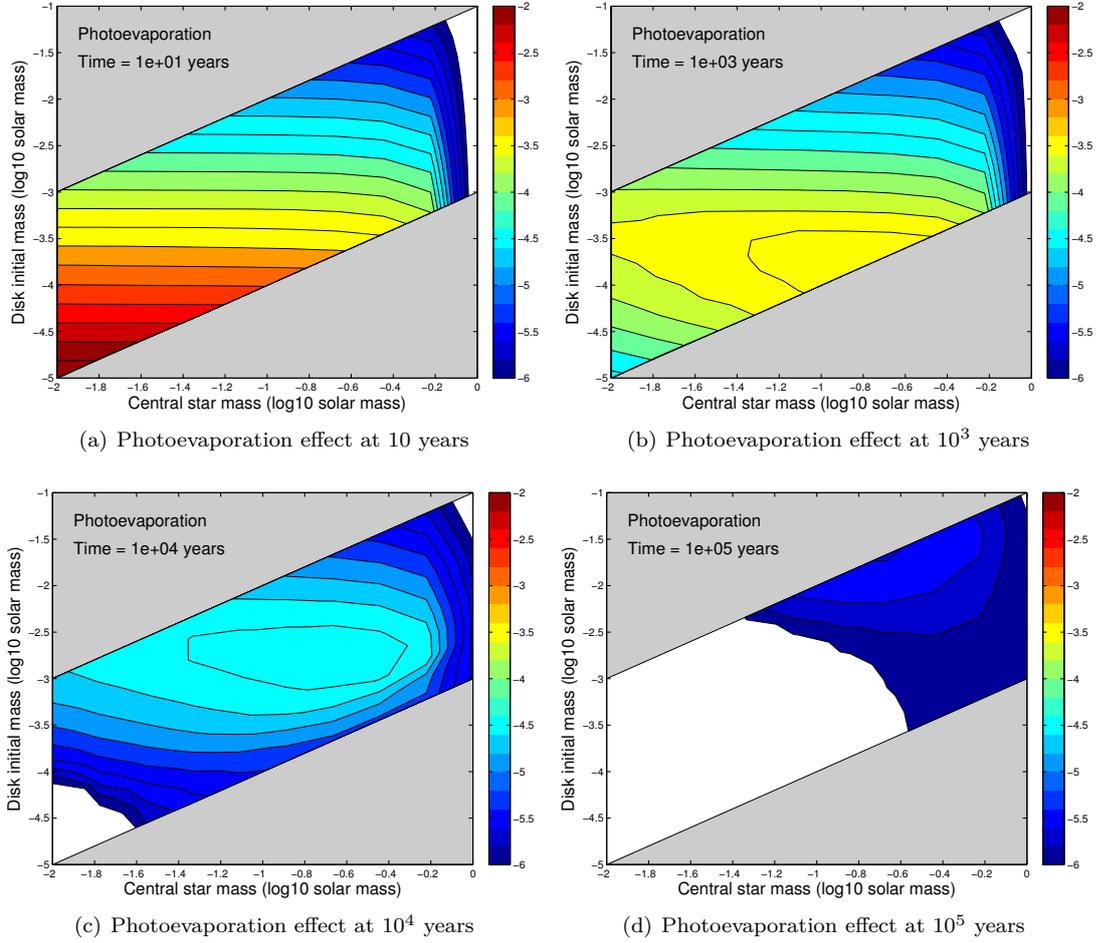

Figure C.4: Mass-loss rates caused by photoevaporation over disk masses, $\dot{M}_{\mathrm{evap}}/M_{\mathrm{disk}}$, for a grid of star and disk masses. The color filled contour and white area show the range modeled, i.e. star mass between $10^{-2}$ and $1\,\mathrm{M}_\odot$ with a disk mass between $10^{-3}$ and $10^{-1}\,\mathrm{M}_\odot$. Each color indicates a given range of values while white means any value below the range of the color bar. The grey regions are regions not modeled here.

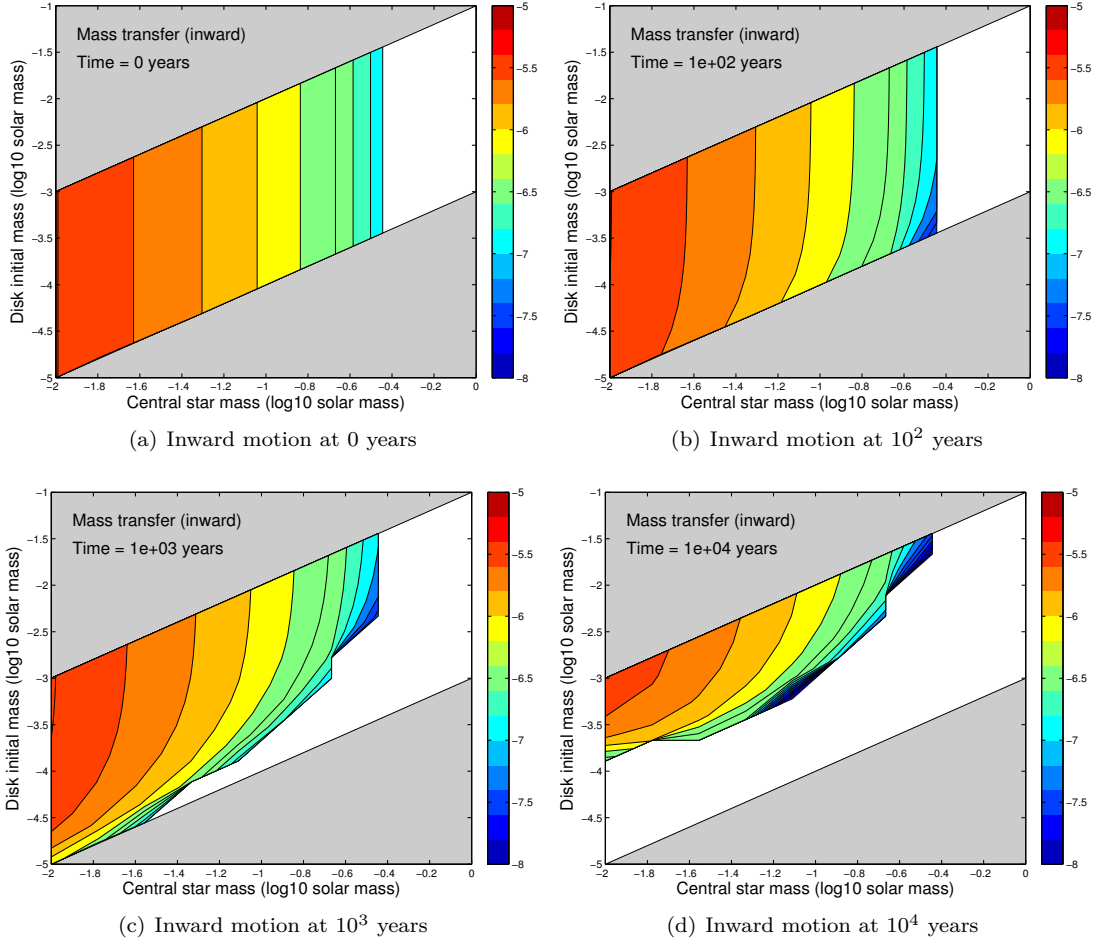

Figure C.5: Inward motion over disk masses, $\dot{M}_{\rm evap}/M_{\rm disk}$, for a grid of star and disk masses. The color filled contour and white area show the range modeled, i.e. star mass between $10^{-2}$ and $1\,M_\odot$ with a disk mass between $10^{-3}$ and $10^{-1}\,M_\odot$. Each color indicates a given range of values while white means any value below the range of the color bar. The grey regions are regions not modeled here.

For the inner disk, it remains the accretion. From the formulation of the accretion rate, $\dot{M}_{\rm acc} = 2\pi r_{\rm in} v_{\rm r}\left(r_{\rm in}\right) \Sigma\left(r_{\rm in}\right)$ (when the loss is considered positive), it is easy to derive the relation with the masses as $\dot{M}_{\rm acc} \propto M_{\rm disk}/\sqrt{M_*}$, so that the fraction of disk mass lost by accretion is only depending on the central star mass with $\dot{M}_{\rm acc}/M_{\rm disk} \propto 1/\sqrt{M_*}$ (cf. Fig. C.7(a)).Without photoevaporation, theses rates evolve slightly to decrease as the disk spread. With photoevaporation, the evolution is similar for thousands of years but when the first disks have been truncated, i.e. the low-mass disks around the least massive stars, their inner disks have been significantly altered so that this is visible on the accretion rate with an important drop (Fig. C.7(b)). With time, this effect propagates for disks around the medium-mass stars at $10^5$ years (Fig. C.7(c)) and has started to impact the ones around solar-mass stars about $10^6$ years (Fig. C.7(d)).

Fig. C.8 summarises the effect of the important parameters driving the dynamical evolution, and thus the lifetime, of the inner disk: the central star mass and the viscosity. Models are here run with an initial disk mass $M_{\rm disk} = 10^{-3}$ M$_\odot$, but is representative of all the previously studied cases according to the non-significant correlation between the lifetime and the disk initial mass (cf. Fig. 8.5).

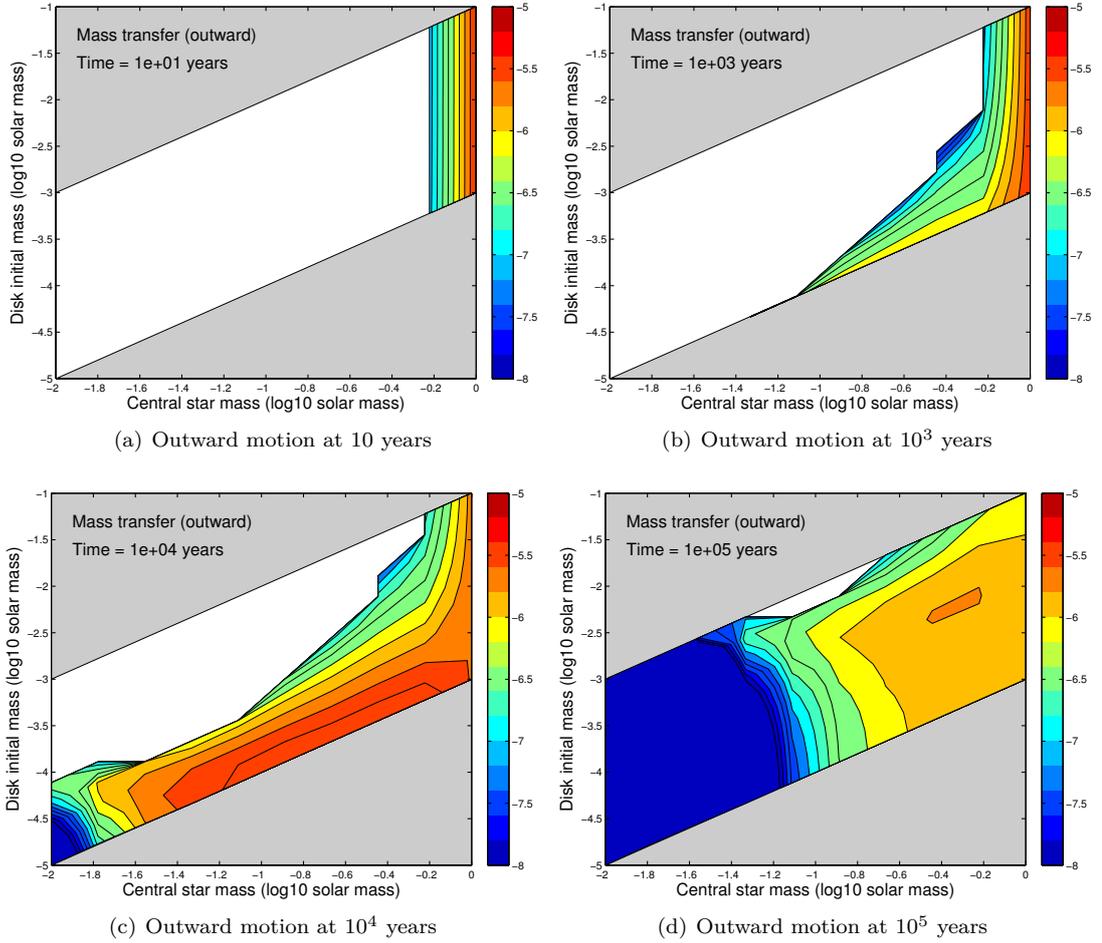

Figure C.6: Outward motion over disk masses, $\dot{M}_{\mathrm{evap}}/M_{\mathrm{disk}}$, for a grid of star and disk masses. The color filled contour and white area show the range modeled, i.e. star mass between $10^{-2}$ and 1 $M_\odot$ with a disk mass between $10^{-3}$ and $10^{-1}$ $M_\odot$. Each color indicates a given range of values while white means any value below the range of the color bar. The grey regions are regions not modeled here.

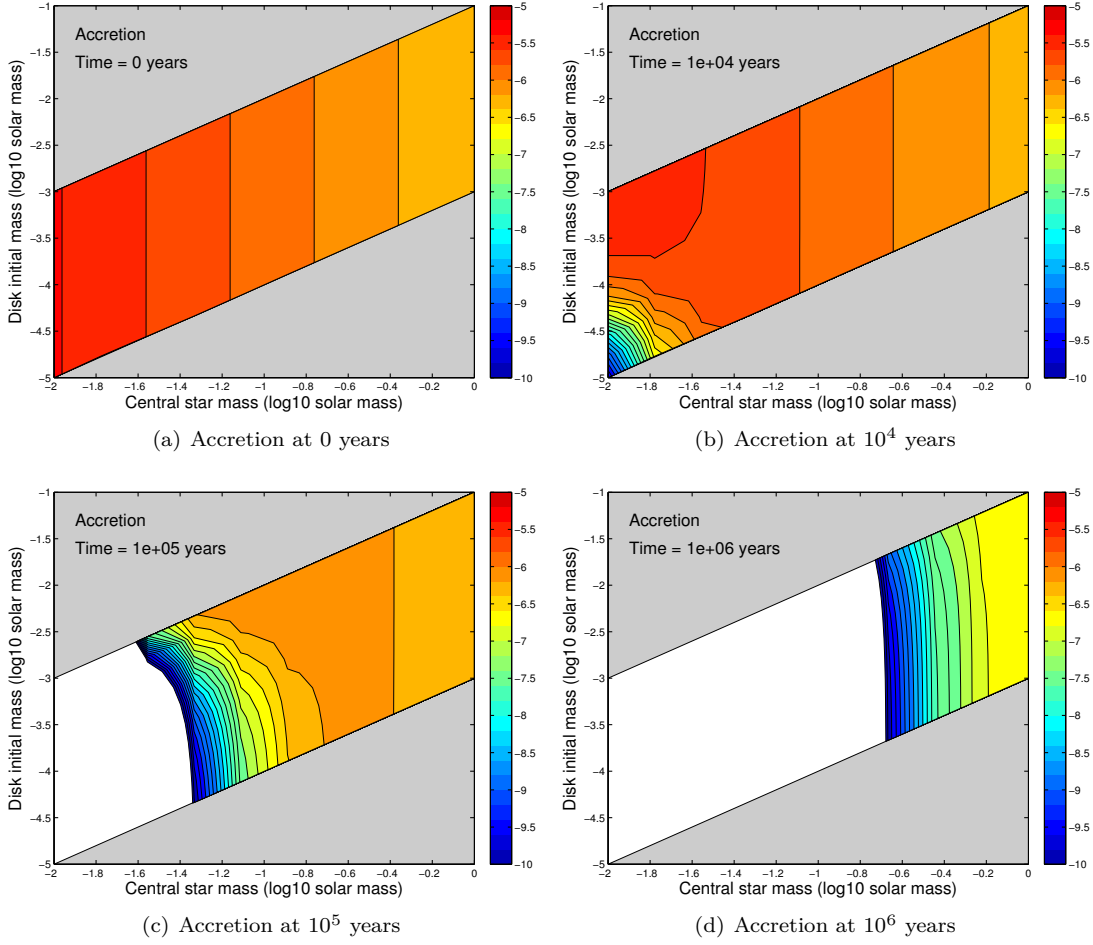

Figure C.7: Accretion rates over disk masses, $\dot{M}_{\mathrm{evap}}/M_{\mathrm{disk}}$, for a grid of star and disk masses. The color filled contour and white area show the range modeled, i.e. star mass between $10^{-2}$ and 1 $M_\odot$ with a disk mass between $10^{-3}$ and $10^{-1}$ $M_\odot$. Each color indicates a given range of values while white means any value below the range of the color bar. The grey regions are regions not modeled here.

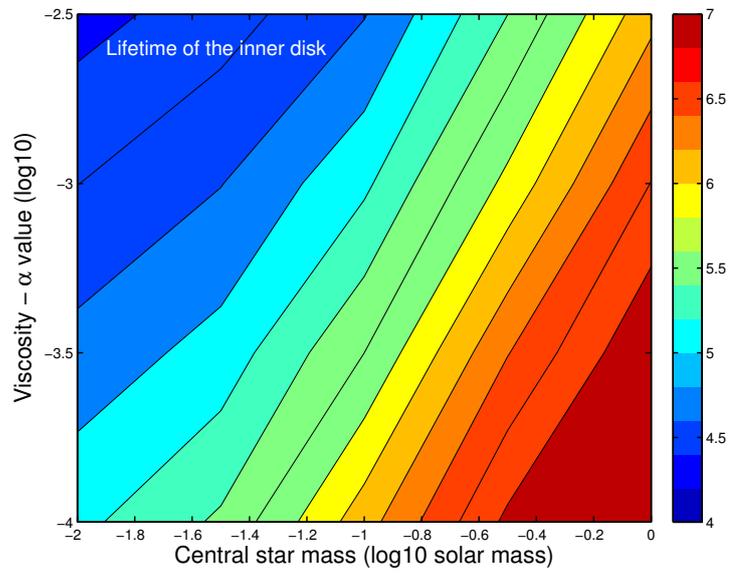

Figure C.8: Lifetime of the inner disk, given as the log10 of years for a disk initial mass $M_{\text{disk}} = 10^{-3}$ M$_\odot$, a grid of star mass ranging from $10^{-2}$ and 1 M$_\odot$ and a viscosity efficiency $\alpha$ from $10^{-4}$ to $10^{-2.5}$. Each color indicates a given range of values while the grey regions are regions not modeled here.